\PassOptionsToPackage{unicode}{hyperref}
\PassOptionsToPackage{hyphens}{url}
\documentclass[
]{article}
\usepackage{amsmath,amssymb}
\usepackage{lmodern}
\usepackage{ifxetex,ifluatex}
\ifnum 0\ifxetex 1\fi\ifluatex 1\fi=0 
  \usepackage[T1]{fontenc}
  \usepackage[utf8]{inputenc}
  \usepackage{textcomp} 
\else 
  \usepackage{unicode-math}
  \defaultfontfeatures{Scale=MatchLowercase}
  \defaultfontfeatures[\rmfamily]{Ligatures=TeX,Scale=1}
\fi
\IfFileExists{upquote.sty}{\usepackage{upquote}}{}
\IfFileExists{microtype.sty}{
  \usepackage[]{microtype}
  \UseMicrotypeSet[protrusion]{basicmath} 
}{}
\makeatletter
\@ifundefined{KOMAClassName}{
  \IfFileExists{parskip.sty}{%
    \usepackage{parskip}
  }{
    \setlength{\parindent}{0pt}
    \setlength{\parskip}{6pt plus 2pt minus 1pt}}
}{
  \KOMAoptions{parskip=half}}
\makeatother
\usepackage{xcolor}
\IfFileExists{xurl.sty}{\usepackage{xurl}}{} 
\IfFileExists{bookmark.sty}{\usepackage{bookmark}}{\usepackage{hyperref}}
\hypersetup{
  pdftitle={Accounting for the Fraction of Carcasses outside the Searched Area and the Estimation of Bird and Bat Fatalities at Wind Energy Facilities},
  hidelinks,
  pdfcreator={LaTeX via pandoc}}
\usepackage[margin=1in]{geometry}
\usepackage{color}
\usepackage{fancyvrb}

\DefineVerbatimEnvironment{Highlighting}{Verbatim}{commandchars=\\\{\}}
\usepackage{framed}
\definecolor{shadecolor}{RGB}{248,248,248}
\newenvironment{Shaded}{\begin{snugshade}}{\end{snugshade}}

\newcommand{\AttributeTok}[1]{\textcolor[rgb]{0.77,0.63,0.00}{#1}}

\newcommand{\CommentTok}[1]{\textcolor[rgb]{0.56,0.35,0.01}{\textit{#1}}}

\newcommand{\ConstantTok}[1]{\textcolor[rgb]{0.00,0.00,0.00}{#1}}

\newcommand{\DecValTok}[1]{\textcolor[rgb]{0.00,0.00,0.81}{#1}}
\newcommand{\DocumentationTok}[1]{\textcolor[rgb]{0.56,0.35,0.01}{\textbf{\textit{#1}}}}

\newcommand{\FloatTok}[1]{\textcolor[rgb]{0.00,0.00,0.81}{#1}}
\newcommand{\FunctionTok}[1]{\textcolor[rgb]{0.00,0.00,0.00}{#1}}

\newcommand{\NormalTok}[1]{#1}

\newcommand{\OtherTok}[1]{\textcolor[rgb]{0.56,0.35,0.01}{#1}}

\newcommand{\SpecialCharTok}[1]{\textcolor[rgb]{0.00,0.00,0.00}{#1}}

\newcommand{\StringTok}[1]{\textcolor[rgb]{0.31,0.60,0.02}{#1}}

\usepackage{graphicx}
\makeatletter
\def\maxwidth{\ifdim\Gin@nat@width>\linewidth\linewidth\else\Gin@nat@width\fi}
\def\maxheight{\ifdim\Gin@nat@height>\textheight\textheight\else\Gin@nat@height\fi}
\makeatother
\setkeys{Gin}{width=\maxwidth,height=\maxheight,keepaspectratio}
\makeatletter
\def\fps@figure{htbp}
\makeatother
\providecommand{\tightlist}{%
  \setlength{\itemsep}{0pt}\setlength{\parskip}{0pt}}
\usepackage[singlelinecheck = false, labelfont = large, labelfont = bf]{caption}
\usepackage{float}
\usepackage{longtable}
\usepackage{graphicx}
\usepackage{threeparttable}
\usepackage{enumitem}
\usepackage[title]{appendix}
\ifluatex
  \usepackage{selnolig}  
\fi
\DeclareMathAlphabet{\mathpzc}{OT1}{pzc}{m}{it}

\title{Accounting for the Fraction of Carcasses outside the Searched
Area and the Estimation of Bird and Bat Fatalities at Wind Energy
Facilities}
\author{}
\date{\vspace{-2.5em}Created 2020-11-15; revised 2023-05-18}

\begin{document}
\maketitle

\setcounter{tocdepth}{2}

\begin{center}
  {\large by 
    Daniel Dalthorp\footnote{United States Geological Survey (USGS), Corvallis, Oregon},
    Manuela Huso\footnotemark[1],
    Mark Dalthorp\footnote{Cornell University, Ithaca, New York},
    Jeff Mintz\footnotemark[1]
  }
\end{center}

\newcommand{\mwin}{m_\textrm{in}}
\newcommand{\mem}{\mathfrak{y}}
\newcommand{\w}{\mathpzc{w}}

\emph{This manuscript has been accepted by the U.S. Geological Survey (USGS) for publication to appear in
	"Techniques and Methods" after some minor formatting changes.}

\newpage

\tableofcontents

\listoffigures

\listoftables

\newpage

\hypertarget{abbreviations}{%
\section*{Abbreviations}\label{abbreviations}}
\addcontentsline{toc}{section}{Abbreviations}

\begin{table}[!htbp]
  \begin{tabular}{ll}
    AICc & Akaike information criterion (corrected for sample size)\\
    $\Delta \textrm{AICc}$ & difference between a model's AICc and the lowest AICc among the models tested\\
    $dwp$ & density weighted proportion, fraction of carcasses lying within the searched area\\
    $g$ & probability of carcass discovery within the searched area\\
    m & meters\\
    $M$ & mortality, number of fatalities\\
    $SE$ & searcher efficiency, probability of finding a carcass that is present at time of search\\
    $\psi$ & probability that carcasses lie in the searched area\\
    xep$x$ & specification of carcass distribution model (Appendix \ref{app:modbrief})\\
   \end{tabular}
\end{table}

NOTE: A hat ( \^{} ) over a parameter signifies an estimate or estimator
of the parameter rather than its true (but unknown) value. For example,
the actual number of fatalities, \(M\), is unknown, and \(\widehat{M}\)
would denote the estimated number of fatalities.

\newpage

\hypertarget{abstract}{%
\section{Abstract}\label{abstract}}

Accurate estimation of bird and bat mortality at wind energy facilities
requires accounting for carcasses that lie outside the search plots,
because they lie beyond the search radius or in areas within the search
radius that remain unsearched due to sub-optimal search conditions such
as thick vegetation, rough or dangerous ground, water, or restricted
access to the land. The \emph{density-weighted proportion} approach (or
\emph{dwp}) to estimating the fraction of carcasses lying in unsearched
areas involves tallying the carcasses found in concentric rings centered
at the turbine, fitting a curve to the carcass densities in the rings,
and dividing the integral of the curve over the area searched by the
integral over the total area. Accounting for unsearched area presents
special difficulties such as extrapolation beyond the search radius,
spatial prediction, and model selection, which are frequently ignored or
under-appreciated, potentially resulting in substantial estimation
errors.

A powerful new R software package (\(\texttt{dwp}\)) is available to
perform the calculations, given the distances at which carcasses were
found from turbines and a map of the searched area to be able to discern
the fraction of the ground searched at each distance. If all ground
within a given search radius has been searched, the map is simply the
search radius. For more complicated search plots, other kinds of maps
may be used: R polygons for plots that can be readily delineated into
searched and not-searched areas (for example, searches restricted to
access roads and turbine pads), GIS shape files for complicated search
patterns (for example, non-uniform vegetation or ground texture
resulting in spatially varying search conditions), or raster files for
complicated search patterns coupled with carcass spatial distribution
depending on both distance and direction from turbines.

This study discusses estimation and interpretation \emph{dwp} in the
context of several realistic examples, provides guidance for use of the
\(\texttt{dwp}\) software for doing the analyses, and addresses
questions of extrapolation, spatial prediction, and model selection.

\hypertarget{overview}{%
\section{Overview}\label{overview}}

Estimation of bird and bat mortality at wind energy facilities typically
involves periodic searches for carcasses on the ground near turbines and
adjustment of the carcass count to account for the estimated fraction of
carcasses missed in the searches. In order for a carcass to be observed,
it must lie inside the area searched, persist until the time of search,
and be found by searchers. Conceptually, the total mortality (\(M\)) is
estimated as \(\hat{M} = \frac{x}{dwp \cdot r \cdot SE}\), where \(x\)
is the number of carcasses observed, \(dwp\) is the fraction of
carcasses lying the searched area, \(r\) is the probability a carcass
persists until the search, and \(SE\) is the \emph{searcher efficiency}
or probability that a carcass is found given that it is present in the
searched area at the time of search. Searcher efficiency and carcass
persistence are typically estimated via field trials and combined into a
detection probability, \(g\). The number of carcasses lying in the
searched area is then estimated as
\(\hat{M}_\textrm{inside} = x/\hat{g}\).

The techniques and models for estimating \(g\) tend to be complicated,
but there are several convenient R packages available to aid in the
estimation of \(SE\), \(r\), \(g\), and \(M\mid\{x,\,\hat{g}\}\). The
most flexible and powerful is \texttt{GenEst} (Dalthorp et al.~2018),
while \texttt{eoa} (Dalthorp et al.~2017) is specially designed for
accuracy and utility when carcass counts are very small or even 0.
Several other packages are available as well (Huso et al.~2012, Wolpert
and Coleman 2015, Korner-Nievergelt et al.~2016). However, in mortality
estimation, \(dwp\) is as influential as \(SE\) and \(r\), but
estimation of \(dwp\) has received relatively little attention in the
literature and in mortality estimation protocols despite its importance.
Mortality estimates that are derived from careful and accurate
estimation of \(SE\) and \(r\) and from careful and accurate modeling of
\(g\) can still be grossly inaccurate unless \(dwp\) is also carefully
and accurately accounted for. Shortcuts in \(dwp\) estimation are likely
to result in unreliable mortality estimates. Estimation of \(dwp\)
presents special difficulties such as extrapolation beyond the search
radius, spatial prediction and model selection, which are discussed in
this report and addressed by the \texttt{dwp} software package for R
(Dalthorp et al.~2021).

We refer to the fraction of carcasses lying in the searched area as the
\emph{density-weighted proportion} (\texttt{dwp}) after Huso et
al.~2012. It is sometimes also referred to as the \emph{area correction}
factor (\texttt{AC}) (Studyvin and Rabie, 2018), or \emph{spatial
coverage} (\(a\)) (Dalthorp et al.~2017). Estimates of \(dwp\) (or
\texttt{AC} or \(a\)) can then be used with \texttt{GenEst} (as the
\(dwp\) parameter) or \texttt{eoa} (as the \(a\) parameter) in
estimating mortality. In addition, the package provides an extensive set
of tools for analyzing carcass density as a function of distance from
turbine which may be of interest beyond estimating mortality at a
particular site and may be used in more general studies of carcass
dispersion at wind energy sites under a variety of conditions.

We begin with a brief discussion of terminology, general concepts,
difficulties, special techniques used in \(dwp\) analysis, and the array
of command line tools that \(\texttt{dwp}\) (1.0) provides for doing the
analyses. Section \ref{sec:principles} gives a brief introduction to the
main tasks and a discussion of some critical issues to consider in
estimating \(dwp\). More detailed discussion of data requirements and
formatting is included in sections \ref{sec:raw}-\ref{sec:formatted},
example analyses in section \ref{sec:examples}. The second half of the
document is a set of technical appendices, discussing details about the
models used in \(\texttt{dwp}\) (Appendix \ref{app:modbrief}), simulated
carcass distributions used in testing the models (Appendix
\ref{app:deqmod}), model performance in the simulation scenarios
(Appendix \ref{app:fit2deq}), and additional techical details about
\(dwp\) and the accuracy of confidence intervals for \(\psi\)
(probability that a carcass falls in the searched area), \(dwp\) and
\(M\) (Appendices \ref{app:psivdwp} and \ref{app:vardwp}).

\hypertarget{general-principles}{%
\section{\texorpdfstring{General Principles
\label{sec:principles}}{General Principles }}\label{general-principles}}

\hypertarget{carcass-dispersion}{%
\subsection{\texorpdfstring{Carcass Dispersion
\label{sec:dispersion}}{Carcass Dispersion }}\label{carcass-dispersion}}

After being struck and killed by a turbine blade, where a carcass will
land is a function of many factors, such as wind velocity, turbine blade
speed, strike angle on the blade, carcass size, carcass cross-sectional
area presented to the wind, wind shear, initial velocity of the carcass
after being struck, and others. After landing, the carcass may be
susceptible to being moved by scavengers, gravity, or wind, or---in
cases where the animal is not immediately killed upon impact---it may
hobble some distance seeking shelter.

The factors effecting carcass location are numerous and complex.
Regardless of which combination of factors are in operation and in what
ways, a general principle of carcass dispersion remains the same,
namely, that carcass densities on the ground decrease with distance
because 1) fewer carcasses fall at great distances from turbines than
near turbines, and 2) carcasses are spread over greater areas at greater
distances from turbines. Although the change in density with distance is
not always monotone, and near the turbine density may actually increase
with distance over a short range, there is a limit to how far carcasses
will fall from a turbine, and the density of carcasses generated by a
turbine will inevitably decrease to zero at great distances from the
turbine.

\begin{figure}
\centering
\includegraphics{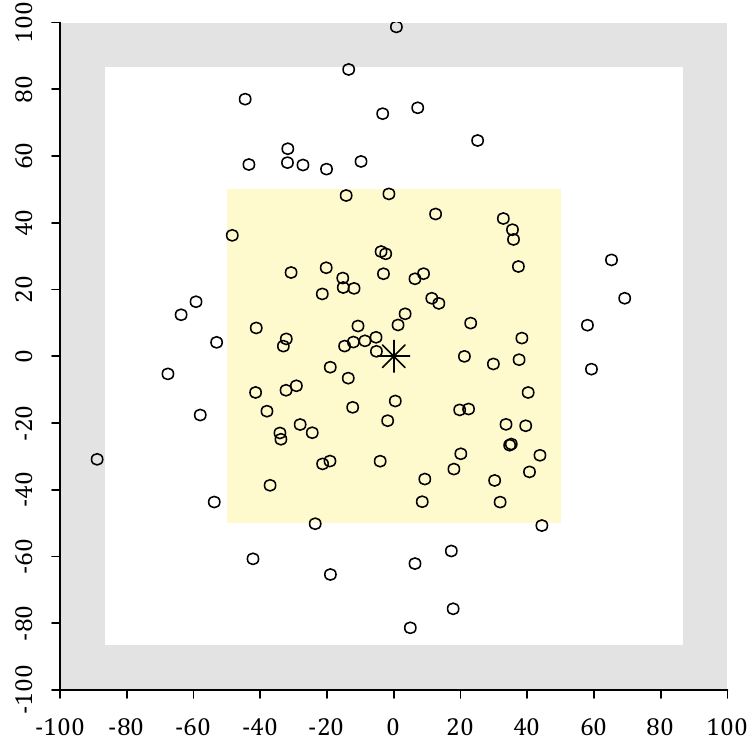}
\caption{\label{fig:carcass_locations}Carcass locations relative to
turbine. Gray and yellow regions each represent 25\% of the area in a
200 m square plot centered at the turbine.}
\end{figure}

Because carcass density varies, a simple area correction that divides
the carcass count by the fraction of the area sampled to get, for
example, \(\hat{M} = x/0.75\) because 75\% of the area was sampled, is
highly prone to error (Huso et al.~2011; Huso and Dalthorp, 2014). For
example, Fig. \ref{fig:carcass_locations} illustrates the locations of
100 carcasses with a gradual decrease in carcass density with increasing
distance from the turbine. The designated search plot area is a square,
200 m on a side, centered at the turbine. If 25\% of the area is not
searchable, the area correction depends strongly on which 25\% is not
searched. For example, the yellow and gray areas each comprise 25\% of
the total area in the \(200\textrm{m} \times 200 \textrm{m}\) square. If
the gray area near the outer edge of the plot is the 25\% that is not
searched, then the simple area correction would estimate \(M\) as
\(\hat{M} = 98/0.75 = 130.7\), which substantially overestimates the
actually mortality of \(M = 100\). If, instead, the yellow region in the
center is the 25\% not searched, then \(\hat{M} = 31/0.75 = 41.3\),
which vastly underestimates \(M\). Clearly, a more reliable solution
than simple area correction is needed.

\hypertarget{accounting-for-unsearched-area}{%
\subsection{\texorpdfstring{Accounting for Unsearched Area
\label{sec:unsearched}}{Accounting for Unsearched Area }}\label{accounting-for-unsearched-area}}

The \texttt{dwp} package fits statistical models of the patterns of
change in carcass density with distance from turbines and uses the
models to predict what fraction of carcasses lie in the searched area.
Given a list of the locations at which carcasses were discovered and a
map of the searched area, the \texttt{prepRing} function\footnote{This
  introductory section is limited to a simple outline of the main
  processes involved in a \texttt{dwp} analysis. Details for formatting
  and importing data and using the functions in the \texttt{dwp} package
  are discussed in detailed examples in later sections. The figures
  shown in this section are meant as simple heuristics, and the
  \texttt{dwp} package does not produce similar pictures.} tallies the
carcasses in concentric 1 m rings and estimates the carcass density in
each ring by dividing the carcass count by the area sampled at that
distance (fig.~\ref{fig:volcano}a).

Then, \texttt{ddFit} is used to fit a curve to the carcass densities
(fig.~\ref{fig:volcano}a, blue line) and rotate the curve around the
turbine to create a volcano-like surface of relative carcass densities
(fig.~\ref{fig:volcano}b)\footnote{The rotation is done mathematically,
  and there is no \texttt{dwp} function for creating such ``volcano
  plots''.}. After proper normalization (which is done automatically by
\texttt{ddFit}), the volcano surface is a probability density function
that shows the relative carcass density at each point. The total volume
of the volcano represents 100\% of the carcasses, and the volume under a
specific area represents the probability that a carcass will lie in that
area.

For example, the searched area at the turbine in fig.~\ref{fig:volcano}
is an irregularly shaped stripe down the middle. The probability of a
carcass lying in the searched area (\(\psi\)) is the volume of the
volcano above that area (fig.~\ref{fig:volcano}d) and is calculated
using \texttt{estpsi}.

Although the basic idea is fairly straightforward, there are some
nuances that are easy to overlook or whose importance is easy to
discount. In particular, estimating how many carcasses lie outside the
search radius requires extrapolation well beyond the range of the data,
which is always hazardous. Furthermore, constructing proper confidence
intervals for the fraction of carcasses within the searched area
(\(dwp\)) requires accounting for both the uncertainty in the
probability that a carcass lies in the searched area (\(\psi\)) and the
uncertainty in the actual number of carcasses in the searched area given
\(\psi\). The complications involved in extrapolation are discussed in
general in section \ref{sec:extrapolation}, examples and discussion of
tools for resolving the complications are given in sections
\ref{sec:exeagle}-\ref{sec:exshape}, and an extended theoretical
discussion is presented in Appendix \ref{app:fit2deq}. For the reader
who is curious about the technical details of how the estimation
uncertainties are accounted for and accurate confidence intervals are
constructed by \texttt{estpsi} and \texttt{estdwp}, the theory is
presented and tested in Appendix \ref{app:vardwp}.

\hypertarget{model-of-carcass-density}{%
\subsection{\texorpdfstring{Model of Carcass Density
\label{sec:density}}{Model of Carcass Density }}\label{model-of-carcass-density}}

In the \texttt{dwp} package, estimation of the fraction of carcasses
that lie within the searched area involves 3 general tasks:

\begin{enumerate}
  \item modeling carcass density as a function of distance from the turbine and integrating over the searched area, 
   \item extrapolating carcass density models to areas beyond the search radius, and 
   \item incorporating the uncertainties that are inherent to the estimation process into the final estimates of $dwp$. 
 \end{enumerate}

The choice of models to use is influenced by the need to prevent over-
or under-estimation of tail probabilities during extrapolation.
\(\texttt{dwp}\) provides several checks to help avoid selection of
unrealistic models: extensibility, tail plausibility, and high influence
filters (discussed in section \ref{sec:selectiontools}). The models
remaining after filtering can be further compared using AICc, however
AICc must be used with discretion as it is an unreliable measure of
model performance outside the search radius, as demonstrated in appendix
\ref{app:fit2deq}.

The model selection and filtering process are guided by several
principles or assumptions:\\

\begin{enumerate}
  \item the distribution of carcass distances is in the exponential family of distributions (appendix \ref{app:modbrief});
  \item the number of animals killed by a turbine is finite, and consequentially there exists a radius beyond which very few animals will be found; and
  \item models should not be overly sensitive to any single observation or data point.
\end{enumerate}

Together, these principles guide us toward a collection of distributions
and rules by which to select from the collection for further comparison.

A set of homogeneity assumptions about carcass dispersion ensure
accuracy of the estimates and inferences:

\begin{enumerate}
  \item fatality rates do not vary among turbines,
  \item carcass distributions do not vary among turbines, and
  \item carcass distributions do not depend on direction ($isotropy$).
\end{enumerate}

These assumptions allow for straightforward pooling of the carcasses
across turbines to fit a single carcass distance model that can be
applied to an entire site. Once a model for a site is selected, \(dwp\)
can be estimated for each turbine, with its particular search plot and
carcass count. Even if the same distance model is used for the entire
site, \(dwp\) will typically vary among turbines, depending on the size
and shape of the search plots, and the degree of uncertainty depends on
the carcass counts. If the homogeneity assumptions are not met, the
analysis may require additional steps to ensure accuracy.

\hypertarget{homogeneity-of-fatality-rates-among-turbines}{%
\subsubsection{Homogeneity of Fatality Rates among
Turbines}\label{homogeneity-of-fatality-rates-among-turbines}}

Homogeneity of rates does not mean that the actual number of fatalities
must be identical at all turbines but that the numbers do not vary
significantly. In this context it is difficult to precisely define
``significantly'' because the effect that variation in rates has on the
model depends on the sizes, shapes, and characteristics of the search
plots, and tolerance for potential bias depends on the objectives of the
research. In theory, variation in rates can adversely affect accuracy.
In practice, however, even substantial variation in mortality rates
among turbines will rarely cause significant inaccuracies in estimating
\(dwp\).

\begin{itemize} 
  \item If the sizes and shapes of the search plots are identical among the turbines and the detection probabilities are the same, then homogeneity of rates is not necessary. The resulting fitted model will be the same regardless of the degree of variation in rates among turbines.
  \item If search plot shapes vary widely among turbines and plot shapes are correlated with fatality rates, then there is potential for modest bias due to heterogeneity of fatality rates. An example of how this might occur would be if turbines that are known or strongly suspected to have lower than average fatality rates are searched on roads and pads only, while turbines with relatively high fatality rates are searched on cleared plots with a large search radius. 
  \item If search plot size and shape does not correlate with mortality rate, then the potential for bias due to lack of homogeneity fatality rates is greatly reduced.
\end{itemize}

\hypertarget{homogeneity-of-carcass-distributions-among-turbines}{%
\subsubsection{Homogeneity of Carcass Distributions among
Turbines}\label{homogeneity-of-carcass-distributions-among-turbines}}

Variation in wind patterns, turbine types, turbine operations (for
example, curtailed versus uncurtailed), carcass type, and other factors
may result in variation in carcass dispersion patterns. We refer to
factors that vary among turbines or carcasses and that influence the
actual carcass distributions as \emph{interacting covariates}. These
factors can be explicitly accounted for in the carcass distance models
when the search plot layouts are imported from shape files or as
\emph{x-y} coordinates on a grid. With the simpler data types (list of
distances, simple geometry, or R polygons), interacting covariates can
be accounted for by fitting seperate models for different levels of the
covariate. For example, bats and eagles may each require their own
models; likewise for curtailed and uncurtailed turbines, large turbines
and small turbines, site A and site B, etc.

\hypertarget{homogeneity-of-carcass-distributions-in-each-direction}{%
\subsubsection{Homogeneity of Carcass Distributions in Each
Direction}\label{homogeneity-of-carcass-distributions-in-each-direction}}

Variation in carcass distributions by direction from a turbine is
referred to as \emph{anisotropy}. If not accounted for, anisotropy can
be a significant source of bias when the shapes of the search plots also
vary by direction. Accounting for anisotropy introduces a considerable
degree of statistical complexity along with more extensive data
requirements. Perhaps the most straightforward and flexible way to
account for anisotropy would be via Poisson regression of carcass counts
on a coordinate grid, as discussed in Maurer et al (2020). The
\(\texttt{dwp}\) package (version 1.0) does not include special
functions for handling anisotropy.

\begin{figure}
\includegraphics[width=1\linewidth]{dwp_estimation11_files/figure-latex/volcano_full} \caption{\label{fig:volcano}Estimation of fraction of carcasses lying within the searched area. a) carcass density (per $m^2$) as a function of distance from the turbine; b) relative carcass density in the area around the turbine, normalized so that it integrates to 1; c) area searched; d) relative density for the area searched. The fraction of carcasses lying in the searched area is the normalized relative carcass density integrated over the searched area.}\label{fig:unnamed-chunk-2}
\end{figure}

\hypertarget{extrapolation-beyond-the-searched-area}{%
\subsection{\texorpdfstring{Extrapolation Beyond the Searched Area
\label{sec:extrapolation}}{Extrapolation Beyond the Searched Area }}\label{extrapolation-beyond-the-searched-area}}

There is a wide variety of possible approaches that could be used to
model carcass density as a function of distance from turbine. Any
approach to fitting a distribution will need to account for carcasses
that lie outside the search radius and beyond the scope of data, and
will thus require extrapolation. All approaches suffer from the inherent
difficulty that model performance within the range of data tells little
to nothing about how well it will perform outside the range of the data.
Traditional model selection criteria such as AIC (Akaike 1974; Burnham
and Anderson 2002), likelihood ratio tests (Vuong 1989), stepwise
regression (Efroymson 1960), cross-validation (Allen 1974) and others
have limited utility when a significant fraction of carcasses lie beyond
the search radius. In fact, two models, A and B, that are equally
plausible in theory and are virtually indistinguishable within the range
of data may be radically different outside the range of data, giving
wildly different estimates of the fraction of carcasses within the
searched area.

The \texttt{dwp} package fits a large number of probability
distributions in the exponential family (Andersen 1970), including the
well-known and named exponential, gamma, normal, lognormal, Rayleigh,
Maxwell-Boltzmann, inverse gamma, inverse Gaussian, normal-gamma,
Pareto, and chi-squared models and several unnamed models, which we
christen with names like ``xep1'' and ``xep012''. Detailed knowledge of
the models is not required for understanding the difficulties inherent
in extrapolation beyond the search radius, but for the curious reader,
further details about the models and our techniques for fitting them are
discussed in Appendix \ref{app:modbrief}.

The example discussed in this section illustrates some important general
concepts and issues involved in determining the fraction of carcasses
within the searched area. Most notably,

\begin{enumerate}
  \item predicting the number or fraction of carcasses lying outside the search radius requires extrapolation;
  \item two (or more) models that are in close agreement within the range of the data may 
     have radically different behaviors outside the search radius;
   \item criteria that draw on knowledge about carcass ecology (for example, the number of 
     carcasses is finite and a very small fraction of carcasses will lie beyond 150 or 200 m from their turbine)
     are a necessary complement to traditional criteria (like AICc, high-influence points) for assessing 
     model suitability; and
     \item the $\texttt{modelFilter}$ function performs a suite of diagnostic tests and ranks the models by
     suitability.
\end{enumerate}

\hypertarget{example-data}{%
\subsubsection{Example Data}\label{example-data}}

\begin{figure}
\centering
\includegraphics{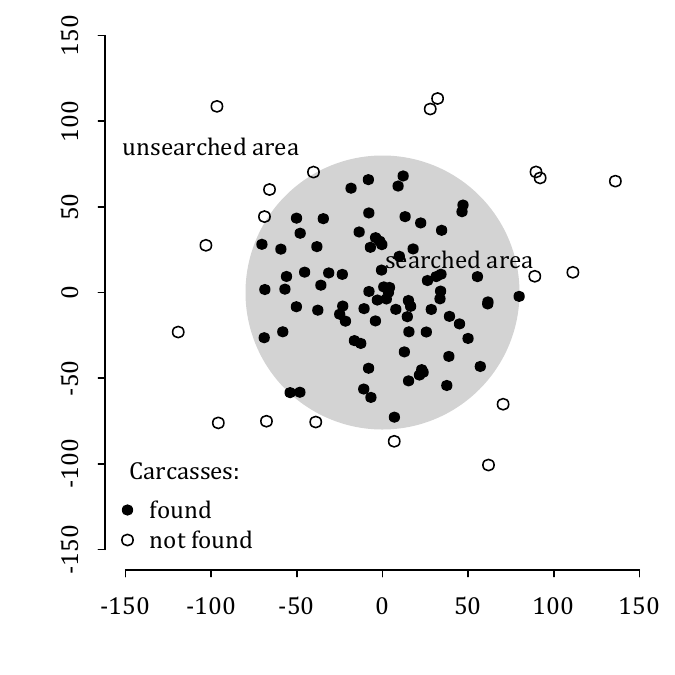}
\caption{\label{fig:ex1dat}Simulated carcass locations. Turbine is at
the center of a circular search plot with a radius of 80 m. There are 81
carcasses within the search radius (all were found) and 19 outside the
search radius (none were found).}
\end{figure}

Problems with extrapolating fitted models beyond the search radius are
illustrated clearly in an example data set in which 100 carcass
locations were simulated. Distances were drawn from a gamma distribution
in which half the carcasses are expected to fall within 41 m of the
turbine and 82\% within 80 m. Searches were conducted in a circular plot
centered at the turbine and with radius 80 m. All 81 simulated carcasses
lying within the search radius were found and none of the 19 carcasses
outside the search radius were found (fig.~\ref{fig:ex1dat}). The
\texttt{dwp} package was used to fit 17 models from the exponential
family of distributions.

\hypertarget{slightly-beyond-the-search-radius-radical-divergence-among-models}{%
\subsubsection{Slightly beyond the Search Radius: Radical Divergence
among
Models}\label{slightly-beyond-the-search-radius-radical-divergence-among-models}}

\begin{figure}
\centering
\includegraphics{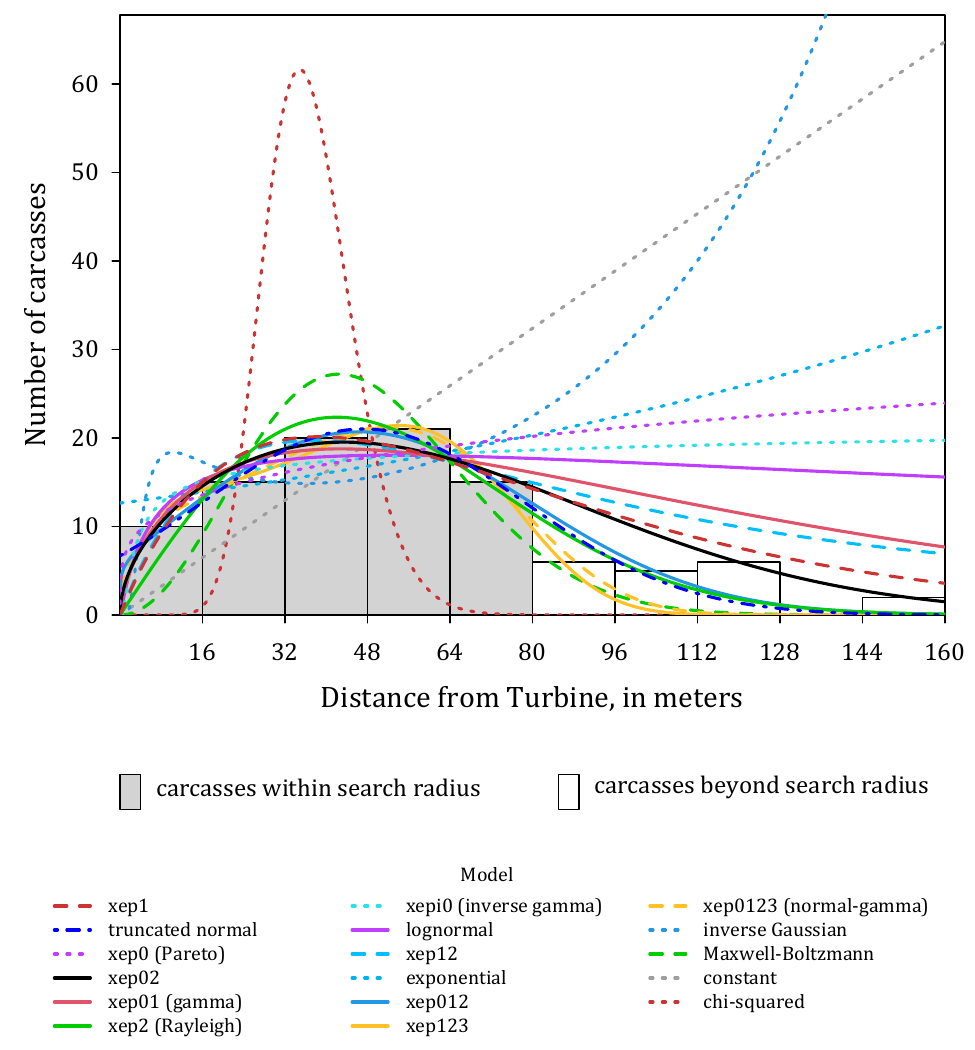}
\caption{\label{fig:allmod}Model spatial predictions of carcasses
outside the search radius}
\end{figure}

Of the 17 models pictured in fig.~\ref{fig:allmod}, all but a few
(constant density, chisquared, Maxwell-Boltzmann, and inverse Gaussian)
appear to provide fairly close fits to the carcass data within the 80 m
search radius (fig.~\ref{fig:allmod}, gray bars), but just beyond the
search radius there is wide divergence among the models
(fig.~\ref{fig:allmod}, lines). Although the vast majority (81\%) of the
carcasses lie within the search radius, there is no guarantee that the
models can accurately predict the final 19\% of carcasses that lie
beyond the search radius. Indeed, the actual fraction of carcasses lying
within the search radius is 81\%, but the models predict anywhere from
0\% to 100\%, depending on model (\(\hat{\psi}\) in table
\ref{tbl:allmod}).

\hypertarget{highly-limited-utility-of-the-akaike-information-criterion-for-model-selection}{%
\subsubsection{Highly Limited Utility of the Akaike Information
Criterion for Model
Selection}\label{highly-limited-utility-of-the-akaike-information-criterion-for-model-selection}}

A traditional staple in model selection is the \emph{Akaike Information
Criterion} or AIC (Burnham and Anderson 2002), which is a measure of how
well a model fits the data given the number of parameters in the model.
We use AICc, which is a slightly modified version of AIC that may be
more accurate than AIC when sample sizes are small (Cavanaugh 1997). To
facilitate easy comparison of models, we define \(\Delta\textrm{AICc}\)
as the difference between a model's AICc score and the minimum of the
AICc scores of all models to be compared. The model with the most
parsimonious fit has \(\Delta\textrm{AICc} = 0\) and serves as the
standard for comparison.

As a rule of thumb, for two models that differ by more than about 7 in
AICc, the one with the smaller AICc is preferable, and models with
scores differing by less than about 2 are often considered statistically
indistinguishable by this measure (Richards 2005). If we only consider
models with \(\Delta\textrm{AICc} < 2\), the range of predicted
proportions of carcasses within the search area narrows only slightly,
to 0\% -- 84.3\%. It is evident that AICc has extremely limited
utility\footnote{That utility may even be exaggerated in this example
  because the \(\Delta\textrm{AICc} < 2\) is far too restrictive,
  eliminating from consideration some models that appear to provide only
  marginally poorer quality fits within the range of data than the
  ``best'' models but without regard to the shapes of the curves outside
  the range of data.} in distinguishing among models and their ability
to extrapolate accurately. AICc measures only the quality of the fit of
a model within the range of data and offers no guidance on performance
outside the range of the data. Additional selection criteria are
essential (section \ref{sec:selectiontools}).

Thus, \(\Delta\textrm{AICc}\) confirms the visual impression that the
constant, chisquared, and Maxwell-Boltzmann models give relatively poor
fits to the data (fig.~\ref{fig:allmod}). Among the remaining models, it
is not clear which provide the best fits to the data within the search
radius. AICc does capture some of the subtle distinctions among the
model fits within the range of data. However, outside the range of the
data, the models diverge radically from each other, with the predicted
fraction of carcasses lying inside the search radius ranging from 0 to
93\% for the models with \(\Delta\textrm{AICc} < 3\). AICc provides no
guidance for assessing model performance beyond the search radius.

\begin{threeparttable}[!htbp]
  \caption{Diagnostic Statistics for Models in Fig. \ref{fig:allmod}}
  \label{tbl:allmod}
  \begin{tabular}{lcccccccc}
    \hline\hline
     & & & & & & \multicolumn{2}{c}{Tail Weight\tnote{d}} & \\
     Model & $k\tnote{a}$ & $\Delta\textrm{AICc}$ & Extensible & $\hat{\psi}\tnote{b}$ & $\hat{M}\tnote{c}$ & R & L & HI\tnote{e}\\
    \hline
     xep1 & 2 & 0.00 & 1 & 0.614 & 131.9 & 0 & 1 & 1 \\ 
      truncated normal & 3 & 0.54 & 1 & 0.843 & 96.1 & 1 & 1 & 1 \\ 
      xep0 (Pareto) & 2 & 0.56 & 0 & 0.000 & Inf & 0 & 1 & 0 \\ 
      xep02 & 3 & 0.87 & 1 & 0.695 & 116.5 & 1 & 1 & 0 \\ 
      xep01 (gamma) & 3 & 1.33 & 1 & 0.451 & 179.6 & 0 & 1 & 0 \\ 
      xep2 (Rayleigh) & 2 & 1.49 & 1 & 0.836 & 96.9 & 1 & 1 & 1 \\ 
      xepi0 (inverse gamma) & 3 & 1.61 & 0 & 0.000 & Inf & 0 & 1 & 0 \\ 
      lognormal & 3 & 1.70 & 1 & 0.033 & 2454.5 & 0 & 1 & 0 \\ 
      xep12 & 3 & 2.13 & 0 & 0.000 & Inf & 0 & 1 & 1 \\ 
      exponential & 2 & 2.64 & 0 & 0.000 & Inf & 0 & 1 & 1 \\ 
      xep012 & 4 & 2.67 & 1 & 0.820 & 98.8 & 1 & 1 & 0 \\ 
      xep123 & 4 & 2.68 & 1 & 0.934 & 86.7 & 1 & 1 & 0 \\ 
      xep0123 (normal-gamma) & 5 & 4.69 & 1 & 0.913 & 88.7 & 1 & 1 & 0 \\ 
      inverse Gaussian & 3 & 6.52 & 0 & 0.000 & Inf & 0 & 1 & 0 \\ 
      Maxwell-Boltzmann & 2 & 18.22 & 1 & 0.931 & 87.0 & 1 & 1 & 1 \\ 
      constant & 1 & 19.82 & 0 & 0.000 & Inf & 0 & 1 & 1 \\ 
      chi-squared & 2 & 380.85 & 1 & 1.000 & 81.0 & 1 & 0 & 1 \\ 
       \hline
  \end{tabular}
  \begin{tablenotes}
    \item[a] Number of parameters in the model.
    \item[b] Estimated probability that a carcass lies in the searched area.
    \item[c] Estimated total number of carcasses.
    \item[d] Indicator of whether the right (R) and left (L) tail probabilities of the distribution are plausible (1) or not (0).
    \item[e] Indicator of whether the model is free of points with high influence (HI = 1). 
    Data points with high influence on a model (HI = 0) cast doubt on the reliability of the model.
  \end{tablenotes}
\end{threeparttable}

\hypertarget{tools-for-model-selection}{%
\subsection{\texorpdfstring{Tools for Model Selection
\label{sec:selectiontools}}{Tools for Model Selection }}\label{tools-for-model-selection}}

With a closer look at the model predictions in fig.~\ref{fig:allmod} and
closer consideration of what is known about carcass dispersion around
turbines, we can filter out most of the obviously poor choices, but we
should do so without reference to the white bars in the histogram
because they are from carcasses outside the search radius, data that we
normally would not have access to.

The \texttt{modelFilter} function performs a series of tests to aid in
model selection and returns the results in a convenient format, greatly
reducing or even eliminating the guesswork. The tests are intuitive,
easy to understand, and easy to implement.

\begin{figure}
\centering
\includegraphics{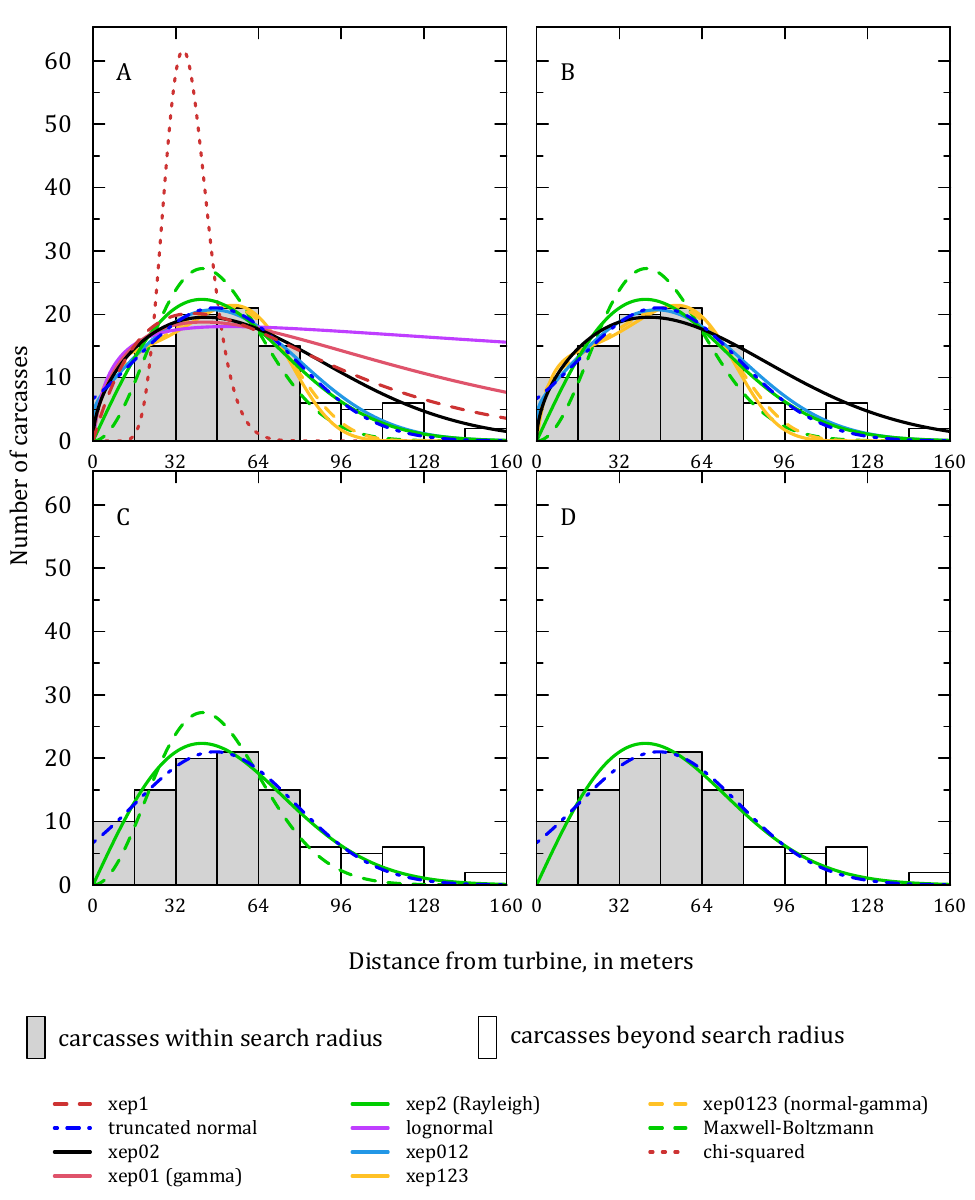}
\caption{\label{fig:filtered_models}Plots showing cCriteria for
filtering out poor model choices for extrapolating beyond the range of
the data and predicting proportion of carcasses outside the search
radius. A. All Extensible models. B. Extensible models, with plausible
tails. C. Extensible models, with plausible tails, and no high influence
points. D. Extensible models, with plausible tails, no high influence
points, and $\Delta \textrm{AICc}$ less than 10. $\Delta \textrm{AICc}$, difference between Akaike
information criterion corrected for finite sample size (AICc) for a
given model and the lowest AICc value among the models tested.}
\end{figure}

\hypertarget{model-filter-extensibility}{%
\subsubsection{Model Filter:
Extensibility}\label{model-filter-extensibility}}

If we assume that the number of carcasses is finite, we can eliminate
from consideration all models that do not converge to zero or do not
converge fast enough to guarantee that the predicted number of carcasses
outside the search area is finite. We refer to models that predict a
finite number of carcasses outside the searched area as
\emph{extensible} because they can be extended outside the search radius
to give properly defined probability distributions\footnote{This is
  discussed further in section \ref{sec:extrapolation}, in the examples
  (sections \ref{sec:exeagle}-\ref{sec:exshape}), and in Appendix
  \ref{app:modbrief}}. Typically, models that are still increasing
beyond the search radius are non-extensible, and models that are
decreasing are extensible (figs. \ref{fig:allmod} and
\ref{fig:filtered_models}a).\footnote{The rule is not hard and fast.
  Sometimes a non-extensible model decreases beyond the search radius
  but at too slow a rate; sometimes an extensible model may be still be
  increasing at the search radius but decline rapidly shortly
  thereafter. These are exceptional situations. They may be rare but
  they are not impossible.}

Among the extensible models---which, for this data, include the
truncated normal, gamma, Rayleigh, lognormal, Maxwell-Boltzmann,
normal-gamma, chi-squared, xep1, xep02, xep12, xep012, and xep123---the
predicted fraction of carcasses beyond the search radius narrows
slightly, from an impossible 0\% -- 100\% to a merely implausible 3.3\%
-- 100\% (table \ref{tbl:allmod}).

Extensibility is a relatively permissive criterion that excludes only
the most grossly inadequate models. Several \texttt{dwp} package
functions automatically test for extensibility. Most notably,
\texttt{ddFit} tags all fitted models as either extensible
(\texttt{\$extensible\ =\ 1}) or not extensible
(\texttt{\$extensible\ =\ 0}), and \texttt{modelFilter} returns a table
of test results (including extensibility) for arrays of models to be
compared.

\hypertarget{model-filter-tail-plausibility}{%
\subsubsection{Model Filter: Tail
Plausibility}\label{model-filter-tail-plausibility}}

Because gravity acts so inexorably on carcasses after turbine strikes,
few carcasses are expected beyond 150 or 200 meters. However, because
the search plots are inevitably finite and the models are fit with
spatially limited data, sometimes a fitted model will perform well
within the search radius but have implausible tails, predicting far too
many carcasses at great or short distances. A rough filter for tail
plausibility identifies obviously inappropriate models. By default,
models that predict that more than 1\% of carcasses lie beyond 200 m or
more than 5\% beyond 150 m fail the \texttt{modelFilter} for a plausible
right tail.\footnote{If desired, users may override these default
  settings.} For this data set with the default settings, the filter
eliminates the gamma, lognormal, and xep1 models (table
\ref{tbl:allmod}), which were extensible but predict too many carcasses
at great distances from the turbine to be plausible. For example the
lognormal predicts that 93.8\% of the carcasses fall beyond 150 m, while
the gamma and xep1 models predict 25.3\% and 90\%, respectively.

Because of the great size and height of turbines, we would expect a
substantial fraction of carcasses to lie beyond 20 m and a significant
fraction beyond 50 m. By default, \texttt{modelFilter} identifies models
that predict that over 50\% of the carcasses lie within 20 m or over
90\% within 50 m as having implausibly heavy left tails. By this
criterion, the chisquared model is eliminated from further consideration
for this data set (table \ref{tbl:allmod}).

The tests of tail plausibility further narrow the range of the predicted
proportion of carcasses within the search radius from 3.3\%--100\% to
69.5\%--93.4\% (table \ref{tbl:allmod}), which is still a wide range but
a vast improvement over using the extensibility criterion by itself
(fig.~\ref{fig:filtered_models}).

\hypertarget{model-filter-high-influence-points}{%
\subsubsection{Model Filter: High Influence
Points}\label{model-filter-high-influence-points}}

In some cases, the apparently good fit of model may strongly depend on
one or a small set of data points, so that removing one or a small
number of data points has a strong impact on the shape of the fitted
model. The presence of these \emph{high influence} points (Davison and
Hinkley 1997; Canty and Ripley 2021) casts considerable doubt on the
reliability of the model for the given data set. Extrapolation would be
especially worrisome for such models because their shapes are strongly
dependent on one or a small number of fickle data points, and the
effects would be felt well beyond the range of the data.

Several of the extensible models with plausible tails have points with
high influence---the xep02, xep012, xep123, and normal-gamma models
(table \ref{tbl:allmod}). Removing these from consideration further
narrows the range of predicted proportions of carcasses within the
searched area to \(\psi \in (0.836, 0.931)\) (table \ref{tbl:allmod})
and eliminates much of the remaining clutter from the graphs of models
still under consideration (fig.~\ref{fig:filtered_models}b, c).

\hypertarget{model-filter-aicc}{%
\subsubsection{Model Filter: AICc}\label{model-filter-aicc}}

After models with implausible extrapolations or worrisome instabilities
have been filtered out, criteria like AICc that measure the quality of a
model's fit strictly within the range of the data are useful for
distinguishing among plausible models. AICc and other internal criteria
are especially valuable when there are substantial unsearched areas
within the search radius, as would be the case with road and pad
searches (Maurer et al.~2020) or with sites with forest or other thick
vegetation, water, cliffs, poisonous snakes, or other features that
render much of the area around turbines practically unsearchable.
Predicting the proportion of carcasses in unsearched areas within the
search radius is an interpolation problem rather than extrapolation, and
the quality of the fit to the data within the search radius is directly
relevant in a way that it is not when extrapolating beyond the search
radius.

In an extensive simulation study of the accuracy of carcass distribution
models (appendices \ref{app:deqmod} and \ref{app:fit2deq}), AICc was
found to be a generally poor guide for gauging the relative adequacy of
models for extrapolating beyond the search radius. Under some of the
conditions tested, models with lower AICc correlated with more accurate
extrapolations; under other conditions, lower AICc was correlated with
\emph{less} accurate extrapolations. However, under most conditions,
there was no apparent correlation between AICc and suitability for
extrapolation. One exception was that models with
\(\Delta\textrm{AICc} > 10\) frequently performed worse than models with
\(\Delta\textrm{AICc} < 10\). For this reason, \texttt{modelFilter}
compares model AICc scores and, by default\footnote{As with all the
  \texttt{modelFilter} parameters, user has the option of overriding the
  default values}, identifies and filters out models with
\(\Delta\textrm{AICc} > 10\).

After filtering out the lone extensible model with plausible tails, no
high influence points, and \(\Delta\textrm{AICc} > 10\) (namely, the
Maxwell-Boltzmann model), the two remaining models---truncated normal
and Rayleigh---predict that, respectively, 84.3\% and 83.6\% of the
carcasses lie within the searched area. Both are in good agreement with
the reality of 81\% of the (simulated) carcasses lying within the
searched area. The fitted models closely match each other and appear to
be good fits to the data (fig.~\ref{fig:filtered_models}). Also, the
difference in their AICc scores is only 0.95, which is too small to make
a statistically meaningful distinction between the models.

\hypertarget{standard-models}{%
\subsubsection{Standard Models}\label{standard-models}}

Several of the distributions are not well suited for modeling carcass
distances. These include the inverse gamma (xepi0), inverse Gaussian,
chisquared, Pareto (xep0), and the exponential distributions. Because
these distributions would rarely (if ever) be among the most appropriate
models for carcasses, they are excluded from the \texttt{dwp} package's
list of 12 ``standard'' distributions but can still be used if
specifically invoked. The standard models, which include xep1, xep01
(gamma), xep2 (Rayleigh), xep02, xep12, xep012, xep123, xep0123
(normal-gamma), truncated normal, Maxwell-Boltzmann, lognormal,
constant, are the primary focus of this report. All the models are
defined and briefly discussed in Appendix \ref{app:modbrief}.

\hypertarget{summary}{%
\subsubsection{Summary}\label{summary}}

In general, when a significant fraction (for example, \textgreater10\%)
of carcasses are expected to lie beyond the search radius and
extrapolation is necessary to estimate the fraction of carcasses lying
in the searched areas, it is essential to consider the suitability of
models for extrapolation, using broad criteria like extensibility, tail
plausibility, and stability (for example, the presence of high-influence
points) before considering strictly internal criteria like AICc. If the
search radius is long enough so that few carcasses are expected beyond
the search radius and a substantial fraction of the area within the
search radius is unsearched, then AICc and high-influence points become
the critical criteria.

To determine which models are likely to perform well or poorly in
predicting carcass numbers outside the search radius, it is necessary to
step beyond traditional model selection criteria and make use of more
general information about the distribution of carcasses. Most notably,
the number of carcasses generated by a turbine will always be finite,
and models that predict infinite carcasses beyond the search radius
should be eliminated from consideration. We refer to models that can be
extended beyond the search radius to infinity and still predict finite
numbers of carcasses as ``extensible.'' Among the original 17 models, 11
are extensible for the example data. Although many of the remaining,
non-extensible models {[}xep0 (Pareto), xepi0 (inverse gamma), xep12,
exponential, inverse Gaussian, constant{]} perform well according to
AICc (table \ref{tbl:allmod}) and provide good fits to the data, they
not suitable for extrapolation beyond the search radius and should not
be used in estimating the fraction of carcasses within the searched
area.

The example illustrates some of the power and utility of the
diagnostics. However, it remains a single example, and results specific
to this particular data set may not be generalizable. For example, the
lognormal model fits relatively well within the range of this data but
was implausible for extrapolation because it predicted that 97\% of the
carcasses lay beyond the 80 m search radius. For this data set,
lognormal is implausible, but for other data sets it may be the best
choice.

\hypertarget{uncertainties-and-confidence-intervals}{%
\subsection{Uncertainties and Confidence
Intervals}\label{uncertainties-and-confidence-intervals}}

In estimating the total mortality (\(M\)), it is necessary to account
for carcasses that lie outside the searched area. If the probability
that a carcass lies in the searched area is \(\psi\), then a reasonable
point estimate for \(M\) would be \(\hat{M} = m_{\textrm{in}}/\psi\),
where \(m_\textrm{in}\) is the number of carcasses lying within the
searched area. However, even if we know precisely the probability that a
carcass will lie in the searched area and we know the number of
carcasses that did lie in the searched area, there will still be
uncertainty about the total number of carcasses.

The problem is akin to coin tossing, where we know the probability of
heads is \(\psi = 0.5\). If someone tosses a coin some unknown number of
times (\(M\)), tells us that there were \(m_{\textrm{in}} = 5\) heads,
and asks us to guess the total number of tosses, a good guess would be
\(\hat{M} = 5/0.5 = 10\), which is the guess with the maximum
likelihood. It is possible that the number of tosses was 10, but it
could easily have been 11 or 15 or 8 instead, but probably not as few as
5 or 6 or as many as 100 or even 25. That binomial uncertainty must be
properly accounted for if we are to create accurate confidence intervals
of either the fraction of carcasses lying in the searched area or the
total mortality.

The probability that a carcass will lie in the searched area, \(\psi\),
is like the theoretical probability of H in a coin toss, while \(dwp\)
is like the fraction of Hs in \(M\) actual coin tosses. The probability
that a carcass will lie in the searched area (\(\psi\)) is unknown. The
\texttt{dwp} package estimates \(\psi\) by integrating (\texttt{estpsi})
a fitted carcass probability distribution (\texttt{ddFit} and
\texttt{modelFilter}) over the searched area (\texttt{prepRing} and
\texttt{addCarcass}). Once \(\psi\) is estimated, \texttt{estdwp}
accounts for the uncertainty about actual fraction of carcasses lying in
each turbine's searched area, given the number of carcasses observed at
each turbine (\(X_i\)) and the estimated probability that a carcass lies
in the searched area (\(\hat{\psi}_i\)).

The \(dwp\) is estimated in a way that is fully compatible with the
\texttt{GenEst} mortality estimator. The distinction between \(dwp\) and
\(\psi\) and their roles in estimation are discussed in more detail in
the technical appendices (Appendices \ref{app:psivdwp} and
\ref{app:vardwp}). With the \texttt{eoa} estimator, an single number is
entered as an aggregate \(dwp\) for the site as a whole. Typically, for
\texttt{eoa} a user would take the average ``total'' \(dwp\), or
\texttt{mean(dwp{[},\ "total"{]})}, where \texttt{dwp} is the \(dwp\)
array estimated from the \texttt{estdwp} function.

\hypertarget{the-dwp-package}{%
\section{\texorpdfstring{The \texttt{dwp}
Package}{The dwp Package}}\label{the-dwp-package}}

Doing basic analyses with the \texttt{dwp} package does not require
extensive prior experience with R, but basic familiarity with the
workflow of entering/loading data, entering commands, and using R's
``help'' features at a basic level is assumed. This tutorial covers a
number of examples and briefly discusses some of the main functions used
in the data management and analysis. Additional information on
formatting, usage, and additional options is available for any function
by entering the function name (case-sensitive) preceded by a question
mark. A list of all functions, grouped by utility and linked to
function-specific help files, can be seen by entering \texttt{?dwp} in
the R command line. A complete reference manual with functions arranged
alphabetically can be found at
\texttt{https://cran.r-project.org/web/packages/dwp/index.html}.

The general work flow involves several steps, and \texttt{dwp} provides
easy-to-use functions to help accomplish each step:

\begin{enumerate}
  \item importing site layout data and formatting for analysis (\texttt{initLayout}),
  \item adding carcass data to the site layout (\texttt{readCarcass} and \texttt{addCarcass}),
  \item fitting carcass distance models (\texttt{ddFit}),
  \item selecting a carcass distance model to use in estimating $dwp$ (\texttt{modelFilter}),
  \item estimating probability of carcass lying in searched area (\texttt{estpsi}), 
  \item estimating fraction of carcasses lying in the searched areas (\texttt{estdwp}), and
  \item exporting to \texttt{GenEst} (\texttt{exportGenEst}).
\end{enumerate}

All data types require a quantitative description of the searched area.
The description may be as simple as the search radius (with the
assumption that all ground within that search radius is searched and no
ground outside that search radius is searched) or as complex as a
detailed GIS-generated map. Altogether there are five options for site
layout data format, including search radius, simple geometries, R
polygons, GIS shape files, and rasters (that is, search plots laid out
on an \(xy\)-grid). The data processing in the first two steps varies,
depending on the data type. However, for many data types the
calculations are all done automatically in the background by
\texttt{dwp} functions, and the user interface for each data type is
similar.

\hypertarget{site-layout-raw-data}{%
\subsection{\texorpdfstring{Site Layout: Raw Data
\label{sec:raw}}{Site Layout: Raw Data }}\label{site-layout-raw-data}}

The \texttt{dwp} package can accommodate several different data formats
for characterizing site layout, including GIS shape files, R polygons,
simple geometric descriptions (square, circular, or road and pad
searches with user-specified dimensions for each turbine), a list of
carcass distances with a search radius, or coordinates on an
\(xy\)-grid.

\begin{enumerate}
  \item \textbf{simple search radius} can accommodate a simple vector of distances from turbine to carcass. It assumes circular search plots with the same radius at each turbine with no unseached area within the plots and the same detection probability in each plot.
  \item \textbf{simple geometry} can accommodate simple variation in search geometries as circular, square, or road \& pad with specific parameters (shape, search radius, size of turbine pad, width of roads, number of roads) for each turbine individually. This option does not allow for multiple search classes.\footnote{A \emph{search class} is an area where the carcass detection probability is expected to be the same in all parts.}
    \item \textbf{R polygons} can accommodate moderately complex search geometries that do not involve multiple search classes with different detection probabilities\footnotemark[\value{footnote}]. 
  \item \textbf{GIS shape files} can accommodate highly complex search geometries, multiple carcass classes (for example, large, medium, small, bats) and multiple search classes\footnotemark with different detection probabilities corresponding to different ground types, search schedules (for example 1-day and 14-day), or other variables that affect detection probability.
  \item \textbf{$xy$-grid} can accommodate highly complex search geometries and multiple covariates and is suitable for more sophisticated custom analyses, such as accounting for anisotropy or interacting covariates. Although it opens the door to more sophisticated analyses, gridded data is substantially more computationally demanding than other data formats, and the models run much more slowly. In addition, the $\texttt{dwp}$ package does not provide automated tools for doing complex, custom analyses.
\end{enumerate}

More complete definitions of each data type are given in worked examples
in sections \ref{sec:exeagle}-\ref{sec:exshape}.

\hypertarget{site-layout-formatted-data}{%
\subsection{\texorpdfstring{Site Layout: Formatted Data
\label{sec:formatted}}{Site Layout: Formatted Data }}\label{site-layout-formatted-data}}

To estimate relative carcass density as a function of distance from a
turbine, the package requires a description of the layout of the areas
searched at a site and a list of locations where carcasses were
discovered. The standard analysis divides the area around each turbine
into 1 meter rings and tallies the number of carcasses and the total
area searched in each ring. The \(xy\)-grid data are processed and
modeled on a grid rather than rings. The grid approach is ultimately
more flexible and powerful than the rings but substantially more
computationally intensive and slightly less accurate.

\hypertarget{ring-structure}{%
\subsubsection{\texorpdfstring{Ring Structure
\label{sec:rings}}{Ring Structure }}\label{ring-structure}}

With the exception of the \(xy\)-grid data, all the data types are
converted to a common ring structure for analysis. A \texttt{rings} data
structure is a list with six elements:

\begin{enumerate}
  \item $\texttt{\$rdat}$ list of data frames---one for each turbine and one for the site 
    as a whole---with 3 or 4 columns. $\texttt{rdat}$ is used in fitting the distance models.
    \begin{itemize}
      \item $\texttt{r}$ = outer radius of 1 m ring
      \item $\texttt{Class}$ = search class (may be $\texttt{NULL}$ or a name other than $\texttt{Class}$)
      \item $\texttt{exposure}$ = area ($m^2$) in the ring represented the given search class
      \item $\texttt{ncarc}$ = number of carcasses in the given ring and search class
    \end{itemize}
  \item $\texttt{\$rpA}$ list of data frames---one for each turbine and one for the site 
   as a whole---giving the proportion of the area searched in each ring. $\texttt{rpA}$ 
   is used in calculating estimated $\texttt{dwp}$.
    \begin{itemize}
      \item $\texttt{r}$ = outer radius of 1 m ring
      \item $\texttt{pinc}$ = proportion of area searched in the ring
    \end{itemize}
 \item $\texttt{\$srad}$ = maximum search radius at any of the turbines searched
 \item $\texttt{\$ncarc}$ = vector of carcass counts at each turbine
 \item $\texttt{\$scVar}$ = name of the search class variable (or  $\texttt{NULL}$)
 \item $\texttt{\$tcenter}$ = array of turbine locations with columns $x$ and $y$ and row names are turbine IDs.
\end{enumerate}

The analysis of \texttt{rings} data begins with the fitting of distance
models. By default, the \texttt{ddFit} function fits 12 different
generalized linear models (GLMs) to the data. Specifically, the models
are Poisson regressions of carcass counts in 1 meter rings, using the
searched area (or the searched area in a give carcass class) as the
exposure and the natural logarithm of the exposure as an offset. The
fitted GLMs are then converted to parametric distributions. The models
are discussed in greater detail in Appendix \ref{app:modbrief}.

The data structure for layouts defined on coordinate grids rather than
shapes are somewhat different, but the analyses are similar.

\hypertarget{coordinate-grid-x-y}{%
\subsubsection{\texorpdfstring{Coordinate Grid
\((x, y)\)}{Coordinate Grid (x, y)}}\label{coordinate-grid-x-y}}

A layout of the site on a grid of \((x, y)\) coordinates marking the
centers of 1 \(m^2\) or 2 m $\times$ 2 m quadrats are the most flexible
for modeling, but they are difficult to work with because of the
computer memory requirements and slow processing speeds. A properly
formatted, small \((x, y)\) gridded layout data set
(\texttt{layout\_xy}) is available in \texttt{dwp}. It is a standard
data frame with \texttt{x} and \texttt{y} coordinates on 1- or 2-meter
grids overlaying each turbine (\texttt{unitCol}) at the site. The
coordinates may either be relative to the turbine location (that is,
each turbine is assumed to be located at (0, 0) according to the grid
coordinates listed for the turbine), or the coordinates are UTMs with
the turbine locations given in a separate file (\texttt{file\_turbine}).
The number of carcasses found in each grid cell must also be provided
(\texttt{ncarcCol}).

The importing and formatting is accomplished via \texttt{initLayout},
and there is no need for further processing before proceeding to the
analysis, as discussed in section \ref{sec:exxy}.

\hypertarget{examples}{%
\section{\texorpdfstring{Examples
\label{sec:examples}}{Examples }}\label{examples}}

The remainder of the body of this guide (prior to the appendices) is
devoted to examples of increasingly complicated and detailed data
analyses based on a variety of data formats seen in fatality monitoring
programs.

\hypertarget{a-terse-analysis}{%
\subsection{A Terse Analysis}\label{a-terse-analysis}}

The first example is a highly streamlined introduction to the main
\texttt{dwp} functions and main tasks of a basic analysis. It serves
strictly as an outline. More detailed and annotated analyses that
address some of the difficulties that arise in a more thorough analysis
are presented in examples \ref{sec:exeagle}-\ref{sec:exshape}.

\hypertarget{import-and-process-the-data}{%
\subsubsection{Import and Process the
Data}\label{import-and-process-the-data}}

In this example, the site layout and carcass location data are imported
from shape files. Detection probability varies with search class on the
ground, as specified in the \texttt{"Class"} column in the shape file.
It is not necessary to know or estimate the detection probabilities in
each search class. Rather, \texttt{Class} can be specified as a
covariate in the model, and the relative detection probabilities will be
implicitly accounted for in the model.

The first steps are to import the data from shape files
(\texttt{initLayout}), format the data into rings (\texttt{prepRing}),
and tally the carcasses discovered in each ring (\texttt{addCarcass}).

\begin{Shaded}
\begin{Highlighting}[]
\CommentTok{\# import site map}
\NormalTok{layout\_shape }\OtherTok{\textless{}{-}} \FunctionTok{initLayout}\NormalTok{(}\AttributeTok{data\_layout =} \StringTok{"searchpoly.shp"}\NormalTok{,}
  \AttributeTok{file\_turbine =} \StringTok{"turbine\_pt.shp"}\NormalTok{, }\AttributeTok{unitCol =} \StringTok{"Turbine"}\NormalTok{)}

\CommentTok{\# format the site map for analysis}
\NormalTok{rings\_shape }\OtherTok{\textless{}{-}} \FunctionTok{prepRing}\NormalTok{(layout\_shape, }\AttributeTok{scVar =} \StringTok{"Class"}\NormalTok{, }\AttributeTok{notSearched =} \StringTok{"Out"}\NormalTok{)}
  
\CommentTok{\# import carcass locations}
\NormalTok{cod }\OtherTok{\textless{}{-}} \FunctionTok{readCarcass}\NormalTok{(}\StringTok{"carcasses.shp"}\NormalTok{, }\AttributeTok{unitCol =} \StringTok{"Turbine"}\NormalTok{)}

\CommentTok{\# add carcasses to the site layout}
\NormalTok{rings\_shape }\OtherTok{\textless{}{-}} \FunctionTok{addCarcass}\NormalTok{(cod, }\AttributeTok{data\_ring =}\NormalTok{ rings\_shape, }\AttributeTok{plotLayout =}\NormalTok{ layout\_shape)}
\end{Highlighting}
\end{Shaded}

\hypertarget{fit-models-and-select-one}{%
\subsubsection{Fit Models and Select
One}\label{fit-models-and-select-one}}

Once the site layout and carcass location data have been imported and
formatted, use \texttt{ddFit} to fit a wide array of possible carcass
distribution models to the data and use \texttt{modelFilter} as an aid
in selecting an appropriate model for estimating \texttt{dwp}. Detailed
help on any of the functions can be found by entering, for example,
\texttt{?ddFit} or \texttt{?modelFilter} in R.

\begin{Shaded}
\begin{Highlighting}[]
\NormalTok{dmod\_shape }\OtherTok{\textless{}{-}} \FunctionTok{ddFit}\NormalTok{(rings\_shape, }\AttributeTok{scVar =} \StringTok{"Class"}\NormalTok{) }\CommentTok{\# scVar is an optional covariate}
\NormalTok{models }\OtherTok{\textless{}{-}} \FunctionTok{modelFilter}\NormalTok{(dmod\_shape, }\AttributeTok{quiet =} \ConstantTok{TRUE}\NormalTok{)}
\NormalTok{best\_mod }\OtherTok{\textless{}{-}}\NormalTok{ models}\SpecialCharTok{$}\NormalTok{filtered}
\end{Highlighting}
\end{Shaded}

\hypertarget{estimate-probability-carcass-lies-in-searched-area}{%
\subsubsection{Estimate Probability Carcass Lies in Searched
Area}\label{estimate-probability-carcass-lies-in-searched-area}}

Now that a model has been selected, use \texttt{estpsi} to estimate
\(\psi\), the probability that a carcass lies in the searched area. The
Greek letter \(\psi\) is ``psi'' in English and short for
``\textbf{p}robability that \textbf{s}earch area \textbf{i}ncludes
carcass''. \texttt{estpsi} estimates \(\psi\) at every turbine and
accounts for the uncertainty in estimating the model parameters.

\begin{Shaded}
\begin{Highlighting}[]
\NormalTok{psihat }\OtherTok{\textless{}{-}} \FunctionTok{estpsi}\NormalTok{(rings\_shape, }\AttributeTok{model =}\NormalTok{ best\_mod)}
\end{Highlighting}
\end{Shaded}

\hypertarget{estimate-dwp-and-export-to-file-for-reading-into-genest}{%
\subsubsection{\texorpdfstring{Estimate \texttt{dwp} and Export to File
for Reading into
GenEst}{Estimate dwp and Export to File for Reading into GenEst}}\label{estimate-dwp-and-export-to-file-for-reading-into-genest}}

The ``density-weighted proportion'' (\textit{dwp}) is the fraction of
carcasses falling in the searched area. The expected \textit{dwp} is
\(\psi\), but the uncertainty in estimating \texttt{dwp} is greater than
the uncertainty in \(\hat{\psi}\) due to variation in the actual
fraction of carcasses falling in the searched area given the probability
of a carcass falling in the searched area. The problem is similar to
flipping a coin 10 times. We know the probability of heads is 0.5, but
the actual number of heads in 10 flips varies. \texttt{estdwp} accounts
for that uncertainty.

After \textit{dwp} is estimated, the result can be formatted\footnote{For
  \texttt{dwp}, GenEst requires a .csv file with a column for
  turbine name and a column for \textit{dwp} estimates for each turbine. The
  \textit{dwp} estimates for each turbine may be either point estimates
  (single number), or simulated values that account for the
  uncertainty in the estimates. In the latter case, the turbine and dwp
  columns will be of length \texttt{nturb} $\times$ \texttt{nsim}, where
 \texttt{nturb} is the number of turbines and \texttt{nsim} is
  the number of simulation reps.} 
and used as a component in GenEst for estimating mortality.

\begin{Shaded}
\begin{Highlighting}[]
\NormalTok{dwphat }\OtherTok{\textless{}{-}} \FunctionTok{estdwp}\NormalTok{(psihat, }\AttributeTok{ncarc =} \FunctionTok{getncarc}\NormalTok{(rings\_shape))}
\FunctionTok{exportGenEst}\NormalTok{(dwphat, }\AttributeTok{file =} \StringTok{"dwp.csv"}\NormalTok{)}
\end{Highlighting}
\end{Shaded}

A series of more detailed examples that demonstrate how to run analyses
on different data types and to highlight some of the difficulties that
may be encountered in a \(dwp\) analysis are given in sections
\ref{sec:exeagle}-\ref{sec:exshape}.

\hypertarget{vector-of-distances-for-eagle-data}{%
\subsection{\texorpdfstring{Vector of Distances for Eagle Data
\label{sec:exeagle}}{Vector of Distances for Eagle Data }}\label{vector-of-distances-for-eagle-data}}

The simplest scenario is a vector of carcass distances coupled with a
search radius. This framework assumes that all ground within the search
radius is searched at every turbine, the detection probability is the
same at each turbine, and nothing else is searched. Bundled with the
package is a data set compiled from eagle carcass searches over several
years at a large site. The searches were conducted with dogs out to a
radius of approximately 100 meters. We defined the search radius to be
100 meters and assumed the plots were circular.

\hypertarget{importing-and-formatting-distance-vector-data}{%
\subsubsection{Importing and Formatting Distance Vector
Data}\label{importing-and-formatting-distance-vector-data}}

The data set is formatted as a data frame, \texttt{layout\_eagle}, which
consists of three columns: \texttt{DateFound}, \texttt{turbine}, and
\texttt{r} (which gives the distance from the carcass to the nearest
turbine). The data frame structure is optional. All that is required is
a vector of distances and a known search radius.

\begin{Shaded}
\begin{Highlighting}[]
\CommentTok{\# the first few lines of the eagle data set}
\FunctionTok{head}\NormalTok{(layout\_eagle) }
\CommentTok{\#\textgreater{}   DateFound turbine  r}
\CommentTok{\#\textgreater{} 1 03{-}Aug{-}05     t21 23}
\CommentTok{\#\textgreater{} 2 10{-}Oct{-}05     t43 46}
\CommentTok{\#\textgreater{} 3 31{-}Oct{-}05     t14 25}
\CommentTok{\#\textgreater{} 4 09{-}Apr{-}06     t25 30}
\CommentTok{\#\textgreater{} 5 28{-}Apr{-}06     t16 65}
\CommentTok{\#\textgreater{} 6 03{-}May{-}06     t21 42}
\end{Highlighting}
\end{Shaded}

\begin{itemize}
  \item Required
    \begin{enumerate}
      \item vector of distances
      \item search radius
    \end{enumerate}
  \item Assumptions
    \begin{enumerate}
      \item circular search plots
      \item no unsearched area within search radius
      \item same search radius at all turbines
      \item isotropic carcass distribution (same in all directions from the turbine)
     \end{enumerate} 
\end{itemize}

After the raw data have been defined in a vector or a column in a data
frame or array, they need to be formatted as a \texttt{rings} object
(section \ref{sec:rings}) .

\begin{Shaded}
\begin{Highlighting}[]
\CommentTok{\# format by feeding the bare vector of distances to prepRing (with search radius)}
\NormalTok{rings\_eagle }\OtherTok{\textless{}{-}} \FunctionTok{prepRing}\NormalTok{(layout\_eagle}\SpecialCharTok{$}\NormalTok{r, }\AttributeTok{srad =} \DecValTok{100}\NormalTok{)}
\end{Highlighting}
\end{Shaded}

\hypertarget{fitting-carcass-distance-models}{%
\subsubsection{Fitting Carcass Distance
Models}\label{fitting-carcass-distance-models}}

After the data have been formatted as a \texttt{rings} object, use
\texttt{ddFit} to fit parametric distributions to the data.

\begin{Shaded}
\begin{Highlighting}[]
\NormalTok{eaglemod }\OtherTok{\textless{}{-}} \FunctionTok{ddFit}\NormalTok{(rings\_eagle)}
\CommentTok{\#\textgreater{} Extensible models:}
\CommentTok{\#\textgreater{}   xep1 }
\CommentTok{\#\textgreater{}   xep01 }
\CommentTok{\#\textgreater{}   xep2 }
\CommentTok{\#\textgreater{}   xep02 }
\CommentTok{\#\textgreater{}   xep12 }
\CommentTok{\#\textgreater{}   xep0123 }
\CommentTok{\#\textgreater{}   tnormal }
\CommentTok{\#\textgreater{}   MaxwellBoltzmann }
\CommentTok{\#\textgreater{}   lognormal }
\CommentTok{\#\textgreater{} }
\CommentTok{\#\textgreater{} Non{-}extensible models:}
\CommentTok{\#\textgreater{}   xep012 }
\CommentTok{\#\textgreater{}   xep123 }
\CommentTok{\#\textgreater{}   constant}
\end{Highlighting}
\end{Shaded}

\begin{figure}
\centering
\includegraphics{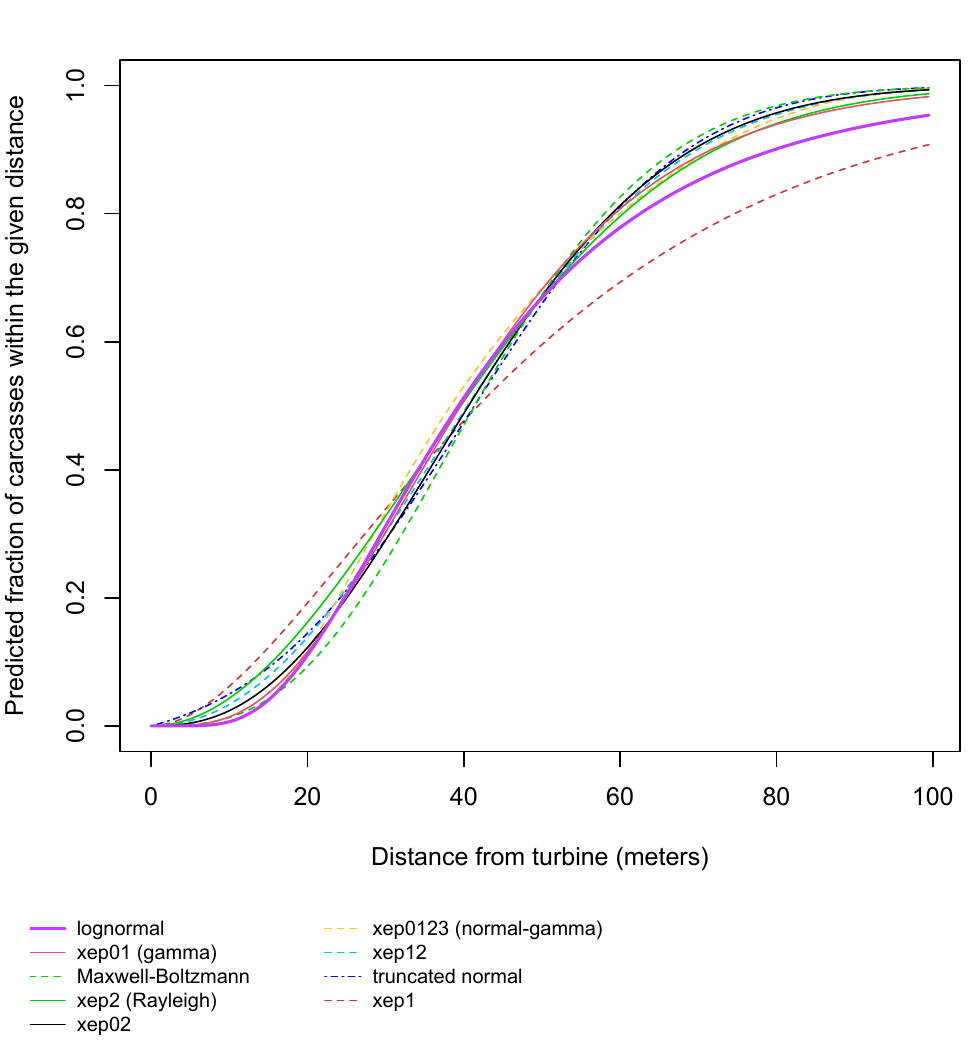}
\caption{\label{fig:eagleCDF}Cumulative distribution functions (CDFs)
for eagle carcass distributions. Figure drawn using the
\(\texttt{plot}\) function from the \(\texttt{dwp}\) package with the
fitted eagle models: \(\texttt{plot(eaglemod)}\).}
\end{figure}

Of the 12 standard models that are fit by default (Appendix
\ref{app:modbrief}), all but 3 are extensible for this data set. With
the exception of xep1 and the lognormal, all the extensible models are
in close agreement with each other and fall within a tight band
(fig.~\ref{fig:eagleCDF}). Since the search coverage within 100 meters
of a turbine is 100\% and there is no unsearched area within the search
radius, the estimated \(dwp\) is equivalent to the fraction of carcasses
falling within 100 meters, which is the \(y\)-value of the CDF at
\(x = 100\) (fig.~\ref{fig:eagleCDF}). Thus, it can be seen that all the
models except for xep1 predict that over 95\% of the carcasses lie
within 100 m, and thus there is little distinction among the models.

\hypertarget{model-selection}{%
\subsubsection{Model Selection}\label{model-selection}}

There appears to be little distinction among the fitted models for the
eagle data beyond xep1 and lognormal predicting somewhat lower \(\psi\)
than the other distributions (fig.~\ref{fig:eagleCDF}). The question is
whether it is the xep1, lognormal, or one of the other 7 models that is
giving the most accurate prediction of \(\psi\) and \(dwp\).

A primary aid in model selection is the function \texttt{modelFilter},
which gives a table of filtering test results, which are discussed in
section \ref{sec:selectiontools} and illustrated in examples in sections
\ref{sec:exeagle}-\ref{sec:exshape}. The filter gives a 0 (fail) or 1
(pass) for each test for each model, along with \(\Delta\textrm{AICc}\).

\begin{Shaded}
\begin{Highlighting}[]
\FunctionTok{modelFilter}\NormalTok{(eaglemod)}
\end{Highlighting}
\end{Shaded}

\begin{verbatim}
#>                  extensible rtail ltail aicc hin deltaAICc
#> lognormal                 1     1     1    1   1  0.000000
#> xep01                     1     1     1    1   1  1.118513
#> MaxwellBoltzmann          1     1     1    1   1  2.286072
#> xep2                      1     1     1    1   1  2.638046
#> xep02                     1     1     1    1   1  3.208478
#> xep0123                   1     1     1    1   1  4.109567
#> xep12                     1     1     1    1   1  4.183636
#> tnormal                   1     1     1    1   1  8.123697
#> xep1                      1     1     1    1   1  8.681678
#> xep012                    0     0     1    1   1  2.863524
#> xep123                    0     0     1    1   0  4.859597
#> constant                  0     0     1    0   1 66.688499
\end{verbatim}

All the extensible models passed all the filtering tests (value = 1).
The lognormal had the lowest AICc, but all the extensible models had
\(\Delta\textrm{AICc} < 10\). When there is little to no unsearched area
within the search radius as with the eagle data, AICc is of little to no
value for distinguishing among models with \(\Delta\textrm{AICc} < 10\)
because the \(dwp\) prediction does not involve any interpolation but is
strictly a matter of extrapolation to the area beyond the search radius.
In general, the heavy-tailed lognormal and xep1 distributions tend to be
less well-suited to eagle carcass distributions and should be used with
caution (Appendix \ref{app:fit2deq}). The other fitted distributions are
remarkably similar for the eagle data, with all predicting at least
98.3\% of the carcasses within the search radius, which is given as
\texttt{p\_win} in the summary \texttt{stats} table for the fitted
models:

\begin{Shaded}
\begin{Highlighting}[]
\FunctionTok{stats}\NormalTok{(eaglemod)}
\CommentTok{\#\textgreater{}                  median  75\%  90\%   95\% mode p\_win deltaAICc}
\CommentTok{\#\textgreater{} lognormal          39.3 57.0 79.7  97.4 29.0 0.954      0.00}
\CommentTok{\#\textgreater{} xep01              39.6 54.9 71.7  83.1 32.5 0.983      1.12}
\CommentTok{\#\textgreater{} MaxwellBoltzmann   41.4 54.5 67.3  75.2 38.1 0.997      2.29}
\CommentTok{\#\textgreater{} xep2               39.6 56.0 72.2  82.4 33.7 0.987      2.64}
\CommentTok{\#\textgreater{} xep02              40.6 55.1 69.3  78.1 36.3 0.994      3.21}
\CommentTok{\#\textgreater{} xep0123            38.3 55.4 71.6  80.3 26.7 0.995      4.11}
\CommentTok{\#\textgreater{} xep12              40.5 55.5 69.9  78.8 36.5 0.993      4.19}
\CommentTok{\#\textgreater{} tnormal            41.4 55.6 68.6  76.3 40.5 0.997      8.13}
\CommentTok{\#\textgreater{} xep1               41.8 67.1 97.0 118.3 24.9 0.908      8.68}
\end{Highlighting}
\end{Shaded}

Thus, regardless of which model we choose (excluding xep1 and
lognormal), \(\widehat{dwp}\) will be practically the same, so we will
select xep01 (gamma) because, among the non-heavy-tailed distributions,
it is the top model in terms of AICc. The fitted parameters of this
model are found in the \(\texttt{parms}\) element of the corresponding
model object. Obtaining confidence intervals for parameters is
demonstrated in section \ref{sec:shapefit}.

\begin{Shaded}
\begin{Highlighting}[]
\NormalTok{eaglemod}\SpecialCharTok{$}\NormalTok{xep01}\SpecialCharTok{$}\NormalTok{parms}
\CommentTok{\#\textgreater{}      shape       rate }
\CommentTok{\#\textgreater{} 4.06980861 0.09448663}
\end{Highlighting}
\end{Shaded}

\hypertarget{psi-and-dwp}{%
\subsubsection{\texorpdfstring{\(\psi\) and
\(dwp\)}{\textbackslash psi and dwp}}\label{psi-and-dwp}}

The probability that a carcass will lie in the searched area can be
predicted using \texttt{estpsi}, which returns a vector of simulated
\(\hat{\psi}\) values that account for the uncertainty in estimation of
\(\psi\), and a confidence interval can be calculated using the R base
function, \texttt{quantile}.

\begin{Shaded}
\begin{Highlighting}[]
\NormalTok{psihat }\OtherTok{\textless{}{-}} \FunctionTok{estpsi}\NormalTok{(rings\_eagle, }\AttributeTok{model =}\NormalTok{ eaglemod[[}\StringTok{"xep01"}\NormalTok{]])}
\CommentTok{\# 90\% confidence interval:}
\FunctionTok{quantile}\NormalTok{(psihat, }\AttributeTok{prob =} \FunctionTok{c}\NormalTok{(}\FloatTok{0.05}\NormalTok{, }\FloatTok{0.95}\NormalTok{))}
\CommentTok{\#\textgreater{}       5\%      95\% }
\CommentTok{\#\textgreater{} 0.943892 0.994859}
\end{Highlighting}
\end{Shaded}

Then, \(\widehat{dwp}\) is calculated from the estimated \texttt{psihat}
and the number of carcasses using \texttt{estdwp}. A confidence interval
can once again be calculated using \texttt{quantile}.

\begin{Shaded}
\begin{Highlighting}[]
\NormalTok{dwphat }\OtherTok{\textless{}{-}} \FunctionTok{estdwp}\NormalTok{(psihat, }\AttributeTok{ncarc =} \FunctionTok{getncarc}\NormalTok{(rings\_eagle))}
\CommentTok{\# 90\% confidence interval:}
\FunctionTok{quantile}\NormalTok{(dwphat, }\AttributeTok{prob =} \FunctionTok{c}\NormalTok{(}\FloatTok{0.05}\NormalTok{, }\FloatTok{0.95}\NormalTok{))}
\CommentTok{\#\textgreater{}    5\%   95\% }
\CommentTok{\#\textgreater{} 0.923 1.000}
\end{Highlighting}
\end{Shaded}

As expected (Appendix \ref{app:psivdwp}), the confidence interval for
\(dwp\) is wider than the CI for \(\psi\). The \(\widehat{dwp}\) can now
be formatted for \texttt{GenEst} and exported.

\begin{Shaded}
\begin{Highlighting}[]
\NormalTok{eagle\_frm }\OtherTok{\textless{}{-}} \FunctionTok{formatGenEst}\NormalTok{(dwphat)}
\FunctionTok{exportGenEst}\NormalTok{(eagle\_frm, }\AttributeTok{file =} \StringTok{"dwp\_eagles.csv"}\NormalTok{)}
\end{Highlighting}
\end{Shaded}

\hypertarget{simple-geometry}{%
\subsection{\texorpdfstring{Simple Geometry
\label{sec:exsimple}}{Simple Geometry }}\label{simple-geometry}}

If the search areas are fairly regular and are amenable to simple
description as square, circular, or road and pad with a few defining
parameters, the data can be imported and processed for analysis using
\texttt{initLayout}, \texttt{prepRing}, and \texttt{addCarcass}. Unlike
the \emph{vector of distances} format, the \emph{simple geometry} format
can accommodate differences in search areas among turbines. However, it
does not allow complex shapes or multiple search classes. A raw
\emph{simple geometry} data set is simply a standard data frame with 3-6
columns, including:

\begin{itemize}
  \item \textbf{turbine:} turbine IDs formatted as syntactically valid R names, which include only letters, numbers,  underscores ( \_ ), and/or periods ( . ) and do not begin with a number.
  \item \textbf{radius:} the search radius. If the search plot is an $m \times m$ square, the radius is its half-width, $m/2$. If the search plot is roads and pad, the radius is the maximum search distance from the turbine.
  \item \textbf{shape:} general descriptor of the shape of the search plot and must be one of \texttt{"square"},
    \texttt{"circular"}, or \texttt{"RP"} (for road and pad searches).
   \item \textbf{padrad:} radius of the turbine pad, which is assumed to be circular; required when \texttt{shape = "RP"},   otherwise optional (and ignored).
   \item \textbf{roadwidth:} width of the access road(s); required when \texttt{shape = "RP"}, otherwise optional (and ignored).
   \item \textbf{n\_road:} number of access roads; required when \texttt{shape = "RP"}, otherwise optional (and ignored).
\end{itemize}

\hypertarget{importing-and-formatting-site-layouts-with-simple-geometry}{%
\subsubsection{Importing and Formatting Site Layouts with Simple
Geometry}\label{importing-and-formatting-site-layouts-with-simple-geometry}}

\begin{figure}
\centering
\includegraphics{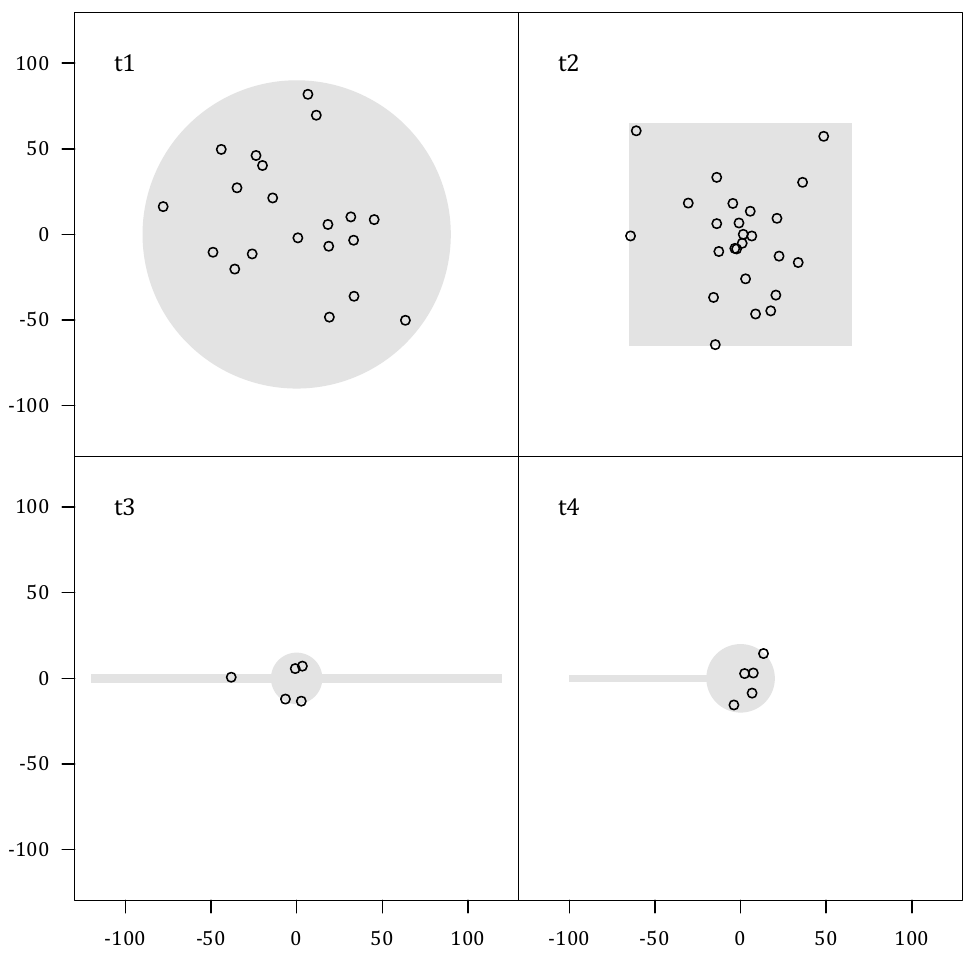}
\caption{\label{fig:simpleLayout}Site layout with simple geometry. Axes
indicate coordinate grids in meters relative to turbine at (0, 0).
Shading represents searched areas, and small circles represent locations
of carcass discoveries.}
\end{figure}

Simple layouts can be read from .csv files using \texttt{initLayout} or
can be entered by hand and then preformatted using \texttt{initLayout}.
There is an example of a simple geometry site description
(\texttt{layout\_simple}) bundled into the package. The data set has
search plots with different shapes at each of four turbines
(fig.~\ref{fig:simpleLayout}).

\begin{Shaded}
\begin{Highlighting}[]
\FunctionTok{data}\NormalTok{(layout\_simple)}
\NormalTok{layout\_simple}
\CommentTok{\#\textgreater{}    turbine radius    shape padrad roadwidth n\_road}
\CommentTok{\#\textgreater{} t1      t1     90 circular     NA        NA     NA}
\CommentTok{\#\textgreater{} t2      t2     65   square     NA        NA     NA}
\CommentTok{\#\textgreater{} t3      t3    120       RP     15         5      2}
\CommentTok{\#\textgreater{} t4      t4    100       RP     20         4      1}
\end{Highlighting}
\end{Shaded}

The \texttt{layout\_simple} raw data set is formatted as a data frame
and needs to be preformatted using \texttt{initLayout} to convert it
into a \texttt{simpleLayout} object which can be recognized by other
\texttt{dwp} functions.

\begin{Shaded}
\begin{Highlighting}[]
\CommentTok{\# initial format:}
\FunctionTok{class}\NormalTok{(layout\_simple)}
\CommentTok{\#\textgreater{} [1] "data.frame"}

\CommentTok{\# convert to simpleLayout:}
\NormalTok{layout\_simple }\OtherTok{\textless{}{-}} \FunctionTok{initLayout}\NormalTok{(layout\_simple)}
\FunctionTok{class}\NormalTok{(layout\_simple)}
\CommentTok{\#\textgreater{} [1] "simpleLayout" "data.frame"}
\end{Highlighting}
\end{Shaded}

Once the site layout data have been preformatted using
\texttt{initLayout}, use the \texttt{prepRing} function to process the
layout data into 1 m concentric rings with the area (\(m^2\)) searched
in each ring at each turbine.

\begin{Shaded}
\begin{Highlighting}[]
\NormalTok{rings\_simple }\OtherTok{\textless{}{-}} \FunctionTok{prepRing}\NormalTok{(layout\_simple)}
\end{Highlighting}
\end{Shaded}

After formatting the site layout data into rings, use
\texttt{addCarcass} to enter the number of carcasses discovered in each
ring at each turbine. The carcass data for simple geometry layouts
should be organized in a data frame with columns for the turbine ID and
the distances from the turbine at which the carcasses were found.

\begin{Shaded}
\begin{Highlighting}[]
\CommentTok{\# carcass data}
\FunctionTok{head}\NormalTok{(carcass\_simple)}
\CommentTok{\#\textgreater{}   turbine     r}
\CommentTok{\#\textgreater{} 1      t2 47.99}
\CommentTok{\#\textgreater{} 2      t2 75.07}
\CommentTok{\#\textgreater{} 3      t1 19.97}
\CommentTok{\#\textgreater{} 4      t1 51.84}
\CommentTok{\#\textgreater{} 5      t4  8.06}
\CommentTok{\#\textgreater{} 6      t2  8.83}

\CommentTok{\# adding carcasses to the formatted (but bare) site layout data}
\NormalTok{rings\_simple }\OtherTok{\textless{}{-}} \FunctionTok{addCarcass}\NormalTok{(carcass\_simple, }\AttributeTok{data\_ring =}\NormalTok{ rings\_simple)}
\end{Highlighting}
\end{Shaded}

After adding carcasses to the site layout formatted as a \texttt{rings}
object using \texttt{prepRing} and \texttt{add\_carcass}, the carcass
distribution models can be fit using \texttt{ddFit}.

\begin{Shaded}
\begin{Highlighting}[]
\NormalTok{dmod\_simple }\OtherTok{\textless{}{-}} \FunctionTok{ddFit}\NormalTok{(rings\_simple)}
\CommentTok{\#\textgreater{} Extensible models:}
\CommentTok{\#\textgreater{}   xep1 }
\CommentTok{\#\textgreater{}   xep01 }
\CommentTok{\#\textgreater{}   xep2 }
\CommentTok{\#\textgreater{}   xep02 }
\CommentTok{\#\textgreater{}   xep123 }
\CommentTok{\#\textgreater{}   tnormal }
\CommentTok{\#\textgreater{}   MaxwellBoltzmann }
\CommentTok{\#\textgreater{}   lognormal }
\CommentTok{\#\textgreater{} }
\CommentTok{\#\textgreater{} Non{-}extensible models:}
\CommentTok{\#\textgreater{}   xep12 }
\CommentTok{\#\textgreater{}   xep012 }
\CommentTok{\#\textgreater{}   xep0123 }
\CommentTok{\#\textgreater{}   constant}
\end{Highlighting}
\end{Shaded}

\begin{figure}
\centering
\includegraphics{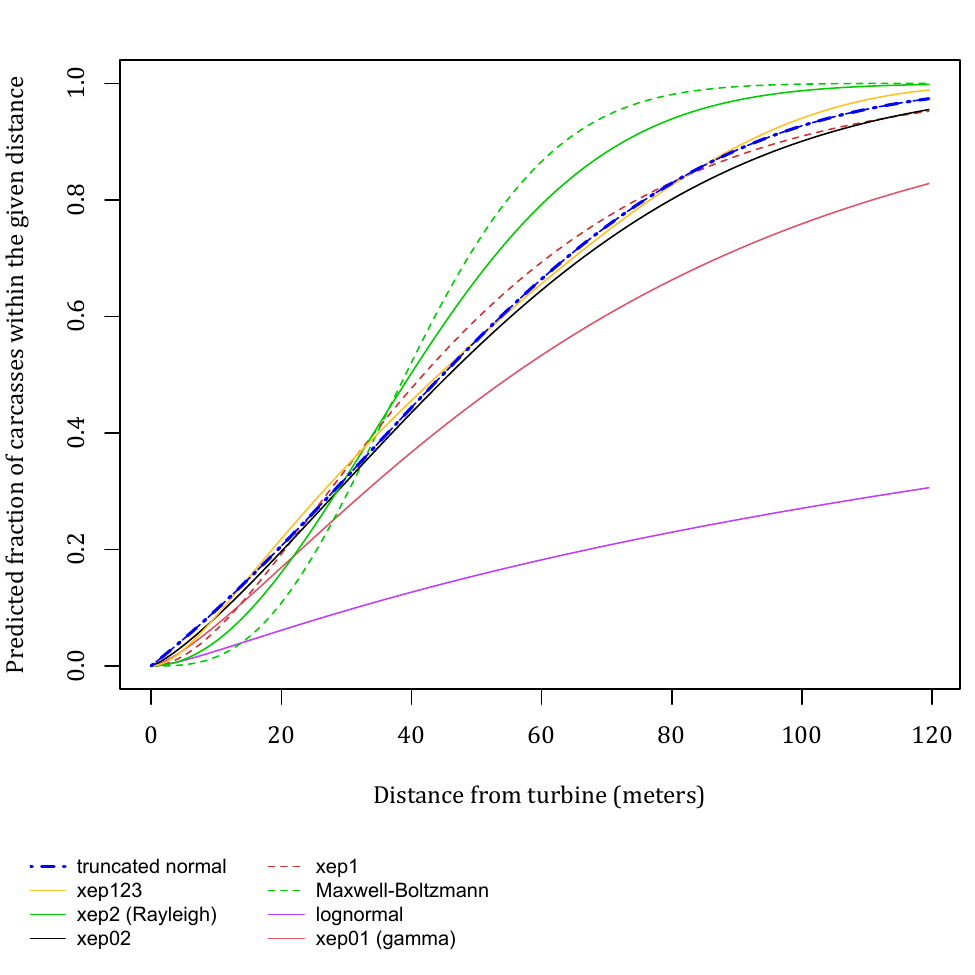}
\caption{\label{fig:simpleCDF}Fitted cumulative distributions of
carcasses at the Simple Geometry site. Only the extensible models are
shown. Figure drawn using the \(\texttt{plot}\) function from the
\(\texttt{dwp}\) package with the fitted models:
\(\texttt{plot(dmod\_simple)}\).}
\end{figure}

The lognormal distribution predicts only 30.6\% of the carcasses to lie
within the maximum search radius among the turbines (120 m), and the
gamma predicts 82.8\%. All the other extensible models predict in excess
of 95\% (fig.~\ref{fig:simpleCDF}) within 120 m. If there were very
little unsearched area within 120 m, then the 6 extensible models other
than lognormal and gamma would give the same 95+\% for
\(\widehat{dwp}\), and there would be no practical difference among
them. However, there is a great deal of unsearched area within the
maximum search radius at the site. The models are quite different from
one another between about 40 and 100 m from the turbine, and the choice
of model is likely to have a significant impact on the resulting \(dwp\)
prediction.

\newpage

\hypertarget{model-selection-for-the-simple-geometry-site}{%
\subsubsection{Model Selection for the Simple Geometry
Site}\label{model-selection-for-the-simple-geometry-site}}

The first step in model selection is to use the \texttt{modelFilter}
function.

\begin{Shaded}
\begin{Highlighting}[]
\FunctionTok{modelFilter}\NormalTok{(dmod\_simple)}\SpecialCharTok{$}\NormalTok{scores}
\end{Highlighting}
\end{Shaded}

\begin{table}[!htbp]
  \caption{Model filter for Simple Geometry site, where 1 = filter test passed, 0 = filter test failed. Tests are described in section “Tools for Model Selection” Extensible = extensibility test, rtail and ltail = heavy r(igth) and l(eft) tail tests, aicc = $\Delta \textrm{AICc} < 10$, hin = high influence test, $\Delta \textrm{AICc}$ = difference between AICc value for a given model and the lowest AICc value among the models tested}
   \label{tbl:simple_scores}
   \begin{tabular}{rcccccc}
 & extensible & rtail & ltail & aicc & hin & deltaAICc \\ 
  \hline
tnormal & 1 & 1 & 1 & 1 & 1 & 1.18 \\ 
  xep123 & 1 & 1 & 1 & 1 & 1 & 2.78 \\ 
  xep2 & 1 & 1 & 1 & 1 & 1 & 8.11 \\ 
  xep02 & 1 & 1 & 1 & 1 & 0 & 0.37 \\ 
  xep1 & 1 & 1 & 1 & 1 & 0 & 0.75 \\ 
  MaxwellBoltzmann & 1 & 1 & 1 & 0 & 1 & 38.47 \\ 
  lognormal & 1 & 0 & 1 & 1 & 1 & 0.00 \\ 
  xep01 & 1 & 0 & 1 & 1 & 0 & 0.10 \\ 
  xep12 & 0 & 0 & 1 & 1 & 1 & 1.05 \\ 
  xep012 & 0 & 0 & 1 & 1 & 0 & 2.22 \\ 
  xep0123 & 0 & 0 & 1 & 1 & 0 & 4.40 \\ 
  constant & 0 & 0 & 1 & 0 & 1 & 43.52 \\ 
   \hline
   \end{tabular}
 \end{table}

The fitted lognormal and gamma (xep01) distributions evidently have
implausibly heavy right tails, as they both fail the \texttt{rtail}
model filter test (table \ref{tbl:simple_scores}). There is at least one
high-influence point for the xep02 and xep1 models, thus failing the
\texttt{hin} test (table \ref{tbl:simple_scores}). Only three
models---the truncated normal (\texttt{tnormal}), \texttt{xep123}, and
Rayleigh (\texttt{xep2})---pass all the diagnostic tests. The truncated
normal and xep123 CDFs are virtually indistinguishable, while the
Rayleigh differs markedly (fig.~\ref{fig:simpleCDF}). Because there is a
great deal of unsearched area at the site, prediction of \(\psi\) and
\(dwp\) relies strongly on interpolation, so a good fit within the range
of the data is crucial, and models with large AICc scores---like the
Rayleigh (xep2)---can be eliminated. The truncated normal has the lowest
AICc score among the models that pass all the diagnostic tests, and that
is the one we select for calculating \(\widehat{dwp}\).

\hypertarget{hatpsi-and-widehatdwp-for-the-simple-geometry-site}{%
\subsubsection{\texorpdfstring{\(\hat{\psi}\) and \(\widehat{dwp}\) for
the Simple Geometry
Site}{\textbackslash hat\{\textbackslash psi\} and \textbackslash widehat\{dwp\} for the Simple Geometry Site}}\label{hatpsi-and-widehatdwp-for-the-simple-geometry-site}}

Having selected the truncated normal distribution (\texttt{tnormal}),
for each turbine we can predict the probability that a carcass lies in
the searched area using \texttt{estpsi}. Since \texttt{estpsi}
integrates the selected model over the searched area (as with the
``volcano'' in fig.~\ref{fig:volcano}), both the searched area
(formatted as \texttt{rings\_simple} and the selected model are required
as arguments to \texttt{estpsi}.

\begin{Shaded}
\begin{Highlighting}[]
\NormalTok{psihat\_simple }\OtherTok{\textless{}{-}} \FunctionTok{estpsi}\NormalTok{(rings\_simple, }\AttributeTok{model =}\NormalTok{ dmod\_simple[}\StringTok{"tnormal"}\NormalTok{])}
\end{Highlighting}
\end{Shaded}

The prediction of \(dwp\) depends not just on the probability of a
carcass lying in the searched area (\(\psi\)) but also on the number of
carcasses discovered in the searched area, so the \texttt{estdwp}
function requires both \(\hat{\psi}\) and the carcass counts.

\begin{Shaded}
\begin{Highlighting}[]
\NormalTok{dwphat\_simple }\OtherTok{\textless{}{-}} \FunctionTok{estdwp}\NormalTok{(psihat\_simple, }\AttributeTok{ncarc =} \FunctionTok{getncarc}\NormalTok{(rings\_simple))}
\end{Highlighting}
\end{Shaded}

The predicted \(dwp\)'s for each turbine can now be plotted and exported
to a \texttt{.csv} file for use in GenEst.

\begin{Shaded}
\begin{Highlighting}[]
\FunctionTok{plot}\NormalTok{(dwphat\_simple)}
\FunctionTok{exportGenEst}\NormalTok{(dwphat\_simple, }\AttributeTok{file =} \StringTok{"dwphat\_simple.csv"}\NormalTok{)}
\end{Highlighting}
\end{Shaded}

\begin{figure}
\centering
\includegraphics{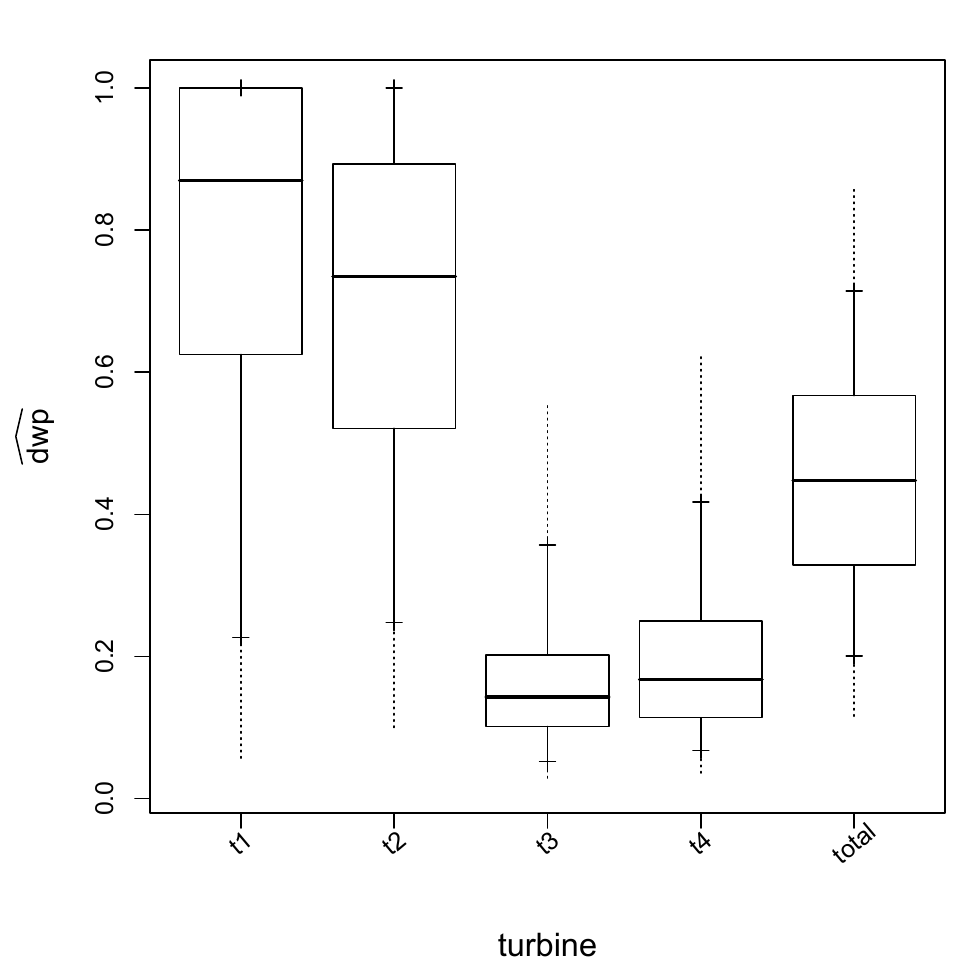}
\caption{\label{fig:simpledwp}Boxplots of predicted dwp for all turbines
and the site as a whole for the simple geometry site. Boxes represent
the interquartile range (25th and 75th percentiles) and medians of
simulated values. Solid whiskers delineate the 90\% confidence
intervals; and the dotted whiskers, the 99\% CIs. Figure drawn using the
\(\texttt{plot}\) function from the \(\texttt{dwp}\) package with the
estimated \(\textit{dwp}\) values: \(\texttt{plot(dwphat\_simple)}\).}
\end{figure}

\newpage

\hypertarget{r-polygons}{%
\subsection{\texorpdfstring{R Polygons
\label{sec:expolygon}}{R Polygons }}\label{r-polygons}}

\hypertarget{r-polygon-site-layouts-data}{%
\subsubsection{R Polygon Site Layouts:
Data}\label{r-polygon-site-layouts-data}}

If the shape of the search plots are somewhat complex, but there is only
one search class, you may import your site layout data in R polygon
format. An R polygon is simply two columns of data giving the \(x\) and
\(y\) coordinates of the vertices of a polygon. For \texttt{dwp}
analysis, the polygons must be associated with turbines in a standard
data frame with three columns:

\begin{enumerate}[leftmargin=2cm]
  \item[\texttt{\textbf{turbine:}}] a turbine ID, which must be a syntactically valid name in R and include letters, numbers, dots ( . ), and underscores ( \_ ) only. Spaces, hyphens, commas, pound signs, and other special characters are not allowed. Syntactically valid names must not begin with a number or with a dot followed by a number.
  \item[\texttt{\textbf{x:}}] $x$ coordinates of the vertices of the polygons delineating search areas at each turbine. All coordinates are assumed to be in meters with the turbine referenced in the `turbine` column assumed to be at (0, 0).
  \item[\texttt{\textbf{y:}}] $y$ coordinates of the vertices of the polygons.
\end{enumerate}

There must be one polygon for each turbine searched, and polygons must
not be self-intersecting. A polygon site layout data set with R polygons
is included as part of the \texttt{dwp} package and can readily be
loaded and viewed:

\begin{Shaded}
\begin{Highlighting}[]
\FunctionTok{data}\NormalTok{(layout\_polygon)}
\NormalTok{layout\_polygon }
\CommentTok{\#\textgreater{}    turbine          x          y}
\CommentTok{\#\textgreater{} 1       t1   4.814164  69.209213}
\CommentTok{\#\textgreater{} 2       t1 {-}12.278842  17.930197}
\CommentTok{\#\textgreater{} 3       t1 {-}64.255532  15.139502}
\CommentTok{\#\textgreater{} 4       t1 {-}54.836937 {-}54.627868}
\CommentTok{\#\textgreater{} 5       t1  22.953680 {-}27.767431}
\CommentTok{\#\textgreater{} 6       t1  70.395491  14.092991}
\CommentTok{\#\textgreater{} 7       t1  20.162985  19.674381}
\CommentTok{\#\textgreater{} 8       t2 {-}19.604416  {-}3.000014}
\CommentTok{\#\textgreater{} 9       t2  {-}6.348615 {-}17.302325}
\CommentTok{\#\textgreater{} 10      t2  14.581596  {-}2.302341}
\CommentTok{\#\textgreater{} 11      t2   6.907185  10.255786}
\CommentTok{\#\textgreater{} 12      t2   2.372306  25.953444}
\CommentTok{\#\textgreater{} 13      t2  {-}1.116063  52.465045}
\CommentTok{\#\textgreater{} 14      t2  {-}5.302105  58.744108}
\CommentTok{\#\textgreater{} 15      t2 {-}17.860231  68.511540}
\CommentTok{\#\textgreater{} 16      t2 {-}23.441621  81.767340}
\CommentTok{\#\textgreater{} 17      t2 {-}35.650911  76.185950}
\CommentTok{\#\textgreater{} 18      t2 {-}25.883479  33.627855}
\CommentTok{\#\textgreater{} 19      t2 {-}35.999748   9.906949}
\end{Highlighting}
\end{Shaded}

\begin{figure}
\centering
\includegraphics{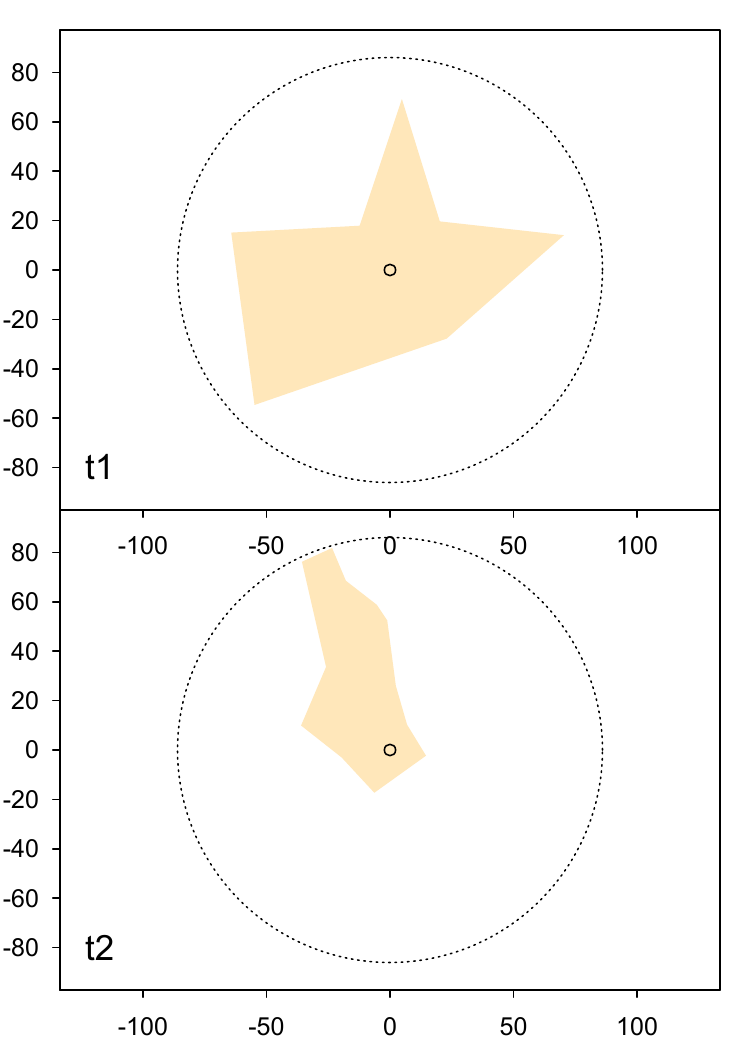}
\caption{\label{fig:polyplot} Search plots for R polygon data set.
Figure drawn using the \(\texttt{plot}\) function from the
\(\texttt{dwp}\) package with the formatted site layout data:
\(\texttt{plot(playout)}\).}
\end{figure}

To begin the analysis, the polygons must be converted to a
\texttt{polygonLayout} object which has been error-checked and converted
to a properly formatted list. This can be done using the
\texttt{initLayout} function with argument
\texttt{dataType\ =\ "polygon"}. Once the polygon layout has been
formatted, it can be plotted (Figure \ref{fig:polyplot}) and prepared
for GLM analysis by tallying the searched area in concentric 1 meter
rings around each turbine using the function \texttt{prepRing}.

\begin{Shaded}
\begin{Highlighting}[]
\CommentTok{\# initial formatting}
\NormalTok{playout }\OtherTok{\textless{}{-}} \FunctionTok{initLayout}\NormalTok{(layout\_polygon, }\AttributeTok{dataType =} \StringTok{"polygon"}\NormalTok{)}

\CommentTok{\# plot}
\FunctionTok{plot}\NormalTok{(playout, }\AttributeTok{las =} \DecValTok{2}\NormalTok{)}
\end{Highlighting}
\end{Shaded}

\begin{Shaded}
\begin{Highlighting}[]

\CommentTok{\# create a ring structure}
\NormalTok{rings\_polygon }\OtherTok{\textless{}{-}} \FunctionTok{prepRing}\NormalTok{(playout)}
\end{Highlighting}
\end{Shaded}

A simulated carcass data set for this polygon site
(\texttt{carcass\_polygon}) is included in the package. It is a simple,
two-column data frame with columns for turbine ID (\texttt{turbine}) and
distances from the turbine (\texttt{r}, in meters) at which carcasses
were discovered in the carcass surveys. The carcass data are added to
the ring structure via \texttt{addCarcass} to complete the data
preparations for analysis.

\begin{Shaded}
\begin{Highlighting}[]
\CommentTok{\# the first few lines of the carcass\_polygon data}
\FunctionTok{head}\NormalTok{(carcass\_polygon)}
\CommentTok{\#\textgreater{}   turbine    r}
\CommentTok{\#\textgreater{} 1      t1 32.5}
\CommentTok{\#\textgreater{} 2      t1 38.2}
\CommentTok{\#\textgreater{} 3      t1 45.1}
\CommentTok{\#\textgreater{} 4      t2  5.6}
\CommentTok{\#\textgreater{} 5      t1 12.4}
\CommentTok{\#\textgreater{} 6      t1 29.5}

\CommentTok{\# add the carcasses from carcass\_polygon data to tha ring structure}
\NormalTok{rings\_polygon }\OtherTok{\textless{}{-}} \FunctionTok{addCarcass}\NormalTok{(carcass\_polygon, }\AttributeTok{data\_ring =}\NormalTok{ rings\_polygon)}
\end{Highlighting}
\end{Shaded}

\hypertarget{r-polygon-site-layouts-carcass-distribution-modeling-and-dwp}{%
\subsubsection{\texorpdfstring{R Polygon Site Layouts: Carcass
Distribution Modeling and
\(dwp\)}{R Polygon Site Layouts: Carcass Distribution Modeling and dwp}}\label{r-polygon-site-layouts-carcass-distribution-modeling-and-dwp}}

After adding carcasses to the formatted \texttt{rings} data, carcass
distance models can be fit using \texttt{ddFit} and assessed using the
functions

\begin{itemize}
  \item $\texttt{stats}$, which gives descriptive statistics for carcass distances,
  \item $\texttt{modelFilter}$, which performs diagnostic tests on the fitted models and ranks them by suitability, 
  \item $\texttt{plot}$, which creates graphs for fitted PDFs or CDFs for fitted carcass distributions, and
  \item $\texttt{aic}$, which calculates AICc scores for fitted distributions.
\end{itemize}

For details on the use of these and other useful functions in the
package, enter \texttt{?dwp}.

\begin{Shaded}
\begin{Highlighting}[]
\CommentTok{\# fit the distance models}
\NormalTok{dmod\_polygon }\OtherTok{\textless{}{-}} \FunctionTok{ddFit}\NormalTok{(rings\_polygon)}
\CommentTok{\#\textgreater{} Extensible models:}
\CommentTok{\#\textgreater{}   xep1 }
\CommentTok{\#\textgreater{}   xep01 }
\CommentTok{\#\textgreater{}   xep2 }
\CommentTok{\#\textgreater{}   xep02 }
\CommentTok{\#\textgreater{}   xep012 }
\CommentTok{\#\textgreater{}   xep123 }
\CommentTok{\#\textgreater{}   tnormal }
\CommentTok{\#\textgreater{}   MaxwellBoltzmann }
\CommentTok{\#\textgreater{}   lognormal }
\CommentTok{\#\textgreater{} }
\CommentTok{\#\textgreater{} Non{-}extensible models:}
\CommentTok{\#\textgreater{}   xep12 }
\CommentTok{\#\textgreater{}   xep0123 }
\CommentTok{\#\textgreater{}   constant}
\FunctionTok{stats}\NormalTok{(dmod\_polygon)}
\CommentTok{\#\textgreater{}                   median     75\%      90\%      95\% mode p\_win deltaAICc}
\CommentTok{\#\textgreater{} xep02               38.5    59.7     81.1     94.7 26.1 0.919      0.00}
\CommentTok{\#\textgreater{} xep01               53.6    93.3    142.1    177.7 22.4 0.712      0.31}
\CommentTok{\#\textgreater{} tnormal             35.3    51.8     67.2     76.6 31.3 0.977      0.61}
\CommentTok{\#\textgreater{} xep1                36.1    58.0     83.8    102.2 21.5 0.906      1.26}
\CommentTok{\#\textgreater{} lognormal        11626.2 61620.5 276438.8 678767.5 25.7 0.023      1.75}
\CommentTok{\#\textgreater{} xep012              37.5    57.3     77.0     89.3 27.2 0.938      2.18}
\CommentTok{\#\textgreater{} xep123              36.9    56.2     72.2     80.9 17.9 0.968      4.30}
\CommentTok{\#\textgreater{} xep2                31.7    44.8     57.7     65.9 26.9 0.994      7.87}
\CommentTok{\#\textgreater{} MaxwellBoltzmann    29.7    39.1     48.2     53.9 27.3 1.000     46.10}
\end{Highlighting}
\end{Shaded}

\begin{threeparttable}[!htbp]
  \caption{Descriptive statistics\tnote{a} for fitted carcass distributions at the R polygon site}
   \label{tbl:polygon_stats}
   \begin{tabular}{rrrrrccc}
      \hline
      \hline
 & median & 75\% & 90\% & 95\% & mode & p\_win & deltaAICc \\ 
  \hline
xep02 & 38.5 & 59.7 & 81.1 & 94.7 & 26.1 & 0.919 & 0.00 \\ 
  xep01 & 53.6 & 93.3 & 142.1 & 177.7 & 22.4 & 0.712 & 0.31 \\ 
  tnormal & 35.3 & 51.8 & 67.2 & 76.6 & 31.3 & 0.977 & 0.61 \\ 
  xep1 & 36.1 & 58.0 & 83.8 & 102.2 & 21.5 & 0.906 & 1.26 \\ 
  lognormal & 11626.2 & 61620.5 & 276438.8 & 678767.5 & 25.7 & 0.023 & 1.75 \\ 
  xep012 & 37.5 & 57.3 & 77.0 & 89.3 & 27.2 & 0.938 & 2.18 \\ 
  xep123 & 36.9 & 56.2 & 72.2 & 80.9 & 17.9 & 0.968 & 4.30 \\ 
  xep2 & 31.7 & 44.8 & 57.7 & 65.9 & 26.9 & 0.994 & 7.87 \\ 
  MaxwellBoltzmann & 29.7 & 39.1 & 48.2 & 53.9 & 27.3 & 1.000 & 46.10 \\ 
   \hline
   \end{tabular}
   \begin{tablenotes}
    \item[a] Statistics are for the estimated quantiles of carcass distances. For example, "median" gives the estimated   median distance of carcasses from the turbine according to the model, "mode" gives the estimated distance that carcasses are most likely to lie from the turbine, and $\texttt{p\_win}$ gives the estimated probability that a carcass will lie within the search radius.
   \end{tablenotes}
 \end{threeparttable}

The extensible models give a broad range of predictions of the
probability of a carcass lying within the search radius, from a low of
2.3\% for the lognormal distribution to a high of 100\% for the
Maxwell-Boltzmann. The \texttt{modelFilter} provides useful guidance on
model selection (table \ref{tbl:polygon_scores}).

\begin{Shaded}
\begin{Highlighting}[]
\FunctionTok{modelFilter}\NormalTok{(dmod\_polygon)}\SpecialCharTok{$}\NormalTok{scores}
\end{Highlighting}
\end{Shaded}

\begin{table}[!htbp]
  \caption{Model filter for Simple Geometry site, where 1 = filter test passed, 0 = filter test failed. Tests are described in section “Tools for Model Selection” Extensible = extensibility test, rtail and ltail = heavy r(igth) and l(eft) tail tests, aicc = $\Delta \textrm{AICc} < 10$, hin = high influence test, $\Delta \textrm{AICc}$ = difference between AICc value for a given model and the lowest AICc value among the models tested}
   \label{tbl:polygon_scores}
   \begin{tabular}{rcccccc}
 & extensible & rtail & ltail & aicc & hin & deltaAICc \\ 
  \hline
tnormal & 1 & 1 & 1 & 1 & 1 & 0.61 \\ 
  xep1 & 1 & 1 & 1 & 1 & 1 & 1.27 \\ 
  xep123 & 1 & 1 & 1 & 1 & 1 & 4.31 \\ 
  xep2 & 1 & 1 & 1 & 1 & 1 & 7.87 \\ 
  xep02 & 1 & 1 & 1 & 1 & 0 & 0.00 \\ 
  xep012 & 1 & 1 & 1 & 1 & 0 & 2.18 \\ 
  MaxwellBoltzmann & 1 & 1 & 0 & 0 & 1 & 46.10 \\ 
  xep01 & 1 & 0 & 1 & 1 & 0 & 0.31 \\ 
  lognormal & 1 & 0 & 1 & 1 & 0 & 1.76 \\ 
  xep12 & 0 & 0 & 1 & 1 & 1 & 2.45 \\ 
  xep0123 & 0 & 0 & 1 & 1 & 0 & 4.01 \\ 
  constant & 0 & 0 & 1 & 0 & 1 & 42.47 \\ 
   \hline
   \end{tabular}
 \end{table}

The lognormal model failed the right tail test because it predicts
implausibly low probabilities of carcasses lying within 80, 120, 150,
and 200 meters, whereas the Maxwell-Boltzmann model failed the left tail
test because it predicts implausibly high probability of carcasses lying
within 50 meters.

\begin{Shaded}
\begin{Highlighting}[]
\FunctionTok{pdd}\NormalTok{(}\FunctionTok{c}\NormalTok{(}\DecValTok{80}\NormalTok{, }\DecValTok{120}\NormalTok{, }\DecValTok{150}\NormalTok{, }\DecValTok{200}\NormalTok{), }\AttributeTok{model =}\NormalTok{ dmod\_polygon[}\StringTok{"lognormal"}\NormalTok{])}
\CommentTok{\#\textgreater{} [1] 0.02202108 0.03217810 0.03924999 0.05018037}
\FunctionTok{pdd}\NormalTok{(}\FunctionTok{c}\NormalTok{(}\DecValTok{20}\NormalTok{, }\DecValTok{50}\NormalTok{), }\AttributeTok{model =}\NormalTok{ dmod\_polygon[}\StringTok{"MaxwellBoltzmann"}\NormalTok{])}
\CommentTok{\#\textgreater{} [1] 0.2167363 0.9184358}
\end{Highlighting}
\end{Shaded}

The xep02 and xep012 models fail the \texttt{hin} test (table
\ref{tbl:polygon_scores}) because they have \emph{high influence}
points, that is, points that are out of line with the others according
to the model but that have a significant impact on the shape of the
fitted curve. This is a red flag that signals an instability in the
model for the given data.

\begin{figure}
\centering
\includegraphics{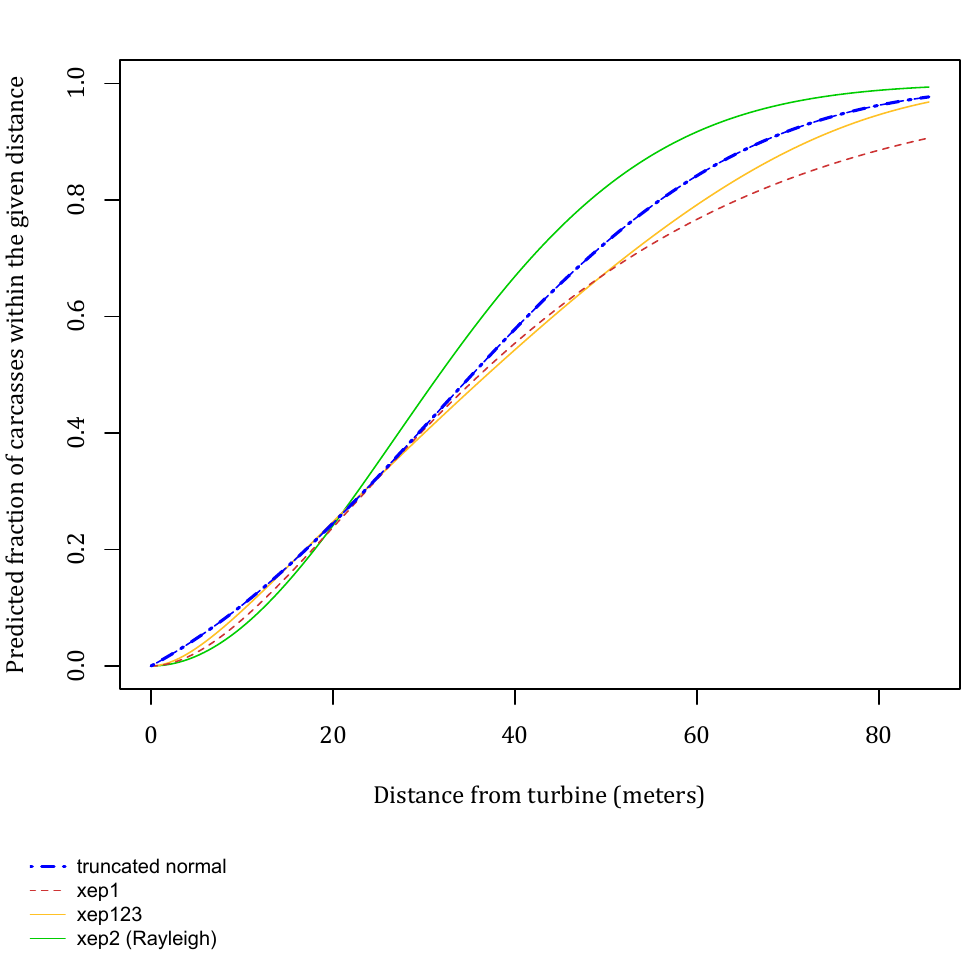}
\caption{\label{fig:polyCDF}Cumulative distribution functions for the
truncated normal, xep1, xep123, xep2 (Rayleigh) models fit to the R
polygon data set. Figure drawn using the \(\texttt{plot}\) function from
the \(\texttt{dwp}\) package with a subset of the fitted models:
\(\texttt{plot(dmod\_polygon[passind]})\), where \(\texttt{passind}\) is
a vector of names of the models to be plotted.}
\end{figure}

The remaining fitted distributions (tnormal, xep1, xep123, xep2) are
broadly comparable according to the diagnostic criteria tested. However,
the distributions differ markedly between 30 m and the maximum search
radius of 85 m (fig.~\ref{fig:polyCDF}), so other considerations come
into play. In particular, there is a tendency for the lighter-tailed
distributions to overestimate \(\psi\) when the search radius is short
or moderate and there is substantial unsearched area within the search
radius as with this example (Appendix \ref{app:fit2deq}), so the
Rayleigh would not be recommended here. Also, when there is substantial
unsearched area, it is important to have a low AICc score, so
\texttt{tnormal} and \texttt{xep1} are to be commended over
\texttt{xep123}.

\begin{figure}
\centering
\includegraphics{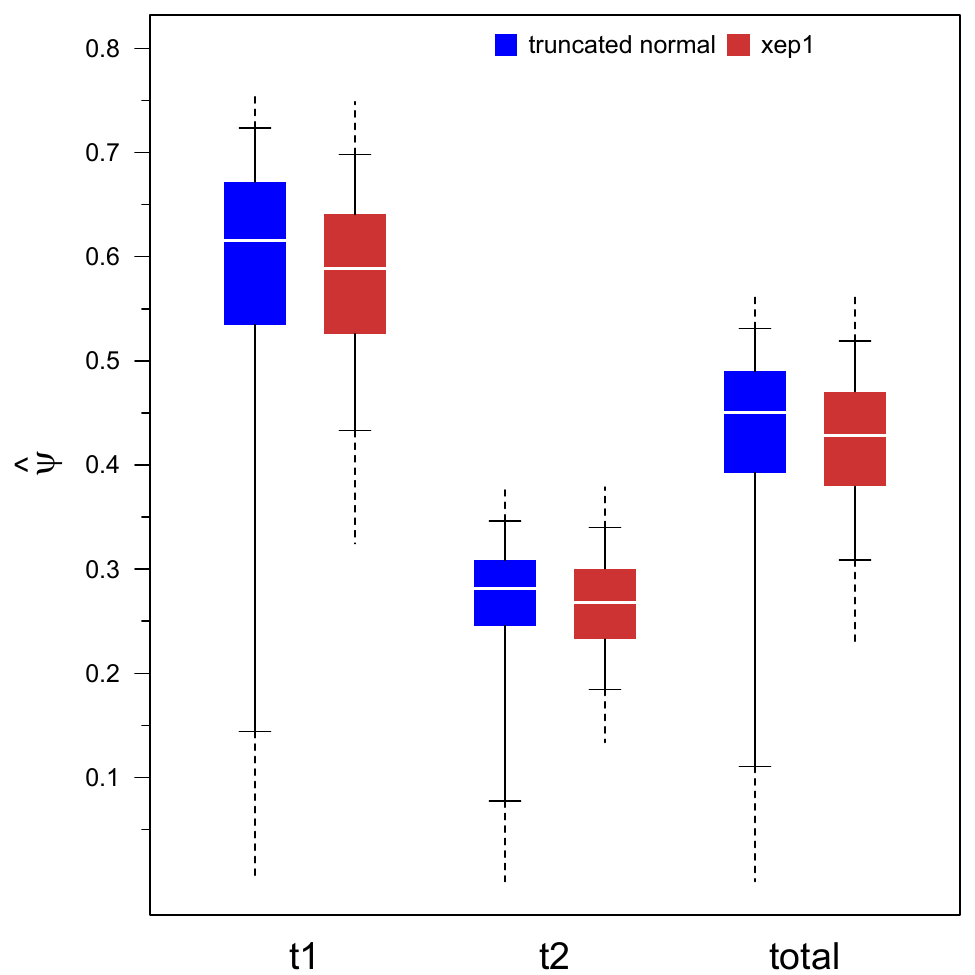}
\caption{\label{fig:psipoly}Boxplots of predicted \(\psi\) for all
turbines and the polygon site as a whole for the truncated normal and
xep1 distributions.}
\end{figure}

The two remaining models (\texttt{tnormal} and \texttt{xep1}) have AICc
scores differing by only 0.65, well below a minimal threshold to be
statistically meaningful. The xep1 distribution has a heavier tail
(fig.~\ref{fig:polyCDF}) and would be the more cautious choice from a
wildlife conservation perspective because it would predict a higher
fraction of carcasses missed in the searches and, therefore, greater
estimated mortality than under the truncated normal. The differences are
slight on average, but the truncated normal has much more frequency of
very small \(\hat{\psi}\) values than does xep1 (fig.~\ref{fig:psipoly})
and therefore a less stable fit.

After selecting the xep1 model, we can calculate \(\widehat{dwp}\) and
export for use in GenEst.

\begin{Shaded}
\begin{Highlighting}[]
\CommentTok{\# estimate probability of carcass lying in searched area}
\NormalTok{psipoly }\OtherTok{\textless{}{-}} \FunctionTok{estpsi}\NormalTok{(rings\_polygon, }\AttributeTok{model =}\NormalTok{ dmod\_polygon[}\StringTok{"xep1"}\NormalTok{])}
\NormalTok{dwp }\OtherTok{\textless{}{-}} \FunctionTok{estdwp}\NormalTok{(psipoly, }\AttributeTok{ncarc =} \FunctionTok{getncarc}\NormalTok{(rings\_polygon))}
\FunctionTok{exportGenEst}\NormalTok{(dwp, }\AttributeTok{file =} \StringTok{"dwp\_poly.csv"}\NormalTok{)}
\end{Highlighting}
\end{Shaded}

\hypertarget{the-x-y-grid-data}{%
\subsection{\texorpdfstring{The \((x, y)\) Grid Data
\label{sec:exxy}}{The (x, y) Grid Data }}\label{the-x-y-grid-data}}

The \texttt{dwp} package includes an example data set with site layout
from a fictitious site with 1 turbine and a road and pad search area.
The site layout data is formatted as an \((x, y)\) grid, which gives the
coordinates (in meters relative to the turbine at (0, 0)) of each square
meter of searched ground at the turbine. The data are stored in a data
frame, \texttt{layout\_xy}, with columns for \(x\) and \(y\)
coordinates, the number of carcasses (\texttt{ncarc}) in the square
meter cell centered at each grid node, the distance from the turbine
(\texttt{r}), and the turbine name.

\begin{Shaded}
\begin{Highlighting}[]
\CommentTok{\# the first few lines of the xy data layout that is bundled with dwp package}
\FunctionTok{head}\NormalTok{(layout\_xy)}
\CommentTok{\#\textgreater{}    x   y ncarc        r turbine}
\CommentTok{\#\textgreater{} 1  0 {-}15     0 15.00000      t1}
\CommentTok{\#\textgreater{} 2 {-}5 {-}14     0 14.86607      t1}
\CommentTok{\#\textgreater{} 3 {-}4 {-}14     0 14.56022      t1}
\CommentTok{\#\textgreater{} 4 {-}3 {-}14     0 14.31782      t1}
\CommentTok{\#\textgreater{} 5 {-}2 {-}14     0 14.14214      t1}
\CommentTok{\#\textgreater{} 6 {-}1 {-}14     0 14.03567      t1}
\NormalTok{xy\_formatted }\OtherTok{\textless{}{-}} \FunctionTok{initLayout}\NormalTok{(layout\_xy, }\AttributeTok{dataType =} \StringTok{"xy"}\NormalTok{)}
\end{Highlighting}
\end{Shaded}

Given gridded site layout data, \texttt{initLayout} returns an
\texttt{xyLayout} object, which is a list with components:

\begin{enumerate}
\item  $\texttt{xydat}$: data frame with  $\texttt{turbine}$, $(x, y)$ grid coordinates 
    relative to turbine, number of carcasses in each grid cell ($\texttt{ncarc}$),
    distance from turbine to center of grid cell ($\texttt{r}$), and (optional)
    columns for covariates or other identifiers;
  \item  $\texttt{tcenter}$: array of turbine locations $(x, y)$, with row names =
    turbine IDs;
  \item  $\texttt{ncarc}$ vector of carcass counts at each turbine, with element
    names = turbine IDs;
  \item  $\texttt{unitCol}$ name of the column in  $\texttt{xydat}$ that has turbine IDs
  \item  $\texttt{tset}$ vector of names of turbines that were searched.
\end{enumerate}

The analysis of gridded data differs slightly from the ring data. The
models are fit on 1 \(\textrm{m}^2\) cells rather than rings. Compared
to the rings format, the grid format is more cumbersome to work with,
runs more slowly, and does not offer as wide of a selection of
pre-formatted models. The advantage of the grid structure is that allows
for more sophisticated modeling options. Of particular interest would be
accounting for \emph{anisotropy}, which is variability in carcass
distribution that depends on direction, as might happen if a prevailing
wind direction strongly affects carcass dispersion patterns. These
anisotropic models are discussed by Maurer et al (2020) and not
addressed here or in the software.

\begin{figure}
\centering
\includegraphics{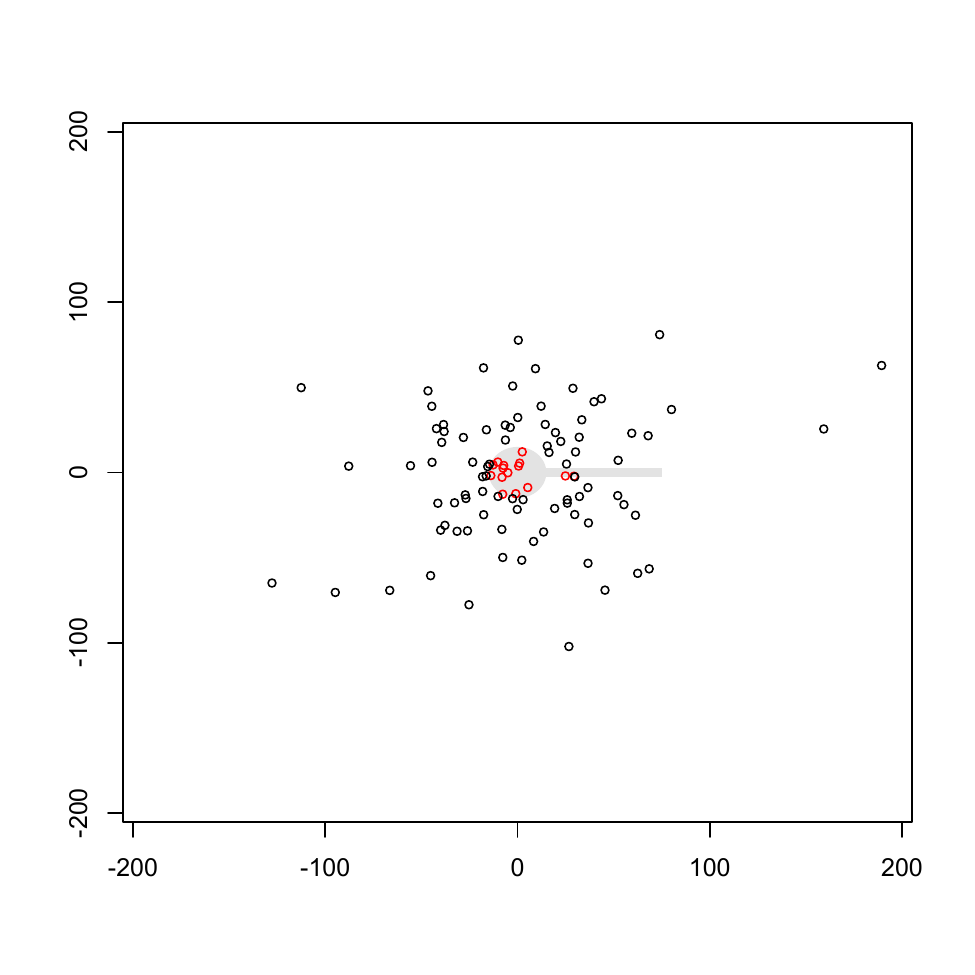}
\caption{\label{fig:xy}Layout of xy grid turbine data with carcasses,
showing the distribution of all carcasses in the simulation, both those
that were found (in the gray searched area) and those that were not
found.}
\end{figure}

The xy-grid data set poses special problems because few carcasses
(\(n = 15\)) were found at a single turbine searched on road and pad
only (fig.~\ref{fig:xy}). Although 65\% of the carcasses fell beyond 30
meters, the maximum distance that any carcass was found at was 29.1
meters. With such a low probability of carcasses lying on roads at
distances beyond 30 meters or so, it is difficult to get a
representative sample with road and pad searches unless the number of
carcasses is large. If the sample includes no carcasses well beyond the
pad radius, then the fitted models will tend to overestimate \(dwp\)
because they have difficulty ``seeing'' the distant carcasses. If, on
the other hand, the sample does include carcasses well beyond the
turbine pad, the fitted models will tend to underestimate \(dwp\)
because they over-estimate the number of carcasses at great distances.
On average, the tendencies to over- or under-estimate balance out, but
on any particular data set, one of the two is likely to be acting. If
there are carcasses at great distance, then the light-tailed models are
preferred because they provide some degree of insurance that the
predicted densities decrease to zero rapidly beyond the range of the
data. If there are no carcasses at great distance, then the heavy-tailed
models are preferred in order to prevent the distribution from
collapsing to zero too rapidly and not properly accounting for carcasses
falling beyond the farthest carcass observed.

The fitted models:

\begin{Shaded}
\begin{Highlighting}[]
\NormalTok{dmod\_xy }\OtherTok{\textless{}{-}} \FunctionTok{ddFit}\NormalTok{(xy\_formatted)}
\CommentTok{\#\textgreater{} Extensible models:}
\CommentTok{\#\textgreater{}   xep1 }
\CommentTok{\#\textgreater{}   xep01 }
\CommentTok{\#\textgreater{}   xep2 }
\CommentTok{\#\textgreater{}   xep02 }
\CommentTok{\#\textgreater{}   xep12 }
\CommentTok{\#\textgreater{}   xep012 }
\CommentTok{\#\textgreater{}   xep123 }
\CommentTok{\#\textgreater{}   xep0123 }
\CommentTok{\#\textgreater{} }
\CommentTok{\#\textgreater{} Non{-}extensible models:}
\CommentTok{\#\textgreater{}  none}
\FunctionTok{plot}\NormalTok{(dmod\_xy)}
\end{Highlighting}
\end{Shaded}

\begin{figure}
\centering
\includegraphics{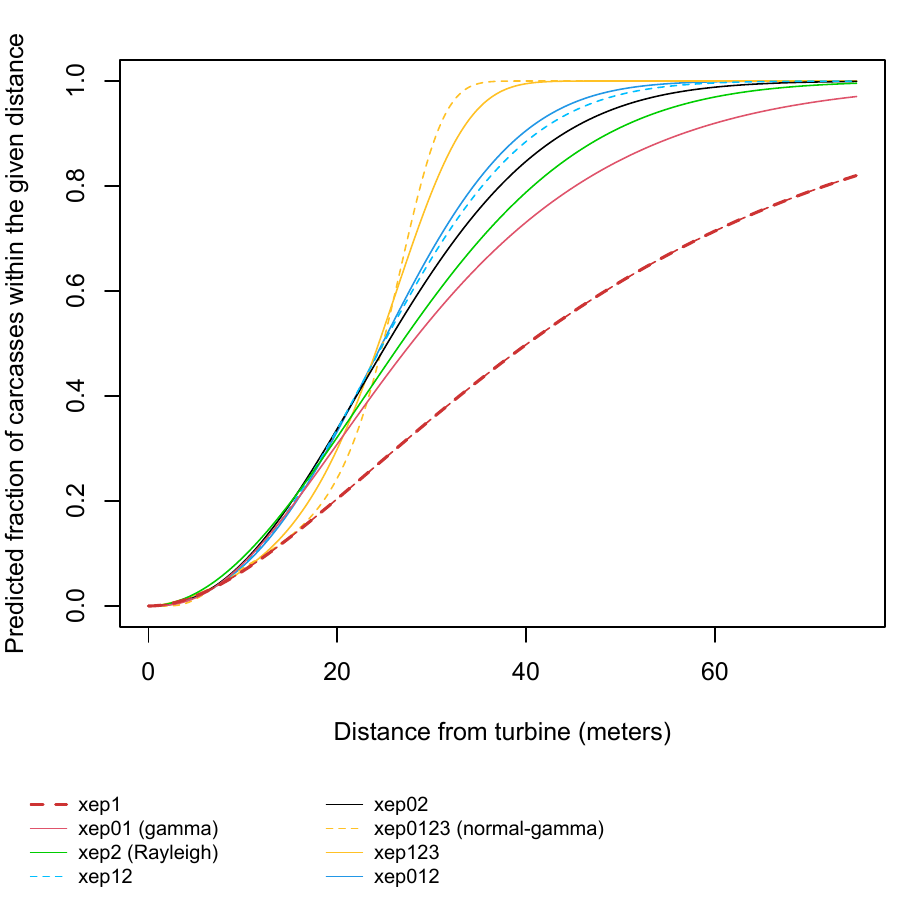}
\caption{\label{fig:xymod}CDFs of fitted models for the xy grid data.
Figure drawn using the \(\texttt{plot}\) function from the
\(\texttt{dwp}\) package with the fitted models:
\(\texttt{plot(dmod\_xy)}\), where \(\texttt{passind}\) is a vector of
names of the models to be plotted.}
\end{figure}

With the gridded data set, there was enormous discrepancy among the
fitted models (fig.~\ref{fig:xymod}), but the \texttt{modelFilter}
winnows out all but two of the models (table \ref{tbl:xy_scores}):

\begin{Shaded}
\begin{Highlighting}[]
\FunctionTok{modelFilter}\NormalTok{(dmod\_xy)}\SpecialCharTok{$}\NormalTok{scores}
\end{Highlighting}
\end{Shaded}

\begin{table}[!htbp]
  \caption{Model filter for site with gridded layout data, where 1 = filter test passed, 0 = filter test failed. Tests are described in section “Tools for Model Selection” Extensible = extensibility test, rtail and ltail = heavy r(igth) and l(eft) tail tests, aicc = $\Delta \textrm{AICc} < 10$, hin = high influence test, $\Delta \textrm{AICc}$ = difference between AICc value for a given model and the lowest AICc value among the models tested}
   \label{tbl:xy_scores}
   \begin{tabular}{rcccccc}
 & extensible & rtail & ltail & aicc & hin & deltaAICc \\ 
  \hline
xep1 & 1 & 1 & 1 & 1 & 1 & 0.90 \\ 
  xep01 & 1 & 1 & 1 & 1 & 1 & 2.11 \\ 
  xep2 & 1 & 1 & 0 & 1 & 1 & 0.00 \\ 
  xep12 & 1 & 1 & 0 & 1 & 1 & 1.61 \\ 
  xep02 & 1 & 1 & 0 & 1 & 1 & 1.74 \\ 
  xep0123 & 1 & 1 & 0 & 1 & 0 & 2.03 \\ 
  xep123 & 1 & 1 & 0 & 1 & 0 & 2.44 \\ 
  xep012 & 1 & 1 & 0 & 1 & 0 & 3.58 \\ 
   \hline
   \end{tabular}
 \end{table}

The xep1 and xep01 distributions are the only ones that do not predict
that over 90\% of the carcasses lie within 50 m of the turbine and thus
pass the \texttt{ltail} filter (table \ref{tbl:xy_scores}). The
predicted 90th percentile of carcasses distances according to the xep01
model is 56.6 meters\footnote{This can be calculated as
  \(\texttt{qdd(0.9, dmod\_xy["xep01"])}\).}, which is just beyond the
cutoff of 50 m for the left-tail test and largely in line with the
other, rejected models. By contrast, the xep1 model has a slightly
better AICc, which, all else being equal, is a reasonable tie-breaker
for model selection.

As a postscript, note that the actual distribution that the carcasses
were generated from (namely,
\(\textrm{gamma}(\alpha = 1.774, \beta = 28.17)\), where \(\beta\) is
the scale parameter) has known average proportion of 79.5\% of carcasses
falling within the search radius of 75 meters. The xep1 model
predicted\footnote{The predicted proportion falling within a given
  radius, \texttt{x}, can be calculated as
  \(\texttt{pdd(x, dmod\_xy["xep1"])}\).} 80\%, which is extremely close
to the expected fraction. None of the other models was even close. All
predicted in excess of 96.5\% within 75 m.

Selecting the xep1 model for estimating \(\psi\) and \(dwp\), we get:

\begin{Shaded}
\begin{Highlighting}[]
\NormalTok{psi\_xy }\OtherTok{\textless{}{-}} \FunctionTok{estpsi}\NormalTok{(xy\_formatted, }\AttributeTok{model =}\NormalTok{ dmod\_xy[}\StringTok{"xep1"}\NormalTok{])}
\NormalTok{dwp\_xy }\OtherTok{\textless{}{-}} \FunctionTok{estdwp}\NormalTok{(psi\_xy, }\AttributeTok{ncarc =} \FunctionTok{getncarc}\NormalTok{(xy\_formatted), }\AttributeTok{forGenEst =} \ConstantTok{TRUE}\NormalTok{)}
\end{Highlighting}
\end{Shaded}

\hypertarget{site-layout-and-carcass-data-from-shape-files-the-casselman-data}{%
\subsection{\texorpdfstring{Site Layout and Carcass Data from Shape
Files: The Casselman Data
\label{sec:exshape}}{Site Layout and Carcass Data from Shape Files: The Casselman Data }}\label{site-layout-and-carcass-data-from-shape-files-the-casselman-data}}

Compared to the simpler data structures (vector, simple geometry, and
polygons), shape files allow for greater power to account for 1)
variation in detection probabilities among search classes and 2)
interacting covariates that affect the distances that carcasses fly but
do require more sophisticated mapping and data management procedures.
However, unlike the xy-grid format, the shape files do not lend
themselves well to accounting for anisotropy.

We use the Casselman data to illustrate some of the analyses that can be
performed using the \texttt{dwp} package with data from shape files. The
Casselman data set was collected at the Casselman Wind Project in
Pennsylvania in 2008 (Arnett et al.~2009a, 2009b). It includes carcass
counts for several bat species found at 22 turbines. Search areas and
locations of carcass discoveries were mapped by a global positioning
system and stored in Geographic Information System (GIS) shape files.
Search areas were classed by the search conditions (\texttt{Easy},
\texttt{Moderate}, \texttt{Difficult}, \texttt{Very\ Difficult}).
Unsearched areas are also delineated in the shape files and classed as
\texttt{Out}. For this analysis, it is assumed that the distances that
carcasses fly after being struck by a turbine blade are not affected by
the search conditions on the ground (Huso and Dalthorp 2014). However,
the probabilities of observing carcasses that fall at a given distance
will differ depending on search class, so the search class must be
included in the models as a non-interacting covariate.

The data set also includes \emph{interacting} covariates that, unlike
search class, affect the distances that carcasses fly. We consider two
of those---turbine operations (curtailed at low wind speeds or not) and
carcass size---in sections \ref{sec:curtail} and \ref{sec:size}.

\hypertarget{shape-files-data}{%
\subsubsection{Shape Files: Data}\label{shape-files-data}}

The \texttt{initLayout} function reads GIS shape files using the
\texttt{st\_read} function from the \texttt{sf} (simple features) R
package for importing and managing GIS data. The importing of the shape
files and subsequent formatting of the data for \texttt{dwp} is done
behind the scenes in the package and does not assume that the user has
any familiarity with the \texttt{sf} package. However, users with a good
working knowledge of \texttt{sf} may be able to use that to their
advantage in managing the data and displaying results.

The importing and formatting of shape file data involves four steps in
R.

\begin{enumerate}
  \item \texttt{initLayout} reads the site layout data into a `shapeLayout` structure;
  \item \texttt{prepRing} pre-processes the site data into a `rings` for analysis;
  \item \texttt{readCarcass} imports carcass observations data from a shape file;
  \item \texttt{addCarcass} adds carcass location data to the `rings` object.
\end{enumerate}

To read shape files, the \texttt{initLayout} function requires ``simple
features'' files, including a polygon or multipolygon shape file that
delineates the area searched at each turbine and a points file that
gives the locations of the turbines on the same coordinate system as the
polygons. The shape files must have their three standard component files
(.shp, .shx, and .dbf) stored in the same folder.

The polygons and turbine point files must have a column with turbine
IDs, which must be syntactically valid R names, i.e., contain
combinations of letters, numbers, dots ( . ), and underscores ( \_ )
only and not begin with a number or a dot followed by a number. Other
characters are not allowed: no spaces, hyphens, parentheses, etc. are
allowed in turbine IDs.

Search areas may be defined for any number of search classes which
affect detection probabilities on the ground but are not expected to
affect the distances that carcasses fly after being struck, for example,
vegetation type or ground texture. Interacting covariates that affect
the actual distances that carcasses fly---like wind speed, turbine
height, carcass size---present special difficulties that are difficult
to model using the kinds of automated procedures used in most of the
convenient \texttt{dwp} functions and are not implemented in the basic
package functions. Savvy R users may be able to use the \texttt{dwp}
functions as a basis for more elaborate analyses, including accounting
for \emph{anisotropy} (or carcass dispersion patterns that vary with
direction from turbines) (Maurer et al.~2020) or analyzing dependence of
dispersion on other interacting variables.

The shape files for site layout should include at a minimum a turbine ID
and a ``geometry'' that gives the coordinates of the vertices for
polygons that delineate the search areas. Other covariates (for example,
search class, species, size, turbine operational constraints) are
optional. When a shape file is read into R using \texttt{sf::st\_read}
(as with \texttt{dwp}'s \texttt{initLayout}), it should look something
like following, showing necessary geographic information in a header and
a summary of the first 10 polygons or ``features'':

\begin{verbatim}
------------------------------------------------------------------------------------------
Simple feature collection with 178 features and 3 fields
geometry type:  MULTIPOLYGON
dimension:      XY
bbox:           xmin: 658608.2 ymin: 4411682 xmax: 662782 ymax: 4415250
projected CRS:  NAD83 / UTM zone 17N
First 10 features:
   OBJECTID Turbine Class                       geometry
1         1       1   Out MULTIPOLYGON (((658822.4 44...
2         2       2   Out MULTIPOLYGON (((658747.7 44...
3         3       3   Out MULTIPOLYGON (((658717 4414...
4         4       4   Out MULTIPOLYGON (((658789.2 44...
5         5       5   Out MULTIPOLYGON (((658847.6 44...
6         6       6   Out MULTIPOLYGON (((658835.4 44...
7         7       7   Out MULTIPOLYGON (((658873 4413...
8         8       8   Out MULTIPOLYGON (((658771.2 44...
9         9       9   Out MULTIPOLYGON (((659276.5 44...
10       10      10   Out MULTIPOLYGON (((659249.2 44...

-------------------------------------------------------------------------------------------
\end{verbatim}

Shape files are read into R using the \texttt{initLayout} function with
the names of the search files (including the file path if they are not
stored in your working directory in R) along with the name of the column
that contains turbine IDs. Other arguments for \texttt{initLayout} are
for reading other file and data formats and are ignored when reading
shape files.

Notice that the turbine IDs (1, 2, \ldots) in the raw data file (shown
above) are not syntactically valid R names. \texttt{initLayout} will
convert these to simple, syntactically valid names, and alert the user
that conversion has taken place. In general, users should use the
\texttt{initLayout} function rather than \texttt{sf::st\_read} directly
for importing the shape files because \texttt{initLayout} employs some
additional error-checks and formats the data for further processing and
analysis:

\begin{Shaded}
\begin{Highlighting}[]
\CommentTok{\# import site layout and carcass observation distances from shape files}
\NormalTok{layout\_shape }\OtherTok{\textless{}{-}} \FunctionTok{initLayout}\NormalTok{(}\AttributeTok{data\_layout =} \StringTok{"searchpoly.shp"}\NormalTok{,}
  \AttributeTok{file\_turbine =} \StringTok{"turbine\_pt.shp"}\NormalTok{, }\AttributeTok{unitCol =} \StringTok{"Turbine"}\NormalTok{)}

\NormalTok{cod }\OtherTok{\textless{}{-}} \FunctionTok{readCarcass}\NormalTok{(}\StringTok{"carcasses.shp"}\NormalTok{, }\AttributeTok{unitCol =} \StringTok{"Turbine"}\NormalTok{)}
\end{Highlighting}
\end{Shaded}

\begin{figure}
\centering
\includegraphics{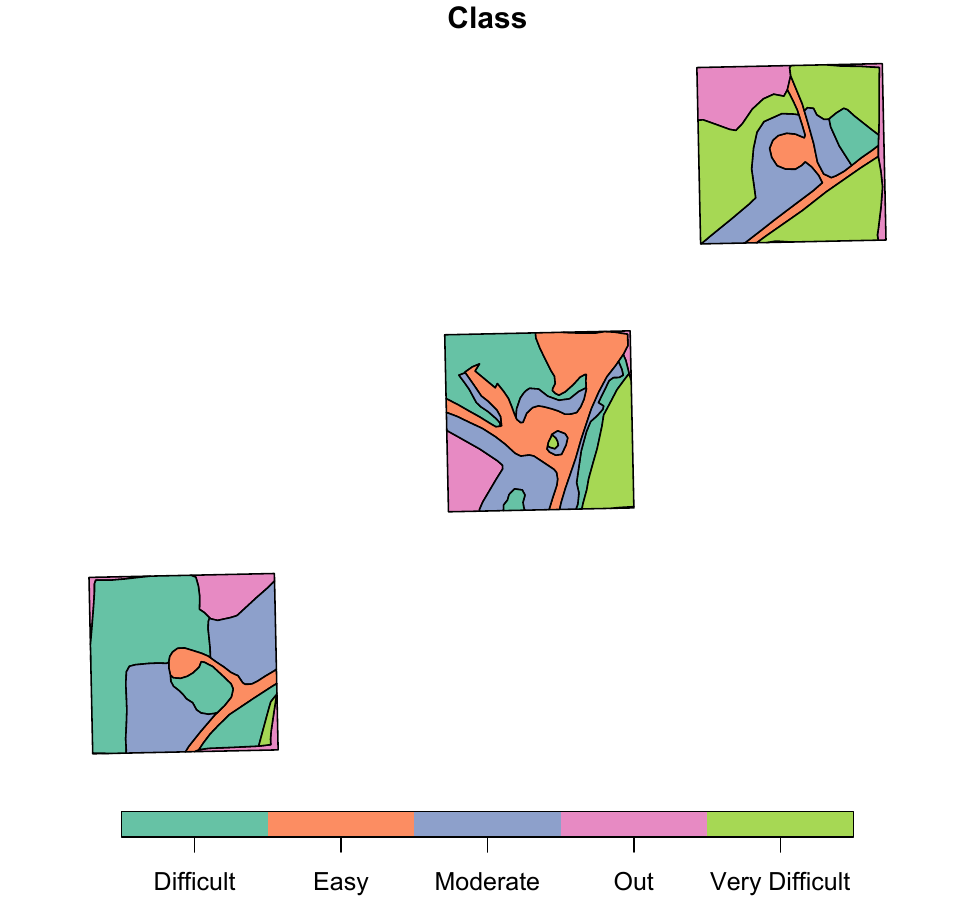}
\caption{\label{fig:shlayout}Maps of search areas for three turbines
with 5 search classes: Easy, Moderate, Difficult, Very Difficult, and
Out. `Easy' is represented primarily by roads and turbine pads, the
other classes represent progressively more challenging search
conditions, and `Out' areas were not searched due to rough terrain or
thick vegetation.Figure drawn using the \(\texttt{plot}\) function from
the \(\texttt{dwp}\) package with formatted site layout data from shape
files.}
\end{figure}

The polygons in the \texttt{searchpoly.shp} shape file are expressed as
collections of (\(x\), \(y\)) coordinates in meters relative to a
reference point. In this example there are a total of 178 polygons (or
``features'') that delineate distinct, non-intersecting search areas at
23 turbines. Search classes include \texttt{Easy}, \texttt{Moderate},
\texttt{Difficult}, \texttt{Very\ Difficult}, and \texttt{Out}. Maps of
the searched areas at three of the turbines are shown in
fig.~\ref{fig:shlayout}.

After the data are imported into R, they must still be formatted for
analysis:

\begin{Shaded}
\begin{Highlighting}[]
\NormalTok{rings\_shape }\OtherTok{\textless{}{-}} \FunctionTok{prepRing}\NormalTok{(layout\_shape, }\AttributeTok{scVar =} \StringTok{"Class"}\NormalTok{, }\AttributeTok{notSearched =} \StringTok{"Out"}\NormalTok{)}
\end{Highlighting}
\end{Shaded}

For a large site with multiple search classes, data formatting may take
several minutes. The result is an S3 object of class \texttt{rings},
which features a characterization of the searched areas at each turbine,
including the total area and number of carcasses in each search class in
each concentric, 1 meter ring from the turbine to the search radius
(stored in \texttt{rdat}, which is used in fitting the distance models)
along with some other data. The ring structure is efficient and
convenient for analysis. A full description of the data structure is
provided in the section on Rings (section \ref{sec:formatted}).

After formatting the search plot configurations into rings for analysis,
carcasses must be added to the rings. If the search plots are delineated
into separate search classes, carcasses must be added by a two-step
process of 1) importing the shape data, and 2) adding the data to the
\texttt{rings}. If there are no search class distinctions, then
carcasses may be added via a data frame with columns for turbine ID
(\texttt{unitCol}) and for the carcass distances from turbine
(\texttt{r}).

\begin{Shaded}
\begin{Highlighting}[]
\CommentTok{\# read carcass data}
\NormalTok{cod }\OtherTok{\textless{}{-}} \FunctionTok{readCarcass}\NormalTok{(}\AttributeTok{file\_cod =} \StringTok{"carcasses.shp"}\NormalTok{, }\AttributeTok{unitCol =} \StringTok{"Turbine"}\NormalTok{)}

\CommentTok{\# add carcasses}
\NormalTok{rings\_shape }\OtherTok{\textless{}{-}} \FunctionTok{addCarcass}\NormalTok{(cod, }\AttributeTok{data\_ring =}\NormalTok{ rings\_shape, }\AttributeTok{plotLayout =}\NormalTok{ layout\_shape)}
\end{Highlighting}
\end{Shaded}

\begin{Shaded}
\begin{Highlighting}[]
\FunctionTok{head}\NormalTok{(cod}\SpecialCharTok{$}\NormalTok{carcasses)}
\CommentTok{\#\textgreater{} Simple feature collection with 6 features and 4 fields}
\CommentTok{\#\textgreater{} Geometry type: POINT}
\CommentTok{\#\textgreater{} Dimension:     XY}
\CommentTok{\#\textgreater{} Bounding box:  xmin: 658875.9 ymin: 4413105 xmax: 659242.4 ymax: 4413770}
\CommentTok{\#\textgreater{} Projected CRS: NAD83 / UTM zone 17N}
\CommentTok{\#\textgreater{}        Date\_ Turbine Species Visibility                 geometry}
\CommentTok{\#\textgreater{} 1 2009{-}08{-}31      t7    PESU       Easy POINT (658875.9 4413770)}
\CommentTok{\#\textgreater{} 2 2009{-}08{-}03      t9    LACI        Mod POINT (659242.4 4413290)}
\CommentTok{\#\textgreater{} 3 2009{-}10{-}07      t9    LANO       Diff POINT (659228.5 4413308)}
\CommentTok{\#\textgreater{} 4 2009{-}08{-}16      t9    PESU        Mod POINT (659195.4 4413340)}
\CommentTok{\#\textgreater{} 5 2009{-}08{-}30     t10    LABO       Easy POINT (659221.9 4413108)}
\CommentTok{\#\textgreater{} 6 2009{-}08{-}24     t10    LACI        Mod POINT (659223.4 4413105)}
\end{Highlighting}
\end{Shaded}

When reading carcass observation data from a shape file into a
\texttt{rings} structure with search class distinctions, \texttt{dwp}
assigns each carcass to the proper ring and search class. The
\texttt{rings} data is required, but there are two ways to insert the
carcass data into the \texttt{rings}. The recommended approach is to
provide the \texttt{plotLayout} data. Then, \texttt{addCarcass}
determines the proper ring and search class to assign each carcass to by
calculating the distances from the carcass \((x, y)\) coordinates in the
carcass observation data to the turbine locations stored in the
\texttt{plotLayout} data, which has been previously read and formatted
via \texttt{initLayout} and \texttt{prepRing}. Search class assignment
is determined by the search class in \texttt{plotLayout} that is
associated with the carcass \((x, y)\) coordinates. If a carcass
location falls in an unsearched area, the \texttt{addCarcass} function
aborts with an error. The calculations are all done safely ``under the
hood'' and directly from the imported shape file data without needing
further user input.

An alternative approach would be leave out the \texttt{plotLayout} data
(\texttt{plotLayout\ =\ NULL}) to add carcasses directly from the the
carcass observation data into the \texttt{rings} data. Carcass are
assigned to rings based on distances calculated from the carcass
locations \((x, y)\) coordinates in the carcass observation data) to the
turbines (\(x\), \(y\)) coordinates in the \texttt{\$tcenter} array in
the \texttt{data\_ring}). Carcasses are assigned to search classes as
indicated by the \texttt{scVar} in the carcass observation file. This
can introduce errors in interpretation, though, because the model
assumes that the ``search classes'' apply to the characteristics of the
general area where a carcass is found (fig.~\ref{fig:shlayout}), but
often the search classes that are tagged to individual carcasses in
carcass observation data apply to the search conditions in the immediate
vicinity of carcasses. For example, if a carcass is found in a small
\texttt{Difficult} patch in a large \texttt{Moderate} area, and the site
map (\texttt{plotLayout}) includes only that \texttt{Moderate} area but
not the \texttt{Difficult} patch, the carcass should be tallied as
\texttt{Moderate}. This potential problem in data interpretation is not
an issue if the \texttt{plotLayout} is included in the
\texttt{addCarcass} function call.

If there are no class distinctions among the search areas, carcasses may
be added to \texttt{rings} from a data frame with columns for turbine ID
and carcass distance. For example the \texttt{cod} argument in the
\texttt{addCarcass} function call may be a data frame rather than an
imported carcass location shape file:

\begin{verbatim}
cod = data.frame(r = c(31.2, 63.9, 18.4, 4.7), turbine = c("t1", "t2", "t2", "t1")
\end{verbatim}

Now that the site layout data has been formatted into rings and the
carcass data has been inserted into the ring structure, the model
fitting and analysis can proceed. For most of the input formats
(excepting the grid data), the analyses follow the same path regardless
of the initial data format. For the Casselman data, we do a basic
analysis first but then indulge some more sophisticated analyses as an
introduction to some of the more advanced questions that can be
addressed using the software.

\hypertarget{model-fitting-and-analysis}{%
\subsubsection{\texorpdfstring{Model Fitting and Analysis
\label{sec:shapefit}}{Model Fitting and Analysis }}\label{model-fitting-and-analysis}}

The basic modeling and analysis is similar, regardless of the initial
data format. Several tools for fitting, interpreting, and evaluating
models are available. However, the shape files allow covariates to be
included in the models, enabling greater power to address more difficult
and varied questions and to account for important factors that the other
data formats cannot easily address.

The analysis begins with the fitting of distance models using
\texttt{ddFit} with ring data or grid data. The basic function call
requires only a formatted data set as returned by the \texttt{prepRing}
function (or, \texttt{initLayout} for grid data). With the shapes data
(\texttt{rings\_shape}, as imported and formatted earlier) in the
example discussed in this section, the search plots are divided into
areas with different detection probabilities or \emph{search classes},
which are stored in the ``\texttt{Class}'' column in the data and should
be included in the model as \texttt{scVar\ =\ "Class"}.

Search class variables (like vegetative cover, for example) affect the
detection probability of carcasses but are not expected to affect how
far carcasses will lie from the turbines. There is no need to estimate
the detection probabilities for the various search classes; the model
automatically estimates relative detection probabilities among search
classes and accounts for the differences. Although the relative
detection probabilities are estimated (for example, the odds of
detection in class A may be 2.52 times as great as in class B), the
absolute detection probabilities (\(g\) in \texttt{GenEst} and
\texttt{eoa}) are not. A number of other options are also available. A
simple example is given below, or enter \texttt{?ddFit} in R for a more
detailed information on arguments and return values.

\begin{Shaded}
\begin{Highlighting}[]
\NormalTok{dmod\_shape }\OtherTok{\textless{}{-}} \FunctionTok{ddFit}\NormalTok{(}\AttributeTok{x =}\NormalTok{ rings\_shape, }\AttributeTok{scVar =} \StringTok{"Class"}\NormalTok{)}
\CommentTok{\#\textgreater{} Extensible models:}
\CommentTok{\#\textgreater{}   xep1 }
\CommentTok{\#\textgreater{}   xep01 }
\CommentTok{\#\textgreater{}   xep2 }
\CommentTok{\#\textgreater{}   xep02 }
\CommentTok{\#\textgreater{}   xep12 }
\CommentTok{\#\textgreater{}   xep012 }
\CommentTok{\#\textgreater{}   tnormal }
\CommentTok{\#\textgreater{}   MaxwellBoltzmann }
\CommentTok{\#\textgreater{}   lognormal }
\CommentTok{\#\textgreater{} }
\CommentTok{\#\textgreater{} Non{-}extensible models:}
\CommentTok{\#\textgreater{}   xep123 }
\CommentTok{\#\textgreater{}   xep0123 }
\CommentTok{\#\textgreater{}   constant}
\end{Highlighting}
\end{Shaded}

\begin{figure}
\centering
\includegraphics{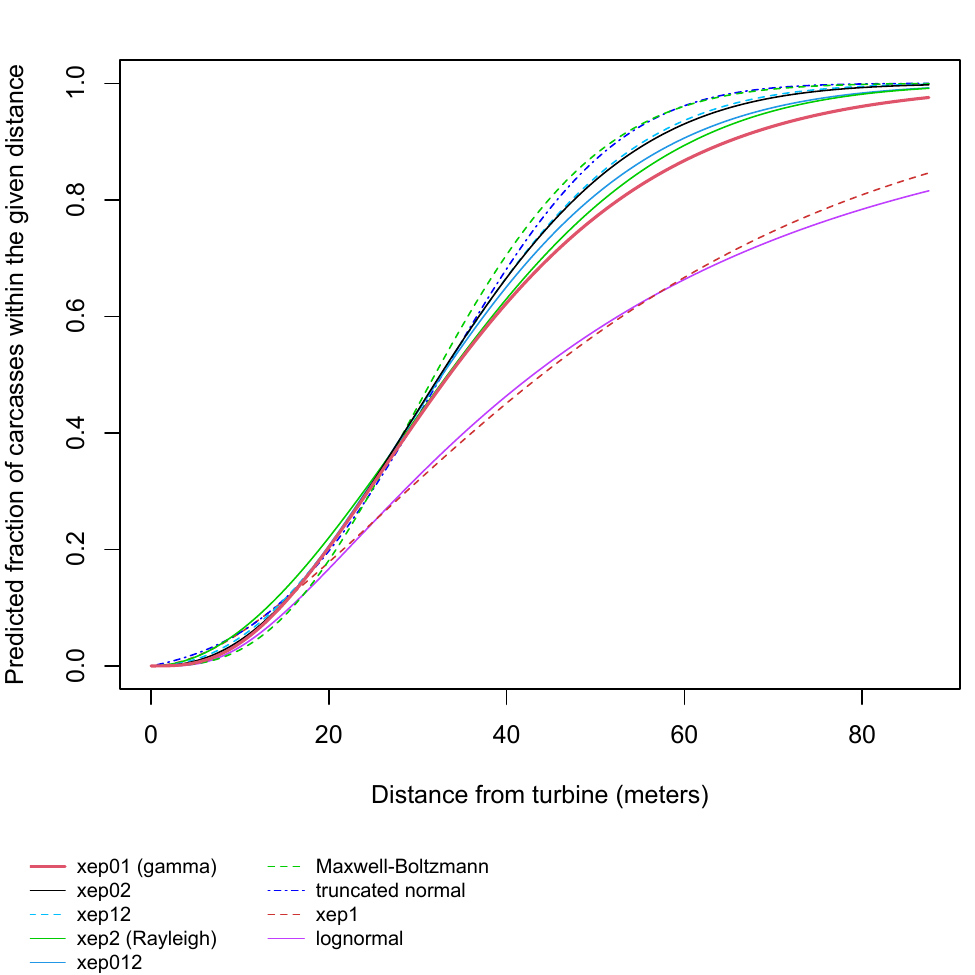}
\caption{\label{fig:CDF}Cumulative distribution functions (CDFs) for
extensible, standard models for the full ``shape'' data set in section
\ref{sec:shapefit}. Figure drawn using the \(\texttt{plot}\) function
from the \(\texttt{dwp}\) package with the fitted models.}
\end{figure}

As the models are fit, their names are listed on the console as either
``extensible'' or ``non-extensible'', depending on whether or not the
model can be extended beyond the search radius and converted to a
probability distribution. Non-extensible models are generally not
appropriate for \texttt{dwp} analysis but may be interesting in their
own right. The \texttt{plot} function can be used to graph the fitted
models (fig.~\ref{fig:CDF}).

\begin{Shaded}
\begin{Highlighting}[]
\FunctionTok{plot}\NormalTok{(dmod\_shape)}
\end{Highlighting}
\end{Shaded}

\begin{figure}
\centering
\includegraphics{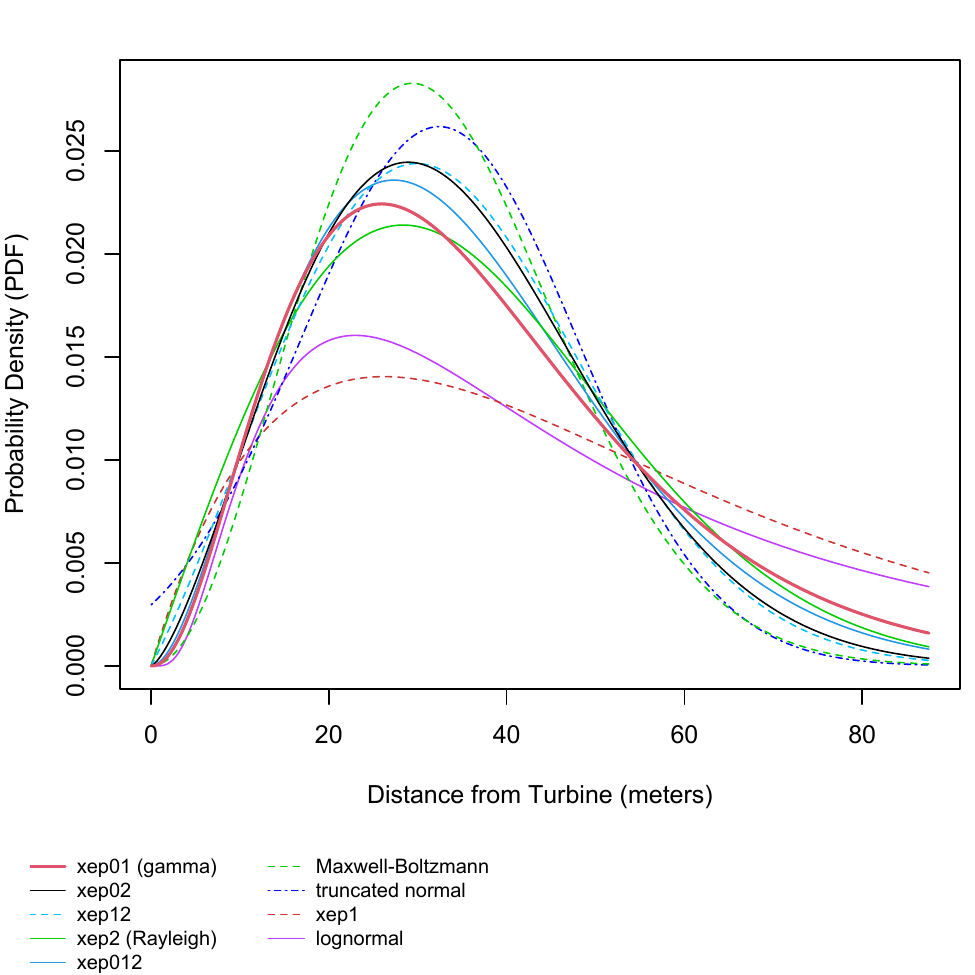}
\caption{\label{fig:PDF}Probability density functions (PDFs) for
extensible, standard models for the full ``shape'' data set in section
\ref{sec:shapefit}. Figure drawn using the \(\texttt{plot}\) function
from the \(\texttt{dwp}\) package with the fitted models.}
\end{figure}

The default for \texttt{plot} is to graph the CDF for each extensible,
fitted model, but the PDFs (\texttt{type\ =\ "PDF"}), distributions
limited strictly to within the search radius
(\texttt{extent\ =\ "win"}), and any subset of models (for example,
\texttt{distr\ =\ c("xep01",\ "xep1")}) may also be plotted
(fig.~\ref{fig:PDF}). By default, the best model according to a default
model filter (\texttt{?modelFilter}) is highlighted, but any model can
be highlighted at the user's discretion by setting the
\texttt{mod\_highlight} argument equal to the name of the model. For
more details on plotting options, enter \texttt{?plot} in R.

\begin{Shaded}
\begin{Highlighting}[]
\FunctionTok{plot}\NormalTok{(dmod\_shape, }\AttributeTok{type =} \StringTok{"PDF"}\NormalTok{)}
\end{Highlighting}
\end{Shaded}

To estimate the probability of a carcass lying in the searched area and
the fraction of carcasses lying in the searched area, a model must be
selected. If it is desired to account for all carcasses lying outside
the areas searched---including areas outside the search radius---an
extensible model is required.

The graph of the CDF shows the estimated probability that a carcass lies
within a given distance from the turbine. In this example, the two
heaviest-tailed models (\texttt{lognormal} and \texttt{xep1}) predict
about 20\% of the carcasses lie outside the search radius of around 88
meters, while the other extensible models predict virtually zero
carcasses outside the search radius. In some data sets, choice of model
can have a profound impact on the estimated fraction of carcasses
falling within the search radius and on \texttt{dwp}. Some of the
important issues involved in model selection are discussed in the
appendices. The \texttt{modelFilter} function is designed to address
many of the issues.

The probability that a carcass lies within the search radius
(\texttt{srad})---or any other specified distance---can be estimated for
any of the extensible, fitted distributions using the \texttt{pdd}
function, which is an analog to \texttt{pnorm} for distance
distributions.

\begin{Shaded}
\begin{Highlighting}[]
\FunctionTok{pdd}\NormalTok{(rings\_shape}\SpecialCharTok{$}\NormalTok{srad, }\AttributeTok{model =}\NormalTok{ dmod\_shape[}\StringTok{"xep01"}\NormalTok{])}
\CommentTok{\#\textgreater{} [1] 0.9764346}
\FunctionTok{pdd}\NormalTok{(rings\_shape}\SpecialCharTok{$}\NormalTok{srad, }\AttributeTok{model =}\NormalTok{ dmod\_shape[}\StringTok{"xep02"}\NormalTok{])}
\CommentTok{\#\textgreater{} [1] 0.997577}
\FunctionTok{pdd}\NormalTok{(rings\_shape}\SpecialCharTok{$}\NormalTok{srad, }\AttributeTok{model =}\NormalTok{ dmod\_shape[}\StringTok{"lognormal"}\NormalTok{])}
\CommentTok{\#\textgreater{} [1] 0.8174281}
\end{Highlighting}
\end{Shaded}

\begin{figure}
\centering
\includegraphics{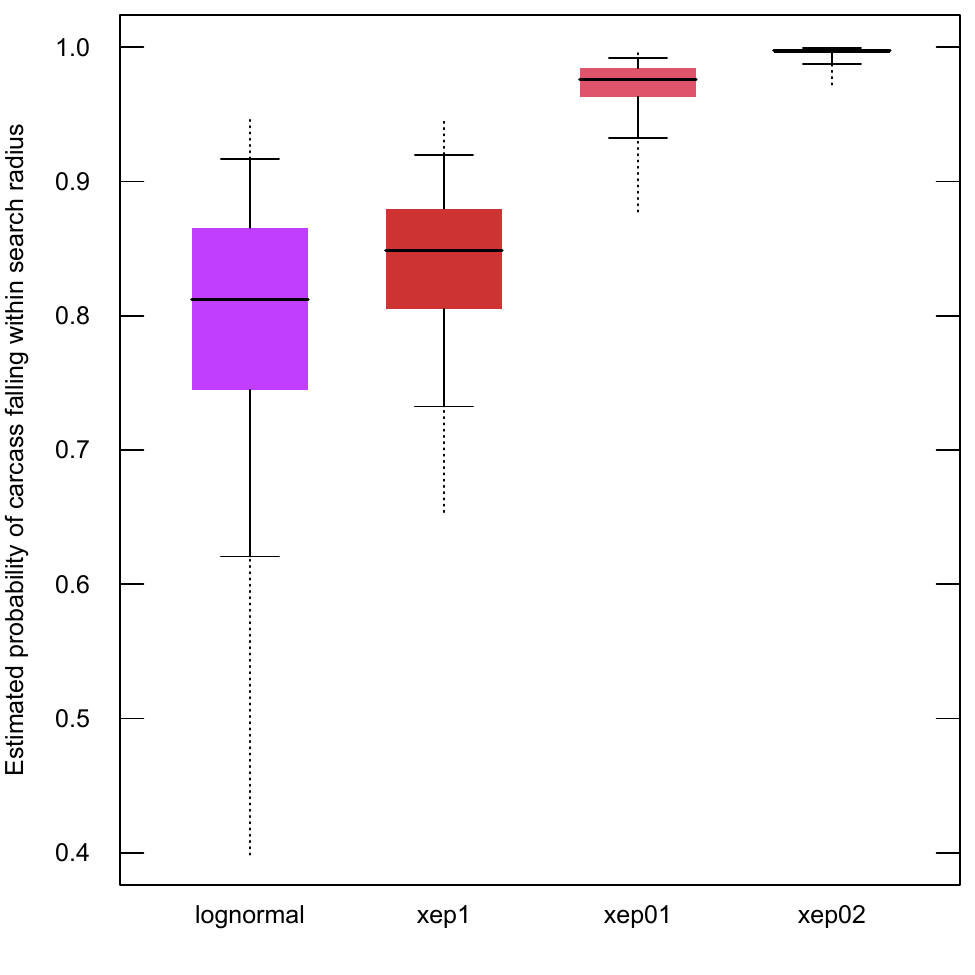}
\caption{\label{fig:boxes}Estimated probability of carcass falling
within search radius according to 4 distributions fit to the full
``shape'' data set.}
\end{figure}

The lines shown in figures \ref{fig:CDF} and \ref{fig:PDF} are for the
maximum likelihood estimators of the models, which represent the most
likely or ``best'' estimate for the given distribution. However, due to
random variation, the distributions are estimated with uncertainty. That
uncertainty can be accounted for by simulating from the asymptotic
distribution of the parameter estimators using the \texttt{ddSim}
(distance distribution simulation) function, and confidence intervals
can be constructed (fig.~\ref{fig:boxes}).

\begin{Shaded}
\begin{Highlighting}[]
\CommentTok{\# simulate parameters to account for uncertaity:}
\NormalTok{simparm01 }\OtherTok{\textless{}{-}} \FunctionTok{ddSim}\NormalTok{(dmod\_shape[}\StringTok{"xep01"}\NormalTok{])}
\NormalTok{simparm1 }\OtherTok{\textless{}{-}} \FunctionTok{ddSim}\NormalTok{(dmod\_shape[}\StringTok{"xep1"}\NormalTok{])}
\NormalTok{simparm02 }\OtherTok{\textless{}{-}} \FunctionTok{ddSim}\NormalTok{(dmod\_shape[}\StringTok{"xep02"}\NormalTok{])}
\NormalTok{simparmln }\OtherTok{\textless{}{-}} \FunctionTok{ddSim}\NormalTok{(dmod\_shape[}\StringTok{"lognormal"}\NormalTok{])}

\CommentTok{\# calculate confidence intervals:}
\NormalTok{srad }\OtherTok{\textless{}{-}}\NormalTok{ rings\_shape}\SpecialCharTok{$}\NormalTok{srad}
\NormalTok{(CI01 }\OtherTok{\textless{}{-}} \FunctionTok{round}\NormalTok{(}\FunctionTok{quantile}\NormalTok{(}\FunctionTok{pdd}\NormalTok{(srad, simparm01), }\AttributeTok{prob =} \FunctionTok{c}\NormalTok{(}\FloatTok{0.05}\NormalTok{, }\FloatTok{0.95}\NormalTok{)), }\DecValTok{3}\NormalTok{))}
\CommentTok{\#\textgreater{}    5\%   95\% }
\CommentTok{\#\textgreater{} 0.938 0.991}
\NormalTok{(CI02 }\OtherTok{\textless{}{-}} \FunctionTok{round}\NormalTok{(}\FunctionTok{quantile}\NormalTok{(}\FunctionTok{pdd}\NormalTok{(srad, simparm02), }\AttributeTok{prob =} \FunctionTok{c}\NormalTok{(}\FloatTok{0.05}\NormalTok{, }\FloatTok{0.95}\NormalTok{)), }\DecValTok{3}\NormalTok{))}
\CommentTok{\#\textgreater{}    5\%   95\% }
\CommentTok{\#\textgreater{} 0.989 0.999}
\NormalTok{(CI1 }\OtherTok{\textless{}{-}} \FunctionTok{round}\NormalTok{(}\FunctionTok{quantile}\NormalTok{(}\FunctionTok{pdd}\NormalTok{(srad, simparm1), }\AttributeTok{prob =} \FunctionTok{c}\NormalTok{(}\FloatTok{0.05}\NormalTok{, }\FloatTok{0.95}\NormalTok{)), }\DecValTok{3}\NormalTok{))}
\CommentTok{\#\textgreater{}    5\%   95\% }
\CommentTok{\#\textgreater{} 0.741 0.912}
\NormalTok{(CIln }\OtherTok{\textless{}{-}} \FunctionTok{round}\NormalTok{(}\FunctionTok{quantile}\NormalTok{(}\FunctionTok{pdd}\NormalTok{(srad, simparmln), }\AttributeTok{prob =} \FunctionTok{c}\NormalTok{(}\FloatTok{0.05}\NormalTok{, }\FloatTok{0.95}\NormalTok{)), }\DecValTok{3}\NormalTok{))}
\CommentTok{\#\textgreater{}    5\%   95\% }
\CommentTok{\#\textgreater{} 0.629 0.917}
\end{Highlighting}
\end{Shaded}

It is also straightforward to calculate confidence intervals for
quantiles of distance distributions using the \texttt{qdd} function,
which is the analog of \texttt{qnorm} but for distance distributions.
For example, to calculate 90\% CIs for the 90th percentile of distances
that carcasses lie, we can use the same sets of simulated parameters
used in estimating the probabilities that carcasses lie within the
search radius, and calculate the 90th percentiles as
\texttt{qdd(0.9,\ ...)}:

\begin{Shaded}
\begin{Highlighting}[]
\CommentTok{\# simulate 90th percentiles:}
\NormalTok{qxep01 }\OtherTok{\textless{}{-}} \FunctionTok{qdd}\NormalTok{(}\FloatTok{0.9}\NormalTok{, }\AttributeTok{model =}\NormalTok{ simparm01)}
\NormalTok{qxep1 }\OtherTok{\textless{}{-}} \FunctionTok{qdd}\NormalTok{(}\FloatTok{0.9}\NormalTok{, }\AttributeTok{model =}\NormalTok{ simparm1)}
\NormalTok{qxep02 }\OtherTok{\textless{}{-}} \FunctionTok{qdd}\NormalTok{(}\FloatTok{0.9}\NormalTok{, }\AttributeTok{model =}\NormalTok{ simparm02)}
\NormalTok{qxepln }\OtherTok{\textless{}{-}} \FunctionTok{qdd}\NormalTok{(}\FloatTok{0.9}\NormalTok{, }\AttributeTok{model =}\NormalTok{ simparmln)}
\CommentTok{\# calculate confidence intervals:}
\NormalTok{(CI01 }\OtherTok{\textless{}{-}} \FunctionTok{round}\NormalTok{(}\FunctionTok{quantile}\NormalTok{(qxep01, }\AttributeTok{prob =} \FunctionTok{c}\NormalTok{(}\FloatTok{0.05}\NormalTok{, }\FloatTok{0.95}\NormalTok{)), }\DecValTok{1}\NormalTok{))}
\CommentTok{\#\textgreater{}   5\%  95\% }
\CommentTok{\#\textgreater{} 56.1 79.5}
\NormalTok{(CI02 }\OtherTok{\textless{}{-}} \FunctionTok{round}\NormalTok{(}\FunctionTok{quantile}\NormalTok{(qxep02, }\AttributeTok{prob =} \FunctionTok{c}\NormalTok{(}\FloatTok{0.05}\NormalTok{, }\FloatTok{0.95}\NormalTok{)), }\DecValTok{1}\NormalTok{))}
\CommentTok{\#\textgreater{}   5\%  95\% }
\CommentTok{\#\textgreater{} 51.3 64.3}
\NormalTok{(CI1 }\OtherTok{\textless{}{-}} \FunctionTok{round}\NormalTok{(}\FunctionTok{quantile}\NormalTok{(qxep1, }\AttributeTok{prob =} \FunctionTok{c}\NormalTok{(}\FloatTok{0.05}\NormalTok{, }\FloatTok{0.95}\NormalTok{)), }\DecValTok{1}\NormalTok{))}
\CommentTok{\#\textgreater{}    5\%   95\% }
\CommentTok{\#\textgreater{}  82.2 131.7}
\NormalTok{(CIln }\OtherTok{\textless{}{-}} \FunctionTok{round}\NormalTok{(}\FunctionTok{quantile}\NormalTok{(qxepln, }\AttributeTok{prob =} \FunctionTok{c}\NormalTok{(}\FloatTok{0.05}\NormalTok{, }\FloatTok{0.95}\NormalTok{), }\AttributeTok{na.rm =}\NormalTok{ T), }\DecValTok{1}\NormalTok{))}
\CommentTok{\#\textgreater{}    5\%   95\% }
\CommentTok{\#\textgreater{}  82.1 227.9}
\end{Highlighting}
\end{Shaded}

\begin{figure}
\centering
\includegraphics{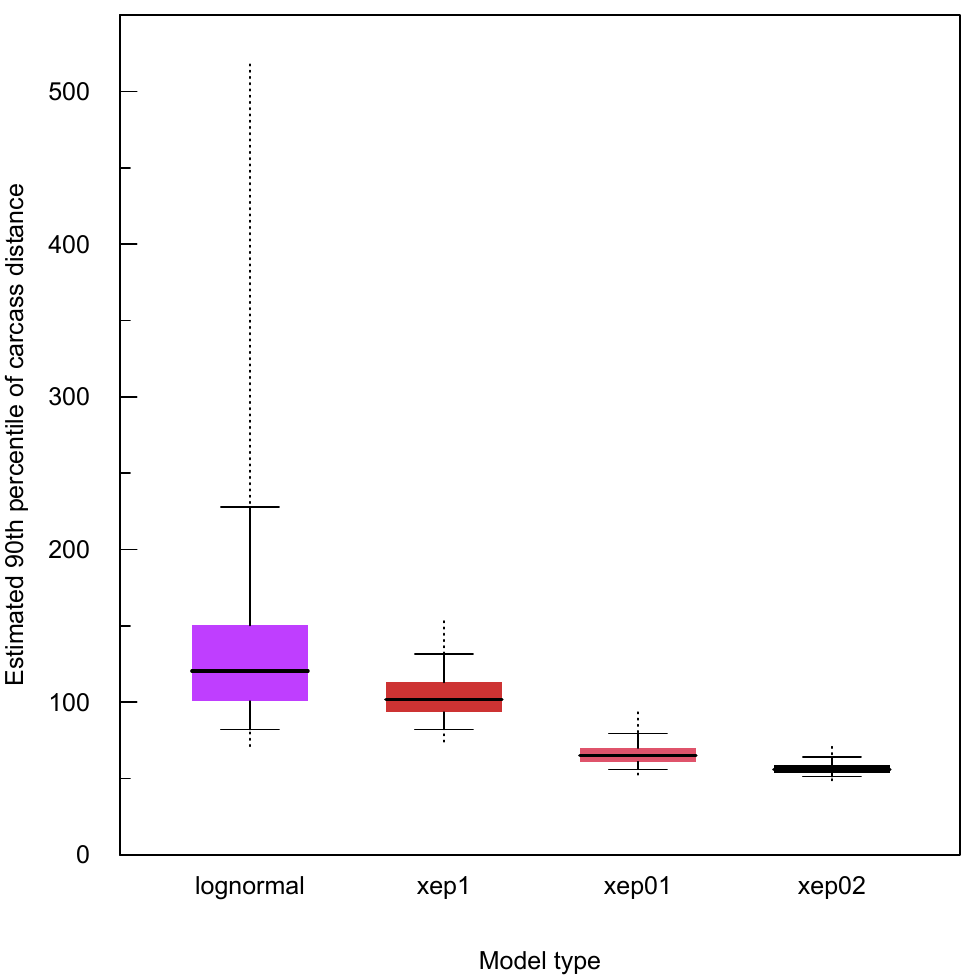}
\caption{\label{fig:boxes_q}Estimated 90th percentile of carcass
distances for 4 distributions fit to the full ``shape'' data set.}
\end{figure}

Thus, according to the \texttt{xep02} model, carcasses had an estimated
90\% probability of falling within
\texttt{qdd(0.9,\ dmod\_shape{[}"xep02"{]})\ =} 56.1 meters of the
turbine with a 90\% confidence interval of {[}51.3, 64.3{]} meters,
whereas the lognormal model estimates a substantially greater distance
of \texttt{qdd(0.9,\ dmod\_shape{[}"lognormal"{]})\ =} 118.5 meters with
a confidence interval of {[}82.1, 227.9{]} meters. The xep01 model was
similar to the xep02, and the xep1 more like the lognormal
(fig.~\ref{fig:boxes_q}).

\begin{verbatim}
#>   5%  95% 
#> 56.1 79.5
#>   5%  95% 
#> 51.3 64.3
#>    5%   95% 
#>  82.2 131.7
#>    5%   95% 
#>  82.1 227.9
#>      0.5%     99.5% 
#>  48.74268 517.79736
\end{verbatim}

\begin{verbatim}
#> null device 
#>           1
\end{verbatim}

\hypertarget{probability-of-carcass-lying-in-the-searched-area}{%
\subsubsection{Probability of Carcass Lying in the Searched
Area}\label{probability-of-carcass-lying-in-the-searched-area}}

\begin{figure}
\centering
\includegraphics{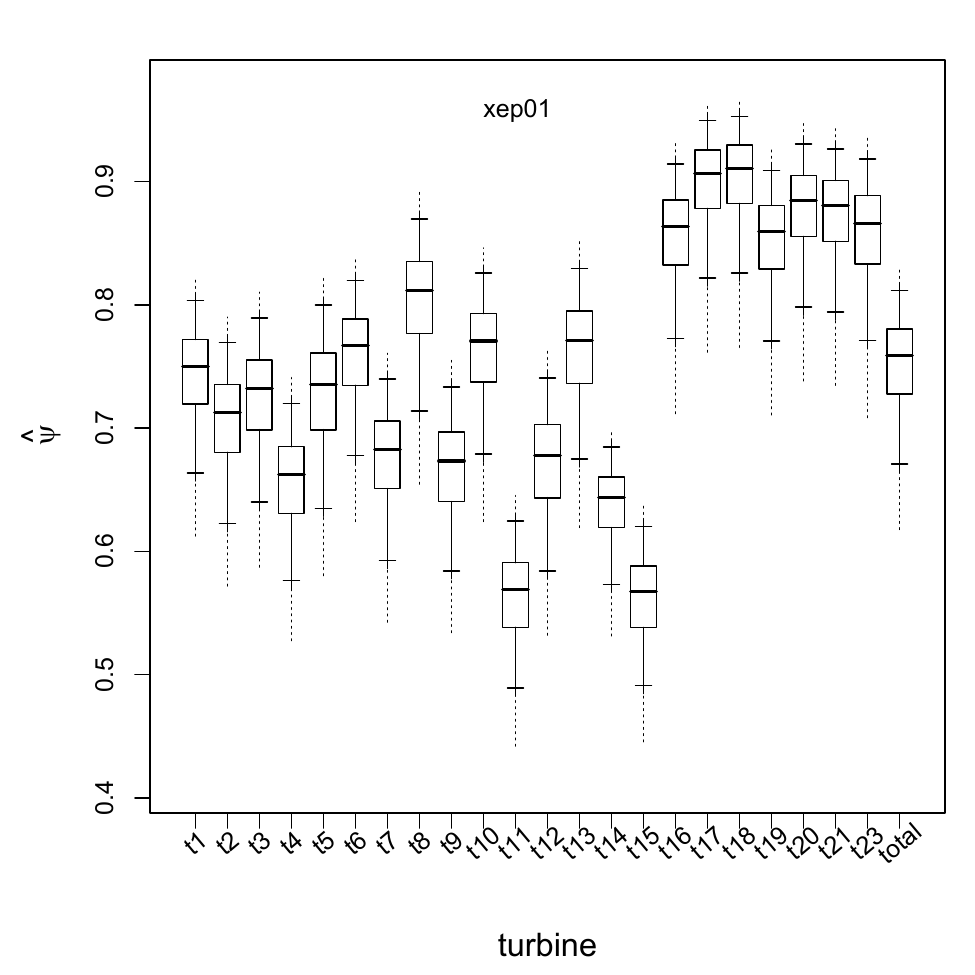}
\caption{\label{fig:psi01}Boxplots of the estimated probability,
\(\hat{\psi}\), of a carcass falling in the searched area. Boxes
represent the IQR with median. Whiskers mark the 95\% and 99\% CIs.
Figure drawn using the \(\texttt{plot}\) function from the
\(\texttt{dwp}\) package with the estimated \(\psi\)'s:
\(\texttt{plot(psi01)}\).}
\end{figure}

The \texttt{pdd} function can be used to estimate the CDF of carcass
distribution as a function of distance from a turbine, which is related
to but distinct from the probability, \(\psi\), that a carcass lies in
the searched area. \(\psi\) is estimated by integrating the PDF of the
fitted carcass distribution over the areas searched at each turbine.
This can be accomplished for any fitted model and site layout using the
\texttt{estpsi} function, and a summary figure can viewed using
\texttt{plot} (fig.~\ref{fig:psi01}).

\begin{Shaded}
\begin{Highlighting}[]
\NormalTok{psi01 }\OtherTok{\textless{}{-}} \FunctionTok{estpsi}\NormalTok{(rings\_shape, }\AttributeTok{model =}\NormalTok{ dmod\_shape[}\StringTok{"xep01"}\NormalTok{])}
\FunctionTok{plot}\NormalTok{(}\FunctionTok{estpsi}\NormalTok{(psi01), }\AttributeTok{main =} \StringTok{"xep01"}\NormalTok{)}
\end{Highlighting}
\end{Shaded}

The probabilities of lying in the searched area ranged from around 55\%
for turbines t11 and t15 to 85-90\% for turbines t17-t23. The combined
total was around 75\%. More precise statistics can be extracted from the
\texttt{psiHat} object returned from the \texttt{estpsi} function. In
this case, that is \texttt{psi01}, which is a matrix with a column of
\texttt{nsim} simulated \(\hat{\psi}\) values for each turbine, with the
uncertainty in the estimation of \(\psi\) at each turbine captured by
the variation in values in the turbine's column.

\begin{figure}
\centering
\includegraphics{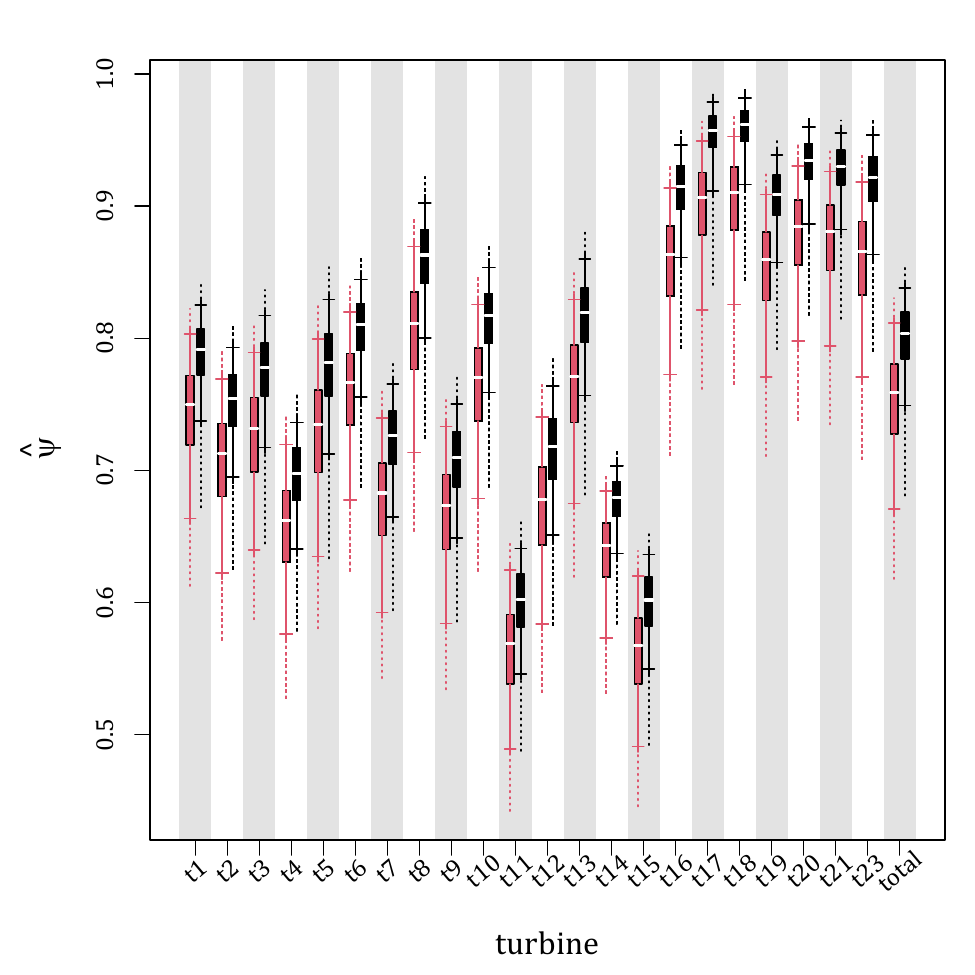}
\caption{\label{fig:psi0102}Comparison of box plots of the estimated
probabilities, \(\hat{\psi}\), of a carcass falling in the searched area
according to the xep01 (black) and xep02 (red) models. Boxes represent
the IQR with median. Whiskers mark the 95\% and 99\% CIs.}
\end{figure}

The \(\hat{\psi}\) values for \texttt{xep01} and \texttt{xep02} varied
by a few percent (fig. \ref{fig:psi0102}), although the 0.005 quantiles
sometimes differed from one another by more than 10\%.

\hypertarget{the-estimated-fraction-of-carcasses-lying-in-searched-areas-dwp}{%
\subsubsection{\texorpdfstring{The Estimated Fraction of Carcasses Lying
in Searched Areas
(\texttt{dwp})}{The Estimated Fraction of Carcasses Lying in Searched Areas (dwp)}}\label{the-estimated-fraction-of-carcasses-lying-in-searched-areas-dwp}}

The \emph{expected} fraction of carcasses lying in the searched area
(that is, the mean \(\widehat{dwp}\)) is equal to the probability of a
carcass lying in the searched area (\(\psi\)), but the uncertainty in
the actual fraction of carcasses lying within the searched area
(\(\widehat{dwp}\)) is greater than the uncertainty in the expected
fraction (\(\hat{\psi}\)). The situation is analogous to a coin toss. We
know the probability of heads is 50\%, and the expected fraction of
times a coin comes up heads in \(n\) coin tosses is \(n/2\). However,
the actual fraction in a sequence of \(n\) flips will usually not be
exactly \(n/2\) but a little more or a little less. That additional,
binomial uncertainty needs to be captured in our estimate of
\texttt{dwp} (Maurer et al.~2020). The distinction between \(\psi\) and
\(dwp\) is discussed further in appendix \ref{app:psivdwp}.

The uncertainties in estimating \(\psi\) are captured in
\texttt{estpsi}, and the uncertainties in estimating \texttt{dwp} given
\(\psi\) are captured using the \texttt{estdwp} function. The function
requires the estimated \(\psi\) and the number of carcasses observed at
each turbine. The number of carcasses is required because the fewer the
carcasses, the greater the relative uncertainty. Again, the coin toss
analogy is illustrative. It would not be surprising to get anything from
0\% to 100\% heads in 4 coin tosses, but it be highly unusual to get
less than 35\% or more than 65\% heads in 40 coin tosses.

The fraction of carcasses lying in the searched area at each turbine is
calculated using \texttt{estdwp}:

\begin{Shaded}
\begin{Highlighting}[]
\NormalTok{dwp01 }\OtherTok{\textless{}{-}} \FunctionTok{estdwp}\NormalTok{(psi01, }\AttributeTok{ncarc =} \FunctionTok{getncarc}\NormalTok{(rings\_shape))}
\end{Highlighting}
\end{Shaded}

The structure of the \texttt{dwp} estimate is the same as that of
\(\hat{\psi}\), namely, a matrix with a column for each turbine,
incorporating the uncertainties in the estimated \texttt{dwp}'s for each
turbine and for the total at the site.

\begin{figure}
\centering
\includegraphics{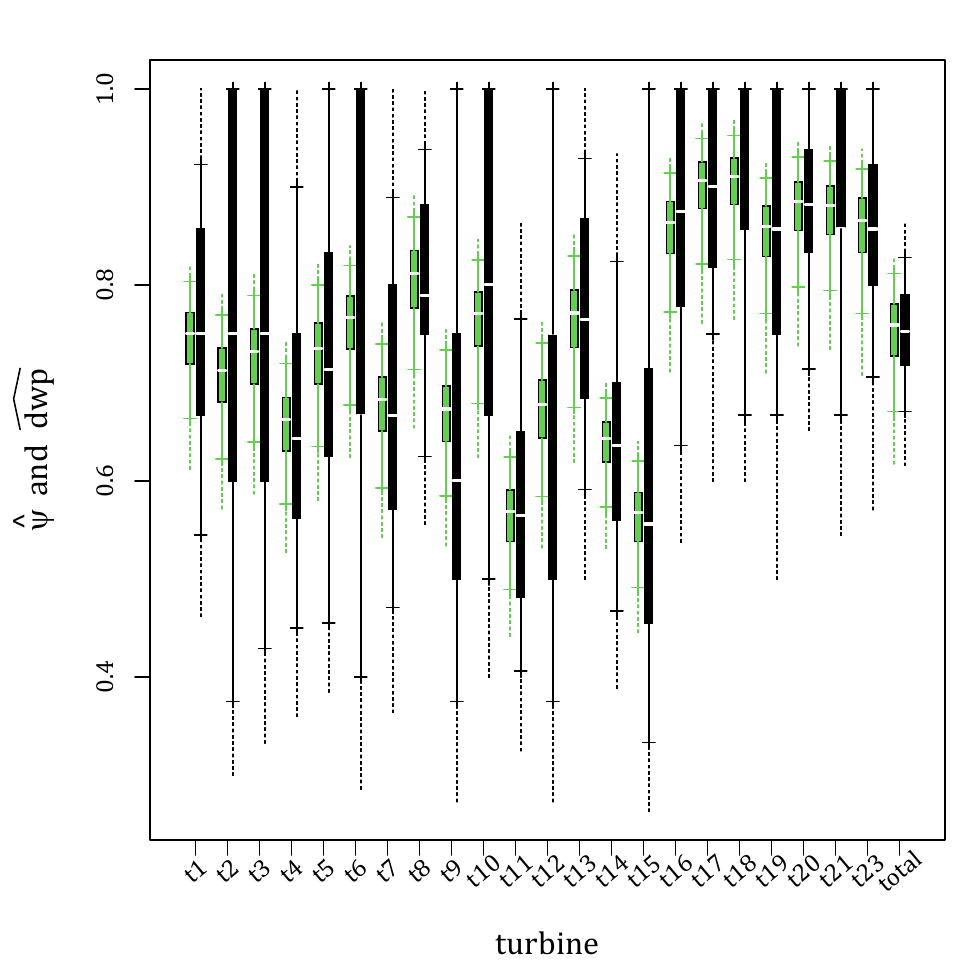}
\caption{\label{fig:psidwp}Comparison of box plots of the estimated
probabilities, \(\hat{\psi}\), of a carcass falling in the searched area
(green) versus the estimated dwp (black) according to the xep01 model.
Boxes represent the IQR with median. Whiskers mark the 95\% and 99\%
CIs.}
\end{figure}

Paired boxplots of \(\hat{\psi}\) and estimated \texttt{dwp} clearly
reveal the greater uncertainty in estimating \texttt{dwp} compared to
\(\psi\) (fig.~\ref{fig:psidwp}).

The estimated \texttt{dwp} can be formatted for use in GenEst either by
setting \texttt{forGenEst\ =\ TRUE} in the argument list for
\texttt{estdwp}, or by using the \texttt{formatGenEst} function on the
estimated \texttt{dwp}.

\begin{Shaded}
\begin{Highlighting}[]
\CommentTok{\# Option 1: format estimated dwp for GenEst in two steps}
\FunctionTok{set.seed}\NormalTok{(}\DecValTok{1942}\NormalTok{) }\CommentTok{\# for repeatability}
\NormalTok{dwp01 }\OtherTok{\textless{}{-}} \FunctionTok{estdwp}\NormalTok{(psi01, }\AttributeTok{ncarc =} \FunctionTok{getncarc}\NormalTok{(rings\_shape))}
\NormalTok{dwp01 }\OtherTok{\textless{}{-}} \FunctionTok{formatGenEst}\NormalTok{(dwp01)}
\FunctionTok{head}\NormalTok{(dwp01)}
\CommentTok{\#\textgreater{}   turbine   dwp}
\CommentTok{\#\textgreater{} 1      t1 0.750}
\CommentTok{\#\textgreater{} 2      t2 0.600}
\CommentTok{\#\textgreater{} 3      t3 0.500}
\CommentTok{\#\textgreater{} 4      t4 0.692}
\CommentTok{\#\textgreater{} 5      t5 1.000}
\CommentTok{\#\textgreater{} 6      t6 1.000}

\CommentTok{\# Option 2: format estimated dwp for GenEst in one step}
\FunctionTok{set.seed}\NormalTok{(}\DecValTok{1942}\NormalTok{)}
\NormalTok{dwp01 }\OtherTok{\textless{}{-}} \FunctionTok{estdwp}\NormalTok{(psi01, }\AttributeTok{ncarc =} \FunctionTok{getncarc}\NormalTok{(rings\_shape), }\AttributeTok{forGenEst =} \ConstantTok{TRUE}\NormalTok{)}
\FunctionTok{head}\NormalTok{(dwp01)}
\CommentTok{\#\textgreater{}   turbine   dwp}
\CommentTok{\#\textgreater{} 1      t1 0.750}
\CommentTok{\#\textgreater{} 2      t2 0.600}
\CommentTok{\#\textgreater{} 3      t3 0.500}
\CommentTok{\#\textgreater{} 4      t4 0.692}
\CommentTok{\#\textgreater{} 5      t5 1.000}
\CommentTok{\#\textgreater{} 6      t6 1.000}
\end{Highlighting}
\end{Shaded}

The formatted \texttt{dwp01} can now be used as
\texttt{data\_dwp\ =\ dwp01} in the argument list in a command line call
to the \texttt{estM} function in GenEst (assuming that the other
required arguments can be supplied as well) or can be exported to a
\texttt{.csv} file for use in the GenEst GUI using the
\texttt{exportGenEst} function.

\begin{Shaded}
\begin{Highlighting}[]
\FunctionTok{exportGenEst}\NormalTok{(dwp01, }\AttributeTok{file =} \StringTok{"my\_dwp.csv"}\NormalTok{)}
\end{Highlighting}
\end{Shaded}

\hypertarget{additional-analysis-for-an-interacting-covariate}{%
\subsubsection{\texorpdfstring{Additional Analysis for an Interacting
Covariate
\label{sec:curtail}}{Additional Analysis for an Interacting Covariate }}\label{additional-analysis-for-an-interacting-covariate}}

In the previous example, we included a covariate to account for
variability in carcass detection probabilities on the ground.
Interacting covariates---that is, variables that affect the actual
distributions of carcasses rather than simply affecting detection
probabilities on the ground---present challenges that are difficult to
overcome in a generic way that would be appropriate for automation in
\texttt{dwp}.

The Casselman data provide an example of an interacting covariate, which
can be analyzed by fitting separate sets of models for different
covariate levels. In particular, the ``shapes'' site has turbines that
were under two different operating protocols, freely operating or
curtailed at low windspeeds. We would expect that the carcasses at the
curtailed turbines would be distributed further from the turbines than
carcasses at the non-curtailed turbines because the curtailed turbines
were operating only under relatively high wind speeds. We can analyze
the data from the two turbine sets separately to see if there are
differences in the distance distributions and consequent \(\hat{\psi}\).

The \texttt{subset} function from the \texttt{dwp} package can be used
to split the \texttt{layout\_shape} data set into two parts: freely
operating turbines (\texttt{layout\_free}) and turbines curtailed at low
wind speeds (\texttt{layout\_curtailed}). The two data sets can then be
analyzed separately and the results compared:

\begin{Shaded}
\begin{Highlighting}[]

\DocumentationTok{\#\#\#\#\#\#\#\# basic setup for freely operating turbines:}
\CommentTok{\# data formatting}
\NormalTok{layout\_free }\OtherTok{\textless{}{-}} \FunctionTok{subset}\NormalTok{(layout\_shape, }\AttributeTok{select =} \StringTok{"Turbine"}\NormalTok{,}
  \AttributeTok{subset =} \FunctionTok{paste0}\NormalTok{(}\StringTok{"t"}\NormalTok{, }\FunctionTok{c}\NormalTok{(}\DecValTok{1}\NormalTok{, }\DecValTok{3}\NormalTok{, }\DecValTok{4}\NormalTok{, }\DecValTok{8}\NormalTok{, }\DecValTok{11}\NormalTok{, }\DecValTok{13}\NormalTok{, }\DecValTok{14}\NormalTok{, }\DecValTok{16}\NormalTok{, }\DecValTok{20}\NormalTok{, }\DecValTok{23}\NormalTok{)))}
\NormalTok{cod\_free }\OtherTok{\textless{}{-}} \FunctionTok{subset}\NormalTok{(cod, }\AttributeTok{select =} \StringTok{"Turbine"}\NormalTok{,}
  \AttributeTok{subset =} \FunctionTok{paste0}\NormalTok{(}\StringTok{"t"}\NormalTok{, }\FunctionTok{c}\NormalTok{(}\DecValTok{1}\NormalTok{, }\DecValTok{3}\NormalTok{, }\DecValTok{4}\NormalTok{, }\DecValTok{8}\NormalTok{, }\DecValTok{11}\NormalTok{, }\DecValTok{13}\NormalTok{, }\DecValTok{14}\NormalTok{, }\DecValTok{16}\NormalTok{, }\DecValTok{20}\NormalTok{, }\DecValTok{23}\NormalTok{)))}
\NormalTok{rings\_free }\OtherTok{\textless{}{-}} \FunctionTok{prepRing}\NormalTok{(layout\_free, }\AttributeTok{scVar =} \StringTok{"Class"}\NormalTok{, }\AttributeTok{notSearched =} \StringTok{"Out"}\NormalTok{, }\AttributeTok{silent =}\NormalTok{ T)}
\NormalTok{rings\_free }\OtherTok{\textless{}{-}} \FunctionTok{addCarcass}\NormalTok{(cod\_free, }\AttributeTok{data\_ring =}\NormalTok{ rings\_free, }\AttributeTok{plotLayout =}\NormalTok{ layout\_free)}

\CommentTok{\# fitting the models}
\NormalTok{mod\_free }\OtherTok{\textless{}{-}} \FunctionTok{ddFit}\NormalTok{(rings\_free, }\AttributeTok{scVar =} \StringTok{"Class"}\NormalTok{)}
\CommentTok{\#\textgreater{} Extensible models:}
\CommentTok{\#\textgreater{}   xep1 }
\CommentTok{\#\textgreater{}   xep01 }
\CommentTok{\#\textgreater{}   xep2 }
\CommentTok{\#\textgreater{}   xep02 }
\CommentTok{\#\textgreater{}   xep12 }
\CommentTok{\#\textgreater{}   tnormal }
\CommentTok{\#\textgreater{}   MaxwellBoltzmann }
\CommentTok{\#\textgreater{}   lognormal }
\CommentTok{\#\textgreater{} }
\CommentTok{\#\textgreater{} Non{-}extensible models:}
\CommentTok{\#\textgreater{}   xep012 }
\CommentTok{\#\textgreater{}   xep123 }
\CommentTok{\#\textgreater{}   xep0123 }
\CommentTok{\#\textgreater{}   constant}

\DocumentationTok{\#\#\#\#\#\#\#\# basic setup for curtailed turbines:}
\NormalTok{layout\_curtailed }\OtherTok{\textless{}{-}} \FunctionTok{subset}\NormalTok{(layout\_shape, }\AttributeTok{select =} \StringTok{"Turbine"}\NormalTok{,}
  \AttributeTok{subset =} \FunctionTok{paste0}\NormalTok{(}\StringTok{"t"}\NormalTok{, }\FunctionTok{c}\NormalTok{(}\DecValTok{2}\NormalTok{, }\DecValTok{5}\NormalTok{, }\DecValTok{6}\NormalTok{, }\DecValTok{7}\NormalTok{, }\DecValTok{9}\NormalTok{, }\DecValTok{10}\NormalTok{, }\DecValTok{12}\NormalTok{, }\DecValTok{15}\NormalTok{, }\DecValTok{17}\NormalTok{, }\DecValTok{18}\NormalTok{, }\DecValTok{19}\NormalTok{, }\DecValTok{21}\NormalTok{)))}
\NormalTok{cod\_curtailed }\OtherTok{\textless{}{-}} \FunctionTok{subset}\NormalTok{(cod, }\AttributeTok{select =} \StringTok{"Turbine"}\NormalTok{,}
  \AttributeTok{subset =} \FunctionTok{paste0}\NormalTok{(}\StringTok{"t"}\NormalTok{, }\FunctionTok{c}\NormalTok{(}\DecValTok{2}\NormalTok{, }\DecValTok{5}\NormalTok{, }\DecValTok{6}\NormalTok{, }\DecValTok{7}\NormalTok{, }\DecValTok{9}\NormalTok{, }\DecValTok{10}\NormalTok{, }\DecValTok{12}\NormalTok{, }\DecValTok{15}\NormalTok{, }\DecValTok{17}\NormalTok{, }\DecValTok{18}\NormalTok{, }\DecValTok{19}\NormalTok{, }\DecValTok{21}\NormalTok{)))}
\NormalTok{rings\_curtailed }\OtherTok{\textless{}{-}} \FunctionTok{prepRing}\NormalTok{(layout\_curtailed, }\AttributeTok{scVar =} \StringTok{"Class"}\NormalTok{, }\AttributeTok{notSearched =} \StringTok{"Out"}\NormalTok{,}
  \AttributeTok{silent =} \ConstantTok{TRUE}\NormalTok{) }\CommentTok{\# "silent" cancels updates on the status of the calculations}
\NormalTok{rings\_curtailed }\OtherTok{\textless{}{-}} \FunctionTok{addCarcass}\NormalTok{(cod\_curtailed, }\AttributeTok{data\_ring =}\NormalTok{ rings\_curtailed,}
  \AttributeTok{plotLayout =}\NormalTok{ layout\_curtailed)}
\NormalTok{mod\_curtailed }\OtherTok{\textless{}{-}} \FunctionTok{ddFit}\NormalTok{(rings\_curtailed, }\AttributeTok{scVar =} \StringTok{"Class"}\NormalTok{)}
\CommentTok{\#\textgreater{} Extensible models:}
\CommentTok{\#\textgreater{}   xep1 }
\CommentTok{\#\textgreater{}   xep01 }
\CommentTok{\#\textgreater{}   xep2 }
\CommentTok{\#\textgreater{}   xep02 }
\CommentTok{\#\textgreater{}   xep12 }
\CommentTok{\#\textgreater{}   xep012 }
\CommentTok{\#\textgreater{}   tnormal }
\CommentTok{\#\textgreater{}   MaxwellBoltzmann }
\CommentTok{\#\textgreater{}   lognormal }
\CommentTok{\#\textgreater{} }
\CommentTok{\#\textgreater{} Non{-}extensible models:}
\CommentTok{\#\textgreater{}   xep123 }
\CommentTok{\#\textgreater{}   xep0123 }
\CommentTok{\#\textgreater{}   constant}
\end{Highlighting}
\end{Shaded}

The descriptive statistics for the fitted models can be summarized using
the function \texttt{stats}, which calculates estimated medians, modes,
and the 75th, 90th, and 95th percentiles according to models, and sorts
the results by \(\Delta\textrm{AICc}\).

\begin{Shaded}
\begin{Highlighting}[]
\FunctionTok{stats}\NormalTok{(mod\_free)}
\CommentTok{\#\textgreater{}                  median  75\%  90\%   95\% mode p\_win deltaAICc}
\CommentTok{\#\textgreater{} xep01              30.6 44.3 59.5  70.1 23.5 0.985      0.00}
\CommentTok{\#\textgreater{} xep2               30.6 43.3 55.9  63.7 26.0 0.996      0.64}
\CommentTok{\#\textgreater{} xep02              30.4 41.9 53.2  60.2 26.8 0.999      1.51}
\CommentTok{\#\textgreater{} xep12              30.5 42.3 53.6  60.7 27.0 0.998      2.35}
\CommentTok{\#\textgreater{} lognormal          35.6 59.1 93.2 122.5 20.3 0.884      2.51}
\CommentTok{\#\textgreater{} MaxwellBoltzmann   30.1 39.7 49.0  54.8 27.7 1.000      5.10}
\CommentTok{\#\textgreater{} xep1               36.8 59.1 85.4 104.1 21.9 0.907      6.41}
\CommentTok{\#\textgreater{} tnormal            30.8 41.3 50.8  56.5 30.3 1.000      9.61}
\FunctionTok{stats}\NormalTok{(mod\_curtailed)}
\CommentTok{\#\textgreater{}                  median   75\%   90\%   95\% mode p\_win deltaAICc}
\CommentTok{\#\textgreater{} MaxwellBoltzmann   35.0  46.1  56.9  63.6 32.2 0.998      0.00}
\CommentTok{\#\textgreater{} xep12              35.7  47.2  58.1  64.7 33.6 0.997      1.02}
\CommentTok{\#\textgreater{} xep02              36.0  48.6  60.9  68.6 32.3 0.993      1.06}
\CommentTok{\#\textgreater{} xep2               38.6  54.5  70.3  80.2 32.8 0.969      1.09}
\CommentTok{\#\textgreater{} xep01              38.8  55.9  74.9  87.9 30.1 0.946      1.95}
\CommentTok{\#\textgreater{} tnormal            35.5  45.7  55.0  60.5 35.3 1.000      2.66}
\CommentTok{\#\textgreater{} xep012             35.7  47.6  59.0  66.0 33.1 0.996      3.06}
\CommentTok{\#\textgreater{} xep1               60.1  96.5 139.4 170.0 35.8 0.695      4.47}
\CommentTok{\#\textgreater{} lognormal          64.2 116.5 199.2 274.6 29.4 0.632      4.93}
\end{Highlighting}
\end{Shaded}

\begin{figure}
\centering
\includegraphics{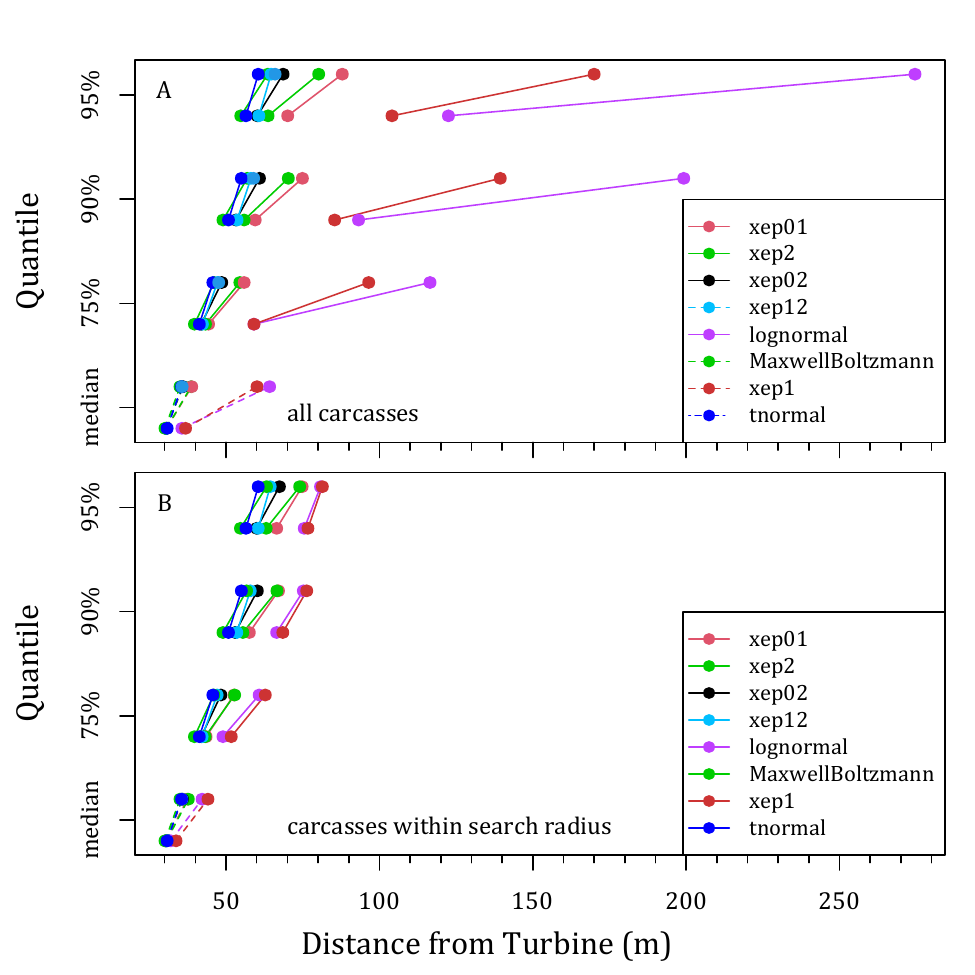}
\caption{\label{fig:fcstat}Comparison of descriptive statistics for the
distance distributions for the curtailed turbines (top row of dots in
each quantile) and the curtailed turbines (bottom row of dots in each
quantile). The lines show the difference between curtailed and free
turbines. A. Estimated quantiles of carcass distances (including
extrapolations to account for carcasses beyond the search radius); B.
estimated quantiles of carcass distances solely within the search
radius.}
\end{figure}

\begin{figure}
\centering
\includegraphics{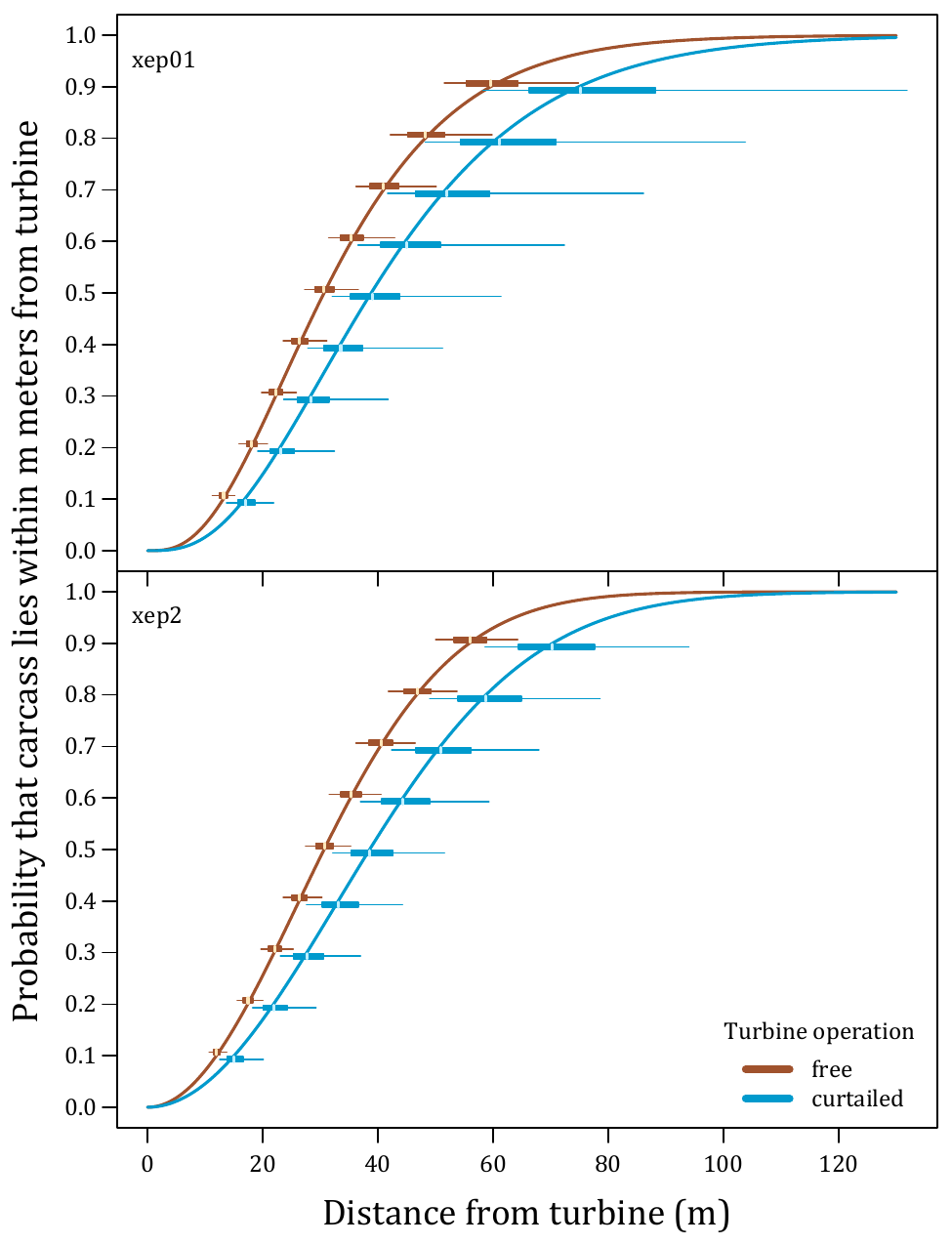}
\caption{\label{fig:cfCDF}Cumulative xep01 and xep2 distribution
functions of carcass distances at curtailed and free turbines at the
Casselman site. Curves show the maximum likelihood esimator. Boxplots
show the interquartile ranges, medians, and 90\% CIs for estimated
distances required to cover 10\%, 20\%, \ldots, 90\% of the carcasses.}
\end{figure}

In general, the carcasses at the curtailed turbines did tend to be
somewhat more distant from turbines than the carcasses at the freely
operating turbines as expected (fig.~\ref{fig:cfCDF}).. The magnitude of
the estimated differences strongly depended on the selected model
(fig.~\ref{fig:fcstat}A).The differences among the models largely vanish
when the models are restricted solely to the range of the data and not
extrapolated beyond the search radius (fig.~\ref{fig:fcstat}B). This
indicates that the discrepancies among models are primarily due to the
extrapolation beyond the search radius and the models' respective
assumptions about the distribution beyond the search radius, that is, in
the tails of the distributions. The heavy-tailed distributions
(lognormal and xep1), predict a relatively large fraction of carcasses
beyond the search radius, while the light-tailed distributions
(truncated normal and Maxwell-Boltzmann) predict relatively few
carcasses beyond the search radius.

The lighter-tailed distributions (Maxwell-Boltzmann and the xep models
with degree 2 or 3) tend to converge rapidly to zero shortly beyond the
most distant carcass and underestimate the total outside the searched
area in comparison with the heavier-tailed distributions (lognormal and
xep models with degree less than 2). For these data, the top-fitting
light-tailed distributions (xep2, xep02, xep12) have similar shapes, and
there is little to distinguish the models. In both the curtailed and
free data sets, the xep01 (gamma distribution) was the highest-ranking,
relatively heavy-tailed distribution and had \(\Delta AICc < 2\). Thus,
the xep01 distribution is an appropriate choice for both types of
turbine, and we compare it with the lighter-tailed xep2 model, which is
also an appropriate choice for both types of turbine.

The estimated probabilities of carcasses lying in the searched areas can
be calculated separately for the two sets of turbines (freely operating
or curtailed):

\begin{Shaded}
\begin{Highlighting}[]
\CommentTok{\# estimated spatial coverage for curtailed turbines...}
\CommentTok{\# ...using distance model fit to data from free turbines}
\NormalTok{psi\_free }\OtherTok{\textless{}{-}} \FunctionTok{estpsi}\NormalTok{(rings\_free, }\AttributeTok{model =}\NormalTok{ mod\_free[}\StringTok{"xep01"}\NormalTok{])}

\CommentTok{\# ...using distance models fit to data from curtailed turbines}
\NormalTok{psi\_curtailed }\OtherTok{\textless{}{-}} \FunctionTok{estpsi}\NormalTok{(rings\_curtailed, }\AttributeTok{model =}\NormalTok{ mod\_curtailed[}\StringTok{"xep01"}\NormalTok{])}
\end{Highlighting}
\end{Shaded}

\begin{figure}
\centering
\includegraphics{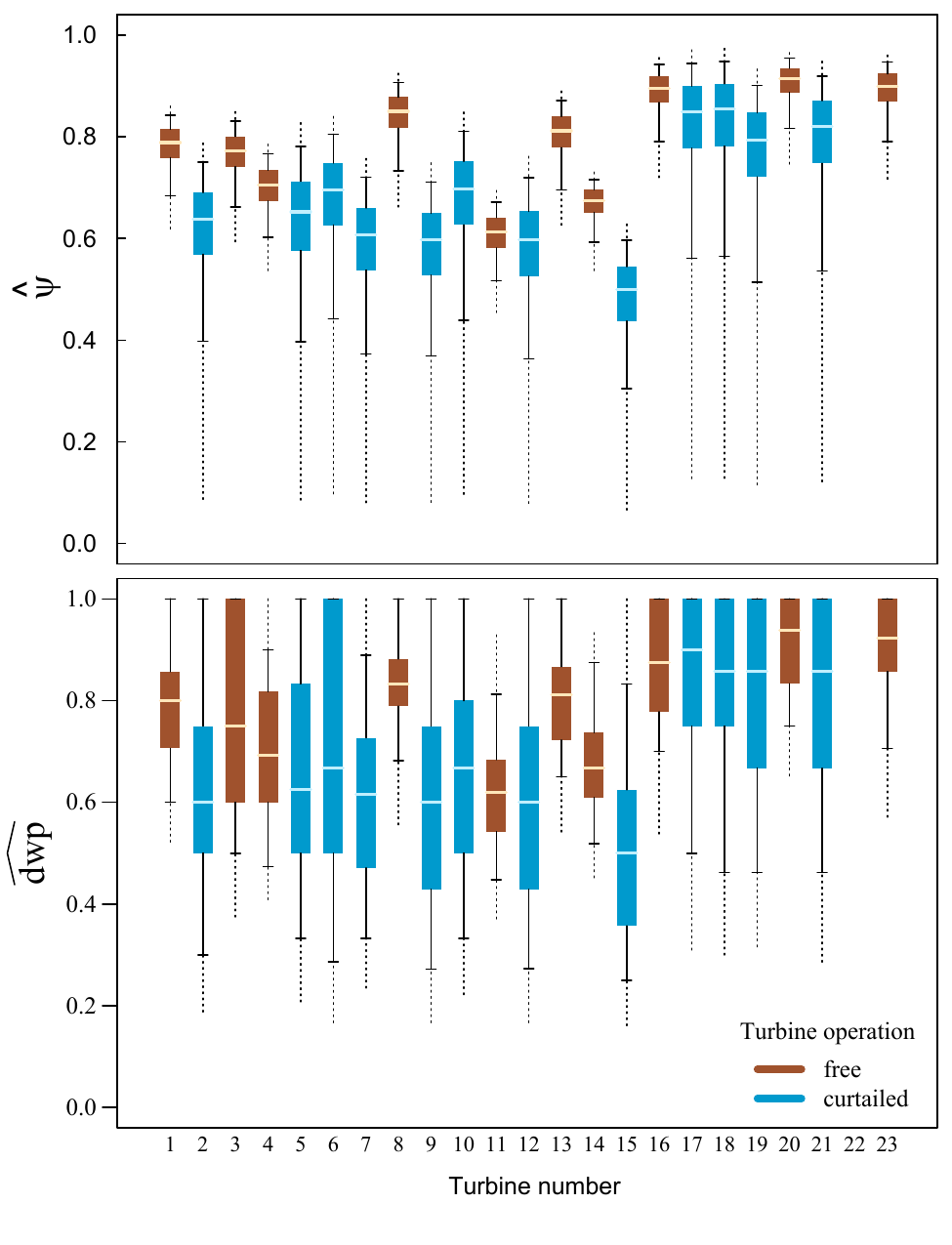}
\caption{\label{fig:cfbox}Boxplots of \(\hat{\psi}\) and
\(\widehat{dwp}\) according to the xep01 distribution for curtailed and
free turbines at the Casselman site.}
\end{figure}

The estimated \(\psi\) and \(dwp\) for the free and curtailed turbines
depends on the configuration of the search plots, which vary by turbine,
so the values are not strictly comparable. However, a few meaningful
patterns can be discerned (fig.~\ref{fig:cfbox}). First, the lower tails
of the distributions for \(\hat{\psi}\) extend much lower for the
curtailed turbines than for the free turbines. This is due to the
presence of more carcasses at greater distances at the curtailed
turbines and a smaller probability that the tail of the distribution is
resolved enough to avoid the possibility that many carcasses lie beyond
the search radius. Next, the uncertainty introduced in estimating
\(dwp\) erases the clarity of the distinction between curtailed and free
turbines.

After calculating the \(\widehat{dwp}\)'s for curtailed and free
turbines separately, the estimates combined into one data frame for
export for GenEst.

\begin{Shaded}
\begin{Highlighting}[]
\NormalTok{ge\_combined }\OtherTok{\textless{}{-}} \FunctionTok{rbind}\NormalTok{(}\FunctionTok{formatGenEst}\NormalTok{(dwp\_curtailed), }\FunctionTok{formatGenEst}\NormalTok{(dwp\_free))}
\FunctionTok{exportGenEst}\NormalTok{(ge\_combined, }\AttributeTok{file =} \StringTok{"dwp\_all.csv"}\NormalTok{)}
\end{Highlighting}
\end{Shaded}

The Rayleigh distribution (xep2) provides an equally high-quality fit
compared to the gamma (xep01) and has the interesting and convenient
property that the ratio of the quantiles of two Rayleighs is equal to
the ratio of their parameters (\(\sigma_1/\sigma_2\)), regardless of the
quantile. Thus, measuring by the fitted Rayleigh distribution, carcasses
at the curtailed turbines were
\(\sigma_\textrm{curtailed}/\sigma_\textrm{free} = 1.258\) times more
distant from the turbines than were the carcasses at the free turbines.
In other words, the median, 90th percentile, 99th percentile, and so on
were all 25.8\% greater at the curtailed turbines than at the free
turbines.

\begin{Shaded}
\begin{Highlighting}[]
\CommentTok{\# estimated spatial coverage for curtailed turbines...}
\CommentTok{\# ...using distance model fit to data from free turbines}
\NormalTok{psi\_free }\OtherTok{\textless{}{-}} \FunctionTok{estpsi}\NormalTok{(rings\_curtailed, }\AttributeTok{model =}\NormalTok{ mod\_free[}\StringTok{"xep01"}\NormalTok{])}

\CommentTok{\# ...using distance models fit to data from curtailed turbines}
\NormalTok{psi\_curtailed }\OtherTok{\textless{}{-}} \FunctionTok{estpsi}\NormalTok{(rings\_curtailed, }\AttributeTok{model =}\NormalTok{ mod\_curtailed[}\StringTok{"xep01"}\NormalTok{])}
\end{Highlighting}
\end{Shaded}

\begin{figure}
\centering
\includegraphics{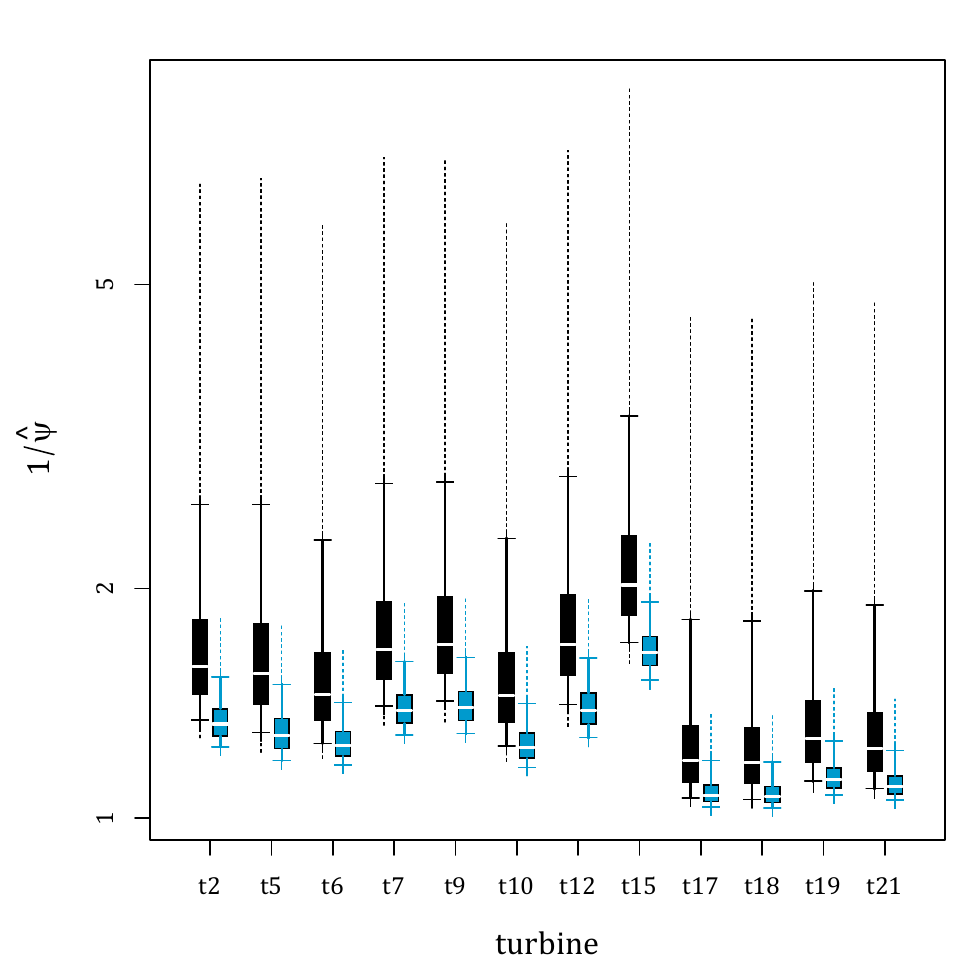}
\caption{\label{fig:fcpsi}Comparison of estimates of the area correction
factor (\(1/\hat{\psi}\)) at curtailed turbines using the correct
distance model (black) based on data from curtailed turbines versus
using an incorrect distance model (blue) derived from data from freely
operating turbines. Boxes show median and IQR. Whiskers show 95\% and
99\% CIs.}
\end{figure}

If not accounted for, the tendency for the carcasses to be more distant
at the curtailed turbines can result in substantial bias when estimating
total mortality. For example, using the distribution that was fit for
the freely operating turbines to estimate mortality at the curtailed
turbines, we would underestimate the mortality by some 40\% on average
at the curtailed turbines due to systematic bias in the estimates of
\(\psi\) (fig.~\ref{fig:fcpsi}).

\hypertarget{carcass-size}{%
\subsubsection{\texorpdfstring{Carcass Size
\label{sec:size}}{Carcass Size }}\label{carcass-size}}

Carcass size may also influence the distances that carcasses may fly,
perhaps even size differences among bats can influence the carcass
dispersion patterns around turbines. The Casselman data include a range
of bat species of different sizes, which we can class as either large
(\texttt{L}) or small (\texttt{S}). The \texttt{dwp} package provides
tools for processing data sets with size class distinctions and
exporting the resulting \texttt{dwp} estimates to GenEst.

The following sections of code show how to split a data set on a carcass
class variable, fit separate distance models for each size class using a
single command, and create GenEst-ready \texttt{dwp} data with carcass
class distinctions.

The first task with this data set is to add a ``Size'' variable to the
carcass data from the freely operating turbines, with LACI, LASE, and
EPFU classed as ``L'' and the other species as small.

\begin{Shaded}
\begin{Highlighting}[]
\NormalTok{cod\_free}\SpecialCharTok{$}\NormalTok{carcasses}\SpecialCharTok{$}\NormalTok{Size }\OtherTok{\textless{}{-}} \FunctionTok{ifelse}\NormalTok{(}
\NormalTok{  cod\_free}\SpecialCharTok{$}\NormalTok{carcasses}\SpecialCharTok{$}\NormalTok{Species }\SpecialCharTok{\%in\%} \FunctionTok{c}\NormalTok{(}\StringTok{"LACI"}\NormalTok{, }\StringTok{"LASE"}\NormalTok{, }\StringTok{"EPFU"}\NormalTok{), }\StringTok{"L"}\NormalTok{, }\StringTok{"S"}\NormalTok{)}
\end{Highlighting}
\end{Shaded}

It is always useful to check whether your command has the desired
effect. Tabulate the carcasses by size and species to verify that the
split was done correctly and that there are sufficient carcasses in each
subset to do the estimation.

\begin{Shaded}
\begin{Highlighting}[]
\FunctionTok{table}\NormalTok{(cod\_free}\SpecialCharTok{$}\NormalTok{carcasses[, }\FunctionTok{c}\NormalTok{(}\StringTok{"Species"}\NormalTok{, }\StringTok{"Size"}\NormalTok{), }\AttributeTok{drop =} \ConstantTok{TRUE}\NormalTok{])}
\CommentTok{\#\textgreater{}        Size}
\CommentTok{\#\textgreater{} Species  L  S}
\CommentTok{\#\textgreater{}    EPFU  3  0}
\CommentTok{\#\textgreater{}    LABO  0 15}
\CommentTok{\#\textgreater{}    LACI 35  0}
\CommentTok{\#\textgreater{}    LANO  0 28}
\CommentTok{\#\textgreater{}    LASE  2  0}
\CommentTok{\#\textgreater{}    MYLU  0 18}
\CommentTok{\#\textgreater{}    PESU  0 12}
\end{Highlighting}
\end{Shaded}

NOTE: The \texttt{drop\ =\ TRUE} is necessary here because cod\_free is
an \texttt{sf} object, which is a specially formatted data frame with
GIS information attached. When that extra information is attached, R
does not recognize the sf object as a data frame, so functions like
\texttt{table} will not work. The GIS information is dropped by using
\texttt{drop\ =\ TRUE} in the subsetting. The result is a standard R
data frame that can be manipulated like any other data frame in R.

To estimate \(\psi\) and calculate \(\widehat{dwp}\) for use in GenEst
when there are distinctions among carcass classes, the \(dwp\)'s must be
calculated for each size class separately, which the \texttt{dwp}
functions, \texttt{estpsi} and \texttt{estdwp}, can do automatically if
the data are properly formatted. Formatting involves adding carcasses
(feeding a carcass data set into \texttt{addCarcass}) to the description
of the site by rings (\texttt{data\_ring}) with a directive to split the
data sets by carcass class (\texttt{ccCol}).

When a \texttt{ccCol} (carcass class column) is included in the argument
list, addCarcass creates separate ring structures for each carcass
class, which in turn are interpreted by \texttt{dwp} functions to
estimate \(\psi\), \(dwp\), and format the results for exporting to
GenEst.

\begin{Shaded}
\begin{Highlighting}[]
\CommentTok{\# split the ring data into separate structures for each size class}
\NormalTok{cod\_free\_size }\OtherTok{\textless{}{-}} \FunctionTok{addCarcass}\NormalTok{(cod\_free, }\AttributeTok{data\_ring =}\NormalTok{ rings\_free,}
  \AttributeTok{plotLayout =}\NormalTok{ layout\_free, }\AttributeTok{ccCol =} \StringTok{"Size"}\NormalTok{)}

\CommentTok{\# fit separate distance models to the different size classes:}
\NormalTok{mod\_size }\OtherTok{\textless{}{-}} \FunctionTok{ddFit}\NormalTok{(cod\_free\_size, }\AttributeTok{scCol =} \StringTok{"Class"}\NormalTok{)}
\CommentTok{\#\textgreater{} Extensible models:}
\CommentTok{\#\textgreater{}   xep1 }
\CommentTok{\#\textgreater{}   xep01 }
\CommentTok{\#\textgreater{}   xep2 }
\CommentTok{\#\textgreater{}   xep02 }
\CommentTok{\#\textgreater{}   xep12 }
\CommentTok{\#\textgreater{}   xep012 }
\CommentTok{\#\textgreater{}   xep123 }
\CommentTok{\#\textgreater{}   xep0123 }
\CommentTok{\#\textgreater{}   tnormal }
\CommentTok{\#\textgreater{}   MaxwellBoltzmann }
\CommentTok{\#\textgreater{}   lognormal }
\CommentTok{\#\textgreater{} }
\CommentTok{\#\textgreater{} Non{-}extensible models:}
\CommentTok{\#\textgreater{}   constant }
\CommentTok{\#\textgreater{} Extensible models:}
\CommentTok{\#\textgreater{}   xep1 }
\CommentTok{\#\textgreater{}   xep01 }
\CommentTok{\#\textgreater{}   xep2 }
\CommentTok{\#\textgreater{}   xep02 }
\CommentTok{\#\textgreater{}   xep12 }
\CommentTok{\#\textgreater{}   tnormal }
\CommentTok{\#\textgreater{}   MaxwellBoltzmann }
\CommentTok{\#\textgreater{}   lognormal }
\CommentTok{\#\textgreater{} }
\CommentTok{\#\textgreater{} Non{-}extensible models:}
\CommentTok{\#\textgreater{}   xep012 }
\CommentTok{\#\textgreater{}   xep123 }
\CommentTok{\#\textgreater{}   xep0123 }
\CommentTok{\#\textgreater{}   constant}

\CommentTok{\# view summary statistics for the resulting models:}
\FunctionTok{stats}\NormalTok{(mod\_size)}
\CommentTok{\#\textgreater{} $L}
\CommentTok{\#\textgreater{}                  median  75\%   90\%   95\% mode p\_win deltaAICc}
\CommentTok{\#\textgreater{} xep2               28.4 40.1  51.7  59.0 24.1 0.999      0.00}
\CommentTok{\#\textgreater{} xep1               30.8 49.4  71.4  87.1 18.4 0.951      0.22}
\CommentTok{\#\textgreater{} xep12              28.4 41.8  55.5  64.4 21.8 0.994      1.66}
\CommentTok{\#\textgreater{} xep01              29.0 44.4  62.1  74.6 19.7 0.976      1.72}
\CommentTok{\#\textgreater{} xep02              28.4 41.0  53.5  61.4 23.3 0.997      1.78}
\CommentTok{\#\textgreater{} lognormal          34.7 62.8 107.0 147.3 16.0 0.854      1.98}
\CommentTok{\#\textgreater{} tnormal            28.7 40.4  51.2  57.7 27.1 1.000      3.26}
\CommentTok{\#\textgreater{} xep123             28.3 40.7  51.6  57.9 23.9 1.000      3.44}
\CommentTok{\#\textgreater{} xep012             28.4 42.2  56.7  66.2 21.1 0.992      3.69}
\CommentTok{\#\textgreater{} xep0123            28.1 41.0  50.7  55.6 13.5 1.000      4.51}
\CommentTok{\#\textgreater{} MaxwellBoltzmann   28.7 37.8  46.7  52.2 26.4 1.000      6.82}
\CommentTok{\#\textgreater{} }
\CommentTok{\#\textgreater{} $S}
\CommentTok{\#\textgreater{}                  median  75\%  90\%   95\% mode p\_win deltaAICc}
\CommentTok{\#\textgreater{} xep2               28.9 40.8 52.6  60.0 24.5 0.998      0.00}
\CommentTok{\#\textgreater{} xep01              28.8 42.1 57.1  67.5 21.6 0.988      0.52}
\CommentTok{\#\textgreater{} xep02              28.9 40.5 51.9  59.1 24.8 0.999      1.97}
\CommentTok{\#\textgreater{} xep12              28.8 41.1 53.3  60.9 24.1 0.998      2.01}
\CommentTok{\#\textgreater{} lognormal          32.1 53.2 83.9 110.1 18.3 0.910      2.04}
\CommentTok{\#\textgreater{} xep1               32.3 51.8 74.9  91.3 19.2 0.941      2.57}
\CommentTok{\#\textgreater{} MaxwellBoltzmann   29.1 38.3 47.3  52.9 26.8 1.000      6.15}
\CommentTok{\#\textgreater{} tnormal            29.2 40.2 50.3  56.3 28.2 1.000      6.63}
\end{Highlighting}
\end{Shaded}

Selection of models to use in the calculations of \(dwp\) must be done
for each size class separately. As with the earlier examples in the
Casselman data (sections \ref{sec:shapefit} and \ref{sec:curtail}), the
xep01 model scores well in the \texttt{modelFilter} for both size
classes:

\begin{Shaded}
\begin{Highlighting}[]
\CommentTok{\# Model filter scores for small carcasses}
\FunctionTok{modelFilter}\NormalTok{(mod\_size[[}\StringTok{"S"}\NormalTok{]])}\SpecialCharTok{$}\NormalTok{scores}
\CommentTok{\#\textgreater{}                  extensible rtail ltail aicc hin  deltaAICc}
\CommentTok{\#\textgreater{} xep2                      1     1     1    1   0  0.0000000}
\CommentTok{\#\textgreater{} xep01                     1     1     1    1   0  0.5211667}
\CommentTok{\#\textgreater{} xep02                     1     1     1    1   0  1.9695832}
\CommentTok{\#\textgreater{} xep12                     1     1     1    1   0  2.0102368}
\CommentTok{\#\textgreater{} lognormal                 1     1     1    1   0  2.0444269}
\CommentTok{\#\textgreater{} xep1                      1     1     1    1   0  2.5680237}
\CommentTok{\#\textgreater{} tnormal                   1     1     1    1   0  6.6354677}
\CommentTok{\#\textgreater{} MaxwellBoltzmann          1     1     0    1   0  6.1499831}
\CommentTok{\#\textgreater{} xep123                    0     0     1    1   0  0.1523155}
\CommentTok{\#\textgreater{} xep0123                   0     0     1    1   0  2.0836871}
\CommentTok{\#\textgreater{} xep012                    0     0     1    1   0  2.4650216}
\CommentTok{\#\textgreater{} constant                  0     0     1    0   0 54.2732536}

\CommentTok{\# Model filter scores for large carcasses}
\FunctionTok{modelFilter}\NormalTok{(mod\_size[[}\StringTok{"L"}\NormalTok{]])}\SpecialCharTok{$}\NormalTok{scores}
\CommentTok{\#\textgreater{}                  extensible rtail ltail aicc hin  deltaAICc}
\CommentTok{\#\textgreater{} xep2                      1     1     1    1   0  0.0000000}
\CommentTok{\#\textgreater{} xep1                      1     1     1    1   0  0.2207812}
\CommentTok{\#\textgreater{} xep12                     1     1     1    1   0  1.6626334}
\CommentTok{\#\textgreater{} xep01                     1     1     1    1   0  1.7142410}
\CommentTok{\#\textgreater{} xep02                     1     1     1    1   0  1.7826877}
\CommentTok{\#\textgreater{} tnormal                   1     1     1    1   0  3.2572103}
\CommentTok{\#\textgreater{} xep123                    1     1     1    1   0  3.4374762}
\CommentTok{\#\textgreater{} xep012                    1     1     1    1   0  3.6869378}
\CommentTok{\#\textgreater{} xep0123                   1     1     1    1   0  4.5038122}
\CommentTok{\#\textgreater{} MaxwellBoltzmann          1     1     0    1   0  6.8218704}
\CommentTok{\#\textgreater{} lognormal                 1     0     1    1   0  1.9813879}
\CommentTok{\#\textgreater{} constant                  0     0     1    0   0 30.2262282}
\end{Highlighting}
\end{Shaded}

The same caveat about the tendency of light-tailed models to
overestimate the \(\psi\)---especially with bats---when the search
radius is not large enough to encompass all the carcasses (and with some
cushion) that is mentioned in sections \ref{sec:exsimple},
\ref{sec:exxy}, and \ref{sec:curtail} and Appendix \ref{app:fit2deq}
applies here, so we again select the xep01 model for both size classes.

There are two ways to call \texttt{estpsi} with data and models that
include carcass class distinctions. They both accomplish the same thing,
namely, to estimate \(\psi\) for each carcass class.

\begin{enumerate}
\def\labelenumi{\arabic{enumi}.}
\tightlist
\item
  Pass the whole array of fitted models (\texttt{mod\_size}, in this
  example) to the \texttt{estpsi} function along with a vector of names
  of which models to select (\texttt{modnames}):
\end{enumerate}

\begin{Shaded}
\begin{Highlighting}[]
\NormalTok{psi\_size }\OtherTok{\textless{}{-}} \FunctionTok{estpsi}\NormalTok{(cod\_free\_size,}
  \AttributeTok{model =}\NormalTok{ mod\_size, }\CommentTok{\# list of models for each carcass class}
  \AttributeTok{modnames =} \FunctionTok{c}\NormalTok{(}\AttributeTok{L =} \StringTok{"xep01"}\NormalTok{, }\AttributeTok{S =} \StringTok{"xep01"}\NormalTok{)) }\CommentTok{\# specify 1 model for each carcass class}
\end{Highlighting}
\end{Shaded}

\begin{enumerate}
\def\labelenumi{\arabic{enumi}.}
\setcounter{enumi}{1}
\tightlist
\item
  Pass a list of models with exactly one model selected for each carcass
  class:
\end{enumerate}

\begin{Shaded}
\begin{Highlighting}[]
\NormalTok{psi\_size }\OtherTok{\textless{}{-}} \FunctionTok{estpsi}\NormalTok{(cod\_free\_size,}
  \AttributeTok{model =} \FunctionTok{list}\NormalTok{(}\AttributeTok{L =}\NormalTok{ mod\_size[[}\StringTok{"L"}\NormalTok{]][}\StringTok{"xep01"}\NormalTok{], }\AttributeTok{S =}\NormalTok{ mod\_size[[}\StringTok{"S"}\NormalTok{]][}\StringTok{"xep01"}\NormalTok{]))}
\end{Highlighting}
\end{Shaded}

Note that it is entirely possible (and easy) to select different model
forms for each class. For example, there would be some justification for
choosing \(\texttt{xep01}\) for \texttt{S} but \(\texttt{xep1}\) for
\texttt{L} based on AICc scores. However, we opted to use the same model
form for each size class because the xep01 model, which was selected as
the ``best'' model for the pooled data, also worked well for each of the
size classes individually.

Once \(\psi\) has been estimated, \texttt{dwp} can be calculated and
formatted for GenEst:

\begin{Shaded}
\begin{Highlighting}[]
\NormalTok{data\_dwp }\OtherTok{\textless{}{-}} \FunctionTok{estdwp}\NormalTok{(psi\_size, }\AttributeTok{ncarc =} \FunctionTok{getncarc}\NormalTok{(cod\_free\_size), }\AttributeTok{forGenEst =}\NormalTok{ T)}
\FunctionTok{dim}\NormalTok{(data\_dwp) }\CommentTok{\# nsim = number of simulated dwp\textquotesingle{}s for each turbine and carcass class}
\CommentTok{\#\textgreater{} [1] 10000     3}

\FunctionTok{head}\NormalTok{(data\_dwp) }\CommentTok{\# the first few rows of the data frame}
\CommentTok{\#\textgreater{}   turbine     L     S}
\CommentTok{\#\textgreater{} 1      t1 1.000 0.875}
\CommentTok{\#\textgreater{} 2      t3 1.000 0.667}
\CommentTok{\#\textgreater{} 3      t4 0.417 0.500}
\CommentTok{\#\textgreater{} 4      t8 1.000 0.923}
\CommentTok{\#\textgreater{} 5     t11 0.400 0.478}
\CommentTok{\#\textgreater{} 6     t13 0.500 0.714}

\CommentTok{\# data\_dwp can be used directly from the command line as the data\_dwp argument}
\CommentTok{\# in GenEst::estM, or it can be exported to a .csv file for importing into the}
\CommentTok{\# GenEst GUI:}
\FunctionTok{exportGenEst}\NormalTok{(data\_dwp, }\AttributeTok{file =} \StringTok{"mydwp.csv"}\NormalTok{)}
\end{Highlighting}
\end{Shaded}

\renewcommand*\appendixpagename{\huge Appendices}
\appendices

\appendixpage

\hypertarget{brief-introduction-to-the-carcass-distributions}{%
\section{\texorpdfstring{Brief Introduction to the Carcass Distributions
\label{app:modbrief}}{Brief Introduction to the Carcass Distributions }}\label{brief-introduction-to-the-carcass-distributions}}

\hypertarget{model-anatomy}{%
\subsection{Model Anatomy}\label{model-anatomy}}

Estimating the expected proportion of carcasses lying in the searched
area involves three steps (fig.~\ref{fig:volcano}):

\begin{enumerate}
  \item modeling carcass density (carcasses per $\textrm{m}^2$) as a function of location or distance from turbine;
  \item normalizing the fitted carcass densities to create a probability density surface; and
  \item integrating the probability density over the searched area.
\end{enumerate}

Each model is a Poisson regression, which is a generalized linear model
(GLM) with log link (McCullagh and Nelder 1983) such that
$y_i \sim \textrm{Poisson}(\lambda_i)$\\ $\log(\lambda_i) = XB$ so
that $\lambda_i = e^{XB}$ where where \(y_i\) is the carcass count
within measurement unit \(i\) at an average distance of \(x_i\) m from
the turbine\footnote{The measurement unit for xy-grid data is a 1
  \(\textrm{m}^2\) patch of ground. For the other data types, it is a 1
  m annulus with inner and outer radii at \(x \pm 0.5\) m.}, and
\(\lambda_i\) is the expected number of carcasses in that unit. In each
model (dropping the index, \(i\)), \(XB\) is of the form
\[p(x) + c + \log(\epsilon(x, c))\] where \(c\) is an indicator variable
for the search class, \(p(x)\) is a polynomial with coefficients to be
estimated in the regression, and the offset term,
\(\log(\epsilon(x, c))\), is the exposure or area (\(\textrm{m}^2\)) in
search class \(c\) of the measurement unit.

Models of this form are distinguished by the form of the polynomial,
whether, for example, it is linear, quadratic, or some other form. It is
not necessary to know or estimate the detection probabilities in each
search class. Rather, specifying \(c\) as a covariate in the model
accounts for the relative detection probabilities among search classes.
Because the search class term serves only to modify the intercept, we
can ignore the search class term, \(c\), in this discussion and consider
the simple case where all the area within the search radius is searched.
Then, the exposure would be proportional to \(x\), and the form
simplifies, to \(\lambda(x) \propto e^{p(x)+log(x)} = xe^{p(x)}\).

We abbreviate such models as \texttt{xep} followed by digits to indicate
the form of their polynomial. For example, \texttt{xep12} would have a
polynomial with linear and quadratic terms and would expand to
\[e^{a + \beta_1 x + \beta_2 x^2 + \log(\varepsilon(x))}\] We represent
a \(\log(x)\) term by ``0'', and a \(1/x\) term by ``i''. Thus
\texttt{xep01} would correspond to
\(p(x) = \beta_0 \log(x) + \beta_1 x\) and \texttt{xepi0} to
\(p(x) = \beta_i 1/x + \beta_0 \log(x)\). In practice, the constant
term, \(a\), is a nuisance parameter that is lost when the model is
normalized to integrate to 1.

Normally, if the leading coefficient\footnote{The leading coefficient is
  the \(\beta\) associated with the highest power of \(x\) in the
  polynomial.} of \(p(x)\) is negative, the model corresponds to a
probability distribution which can be extended beyond the search radius,
and the model can be extrapolated to estimate the fraction of carcasses
falling outside the search radius. However, if the model also has a
\(\log(x)\) term, that coefficient (\(\beta_0\)) must be greater than -2
for the model to yield a proper probability distribution. Similarly, a
\(1/x\) coefficient (\(\beta_i\)) must be negative. In all cases, if the
leading coefficient is positive, the model is not extensible beyond the
search radius and cannot be used to estimate the fraction falling
outside the search radius.

All the extensible fitted Poisson regression models can be converted
into probability distributions in the exponential family simply by
normalizing the fitted model by dividing by its integral from 0 to
\(\infty\). In this way, some of the models give rise to familiar
distributions, like the normal, lognormal, gamma, and exponential. For
example, a normalized \texttt{xep01} model is a gamma distribution, and
a normalized \texttt{xep} model with polynomial
\(p(x) = \beta_0\log(x) + \beta_1 \log(x)^2\) is a lognormal
distribution. However, although all the models do give rise to
distributions in the exponential family, most do not have familiar
names. In all cases, the \texttt{ddFit} function automatically
normalizes the fitted distance functions whenever possible, thereby
converting them to probability distributions. The package also includes
functions for calculating the probability density (PDF), the cumulative
distribution (CDF), quantiles, and random deviates for all the distance
distributions in much the same format as the d/p/q/r family of R
functions for other distributions. For the fitted distance distributions
in the \texttt{dwp} package, the functions \texttt{ddd}, \texttt{pdd},
\texttt{qdd}, and \texttt{rdd} serve as analogs to R's \texttt{dnorm},
\texttt{pnorm}, \texttt{qnorm}, and \texttt{rnorm} for the normal
distribution.

In addition to the \texttt{xep} models which have offset =
\(\log(\textrm{exposure})\) to account for the area searched in each
ring, \texttt{ddFit} will fit a few models that have a modified offset
that serves dual purposes. In addition to accounting for area sampled in
each ring, the modified offset alters the shape of the distribution in a
standard way as the distance changes. These \texttt{xep}-like
distributions with modified offset that are fit by \texttt{ddFit}
include the normal, exponential, \(\chi^2\), Maxwell-Boltzmann, and
inverse Gaussian. Finally, a model that assumes the carcass density is
constant throughout the searched area is fit (table \ref{tbl:formsST}).

\begin{table}[!htbp]
  \caption{Poisson Regressions, Standard Distributions}
  \label{tbl:formsST}
  \renewcommand{\arraystretch}{1.4}
  \begin{tabular}{lll}
    \textbf{Distribution} & \textbf{GLM Form and Density} & \textbf{Extensibility}\\
    \hline\hline \\[-3ex]
    Constant & $y \sim \textrm{offset}(\log(\varepsilon(x)))$ & NA\\[3pt]
    & $f(x) \propto x$ &\\[5pt]
    \hline \\[-3ex]
    \texttt{xep1} & $y \sim x + \textrm{offset}(\log(\varepsilon(x)))$ & $\beta_1<0$\\[3pt]
    & $f(x) \propto x e^{\beta_1x}$ &\\[3pt]
    \hline \\[-3ex]
    \texttt{xep01} (gamma)   & $y \sim \log x + x + \textrm{offset}(\log(\varepsilon(x)))$ & $\beta_0 > -2, \beta_1 < 0$\\[3pt]
    & $f(x) \propto x e^{\beta_0\log x + \beta_1x}$ &\\
    \hline \\[-3ex]
    \texttt{xep2} (Rayleigh) & $y \sim x^2 + \textrm{offset}(\log(\varepsilon(x)))$ & $\beta_2 < 0$\\[3pt]
     & $f(x) \propto x e^{\beta_2x^2}$ &\\
    \hline \\[-3ex]
    \texttt{xep02}           & $y \sim \log x + x^2 + \textrm{offset}(\log(\varepsilon(x)))$& $\beta_0 > -2, \beta_2 < 0$\\[3pt]
     & $f(x) \propto x e^{\beta_0\log x + \beta_2x^2}$ &\\
    \hline \\[-3ex]
    \texttt{xep12}           & $y \sim x + x^2 + \textrm{offset}(\log(\varepsilon(x)))$ & $\beta_2 < 0$\\[3pt]
    & $f(x) \propto x e^{\beta_1x + \beta_2x^2}$ &\\
    \hline \\[-3ex]
    \texttt{xep012}          & $y \sim \log x + x + x^2 + \textrm{offset}(\log(\varepsilon(x)))$ & $\beta_0 > -2, \beta_2 < 0$\\[3pt]
    & $f(x) \propto x e^{\beta_0\log x + \beta_1x + \beta_2x^2}$ &\\
    \hline \\[-3ex]
    \texttt{xep123}          & $y \sim x + x^2 + x^3 + \textrm{offset}(\log(\varepsilon(x)))$ & $\beta_3 < 0$\\[3pt]
    & $f(x) \propto x e^{\beta_1x + \beta_2x^2 + \beta_3x^3}$ &\\
    \hline \\[-3ex]
    \texttt{xep0123} & $y \sim \log x + x + x^2 + x^3 + \textrm{offset}(\log(\varepsilon(x)))$ & $\beta_0 > -2, \beta_3 < 0$\\[3pt]
    & $f(x) \propto x e^{\beta_0\log x + \beta_1x + \beta_2x^2 + \beta_3x^3}$ &\\
    \hline \\[-3ex]
    Lognormal         & $y \sim \log x + \log^2x + \textrm{offset}(\log(\varepsilon(x)))$ & $\beta_1 < 0$\\[3pt]
    & $f(x) \propto x e^{\beta_0\log(x) + \beta_1\log(x)^2}$ &\\
    \hline \\[-3ex]
    Truncated Normal  & $y \sim x + x^2+ \textrm{offset}(\log(\varepsilon(x)) - \log x)$ & $\beta_2 < 0$\\[3pt]
    & $f(x) \propto x e^{\beta_1x + \beta_2x^2}$ &\\
    \hline \\[-3ex]
    Maxwell Boltzmann & $y \sim x^2+ \textrm{offset}(\log(\varepsilon(x)) + \log x)$ &$\beta_2 < 0$\\[3pt]
    & $f(x) \propto x^2 e^{\beta_2x^2}$ & \\
    \hline\hline \\[-3ex]
  \end{tabular}
\end{table}

By default, a standard set of 12 distributions is fit, although others
can be fit as well (table \ref{tbl:formsSU}). Enter \texttt{?ddFit} for
details on the function, including how to specify which distributions to
fit.

\begin{table}[!htbp]
  \caption{Poisson Regressions, Supplementary Distributions}
  \label{tbl:formsSU}
  \begin{tabular}{lll}
    \textbf{Distribution} & \textbf{GLM Form and Density} & \textbf{Extensibility}\\
    \hline\hline \\[-3ex]
    \texttt{xep0} (Pareto) & $y \sim \log x + \textrm{offset}(\log(\varepsilon(x)))$ & $\beta_0 < -2$ and $x \in [1, \infty)$\\[3pt]
     & $f(x) \propto x^{\beta_0 + 1}$ &\\
    \hline \\[-3ex]
    \texttt{xepi0} (inverse gamma) & $y \sim 1/x + \log x + \textrm{offset}(\log(\varepsilon(x)))$ & $\beta_0 < -2$ and $\beta_i < 0$)\\[3pt]
     & $f(x) \propto x^{\beta_0 + 1} e^{\beta_i/x}$ &\\
    \hline \\[-3ex]
    Chisquared  & $y \sim \log(x) + \textrm{offset}(\log(\varepsilon(x)) - x/2)$ & $\beta_0 > -2$\\[3pt]
    & $f(x) \propto x^{\beta_0 + 1} e^{- x/2}$ &\\
    \hline \\[-3ex]
    Exponential & $y \sim x + \textrm{offset}(\log(\varepsilon(x)) -\log(x))$ & $\beta_1 < 0$\\[3pt]
    & $f(x) \propto e^{\beta_1x}$ &\\
    \hline \\[-3ex]
    Inverse Gaussian & $y \sim 1/x + x + \textrm{offset}(\left(\log(\varepsilon(x))-5/2\log(x)\right)$ & $\beta_i, \beta_1 < 0$\\[10pt]
     & $f(x) \propto \sqrt{1/x^3} e^{\beta_1x + \beta_i\cdot1/x}$ &\\
    \hline\hline \\[-3ex]
  \end{tabular}
\end{table}

\hypertarget{model-characteristics}{%
\subsection{Model Characteristics}\label{model-characteristics}}

There are never universally accepted criteria for determining which
model is definitively the ``best.'' When the model must be extrapolated
well beyond the range of data, the problem becomes much more difficult.
How well a model fits within the range of the data may give very little
indication of how well it will fit outside the range of data.
Conventional tools like Akaike Information Criterion (AIC) can be very
useful for comparing how well various models fit the data, but two
models that are virtually indistinguishable within the range of data can
diverge dramatically outside the range of data. However, there are some
useful guidelines for model selection that are discussed briefly here
and in greater detail in the discussions of examples (sections
\ref{sec:exeagle}-\ref{sec:exshape}) and in Appendix \ref{app:fit2deq}.

\begin{enumerate}
  \item Models with lower AICc scores give better fits within the range of the data. Models with AICc scores within about 4 are considered indistinguishable by this measure. AICc scores are less informative about model reliability for extrapolating beyond the search radius, but models with $\Delta\textrm{AICc}$ scores above 10 do not seem to fare as well as models with $\Delta\textrm{AICc} < 5$.

  \item Models with heavy tails tend to overestimate the fraction of carcasses falling outside the search radius when the search radius is long. The weight of the tail is primarily a function of the degree of the polynomial. For example, the tail of an $\texttt{xep}$ distribution with degree 2 will have a lighter tail than one with degree 1. The weight of the tail also depends somewhat on the other terms in the polynomial, with lower order terms contributing to lighter tails.

   \item Road \& pad searches (RP) or other search patterns in which the proportion of area searched per ring decreases with distance from turbine and the search coverage is low but non-zero for a broad range of distances greater than about 40 meters can lead to reduced stability in the models. In this situation:
      \begin{itemize}
        \item there will be a tendency to overestimate the fraction of carcasses within the search radius if no carcasses are found at great distances, and we recommend using a relatively heavy-tailed distribution these situations to account for the possibility of significant numbers of carcasses beyond the search radius, and
        \item there will be a tendency to strongly underestimate the fraction of carcasses within the search radius if carcasses are found at great distances, and we recommend using a relatively light-tailed distribution in these cases.
      \end{itemize}

    \item Distributions of smaller carcasses tend to have heavier tails than distributions of larger carcasses because the smaller carcasses are more strongly affected by wind and are blown farther under high wind conditions and are not flung as far under light wind conditions, as demonstrated in ballistic simulations in appendices \ref{app:deqmod} and \ref{app:fit2deq}.
\end{enumerate}

\hypertarget{mechanistic-model-of-carcass-deposition}{%
\section{\texorpdfstring{Mechanistic Model of Carcass Deposition
\label{app:deqmod}}{Mechanistic Model of Carcass Deposition }}\label{mechanistic-model-of-carcass-deposition}}

To explore properties of carcass density models and their performances
on known data, we generated carcass distributions according to a
ballistics model similar to those used by Hull \& Muir (2010) and by
Prakash and Markfort (2020). Models are fit to simulated carcass
dispersions under several different scenarios with different carcass
types (eagle and bat), wind regimes, and carcass search protocols. Since
the actual (simulated) carcass distributions are known, models can be
compared with each other based on how well they are able to predict the
true data under a variety of conditions.

\hypertarget{the-model}{%
\subsection{The Model}\label{the-model}}

Physics puts a fairly strict limit on how far carcasses may fall from
the turbine. Gravity inexorably pulls them to the ground after a brief
time. Excluding cases where 1) a bird or bat is not killed immediately
but is able to fly or walk a great distance after getting struck by a
turbine blade, 2) a carcass is carried a great distance by a scavenger,
3) a carcass is swept upward or downward by a wind with a strong
vertical component, or 4) a carcass is blown or rolls a great distance
after landing, we can use well-known differential equations from
ballistics modeling to get a qualitative understanding of carcass
spatial distributions around a turbine. Of particular interest is the
general shape of the tails of the carcass distribution, which, in turn,
informs the choice of parametric distribution for empirical modeling of
carcass distributions and estimation of the proportion of carcasses
landing in the searched areas.

Mechanistic differential equation models can be useful in addressing
simple, qualitative questions about carcass dispersion. Which general
shapes of distributions are best in which types of situations? How well
does AICc perform for identifying models that will most accurately
predict the fraction of carcasses landing in the searched area? How do
carcass size and wind speed broadly affect the distribution of carcasses
on the ground?

However, a quantitative understanding that accurately predicts the
actual distributions of carcasses in the field as a function of
environmental covariates such as wind profile and turbine size is much
more difficult and may not be feasible. The precise distribution
critically depends on a number of parameters and assumptions that are
difficult to validate in the field. Factors such as wind speed and
direction at the time of strike, strike location on the blade, carcass
velocity at the time of strike, change in carcass velocity upon strike
(angle of the blade, transfer of momentum from blade to carcass), drag
coefficient and its change during flight as carcass changes shape and
direction, rotor speed at time of collision, carcass fluttering or
spinning after collision, dependence of wind velocity on position
(including wind shear and wake effect), turbulence behind the blade,
displacement by scavengers after landing, and even the type of ground
cover (which affects the wind gradient with height) all have a
significant impact on the resulting carcass distribution on the ground,
but none of them are known with any practical, quantitative degree of
reliability.

If we make a few simplifying assumptions, the ballistics equations are
not difficult. We start by assuming that the animal is killed upon
impact with the turbine blade and that the wind is blowing horizontally
(with no vertical component) and normal to the plane of the blades. Like
Prakash and Markfort (2021) and Hull and Muir (2011), we ignore the wake
effect behind the turbine, but we do include wind shear.

To calculate a carcass landing position after being struck by a turbine,
we first define a coordinate system centered at the base of the turbine
to track carcass position in 3 dimensions, \(s=(x,y,w)\), where \(x\) is
the horizontal position in the plane of the turbine blades, \(y\) is
height above the ground, and \(w\) is the horizontal distance normal to
the plane of the turbine blades in the downwind direction. We then use
ballistics differential equations to describe the trajectory of the
carcass as a function of time.

Define \(x\) and \(y\) to be the horizontal and vertical displacements
of a carcass from the turbine base, and \(w\), the distance downwind
from the turbine. If \(s(t)\) is the position vector of the carcass as a
function of time and \(v(t) = s'(t)\) is the velocity, then the pair
\((s(t), v(t))\) is a 6-dimensional vector which satisfies a set of
first order differential equations: \[s'(t) = v(t)\]
\[v'(t) = -\mathfrak{y}\left(v(t) - \mathpzc{w}(s(t))\right) \lVert v(t) - \mathpzc{w}(s(t)) \rVert - G\]
where \(\mathpzc{w}(s(t))\) is the wind velocity at \(s(t)\),
\(\mathfrak{y}= 9.807/{y'_T}^2\) (with \(y'_T\) = terminal velocity of
the carcass) is an aerodynamic constant associated with each carcass
type, and \(G = [0, 9.807, 0]\) is the acceleration due to gravity.
Given an initial carcass strike point on the blade and an initial
carcass velocity, we track the trajectory of the carcass through space
using the method of lines with \(\Delta t = 0.01\) seconds. The final
position of the carcass on the ground is \((x, w)\) when \(y = 0\).

This set of governing equations is similar to those used by Prakash and
Markfort (2021) but allows wind speed to vary with carcass position,
thereby enabling the accommodation of wind shear and wake effect. Like
Prakash and Markfort (2021) and Hull and Muir (2011), we ignore the wake
effect. We do account for wind shear through the Wind Power Law, or
\(\mathpzc{w}(y) = \mathpzc{w}(y_n)\cdot{y/y_n}^a\), where
\(\mathpzc{w}(y)\) is the wind speed at \(y\) meters above the ground,
\(y_n\) is the height of the nacelle, and \(a\) is the Hellman
coefficient (Counihan 1975).

In addition, we assume that the cross-sectional area of the carcass
presented to the wind (\(A\)) and the drag coefficient (\(C_d\)) are
constant, so the term \(0.5\, C_d A\rho/m\) (Prakash and Markfort
(2021); Hull and Muir (2011)) simplifies to \(9.807/{y'_T}^2\) after
noting that the terminal velocity is the vertical velocity at which the
wind resistance equals the force of gravity so there is no longer any
vertical acceleration. In other words, terminal velocity, \{y'\_T\}, is
the velocity such that \(\rho/2C_dA/m\cdot{y'_T}^2 - g = 0\) from
Prakash and Markfort's (2021) equations 3 and 4 with no acceleration in
the horizontal directions, giving
\(\rho/2C_dA/m = G/{y'_T}^2 = \mathfrak{y}\). Thus, the terminal
velocity, which is relatively easy to measure and is relatively
well-known, is used in lieu of the product of parameters with largely
unknown values (\(C_d\) and \(A\)).

\hypertarget{simulation-parameters}{%
\subsection{\texorpdfstring{Simulation Parameters
\label{app:simparm}}{Simulation Parameters }}\label{simulation-parameters}}

We used the ballistics model described in the previous section (appendix
\ref{app:deqmod}) to explore the effects of carcass type, wind regime,
strike position, and flight speed on carcass distributions on the ground
(appendix \ref{app:deqresults}). Random carcass locations were generated
from the carcass distributions and then sampled to create simulated data
sets. Distributions were fit to the simulated data sets using
\texttt{ddFit}. Statistical properties of the models are explored and
discussed in Appendix \ref{app:fit2deq} to provide guidance on
\(\texttt{dwp}\) model selection.

Parameter values used in the carcass dispersion models and simulations
are a mix of fixed constants (such as turbine specifications) that are
constant across the simulation scenarios and variable parameters that
may differ among carcasses and simulation scenarios (table
\ref{tbl:parms}).

\begin{table}[!htbp]
  \caption{Parameters used in the ballistics model and simulations}
  \label{tbl:parms}
  \begin{tabular}{ll}
     Parameter & Value \\
    \hline\hline
    Fixed across scenarios & \\
    \hline
    nacelle height & 80 m\\
    blade length & $r_0 = 45$ m\\
    tip speed ratio & 6\\
    acceleration due to gravity & $G = 9.807\, \mathrm{m/s^2}$\\
    Hellman coefficient & $a = 0.22$\\
    strike position on the blade & random uniform\\
    \hline
    Varying among scenarios &\\
    \hline
    terminal velocity & $V_t = 8.8$ m/s for bats, and $V_t = 25$ m/s for eagles\\
    wind speed & low, moderate, high\\
    wind profile & constant, varying\\
    carcass initial horizontal velocity & 0, varying (up to 8 m/s)\\
    search plot type & cleared, roads and pads\\
    \hline
  \end{tabular}
\end{table}

The Hellman coefficient governs the wind shear or the tendency for wind
speed to increase with height. It is a measure of the coarseness of the
landscape texture (Kaltschmatt et al.~2007), with smaller values
corresponding to smoother surfaces and smaller changes in wind speed
with height. Values typically range from about 0.05 for very smooth
ground to 0.5 or 0.6 for cities with high-rise buildings.

The terminal velocity is a measure of an object's air resistance and is
an amalgam of the drag coefficient (\(C_d\)), the cross-sectional area
presented to the wind, the air density---all of which are unknown---and
the object's mass. The terminal velocity is used in lieu of the product
of parameters with largely unknown values. We assume a value of 8.8 m/s
for the terminal velocity for bats, which we calculated from data given
by Grodsky et al.~(2011).\footnote{Grodsky et al.~(2011) reported that
  bats took an average of 10--12 seconds to fall 91.4 m in an
  experiment. Terminal velocity is the velocity, \(y'\), at which the
  carcass is no longer accelerating as it falls but is falling at a
  constant rate, that is \(y'' = -G + A(y'_T)^2 = 0\), where \(y'_T\) is
  the terminal velocity, \(A\) is the projected area, and \(G\) is the
  acceleration due to gravity. Solving for \(y\), we get
  \(y(t) = y_0 - \log(\cosh(3.1305 t \sqrt{A}) )/A\). Then, taking the
  initial height as \(y_0 = 91.4\) and the final height after 11 seconds
  as \(y(11) = 0\), we get \(A = 0.1263\) and a terminal velocity of
  \(y'_T = 8.8\) m/s for bats, which is in good agreement with the
  findings of Prakash and Markfort (2020).} We used a terminal velocity
of 25 m/s for eagles, which is slightly less than it is for live cats
(Whitney and Mehlhaff, 1987).

Turbine nacelle height and blade length match those of the 1.8 MW Vestas
V90 turbine. We assume the turbine is operating at a tip speed ratio of
6, that is, that the tips of turbine blades are moving at 6 times the
wind speed, while the wind varies.

Simulation scenarios assume that strike position on the blades is
uniformly distributed and that the initial velocity of the carcass after
being struck is equal to the sum of the blade and carcass velocities at
the point of impact.

Wind speeds behind the turbine and at the height of the nacelle were
either constant at 4, 8, or 12 m/s or varying with low, moderate, or
high average. Winds in the low wind speed regime were assumed to be
Weibull distributed with mean of 5.32 m/s and standard deviation of 2.78
m/s. The high wind speed regime was also assumed to be Weibull
distributed, with mean of 9.18 m/s and standard deviation of 2.1 m/s.
The moderate wind speed regime mimics the winds measured at a site in
the Netherlands, with a mean of 7.5 m/s, standard deviation of 1.25 m/s,
and having a probability density function (PDF) of wind speeds of
\(f(w) = 10^{-0.04(w -7)^2 - 0.8}\), which approximately fits the
moderate wind profile in fig.~2 (row \(\mathrm{W_{80}}\), column JJA) of
He et al.~(2013) . In all cases, no carcasses were produced when wind
speeds were \textless3.5 m/s. Wind direction was assumed to be uniformly
distributed, with no prevailing direction.

Animal flight velocity was assumed to be constant in half the scenarios
and variable in half. In the constant scenarios, carcasses attain the
same velocity as the turbine blade at the point of impact upon being
struck, implying a coefficient of restitution of \(e = 1\), following
Hull and Muir (2010) and Prakash and Markfort (2020). In the variable
scenarios, flight speed relative to the ground was assumed to be 8 m/s
with 0 vertical component but horizontal direction varied uniformly
between 0 and 2\(\pi\), with the component of carcass initial velocity
in the plane of the blades (\(x\)) erased at the moment of impact and
the component normal to the plane of the blades (\(w\)) added to the
wind velocity.

\begin{figure}
\centering
\includegraphics{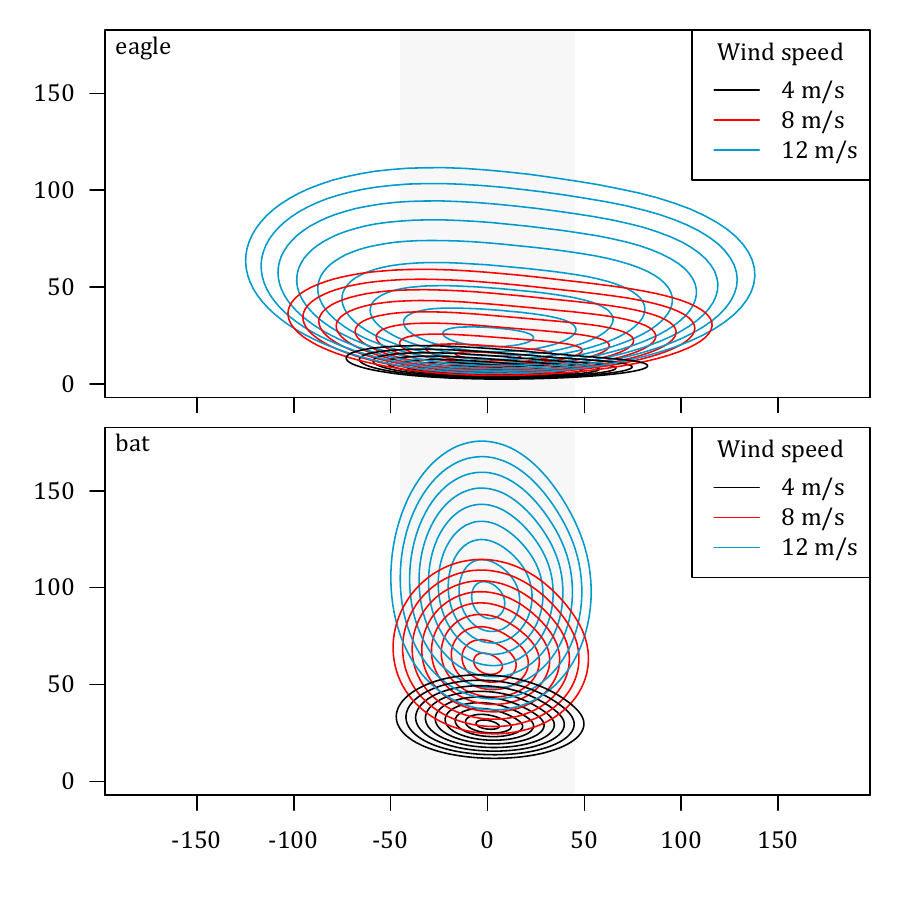}
\caption{\label{fig:contour}Projection of turbine-carcass strike points
onto the ground. Turbine is located at (0, 0), with wind blowing north
(bottom to top) at a constant velocity. Lines are projected images of
turbine-carcass strike points onto the ground. Contours correspond to
strike points on the blades in concentric rings with radii of 5, 10,
\ldots, 45 m. Shading marks the span of the turbine blades in the x
direction.}
\end{figure}

The zone where carcasses are likely to fall would be expected to extend
to a radius at least as great as the turbine blades (\(r_0\)). Carcasses
may land outside that range if the impact from the blade knocks them
beyond \(r_0\) in the plane of the blades or if the wind carries them
beyond \(r_0\) in the windward direction. Figure \ref{fig:contour}
compares the carcass fall zones for eagles and bats at wind speeds of 4,
8, and 12 m/s, showing images of the disk spanned by the turbine blades
projected onto carcass dispersion patterns on the ground. The contours
are mappings of concentric rings around the turbine hub to the position
on the ground where an eagle or bat carcass would be expected to fall
under the given wind speed and carcass type under the conditions of the
simulation scenario (table \ref{tbl:parms}).

Eagle carcasses are large, and, compared to small carcasses, their
trajectories are less affected by the wind. Like a baseball getting
struck by a bat, an eagle directly hit by a turbine blade is flung in
the direction dictated by the blade, with relatively little modification
of its subsequent trajectory by the wind. In the simulations, under low
wind speed conditions (4 m/s) eagle carcasses fell near the line of the
blades. With stronger winds, carcasses landed significantly downwind
(fig.~\ref{fig:contour}).

By contrast bat carcasses are small and may be carried aloft by the wind
for great distances, like popcorn thrown into the air on a windy day. In
the simulations, even at the relatively low wind speed of 4 m/s,
carcasses struck near the top of the swept area were blown over 50
meters, significantly beyond the \(r_0 = 45 \mathrm{m}\) blade length
(fig.~\ref{fig:contour}). Greater windspeeds had little effect on the
distribution of carcasses in the plane of the blades (left-right in
fig.~\ref{fig:contour}) but a substantial effect on the distribution of
carcasses in the downwind direction (bottom to top in
fig.~\ref{fig:contour}). With a wind of 12 m/s, carcasses near the top
of the swept area were blown close to 200 m.

\hypertarget{distributions-of-carcass-distances}{%
\subsection{\texorpdfstring{Distributions of Carcass Distances
\label{app:deqresults}}{Distributions of Carcass Distances }}\label{distributions-of-carcass-distances}}

\begin{figure}
\centering
\includegraphics{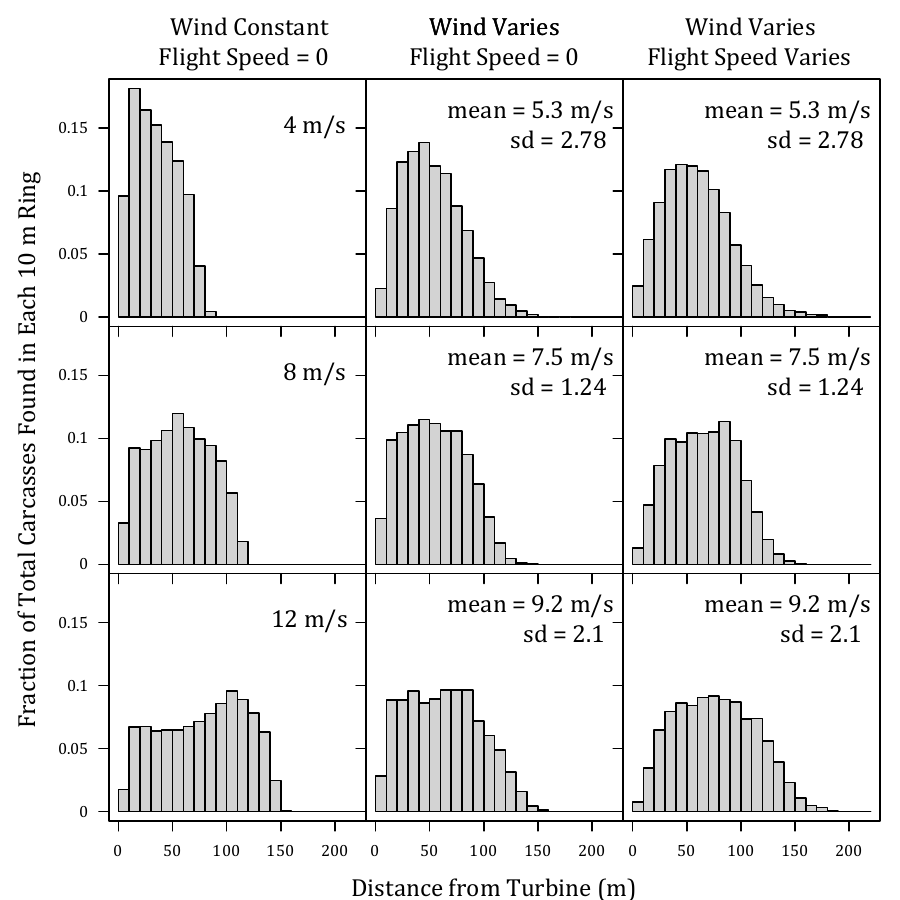}
\caption{\label{fig:eaglehist}Histograms of distances of eagle carcasses
to turbine. Upon collision with a turbine blade, carcass velocity takes
on the velocity of the blade plus the flight speed. Flight speed is
either assumed to be 0, in which case initial carcass velocity equals
turbine blade velocity at point of impact, or variable, in the range of
wind speed ± 8 m/s.}
\end{figure}

\begin{figure}
\centering
\includegraphics{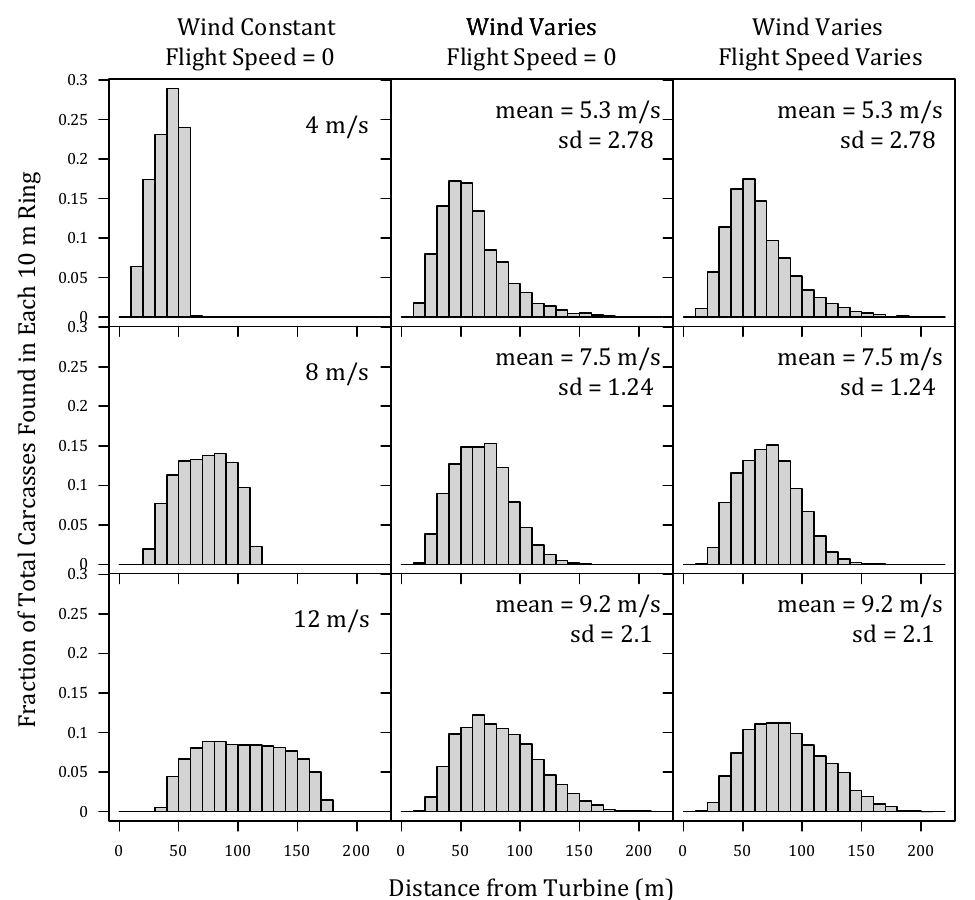}
\caption{\label{fig:bathist}Histograms of distances of bat carcasses to
turbine. Upon collision with a turbine blade, carcass velocity takes on
the velocity of the blade plus the flight speed. Flight speed is either
assumed to be 0, in which case initial carcass velocity equals turbine
blade velocity at point of impact, or variable, in the range of wind
speed ± 8 m/s.}
\end{figure}

Because bats are be more heavily influenced by the wind than eagles are
and are blown farther, a smaller proportion of bats are expected to fall
very close to the turbine, as reflected in the histograms of simulated
carcass distances shown in figs. \ref{fig:eaglehist} and
\ref{fig:bathist}.

The eagle histograms are qualitatively similar to those constructed for
large birds by Hull and Muir (2011, fig.~2). Under most of the simulated
conditions, the distributions have relatively flat tops over much of the
range of distances but then drop precipitously. The most consequential
difference between our model and that of Hull and Muir (2011) is that
the latter neglects the downwind direction. Because eagle carcasses are
typically not blown far by the wind after being struck, the downwind
direction plays a relatively minor role in determining eagle distance
distributions, which explains the similarity between the patterns in
fig.~\ref{fig:eaglehist} and those observed by Hull and Muir (2011).

The bat histograms (fig.~\ref{fig:bathist}) show relatively few
carcasses very near the turbines and more of a classic bell shape
(albeit, somewhat right-skewed) than the eagle histograms. That these
bat histograms bear little resemblance to those of Hull and Muir (2011,
fig.~2) and Prakash and Markfort (2020, fig.~9) is due to the importance
of the downwind direction, which is included in our model but neglected
in the other two.

\hypertarget{fitting-parametric-distributions-to-the-dispersion-patterns}{%
\section{\texorpdfstring{Fitting Parametric Distributions to the
Dispersion Patterns
\label{app:fit2deq}}{Fitting Parametric Distributions to the Dispersion Patterns }}\label{fitting-parametric-distributions-to-the-dispersion-patterns}}

The general shape of the actual distribution of carcasses varies with
the ballistics parameters and environmental conditions. The performance
of a fitted distribution depends on how well the distribution family's
shape conforms to the actual distribution and, perhaps even more
critically, how accurately the fitted model reflects the distribution of
carcasses lying outside the searched area.

In practice, the search radius is finite, and a fitted model does not
``know'' what fraction of carcasses lie within the search radius and has
to ``guess'' based on its general shape. If the general shape happens to
align well with the true distribution, the prediction of the fraction
lying outside the search radius will be accurate; if not, the error may
be substantial.

The ballistics modeling suggests that eagle carcass distributions are
often likely to be relatively flat near the turbine but decline
precipitously with distance at some point (fig.~\ref{fig:eaglehist}), so
that, under many conditions, eagle distributions would be expected to be
light-tailed. Thus, relatively heavy-tailed distributions such as the
lognormal and \texttt{xep1} would usually not be a good match for eagle
distributions. By contrast, the simulations suggest that bat carcass
distributions will tend to have few carcasses very near the turbines,
and, compared with the eagle distributions, a more gradual climb to a
narrower peak (fig.~\ref{fig:bathist}) and have a different suite of
distributions with the best fits.

Although some broad descriptions of model characteristics and their
general applicability can be made, it may not be possible to determine
\emph{a priori} which particular models will provide the best
predictions in a given situation. Factors like wind regime and species
determine the actual distribution of carcasses, but other factors---like
search radius and plot shape (for example, cleared plots or roads \&
pads)---can heavily influence the aptness of a model for predicting the
fraction of carcasses outside the searched area. Selecting a specific
model for a specific situation must be done on a case by case basis.

\hypertarget{model-accuracy-in-predicting-psi}{%
\subsection{\texorpdfstring{Model Accuracy in Predicting \(\psi\)
\label{app:predpsi}}{Model Accuracy in Predicting \textbackslash psi }}\label{model-accuracy-in-predicting-psi}}

The accuracy of a carcass distribution model for predicting the
probability that a carcass lies in the searched area depends both on how
well the model conforms to the shape of the actual distribution and how
the fit within the searched area can be extended outside the searched
area. There are two types of unsearched area: 1) area outside a search
radius, and 2) unsearched area within the search radius. Spatial
prediction for carcasses outside the search radius involves
extrapolation, and accuracy requires making effective use of additional
information, apart from the data itself because a good or even perfect
fit within the range of data gives no guarantee of a good or even
plausible fit outside the range of the data. Prediction for unsearched
areas within the search radius involves interpolation, and the relative
quality of a model's fit to the data within the searched area (for
example, AICc) is a reasonable guide for model selection for
interpolation. Thus, wind conditions, carcass type, search radius, and
plot shape are all important factors in determining which models will
provide the most accurate predictions.

\begin{figure}
\centering
\includegraphics{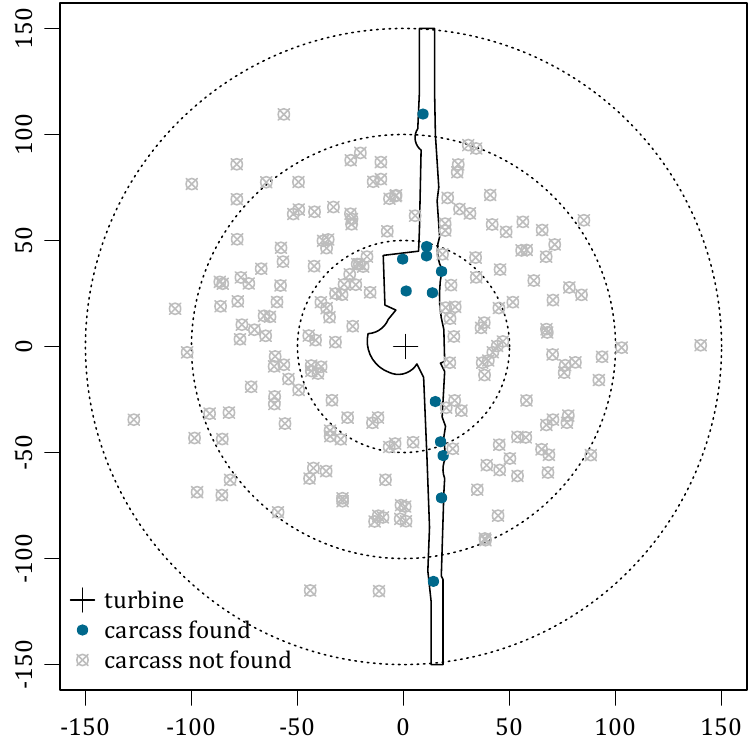}
\caption{\label{fig:searchPlots}Search areas used in the simulations.
Searches were conducted out to radii of 50, 100, or 150 m from the
turbine (circles) and included either all carcasses within the search
radius (``cleared plot'') or only those carcasses on the roads and
turbine pad (``RP''). Figure shows carcasses found on an RP search out
to 150 m.}
\end{figure}

\begin{figure}
\centering
\includegraphics{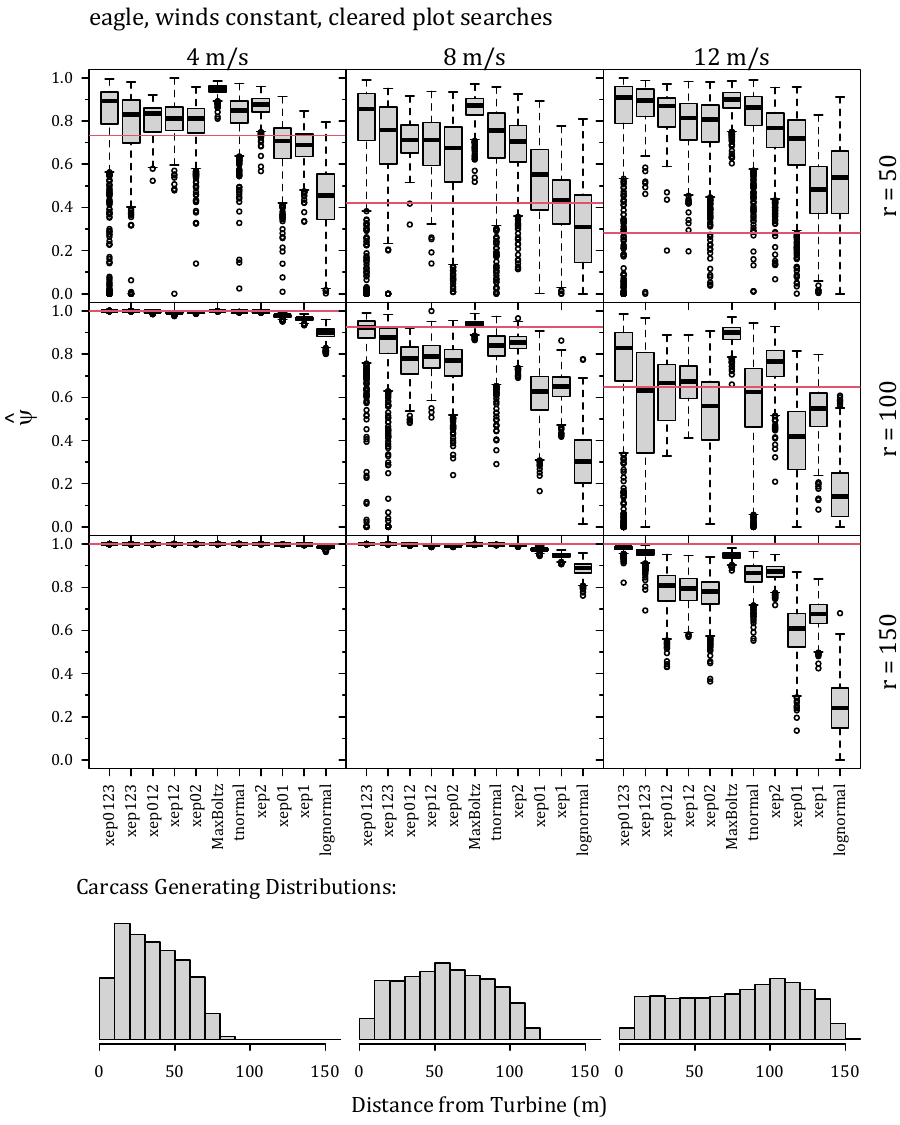}
\caption{\label{fig:psi111}Estimated \(\psi\) for the standard models
with simulated M = 200 eagles with winds constant and cleared plot
searches. Boxes show sample IQR with median; whiskers extend to the most
extreme points within 1.5 IQR of the box; points beyond 1.5 IQR of the
box are shown as small circles. Red lines show the true \(\psi\).}
\end{figure}

\begin{figure}
\centering
\includegraphics{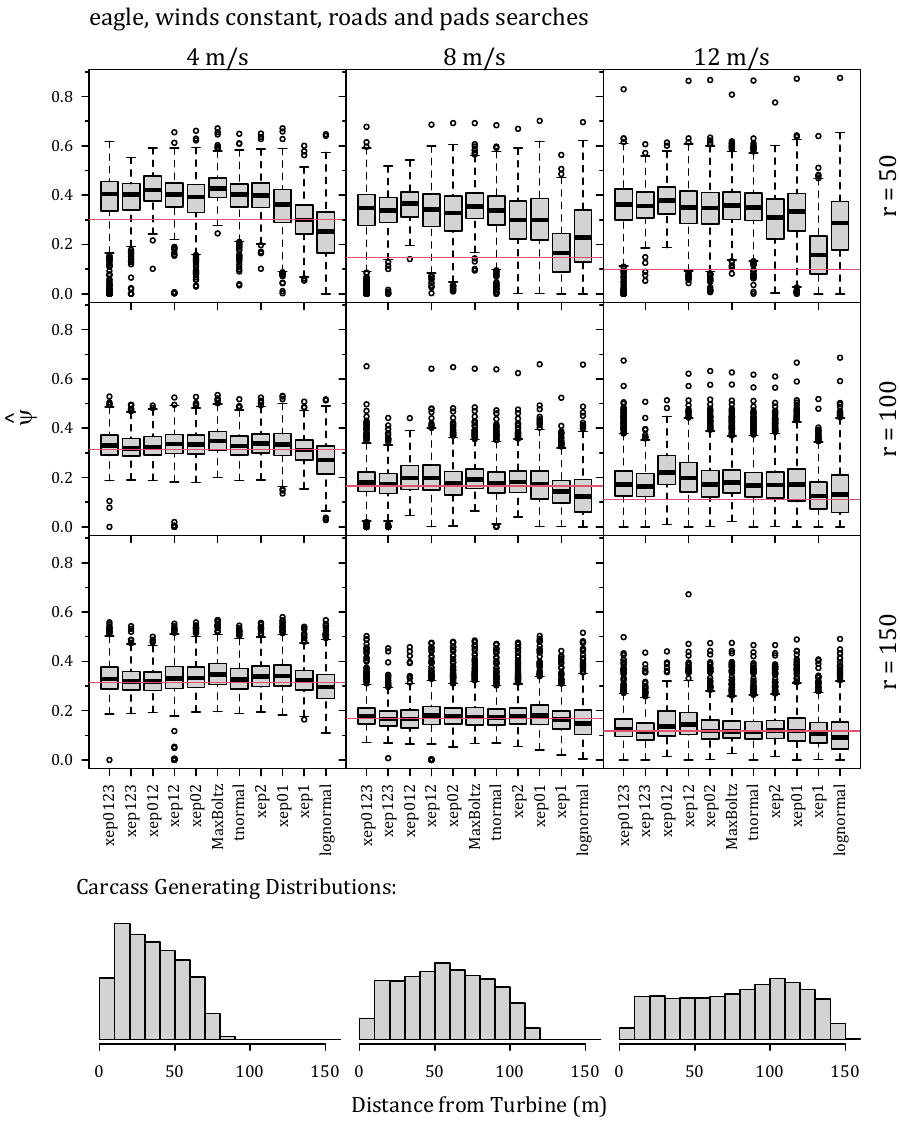}
\caption{\label{fig:psi112}Estimated \(\psi\) for the standard models
with simulated M = 200 eagles with winds constant and roads \& pads
searches. Boxes show sample IQR with median; whiskers extend to the most
extreme points within 1.5 IQR of the box; points beyond 1.5 IQR of the
box are shown as small circles. Red lines show the true \(\psi\).}
\end{figure}

\begin{figure}
\centering
\includegraphics{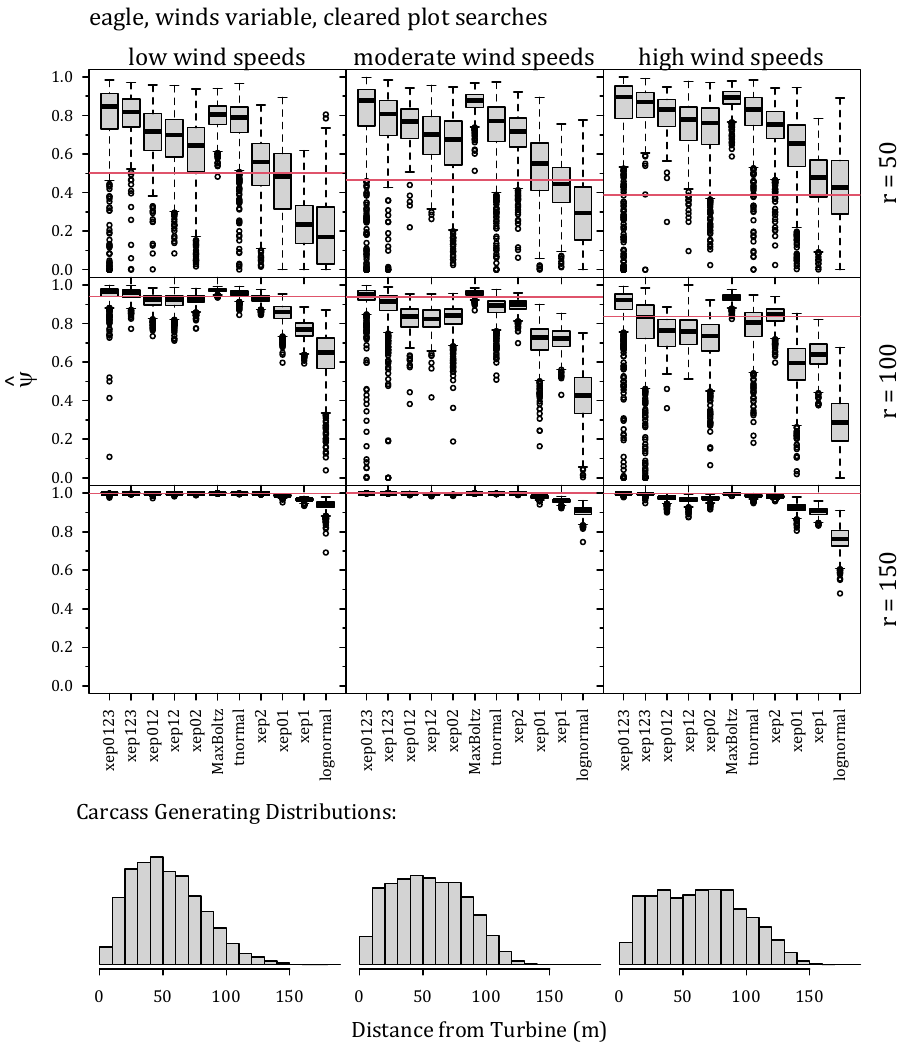}
\caption{\label{fig:psi121}Estimated \(\psi\) for the standard models
with simulated M = 200 eagles with winds variable and cleared plot
searches. Boxes show sample IQR with median; whiskers extend to the most
extreme points within 1.5 IQR of the box; points beyond 1.5 IQR of the
box are shown as small circles. Red lines show the true \(\psi\).}
\end{figure}

\begin{figure}
\centering
\includegraphics{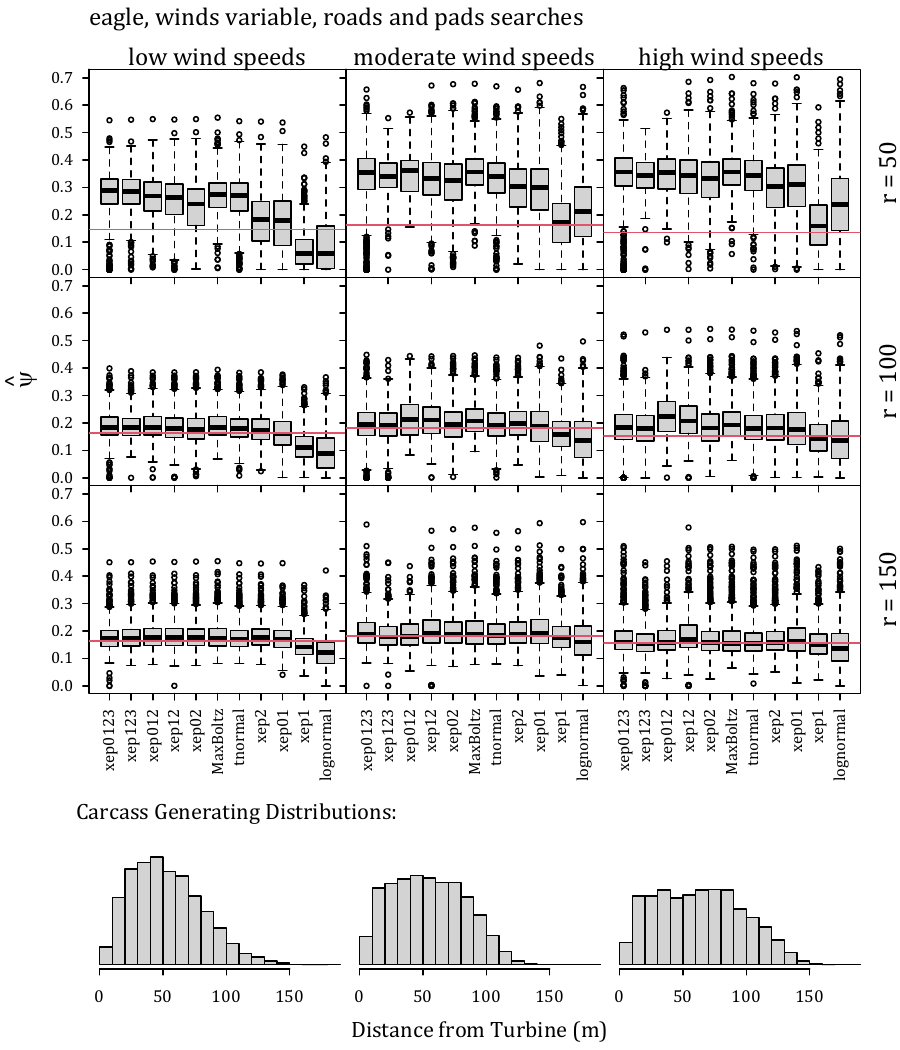}
\caption{\label{fig:psi122}Estimated \(\psi\) for the standard models
with simulated M = 200 eagles with winds variable and roads \& pads
searches. Boxes show sample IQR with median; whiskers extend to the most
extreme points within 1.5 IQR of the box; points beyond 1.5 IQR of the
box are shown as small circles. Red lines show the true \(\psi\).}
\end{figure}

\begin{figure}
\centering
\includegraphics{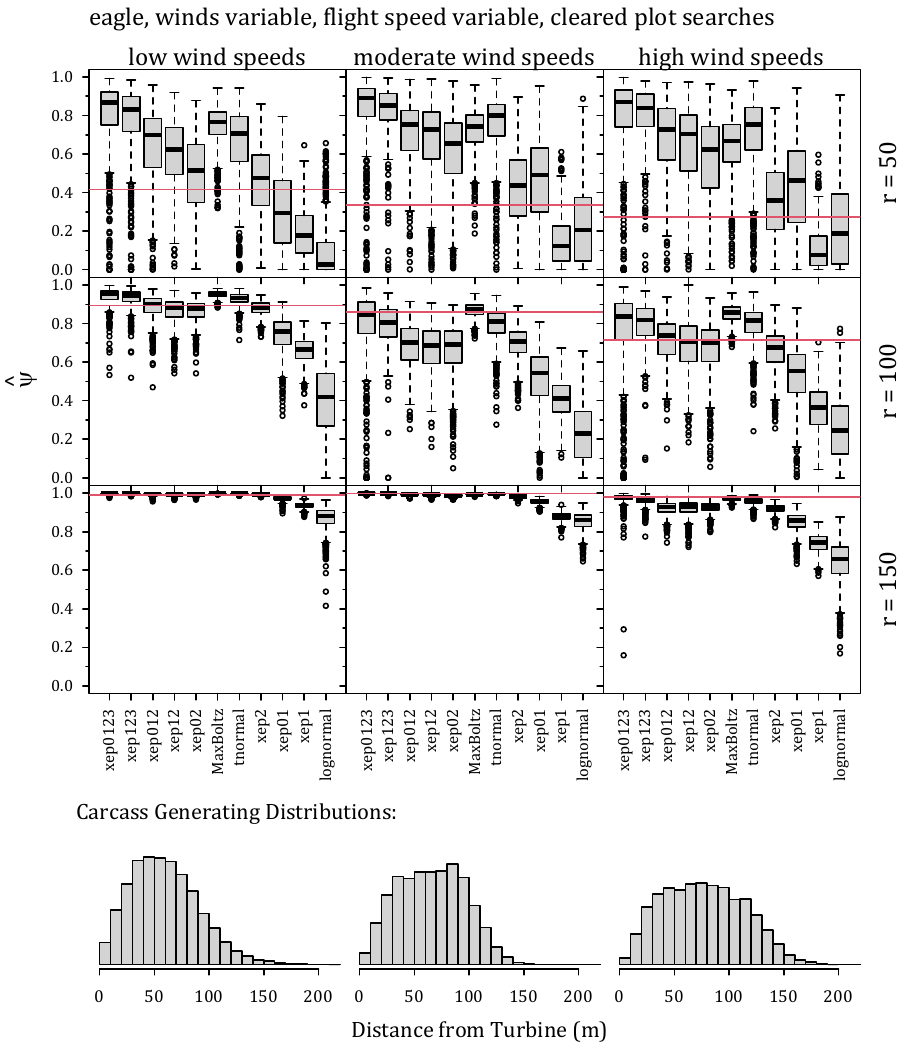}
\caption{\label{fig:psi131}Estimated \(\psi\) for the standard models
with simulated M = 200 eagles with winds variable, flight speed variable
and cleared plot searches. Boxes show sample IQR with median; whiskers
extend to the most extreme points within 1.5 IQR of the box; points
beyond 1.5 IQR of the box are shown as small circles. Red lines show the
true \(\psi\).}
\end{figure}

\begin{figure}
\centering
\includegraphics{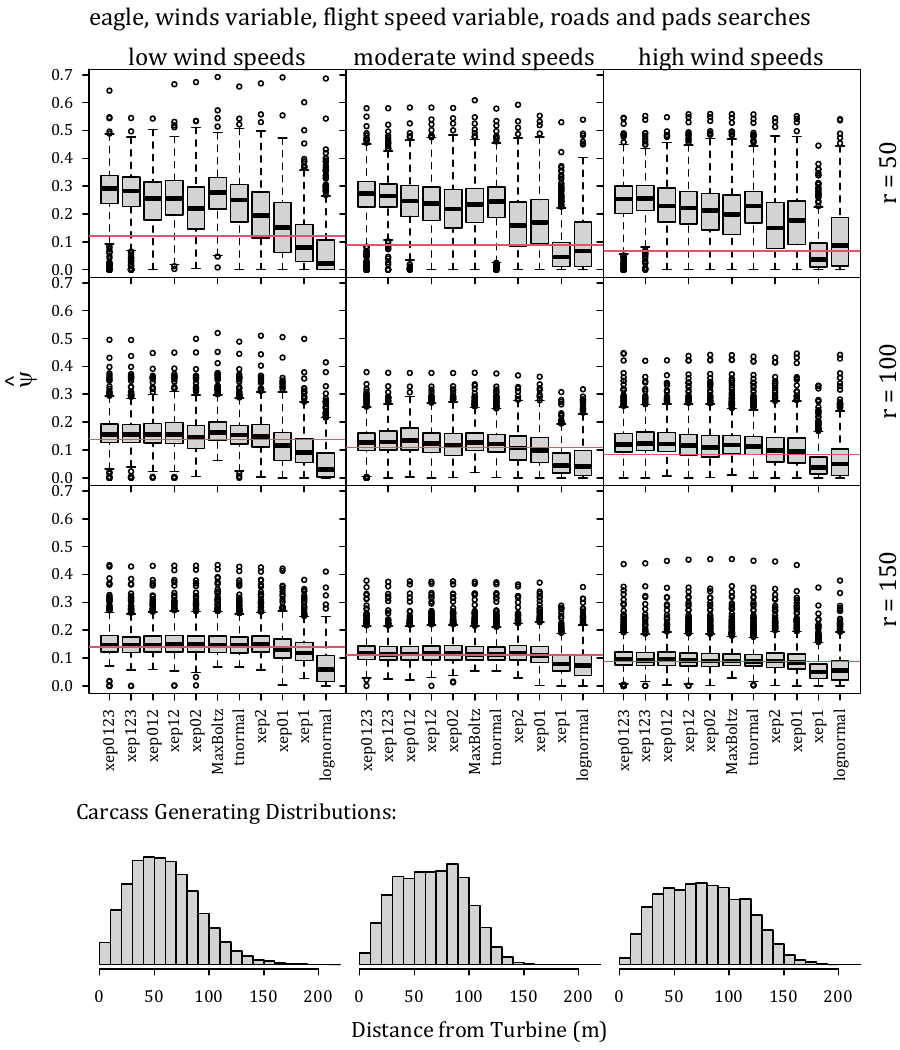}
\caption{\label{fig:psi132}Estimated \(\psi\) for the standard models
with simulated M = 200 eagles with winds variable, flight speed variable
and roads \& pads searches. Boxes show sample IQR with median; whiskers
extend to the most extreme points within 1.5 IQR of the box; points
beyond 1.5 IQR of the box are shown as small circles. Red lines show the
true \(\psi\).}
\end{figure}

\begin{figure}
\centering
\includegraphics{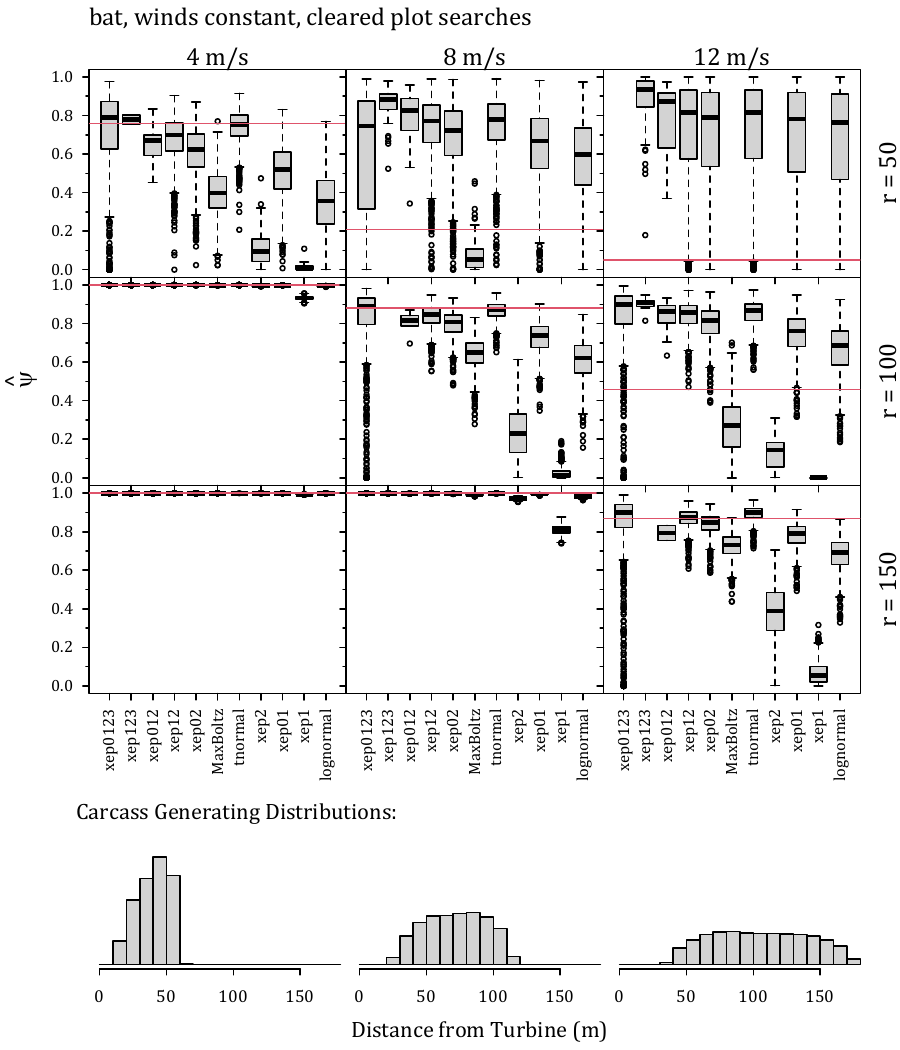}
\caption{\label{fig:psi211}Estimated \(\psi\) for the standard models
with simulated M = 200 bats with winds constant and cleared plot
searches. Boxes show sample IQR with median; whiskers extend to the most
extreme points within 1.5 IQR of the box; points beyond 1.5 IQR of the
box are shown as small circles. Red lines show the true \(\psi\).}
\end{figure}

\begin{figure}
\centering
\includegraphics{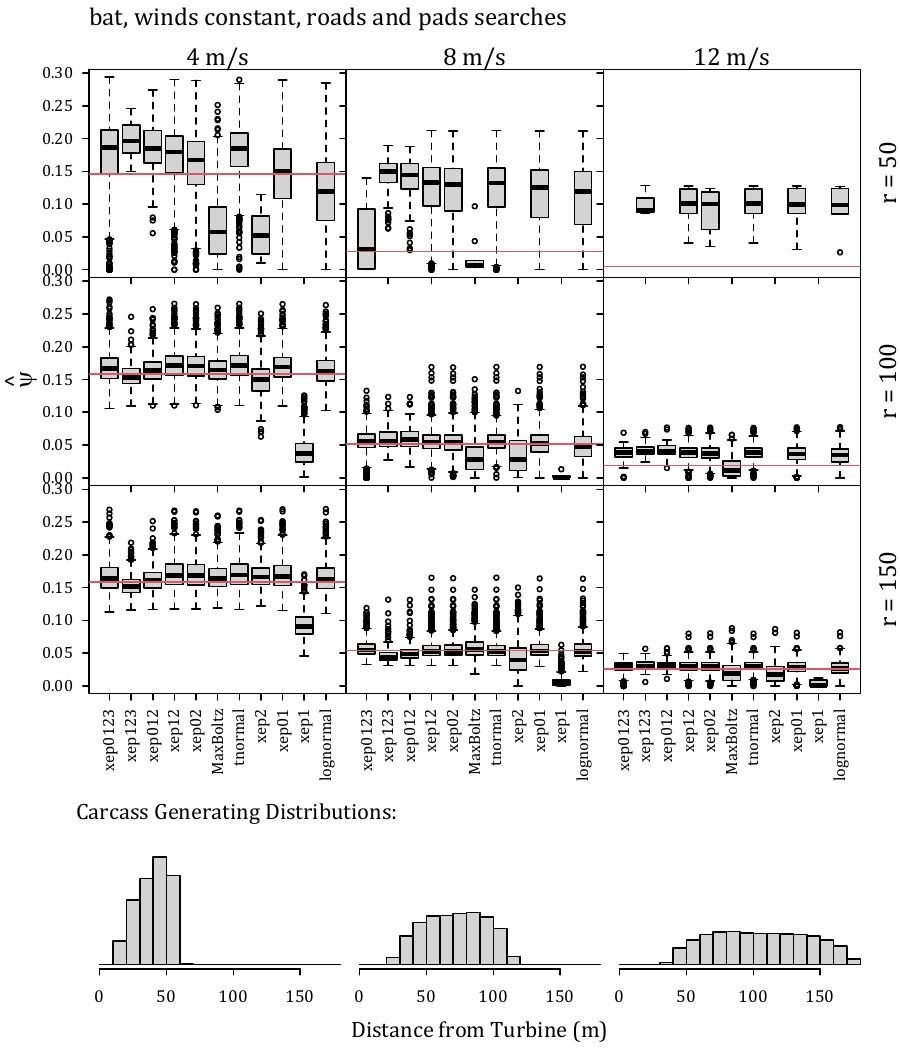}
\caption{\label{fig:psi212}Estimated \(\psi\) for the standard models
with simulated M = 200 bats with winds constant and roads \& pads
searches. Boxes show sample IQR with median; whiskers extend to the most
extreme points within 1.5 IQR of the box; points beyond 1.5 IQR of the
box are shown as small circles. Red lines show the true \(\psi\).}
\end{figure}

\begin{figure}
\centering
\includegraphics{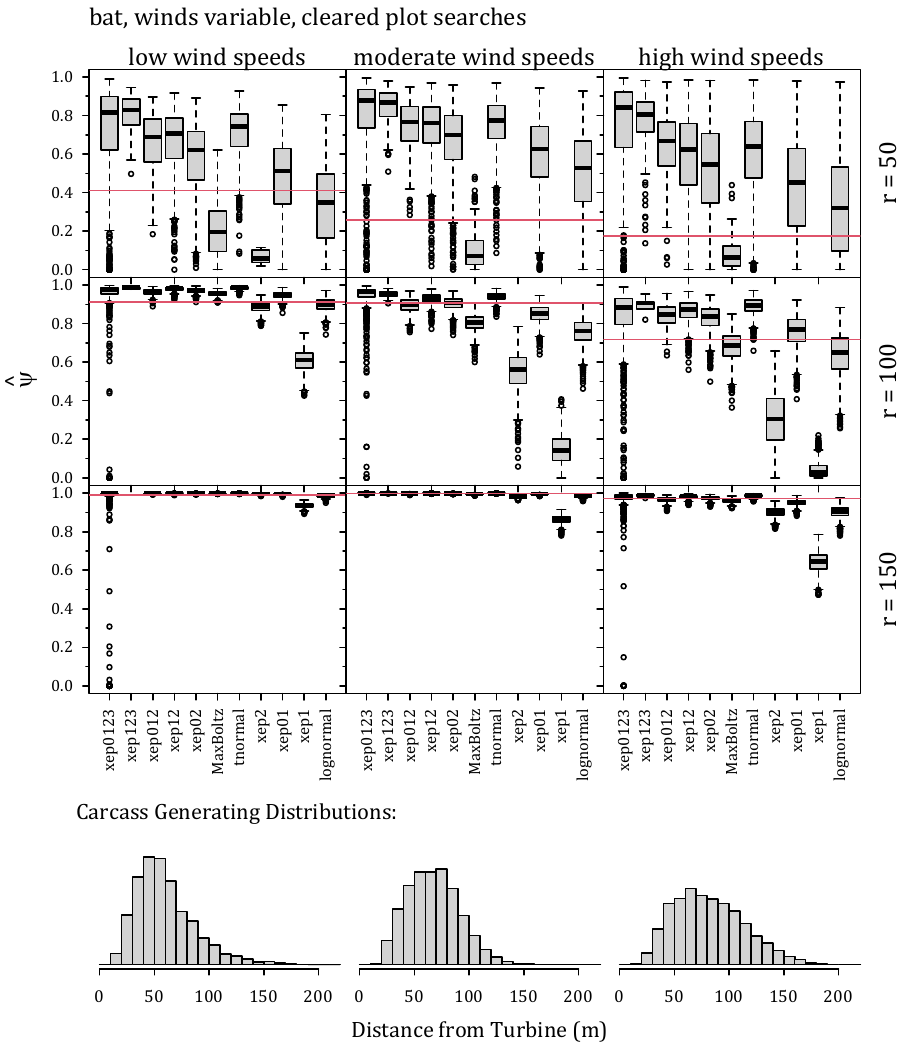}
\caption{\label{fig:psi221}Estimated \(\psi\) for the standard models
with simulated M = 200 bats with winds variable and cleared plot
searches. Boxes show sample IQR with median; whiskers extend to the most
extreme points within 1.5 IQR of the box; points beyond 1.5 IQR of the
box are shown as small circles. Red lines show the true \(\psi\).}
\end{figure}

\begin{figure}
\centering
\includegraphics{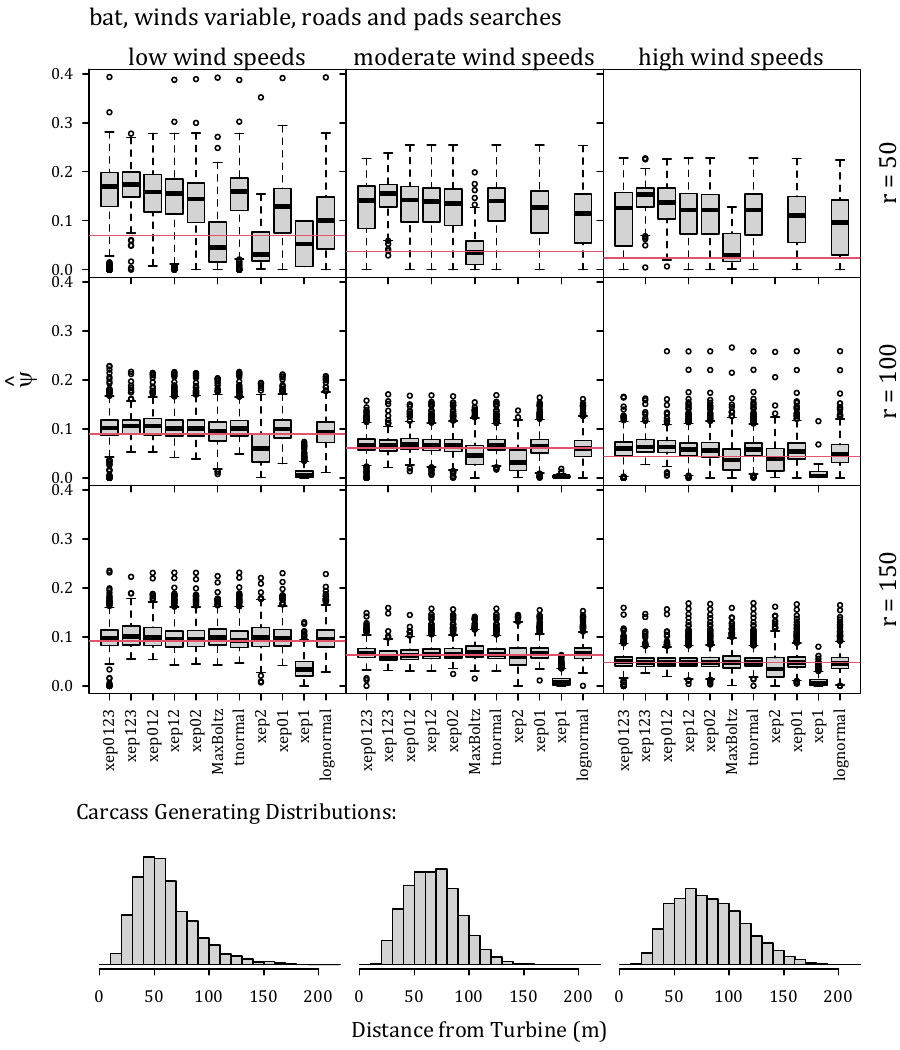}
\caption{\label{fig:psi222}Estimated \(\psi\) for the standard models
with simulated M = 200 bats with winds variable and roads \& pads
searches. Boxes show sample IQR with median; whiskers extend to the most
extreme points within 1.5 IQR of the box; points beyond 1.5 IQR of the
box are shown as small circles. Red lines show the true \(\psi\).}
\end{figure}

\begin{figure}
\centering
\includegraphics{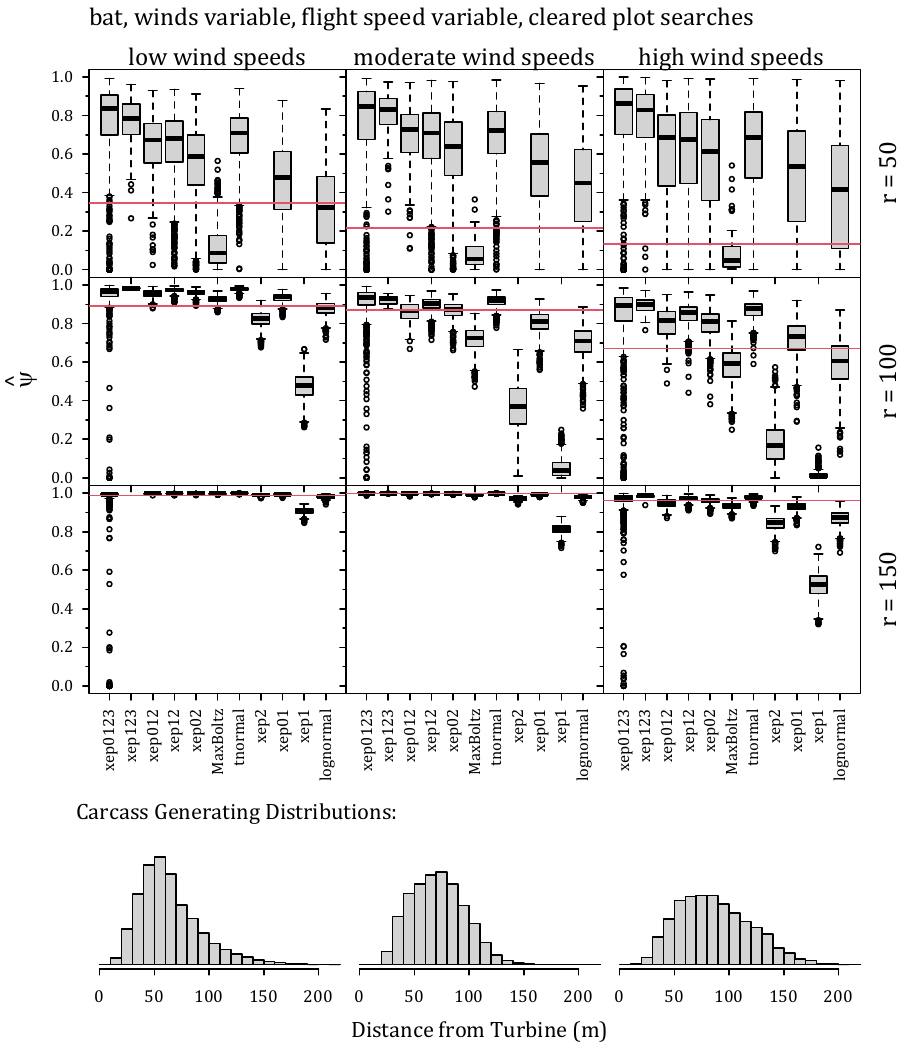}
\caption{\label{fig:psi231}Estimated \(\psi\) for the standard models
with simulated M = 200 bats with winds variable, flight speed variable
and cleared plot searches. Boxes show sample IQR with median; whiskers
extend to the most extreme points within 1.5 IQR of the box; points
beyond 1.5 IQR of the box are shown as small circles. Red lines show the
true \(\psi\).}
\end{figure}

\begin{figure}
\centering
\includegraphics{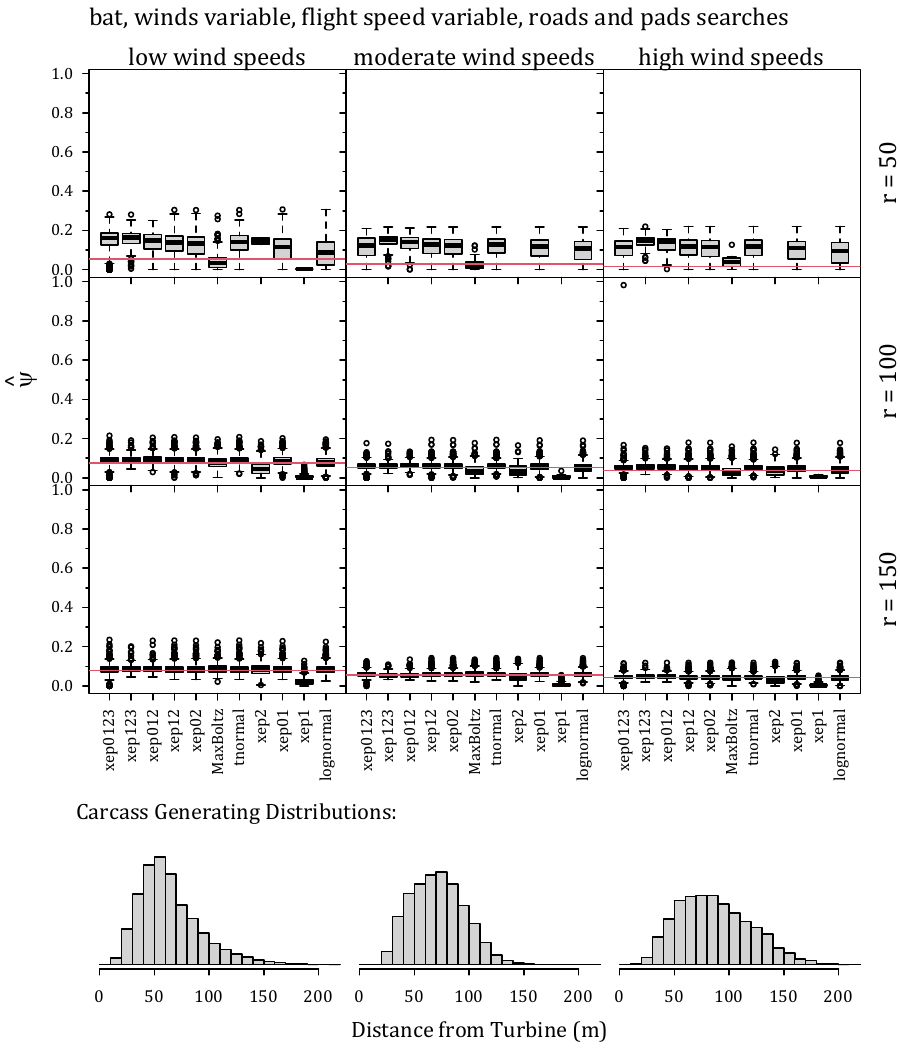}
\caption{\label{fig:psi232}Estimated \(\psi\) for the standard models
with simulated M = 200 bats with winds variable, flight speed variable
and roads \& pads searches. Boxes show sample IQR with median; whiskers
extend to the most extreme points within 1.5 IQR of the box; points
beyond 1.5 IQR of the box are shown as small circles. Red lines show the
true \(\psi\).}
\end{figure}

We tested the model predictions for the standard \(dwp\) models
(Appendix \ref{app:modbrief}) in a variety of simulation scenarios
(table \ref{tbl:parms}), with \(n = 1000\) replicates for each scenario.
For each replicate, a total of 200 random carcass distances were
generated from the distance distributions arising from the ballistics
models for eagles and bats under the wind regimes described in table
\ref{tbl:parms} and Appendix \ref{app:deqresults} and shown in figs.
\ref{fig:eaglehist} and \ref{fig:bathist}. Random directions were
generated as uniform(0, \(2\pi\)) deviates. Carcasses that lay within
the searched area (fig.~\ref{fig:searchPlots}) were found and included
in the data used for fitting the models; carcasses falling outside the
searched area were not included. For replicates in which at least 5
carcasses were found, each of the standard \(dwp\) models was fit and
\(\hat{\psi}\) was calculated using the maximum likelihood estimator for
the model parameters whenever the model was extensible. For each
simulation scenario, box plots of \(\hat{\psi}\) are presented for each
of the standard \(dwp\) models (figs. \ref{fig:psi111} -
\ref{fig:psi232}), ordered by degree of the polynomial in the model
(table \ref{tbl:formsST}).

\hypertarget{accuracy-of-hatpsi-eagles-under-constant-winds-and-cleared-search-plots}{%
\subsubsection{\texorpdfstring{Accuracy of \(\hat{\psi}\): Eagles under
constant winds and cleared search plots
\label{app:fixeagle_cleared}}{Accuracy of \textbackslash hat\{\textbackslash psi\}: Eagles under constant winds and cleared search plots }}\label{accuracy-of-hatpsi-eagles-under-constant-winds-and-cleared-search-plots}}

For eagles under constant wind conditions in the simulations, the
distribution (PDF) of carcasses was fairly flat out to a certain point
but then dropped rapidly to zero (fig.~\ref{fig:psi111}). Under these
conditions, a short search radius misses the precipitous decline in the
probability density, and the empirical models have great difficulty
estimating the fraction of carcasses within the search radius.

At 4 m/s, the carcass PDF (fig.~\ref{fig:psi111}, histogram on bottom
left) begins a long decline as distance increases beyond approximately
10 m. The heavy-tailed lognormal continues the gradual decrease past the
search radius, unable to accommodate the sharp drop beginning at about
75 m. As a result, the lognormal regularly underestimates \(\psi\) under
these conditions as the model implicitly assumes that there is a
substantial number of carcasses past the search radius. By contrast, the
distributions with degree \(\geq\!2\) have an inherently greater
acceleration in the rate of decline in the PDF and tend to predict fewer
carcasses within the search radius than do the heavier-tailed
distributions. With a search radius of 100 m, all carcasses fell within
the search plot, and all the fitted distributions except the lognormal
routinely estimated \(\hat{\psi} > 90\%\).

At 12 m/s (fig.~\ref{fig:psi111}, right column), the situation changes
dramatically. The density of carcasses (PDF) is largely flat but
gradually increasing with distance from the turbine until about 125 m,
when the density drops rapidly to zero. The short search radius
(\(r = 50\) m) entirely misses the decrease, and almost all the fitted
distributions vastly overestimate the fraction of carcasses within the
search area as their implicit assumptions about how the density would
converge to zero are in error, resulting curves that underestimate the
extent of the flat part of the density. The relatively heavy-tailed
lognormal and xep1 distributions are forced to decrease gradually from
their peak, and their degree of overestimation of \(\psi\) in this
scenario is modest compared to the lighter-tailed distributions.

The distributions of carcasses within the first 100 m from a turbine
were similar for winds of 8 and 12 m/s (fig.~\ref{fig:psi111},
histograms), so the \(r = 50\) and 100 m search radii yield similar
fitted distributions (fig.~\ref{fig:psi111}, comparing the 8 m/s panels
with the 12 m/s panels for \(r = 50\) and 100). However, the accuracy of
the estimates depends on the target (fig.~\ref{fig:psi111}, red
horizontal lines).

\hypertarget{accuracy-of-hatpsi-eagles-under-constant-winds-and-constant-flight-speeds-for-road-pad-searches}{%
\subsubsection{\texorpdfstring{Accuracy of \(\hat{\psi}\): Eagles under
constant winds and constant flight speeds for road \& pad searches
\label{app:fixeagle_rp}}{Accuracy of \textbackslash hat\{\textbackslash psi\}: Eagles under constant winds and constant flight speeds for road \& pad searches }}\label{accuracy-of-hatpsi-eagles-under-constant-winds-and-constant-flight-speeds-for-road-pad-searches}}

Under constant wind conditions, predictions across the fitted
distributions were more consistent in the road \& pad searches than in
the cleared plot searches (fig.~\ref{fig:psi112}). In each panel, the
distinctions among the fitted distributions were slight. However, the
lognormal almost always predicted a smaller fraction of carcasses within
the searched area than did all other models. The only exceptions were
for a short search radius was short (\(r = 50\) m) and the wind speed
was 8 or 12 m/s, in which case xep1 had the smallest (and most accurate)
estimates for \(\psi\). When the search radius was not short
(\(r = 100\) or 150 m), there was little distinction among the model
fits.

\hypertarget{accuracy-of-hatpsi-eagles-under-varying-winds-and-constant-flight-speed-for-cleared-plot-and-road-pad-searches}{%
\subsubsection{\texorpdfstring{Accuracy of \(\hat{\psi}\): Eagles under
varying winds and constant flight speed for cleared plot and road \& pad
searches}{Accuracy of \textbackslash hat\{\textbackslash psi\}: Eagles under varying winds and constant flight speed for cleared plot and road \& pad searches}}\label{accuracy-of-hatpsi-eagles-under-varying-winds-and-constant-flight-speed-for-cleared-plot-and-road-pad-searches}}

The eagle distributions and fitted models under variable winds (figs.
\ref{fig:psi121}, \ref{fig:psi122}) were similar to those under constant
winds (figs. \ref{fig:psi111}), \ref{fig:psi112}). Refer to sections
\ref{app:fixeagle_cleared} and \ref{app:fixeagle_rp} for discussion.

\hypertarget{accuracy-of-hatpsi-eagles-under-varying-winds-and-flight-speeds-cleared-plot-searches-road-pad-searches}{%
\subsubsection{\texorpdfstring{Accuracy of \(\hat{\psi}\): Eagles under
varying winds and flight speeds (cleared plot searches, road \& pad
searches)}{Accuracy of \textbackslash hat\{\textbackslash psi\}: Eagles under varying winds and flight speeds (cleared plot searches, road \& pad searches)}}\label{accuracy-of-hatpsi-eagles-under-varying-winds-and-flight-speeds-cleared-plot-searches-road-pad-searches}}

The eagle distributions under variable winds and flight speeds (figs.
\ref{fig:psi131}, \ref{fig:psi132}) were notably less flat than the
eagle scenarios in which the effect of initial flight speed was
nullified upon impact by the turbine blade. The predicted proportion of
carcasses within the searched area generally decreased with the order of
the fitted distribution, with the accuracy increasing with search
radius. With road \& pad searches, there was a marked tendency for
models to overpredict \(\psi\), especially when the search radius was
short (\(r = 50\)). That tendency to overpredict was fully compensated
for in the lognormal and xep1 models, which, as heavy-tailed
distributions, naturally tend to underpredict.

\hypertarget{accuracy-of-hatpsi-bats}{%
\subsubsection{\texorpdfstring{Accuracy of \(\hat{\psi}\):
Bats}{Accuracy of \textbackslash hat\{\textbackslash psi\}: Bats}}\label{accuracy-of-hatpsi-bats}}

Bat carcass distributions tended to have fewer carcasses near the
turbine and a somewhat more elongated right tail than did the eagle
distributions (figs. \ref{fig:psi211}-\ref{fig:psi232}). As with the
eagles, the models tended to overpredict by a substantial margin when
the search radius was short (\(r = 50\) m), variation among model
predictions was much smaller for road \& pad searches than for cleared
plot searches, and \(\hat{\psi}\) tended to decrease with the degree of
the model (as reflected in the decreasing trend in the boxes in figs.
\ref{fig:psi211}-\ref{fig:psi232}).

The same pattern naturally arises with the road \& pad searches. The
sample is likely to not have any carcasses at great distances because
the search coverage is too low there. The light-tailed distributions
drop off quickly, so there's not much probability in the tails and they
overestimate the fraction of carcasses in the search radius. As the
search radius increases, the heavy-tailed distributions tend to think
there's still a lot of carcasses farther out, and consequently they tend
to underestimate the fraction falling in the searched area.

\hypertarget{akaike-information-criterion-model-accuracy-and-psi}{%
\subsection{\texorpdfstring{Akaike Information Criterion, Model
Accuracy, and
\(\psi\)}{Akaike Information Criterion, Model Accuracy, and \textbackslash psi}}\label{akaike-information-criterion-model-accuracy-and-psi}}

\begin{figure}
\centering
\includegraphics{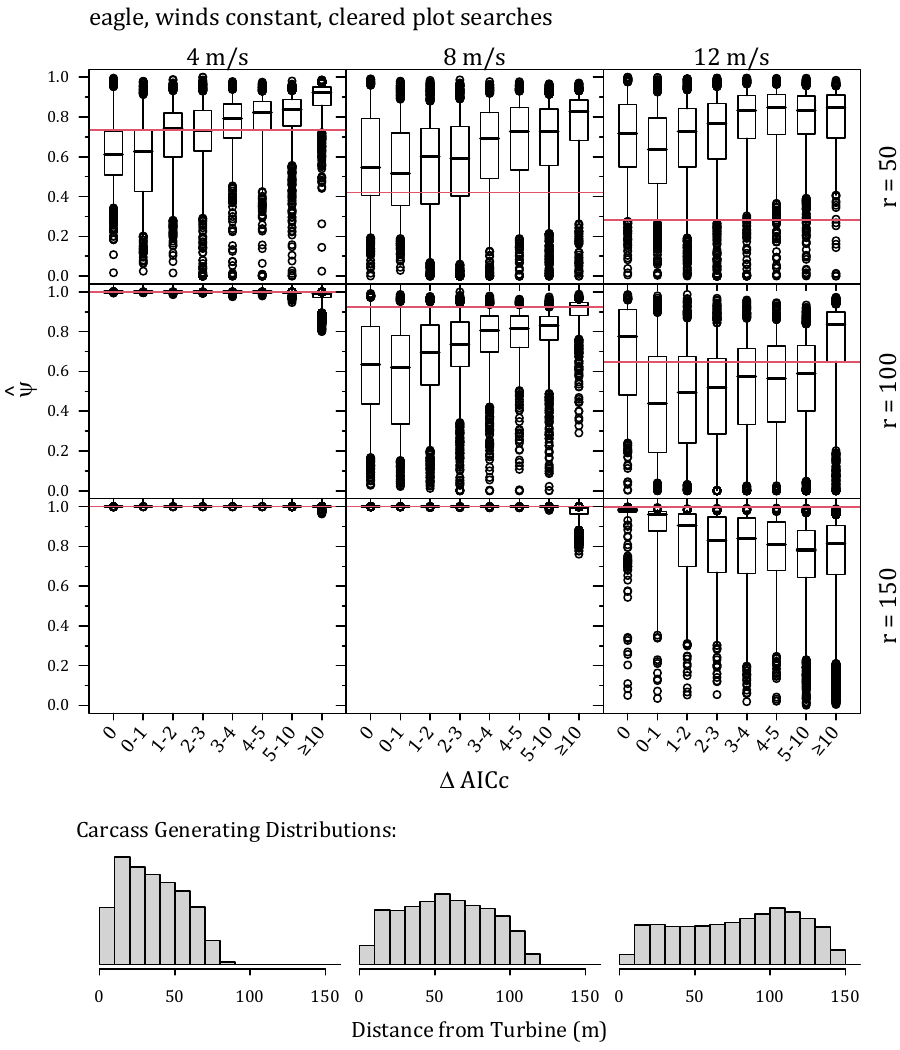}
\caption{\label{fig:psiaic111}AICc and estimated \(\psi\) for the
standard models with simulated M = 200 eagles with winds constant and
cleared plot searches. Boxes show sample IQR with median; whiskers
extend to the most extreme points within 1.5 IQR of the box; points
beyond 1.5 IQR of the box are shown as small circles. Red lines show the
true \(\psi\).}
\end{figure}

The Akaike information criterion (AICc) is a well-tested and commonly
used tool for model selection and has proven to be effective for
distinguishing among models by the relative quality of the fit to the
data. However, its validity is limited to the scope of the data, and
there is no guarantee that models that perform well within the scope of
the data will perform well---or will even be remotely plausible---when
extended outside the scope of the data, as must be done when predicting
the proportion of carcasses that lie outside the searched area
(\(1-\psi\)). To test the utility of AICc for prediction of \(\psi\), we
summarized the results from the simulations in section \ref{app:predpsi}
by \(\Delta\mathrm{AICc}\), binning the \(\hat{\psi}\) values by the
\(\Delta\mathrm{AICc}\) scores of the fitted model. For example, for
eagles under constant wind speeds of 4 m/s and searched on cleared plots
out to a radius of \(r = 50\) m, we generated 1000 simulated data sets,
fit the 11 standard \(dwp\) models, and binned the \(\hat{\psi}\) values
according to their \(\Delta\mathrm{AICc}\) scores
(fig.~\ref{fig:psiaic111}, upper left panel). The boxplot of
\(\hat{\psi}\) values for \(\Delta\mathrm{AICc} = 0\) summarizes the
collection of \(\hat{\psi}\)'s for the best-fitting model in each of the
1000 replicate data sets. The boxplot for \(\Delta\mathrm{AICc}\) = 0--1
is a summary of the \(\hat{\psi}\) values for models with
\(\Delta\mathrm{AICc} \in (0, 1]\), and the rightmost boxplot in each
panel is for models with \(\Delta\mathrm{AICc} \geq 10\).

\hypertarget{eagles}{%
\subsubsection{Eagles}\label{eagles}}

For eagles under constant wind conditions and the initial carcass
position and velocity are taken to be that of the rotor at the point of
impact, the distribution (PDF) of carcasses was fairly flat out to a
certain point but then dropped rapidly to zero
(fig.~\ref{fig:eaglehist}, left panels). Under these conditions, a short
search radius misses the precipitous decline and empirical models have
great difficulty estimating the fraction of carcasses within the search
radius. The key feature of the distribution---namely, the point marking
the beginning of the steep drop---is absent from the data, so measures
of how well the models fit within the range of the data (AICc, for
example) cannot make meaningful distinctions among the models for
extrapolating beyond the search radius.

In general, for cleared plots and fixed wind speeds,
\(\Delta\mathrm{AICc}\) was largely unrelated to accuracy in predicting
\(\psi\). For search radii of \(r = 50, \, 100\) m, \(\hat{\psi}\)
tended to increase with \(\Delta\mathrm{AICc}\)
(fig.~\ref{fig:psiaic111}), presumably because the heavier-tailed models
tended to fit better within the range of data than the lighter-tailed
models. However, the actual simulated carcass distributions are
light-tailed, so \(\Delta\mathrm{AICc}\) gives an unreliable measure of
the aptness of the model fits. For example, at 8 m/s, and a short search
radius (fig.~\ref{fig:psiaic111}, top middle panel), the higher the
\(\Delta\mathrm{AICc}\), the poorer the prediction. However, with a
longer search radius and a corresponding greater fraction of carcasses
within the search radius, the greater the AICc, the better the
prediction, the best-fitting models had by far the least accurate
predictions (fig.~\ref{fig:psiaic111}, center panel).

When the search radius was close to the maximum carcass distance, as in
the high winds, long search radius example, the AICc tended to
successfully select the model with the best estimate of \(\psi\) (bottom
right panel). When windspeeds dropped (4m/s or 8m/s) but search radius
remained long (bottom row left and center, or middle row at left), few
if any carcasses fell beyond the search radius and all models did well
at determinig that \(\psi = 1\). These scenarios are effectiely
interpolation rather than extrapolation given the length of the search
radius and the shape of the generating distributions.

Because \(\Delta\mathrm{AICc}\) applies strictly to the fit of models
within the range of data, we would expect \(\Delta\mathrm{AICc}\) to
perform better when the fraction of carcasses within the search radius
is higher, as illustrated in fig.~\ref{fig:psiaic111} for \(r = 150\),
where close to 100\% of the carcasses lie within the search radius, and,
in the case where there are distinctions among models (lower right
panel), greater \(\Delta\mathrm{AICc}\) values are associated with
poorer fits. However, in the center panel (8 m/s and \(r = 100\)), 90\%
of the carcasses lie within the search plot but the best predictors
tended to be the models with the poorest fits within the range of the
data. In this simulation scenario, the distribution of carcasses was
largely flat within the 100 m search radius but dropped rapidly to zero
just outside the search radius, requiring a light-tailed distribution in
order to capture the abrupt end of the distribution. However, it is the
heavy-tailed distributions that fit best within that first 100 m from
the turbine, hence the inverse relationship between the quality of the
fit within the range of data and the adequacy for prediction outside the
range of the data.

\begin{figure}
\centering
\includegraphics{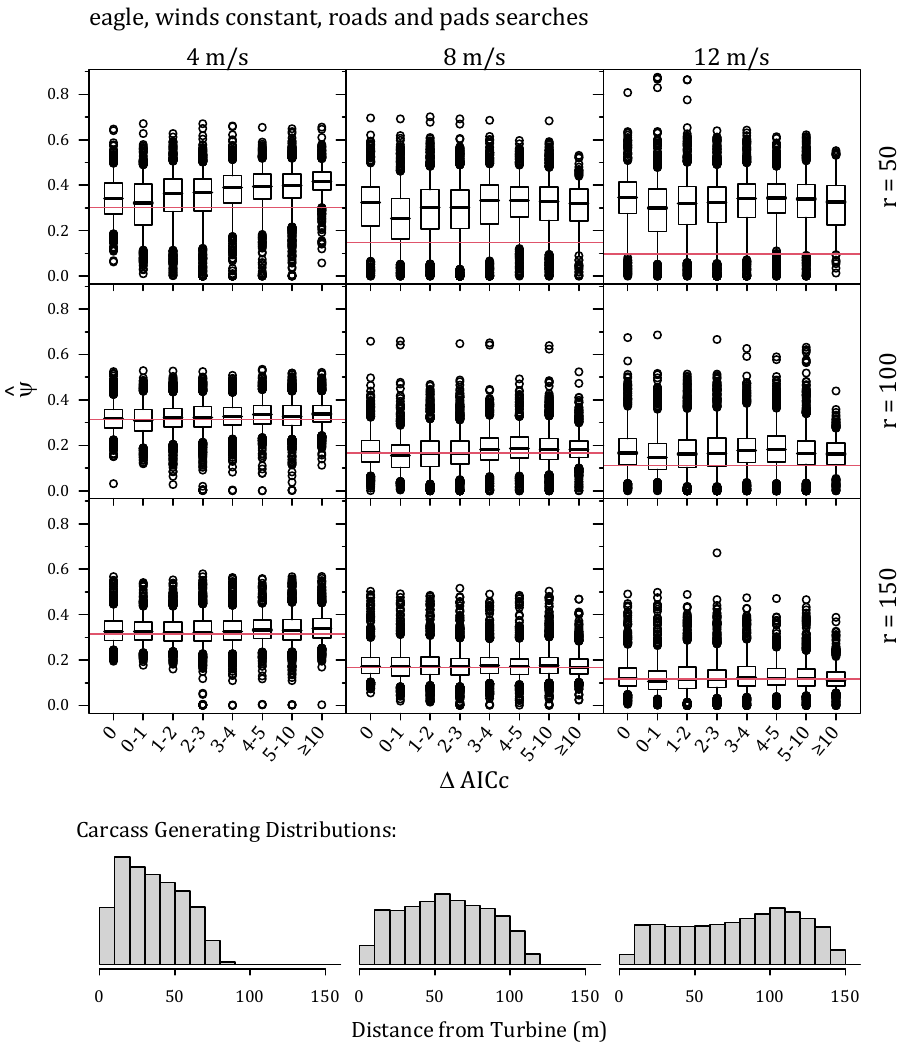}
\caption{\label{fig:psiaic112}AICc and estimated \(\psi\) for the
standard models with simulated M = 200 eagles with winds constant and
roads \& pads searches. Boxes show sample IQR with median; whiskers
extend to the most extreme points within 1.5 IQR of the box; points
beyond 1.5 IQR of the box are shown as small circles. Red lines show the
true \(\psi\).}
\end{figure}

\begin{figure}
\centering
\includegraphics{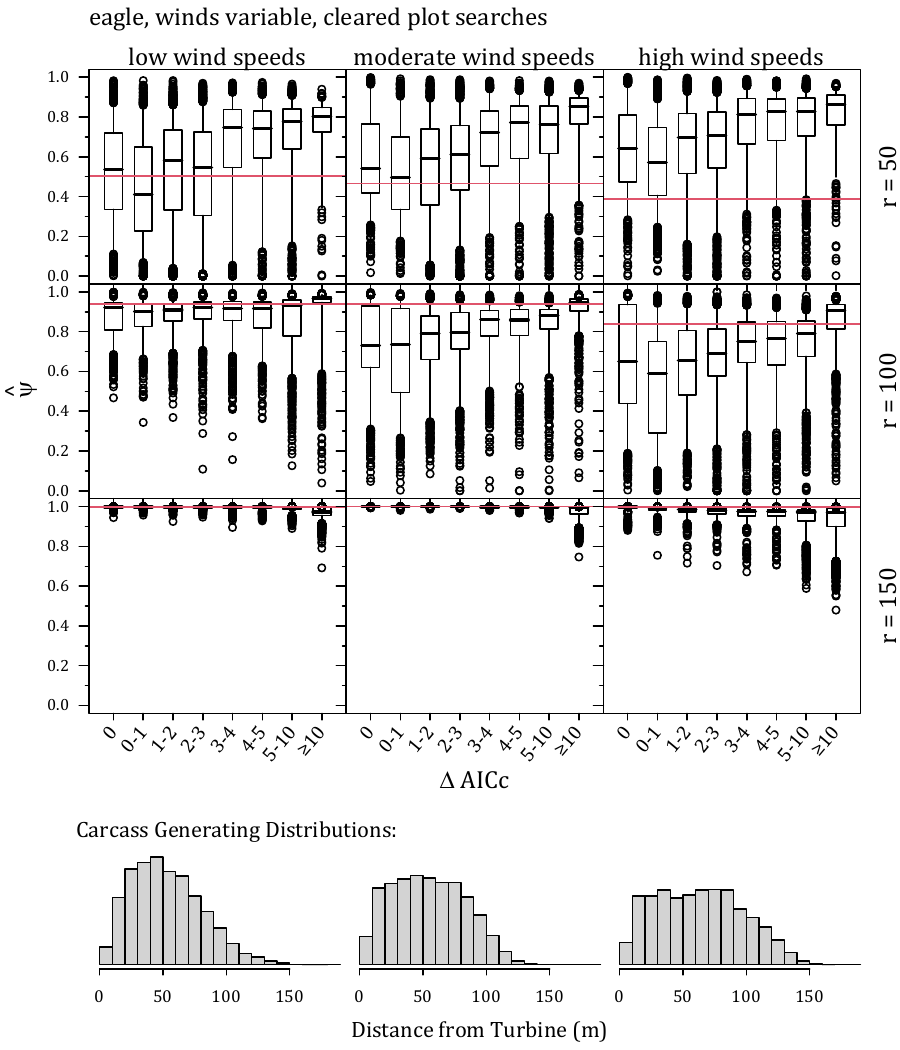}
\caption{\label{fig:psiaic121}AICc and estimated \(\psi\) for the
standard models with simulated M = 200 eagles with winds variable and
cleared plot searches. Boxes show sample IQR with median; whiskers
extend to the most extreme points within 1.5 IQR of the box; points
beyond 1.5 IQR of the box are shown as small circles. Red lines show the
true \(\psi\).}
\end{figure}

\begin{figure}
\centering
\includegraphics{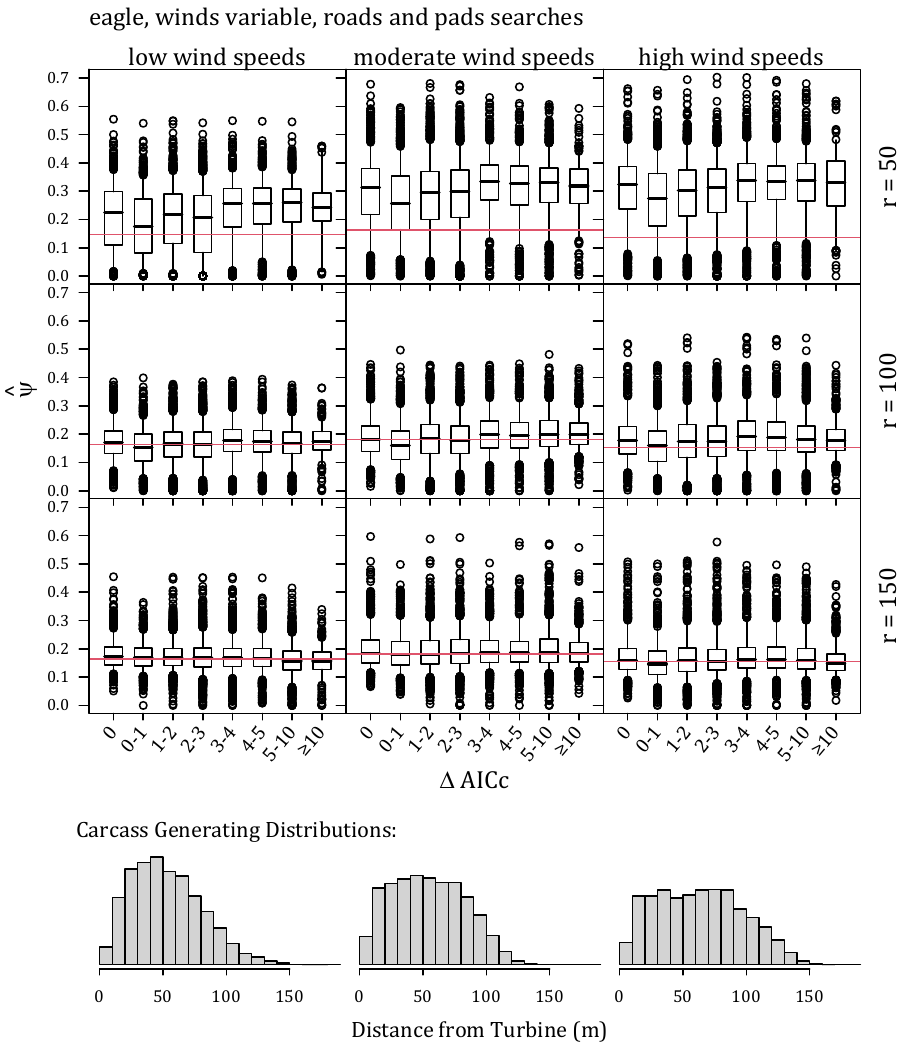}
\caption{\label{fig:psiaic122}AICc and estimated \(\psi\) for the
standard models with simulated M = 200 eagles with winds variable and
roads \& pads searches. Boxes show sample IQR with median; whiskers
extend to the most extreme points within 1.5 IQR of the box; points
beyond 1.5 IQR of the box are shown as small circles. Red lines show the
true \(\psi\).}
\end{figure}

\begin{figure}
\centering
\includegraphics{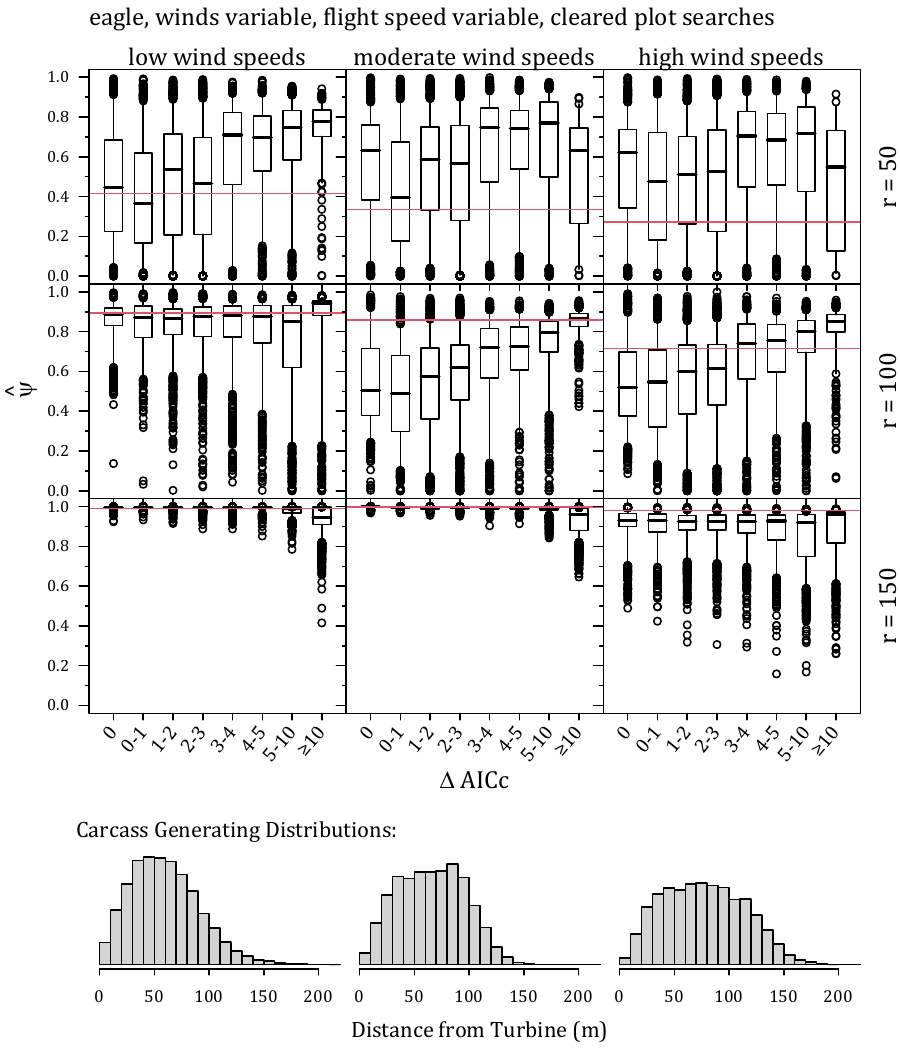}
\caption{\label{fig:psiaic131}AICc and estimated \(\psi\) for the
standard models with simulated M = 200 eagles with winds variable,
flight speed variable and cleared plot searches. Boxes show sample IQR
with median; whiskers extend to the most extreme points within 1.5 IQR
of the box; points beyond 1.5 IQR of the box are shown as small circles.
Red lines show the true \(\psi\).}
\end{figure}

\begin{figure}
\centering
\includegraphics{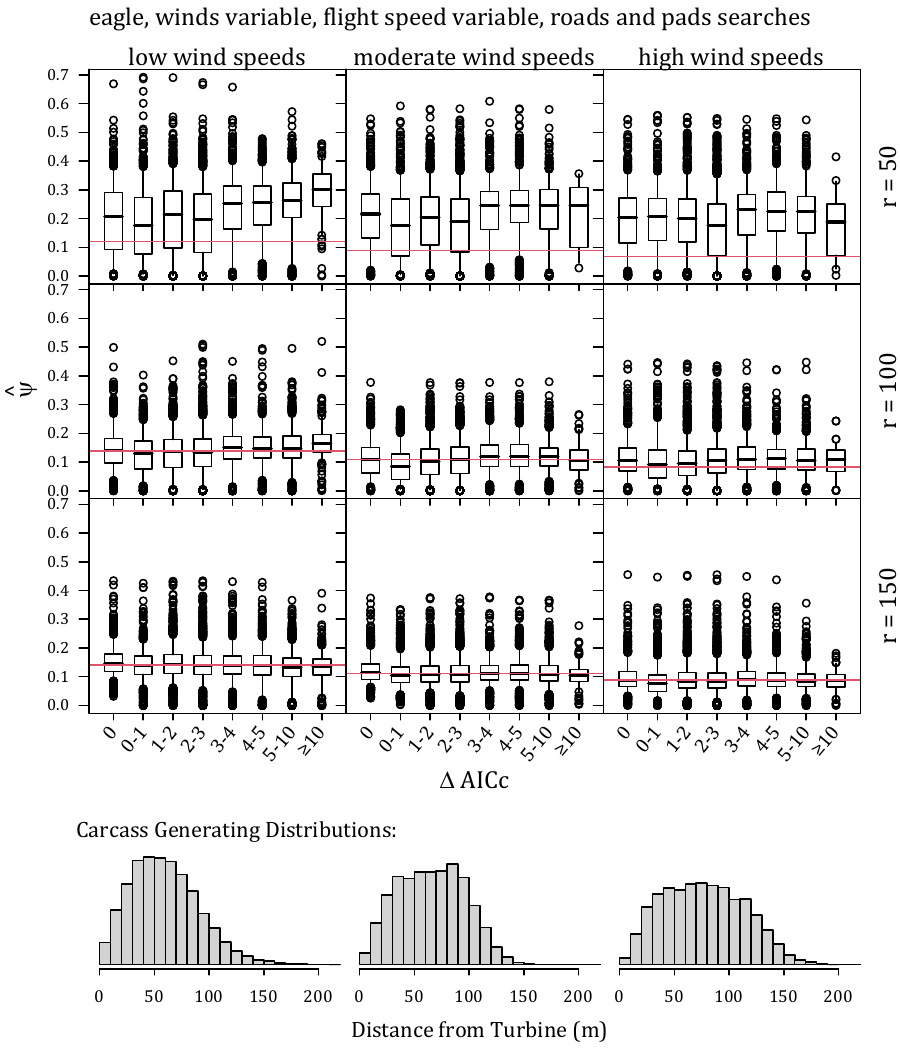}
\caption{\label{fig:psiaic132}AICc and estimated \(\psi\) for the
standard models with simulated M = 200 eagles with winds variable,
flight speed variable and roads \& pads searches. Boxes show sample IQR
with median; whiskers extend to the most extreme points within 1.5 IQR
of the box; points beyond 1.5 IQR of the box are shown as small circles.
Red lines show the true \(\psi\).}
\end{figure}

With road \& pad sampling, much of the spatial prediction is in
extending the models to the unsearched area within the search radius.
Because interpolation to unsearched areas within the range of the data
plays a large role in prediction of \(\psi\) from road \& pad data, we
would expect \(\Delta\mathrm{AICc}\) to be a better indicator of model
accuracy than it was for the cleared plot searches. Indeed, lower
\(\Delta\mathrm{AICc}\) values do seem to be somewhat associated with
more accurate prediction in the road \& pad scenarios
(fig.~\ref{fig:psiaic112}, fig.~\ref{fig:psiaic122} and
fig.~\ref{fig:psiaic132}) than in the corresponding cleared plot
scenarios (fig.~\ref{fig:psiaic111}, \ref{fig:psiaic121}, and
\ref{fig:psiaic131}, respectively).

\hypertarget{bats}{%
\subsubsection{Bats}\label{bats}}

\begin{figure}
\centering
\includegraphics{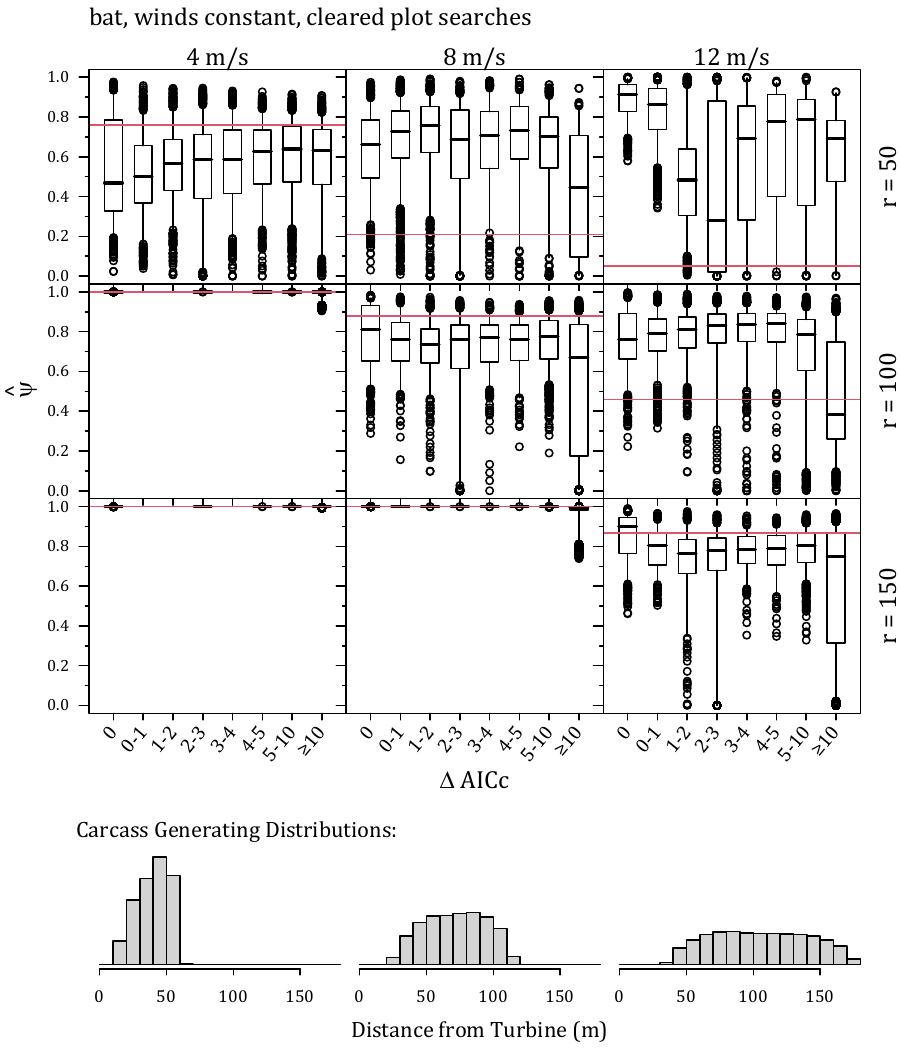}
\caption{\label{fig:psiaic211}AICc and estimated \(\psi\) for the
standard models with simulated M = 200 bats with winds constant and
cleared plot searches. Boxes show sample IQR with median; whiskers
extend to the most extreme points within 1.5 IQR of the box; points
beyond 1.5 IQR of the box are shown as small circles. Red lines show the
true \(\psi\).}
\end{figure}

Fig. \ref{fig:psiaic211} shows bat carcass distributions and estimated
probabilities that carcasses lie in the searched area for constant wind
speeds of 4, 8, and 12 m/s and cleared plot search radii of 50, 100, and
150 m. As with eagles, AICc does not appear to be a useful predictor of
model performance in estimating \(\psi\) except in cases where the
search radius is large enough to encompass nearly the whole of the
distribution.

\begin{figure}
\centering
\includegraphics{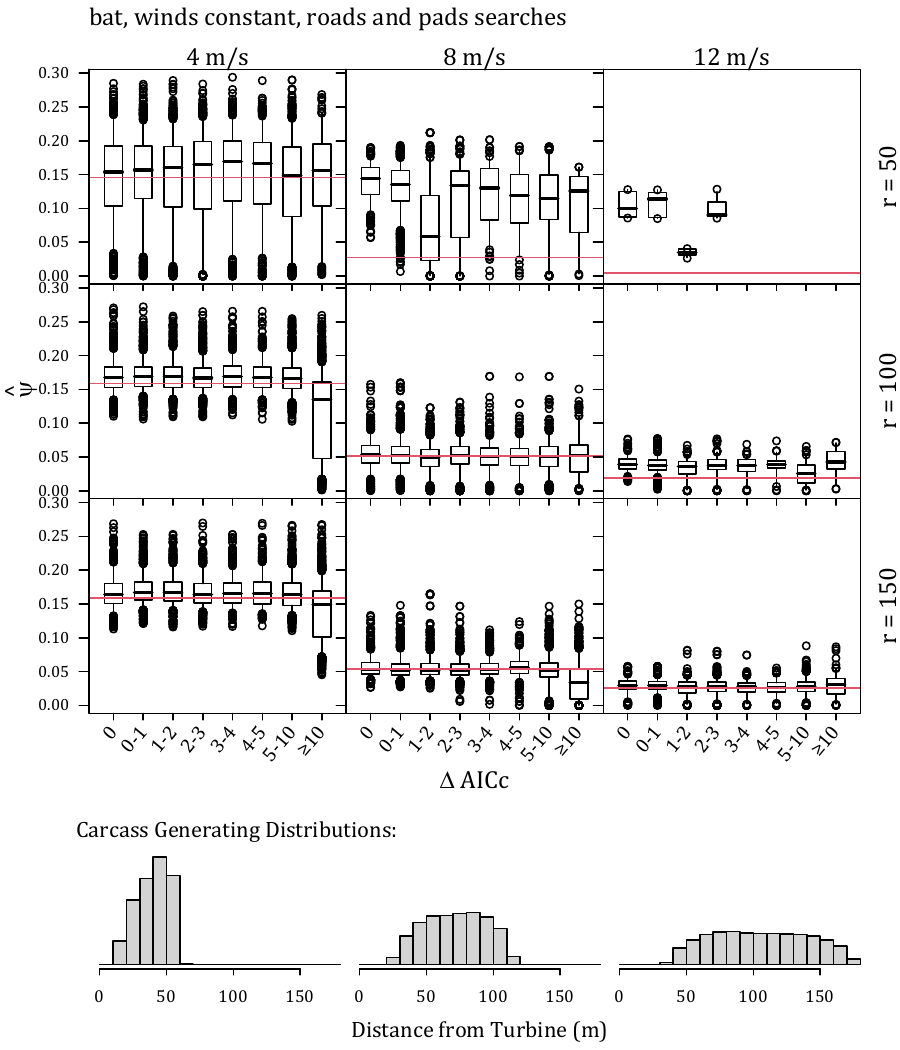}
\caption{\label{fig:psiaic212}AICc and estimated \(\psi\) for the
standard models with simulated M = 200 bats with winds constant and
roads \& pads searches. Boxes show sample IQR with median; whiskers
extend to the most extreme points within 1.5 IQR of the box; points
beyond 1.5 IQR of the box are shown as small circles. Red lines show the
true \(\psi\).}
\end{figure}

\begin{figure}
\centering
\includegraphics{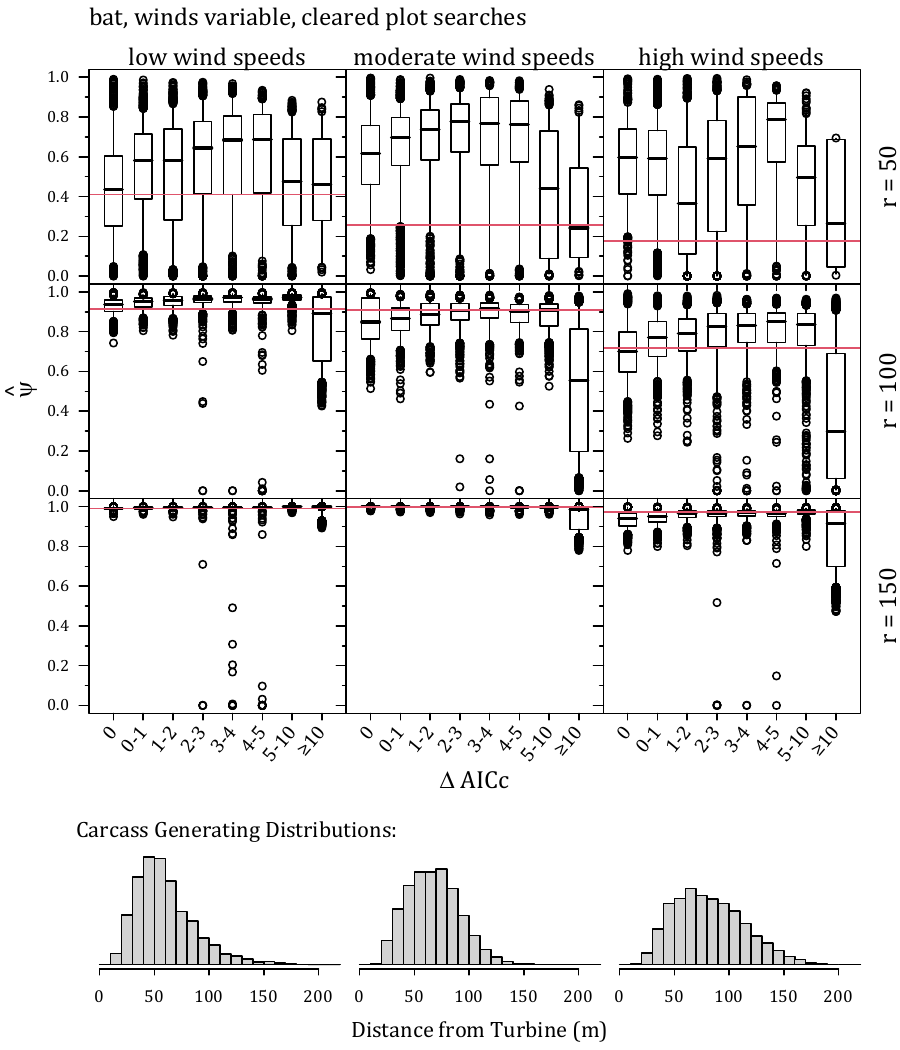}
\caption{\label{fig:psiaic221}AICc and estimated \(\psi\) for the
standard models with simulated M = 200 bats with winds variable and
cleared plot searches. Boxes show sample IQR with median; whiskers
extend to the most extreme points within 1.5 IQR of the box; points
beyond 1.5 IQR of the box are shown as small circles. Red lines show the
true \(\psi\).}
\end{figure}

\begin{figure}
\centering
\includegraphics{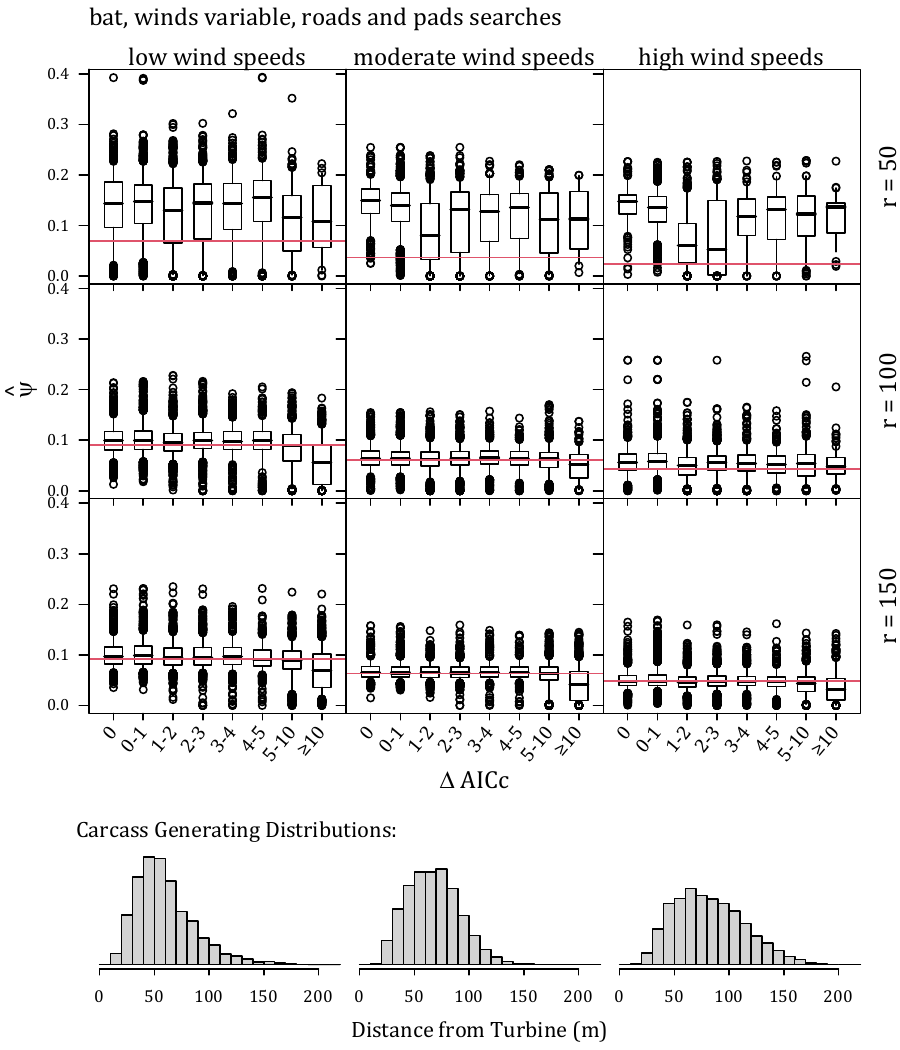}
\caption{\label{fig:psiaic222}AICc and estimated \(\psi\) for the
standard models with simulated M = 200 bats with winds variable and
roads \& pads searches. Boxes show sample IQR with median; whiskers
extend to the most extreme points within 1.5 IQR of the box; points
beyond 1.5 IQR of the box are shown as small circles. Red lines show the
true \(\psi\).}
\end{figure}

\begin{figure}
\centering
\includegraphics{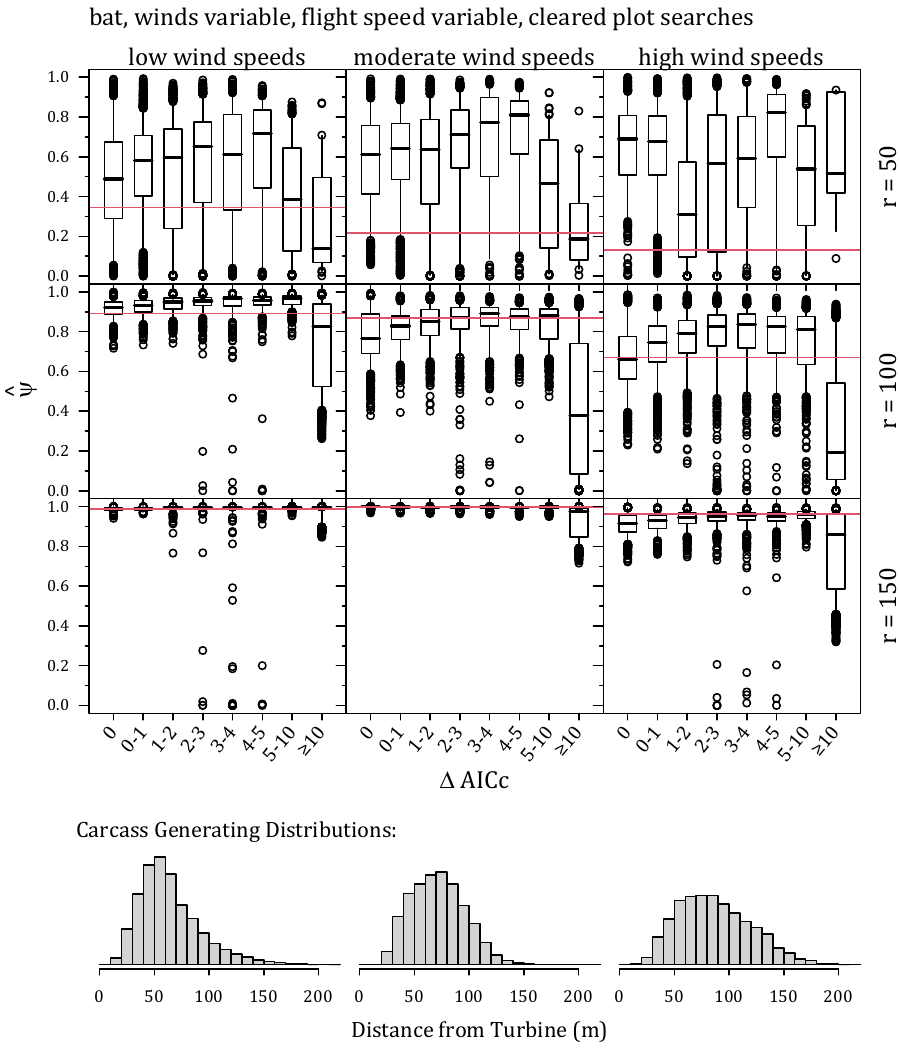}
\caption{\label{fig:psiaic231}AICc and estimated \(\psi\) for the
standard models with simulated M = 200 bats with winds variable, flight
speed variable and cleared plot searches. Boxes show sample IQR with
median; whiskers extend to the most extreme points within 1.5 IQR of the
box; points beyond 1.5 IQR of the box are shown as small circles. Red
lines show the true \(\psi\).}
\end{figure}

\begin{figure}
\centering
\includegraphics{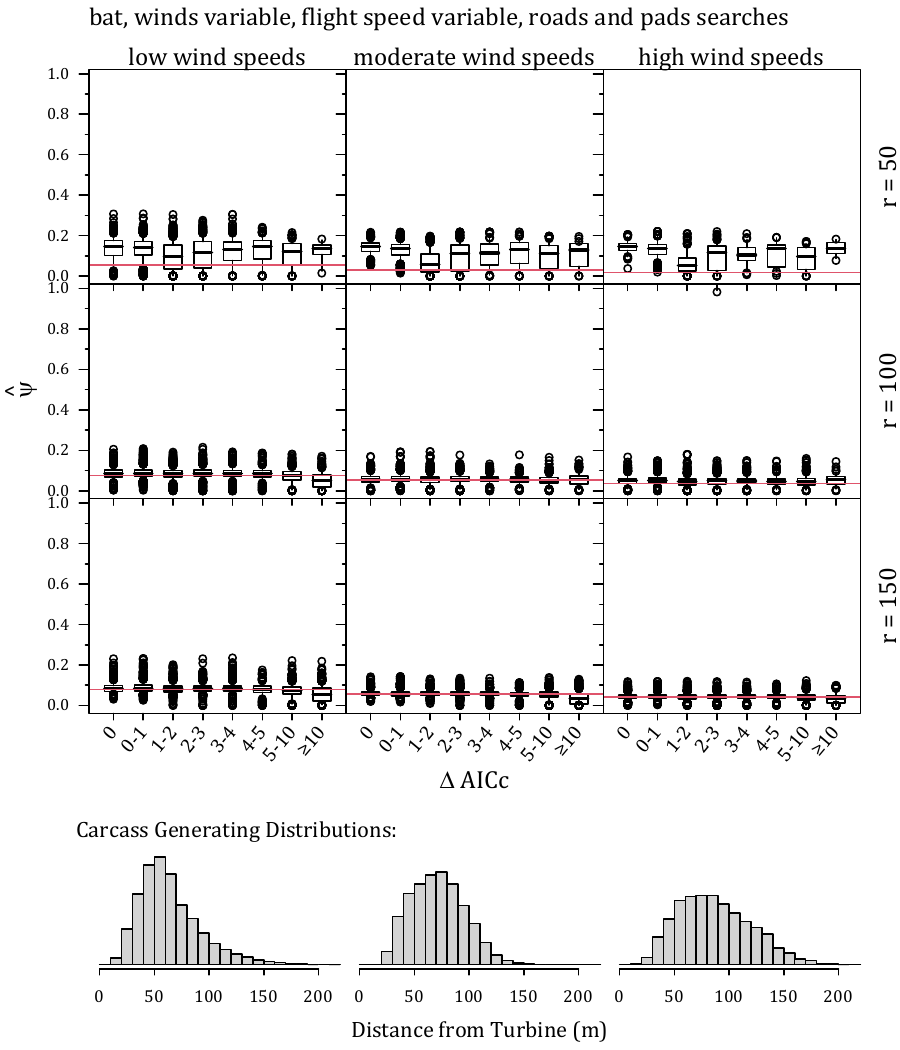}
\caption{\label{fig:psiaic232}AICc and estimated \(\psi\) for the
standard models with simulated M = 200 bats with winds variable, flight
speed variable and roads \& pads searches. Boxes show sample IQR with
median; whiskers extend to the most extreme points within 1.5 IQR of the
box; points beyond 1.5 IQR of the box are shown as small circles. Red
lines show the true \(\psi\).}
\end{figure}

For bat carcasses, \(\Delta\mathrm{AICc}\) appears to be even less
reliable for selecting the most accurate models for \(\psi\) than it was
for eagle carcasses. In some cases, the worse-fitting models (high
\(\Delta\mathrm{AICc}\)) gave more accurate predictions (for example,
fig.~\ref{fig:psiaic211}, upper right panel, with 12 m/s winds and
search radius of \(r = 50\) m); in some cases, the best-fitting models
gave the most accurate predictions (for example,
fig.~\ref{fig:psiaic211}, center panel, with 8 m/s winds and search
radius of \(r = 100\) m); in most cases, though, there was little
discernible relationship between \(\Delta\mathrm{AICc}\) and accuracy
(figs. \ref{fig:psiaic211}-\ref{fig:psiaic232}) .

The road \& pad searches yielded accurate predictions for \(\psi\) for
bats under constant or variable windspeeds (figs.
\ref{fig:psiaic212}\footnote{The missing boxplots in the upper right
  panel of fig.~\ref{fig:psiaic212} are due to so few models being able
  to fit viable distributions to the data because the carcass PDF did
  not start its decline to zero until well beyond the search radius of
  50 m.}, \ref{fig:psiaic222}, and \ref{fig:psiaic232}), except when the
search radius was too short to capture the distribution's dropoff in the
right tail, in which case models tended to vastly overestimate \(\psi\).
The \(\Delta\textrm{AICc}\) had very little predictive value because
\(\Delta\textrm{AICc}\) measures relative fit within the range of the
data but gives no information on how well the model extrapolates beyond
the data or whether a given model happens to have a tail that matches
that of the actual carcass distribution.

Under the conditions of fig.~\ref{fig:psiaic222}, it is remarkable how
performance appears to be almost entirely independent of AICc.

\newpage

\hypertarget{mortality-estimation-and-the-distinction-between-psi-and-dwp}{%
\section{\texorpdfstring{Mortality Estimation and the Distinction
between \(\psi\) and \(dwp\)
\label{app:psivdwp}}{Mortality Estimation and the Distinction between \textbackslash psi and dwp }}\label{mortality-estimation-and-the-distinction-between-psi-and-dwp}}

Accurate estimation of the fraction of carcasses lying in the searched
area depends on both the probability that a carcass lies in the searched
area (\(\psi\)) and the number of carcasses that were found in the
searched area (\(X\)). For example, if \(\psi = 0.2\) and \(X = 2\)
carcasses were found at turbine A, the most likely total number of
carcasses would be \(\hat{M_A} = 2/0.2 = 10\), but \(M_A\) could readily
be anywhere in its 90\% credible interval\footnote{Credible interval is
  a Bayesian analog to the confidence interval of classical statistics.
  Credible intervals for \(M_A\) given \(X_A\) and \(\psi\) can be
  calculated in the \texttt{dwp} package using, for example,
  \texttt{MCI(postM(x\ =\ 2,\ g\ =\ 0.2))}.} of {[}4, 26{]}, and the
fraction of carcasses lying in the searched area at the turbine
(\(dwp_A\)) would likely be in the interval {[}2/26, 2/4{]} or
7.7--50\%. By contrast, if \(\psi = 0.2\) at turbine B too, but
\(X_B = 20\) rather than 2, then the credible interval for \(M_B\) would
be {[}73, 109{]}, and the fraction of carcasses lying in the searched
area at turbine B (\(dwp_A\)) would likely be in a much narrower
interval---{[}20/139, 20/73{]} or 14.4--27.4\%. \(dwp\) properly
reflects the greater relative information content of \(X = 20\) compared
to \(X = 2\), but \(\psi\) does not.

\begin{figure}
\centering
\includegraphics{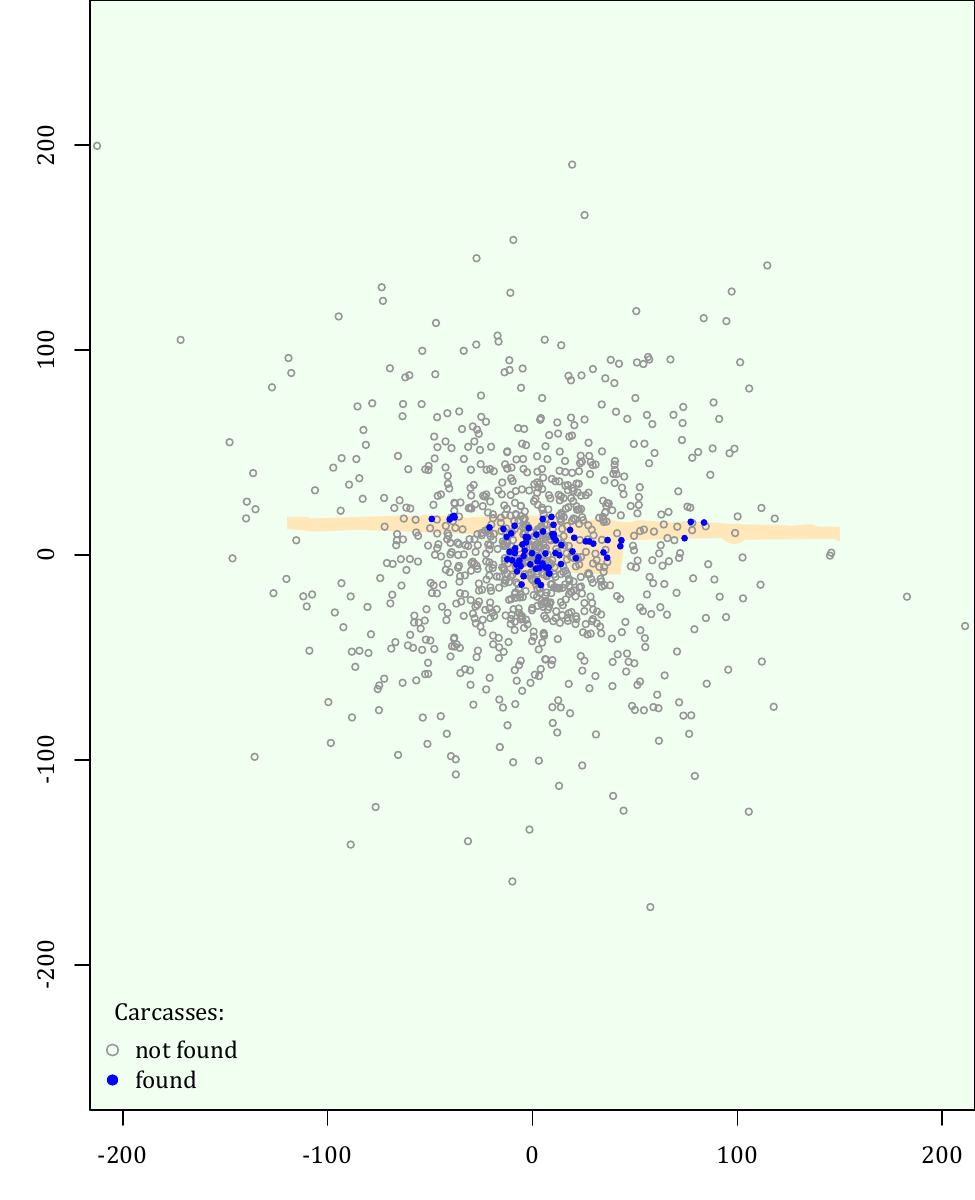}
\caption{\label{fig:dwpsim0}Simulated carcass dispersion and discovery
(n = 1000). The horizontal tan strip in the center represents the
searched area.}
\end{figure}

To illustrate in another, more realistic way and validate the accuracy
of the \(dwp\) estimator, we ran sets of 1000 simulated collisions
between bats and a turbine and the resulting dispersion of carcasses on
the ground near the turbine. Carcass distances were generated according
to a gamma distribution with 60\% of the carcasses lying within 50 m of
the turbine and 90\% within 100 m (that is, a gamma distribution with
\texttt{shape}= 1.7744 and \texttt{rate} = 0.0355). Carcass directions
were uniformly distributed between \(0^\circ\) and \(360^\circ\). The
search area was confined to a fairly large turbine pad and a moderately
wide access road, where the visibility was good and search conditions
were easy. Carcasses were scavenged and removed from the field with
persistence times randomly distributed as Weibull(\texttt{shape} = 0.64,
\texttt{scale} = 1.705), with median persistence time of 1 day and 90\%
of the carcasses removed within \textasciitilde6 days. Searches were
conducted on the road \& pad area out to 150 m from the turbine every
5th day over a period spanning 150 days. Searcher efficiency for a
carcass in the first search after arrival was 80\% and
\(0.8\cdot 0.75^{k -1}\) on the \(k\)th search. Results for one set of
1000 carcasses are displayed in fig.~\ref{fig:dwpsim0}.

For each simulated data set with 1000 carcasses, \texttt{ddFit} from the
\texttt{dwp} package was used to fit a gamma distribution to the
dispersion of the recovered carcasses. Then, a 90\% confidence interval
for the probability that a carcass would lie in the searched area was
calculated using \texttt{estpsi}, a 90\% confidence interval for the
fraction of carcasses lying in the searched area was calculated using
\texttt{estdwp}, and the actual fraction of carcasses lying in the
searched area was recorded. The process was repeated 1000 times to
create 1000 confidence intervals for \(dwp\) calculated using
\texttt{estdwp} and 1000 \(\psi\) using \texttt{estpsi}. Following this
process, each simulated 90\% confidence interval should have a 90\%
chance of including the actual fraction of carcasses in the searched
area.

\begin{figure}
\centering
\includegraphics{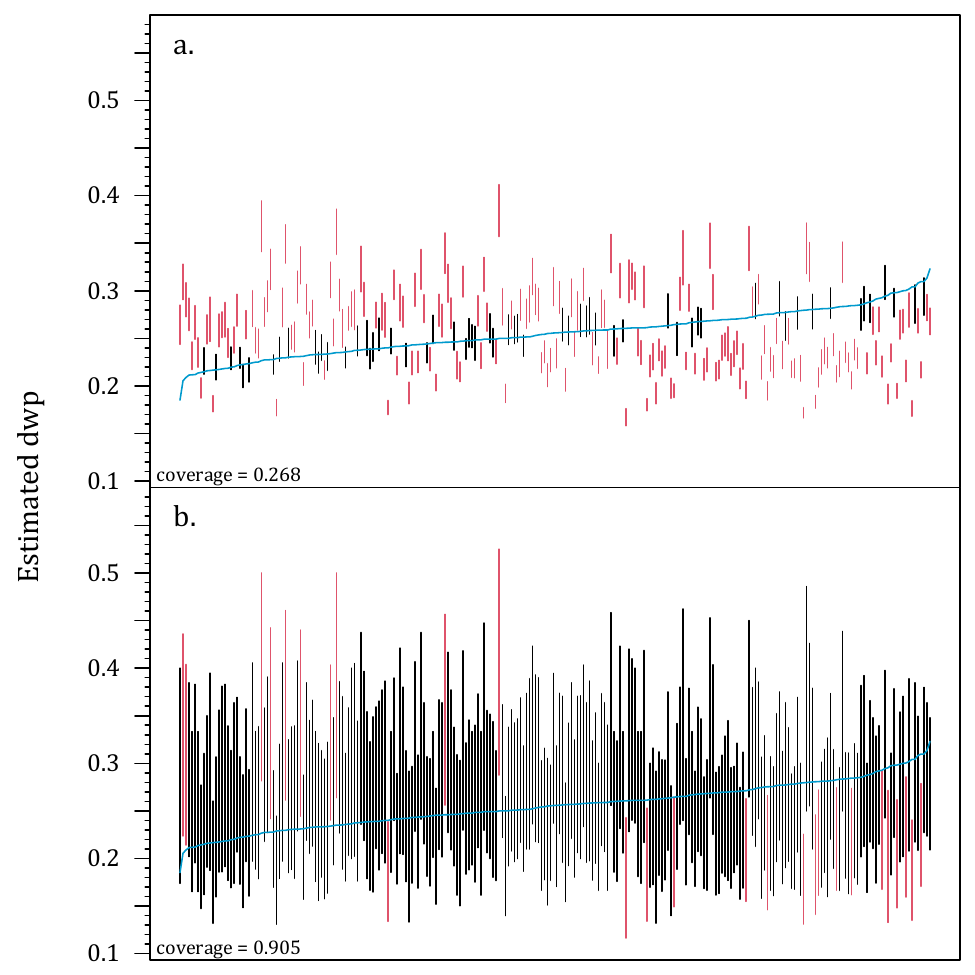}
\caption{\label{fig:CIdwp}Confidence intervals for the fraction of
carcasses lying in the searched area (\(dwp\)) using a. \texttt{estpsi}
or b. \texttt{estdwp} from the \texttt{dwp} package as the estimator.250
confidence interval are plotted for each, with black to indicate CIs
that cover the true \(dwp\) and red to indicate missing the true
\(dwp\). The blue lines in the center show the target, actual \(dwp\)
for each simulated CI. The CIs are ordered by increasing target, actual
\(dwp\). Coverages are the proportion of 1000 simulated CIs that cover
the target \(dwp\).}
\end{figure}

The proportion of intervals that do include (or ``cover'') the actual
parameter (namely, the fraction of carcasses in the searched area) is
referred to as the \emph{coverage} and is an essential test of estimator
performance. Fig. \ref{fig:CIdwp} displays 250 simulated CIs for the
fraction of carcasses lying in the searched area according to
\(\hat{\psi}\) and \texttt{estdwp} as estimators. The CIs based on
\(\hat{\psi}\) appear to be too small and routinely miss the true
\texttt{dwp}. This is wholly expected because \(\psi\) is the
probability of a carcass lying in the searched area rather than the
actual, realized fraction of carcasses lying in the searched area and is
inherently much less variable than \texttt{dwp}. Coverage for the
nominal 90\% CIs was only 26.8\%. This is not a problem with \(\psi\) or
the calculation of \(\hat{\psi}\); rather, it is a confusion of \(\psi\)
as \texttt{dwp} and an attempt to use \(\psi\) to do \texttt{dwp}'s job.
By contrast, coverage for the nominal 90\% CIs based on \texttt{dwp} was
90.5\%, which is as close to 90\% as can be expected for even a perfect
estimator in such a small simulation.

\hypertarget{accounting-for-the-uncertainty-in-dwp-technical-details}{%
\section{\texorpdfstring{Accounting for the Uncertainty in \texttt{dwp}:
Technical Details
\label{app:vardwp}}{Accounting for the Uncertainty in dwp: Technical Details }}\label{accounting-for-the-uncertainty-in-dwp-technical-details}}

To be able to use the software effectively, it is not necessary to
understand the details of how the \texttt{dwp} package accounts for the
uncertainty in \texttt{dwp}. The details are offered here for
completeness.

The uncertainty in \(\widehat{dwp}\) given the number of carcasses in
the searched area (\(m_\textrm{in}\)) and the probability of a carcass
lying in the searched area (\(\psi\)) is accounted for by sampling from
the posterior distribution of \(M\) given \(m_\textrm{in}\) and
\(\psi\)~or, more precisely,
\[\textrm{Pr}(M = m \mid m_\textrm{in},\psi) = \frac{\textrm{Pr}(m_\textrm{in}\mid  M = m, \psi) \cdot \textrm{Pr}(M = m)}{\sum_m\textrm{Pr}(m_\textrm{in}\mid  M = m, \psi) \cdot \textrm{Pr}(M = m)}\]
where \(m_\textrm{in}\sim \textrm{binomial}(M, \psi)\) and
\(\textrm{Pr}(M)\) is the integrated reference prior,
\(M \propto \sqrt(M + 1) - \sqrt(M)\) (Berger et al.~2012; Dalthorp et
al.~2017). In practice, \(\psi\) is not known but estimated using
\texttt{ddFit} from the \texttt{dwp} package. The uncertainties in
\(\hat{\psi}\) and \(M \mid (\psi, m_\textrm{in})\) are assessed via a
three-step process:

\begin{enumerate}
  \item simulate the regression parameter estimates from the fitted GLM as multivariate
normals (Nelder and Wedderburn 1972) to account for uncertainty in the fitted model (using `ddSim`, 
which is called by \texttt{estpsi} but can be called directly if desired),
  \item integrate the simulated models over the searched area (\texttt{estpsi}) to account for uncertainty in $\hat{\psi}$; and 
  \item calculate a simulated $\widehat{dwp}_i = \frac{m_\textrm{in}}{M \mid (m_\textrm{in}, \hat{\psi})}$ for 
each simulated value of $\hat{\psi}$, using a random draw from the posterior distribution of 
$M \mid (m_\textrm{in}, \hat{\psi})$. 
\end{enumerate}

The vector of estimated \texttt{dwp} values generated in this way
incorporates the uncertainty in \(M \mid (m_\textrm{in}, \hat{\psi})\),
yielding accurate CIs for \texttt{dwp}.

\hypertarget{dwp-and-the-estimation-of-m}{%
\subsection{\texorpdfstring{\(dwp\) and the Estimation of
\(M\)}{dwp and the Estimation of M}}\label{dwp-and-the-estimation-of-m}}

After being properly formatted (using the function
\texttt{exportGenEst}), the estimated \(dwp\) can be used with GenEst to
estimate total mortality. At a site with a single search class with a
single turbine, a simplified version of the model\footnote{The estimator
  for \(M\) that is used in GenEst is difficult to write down
  succinctly. Readers interested in finer detail are directed to the
  GenEst technical manual (Dalthorp et al.~2018).} for estimating \(M\)
is \(\hat{M} = X/(\hat{g}\cdot\widehat{dwp})\), where \(X\) is the
number of carcasses found, \(\hat{g}\) is the estimated detection
probability for carcasses lying in the searched area during the span of
the monitoring field season, and \(\widehat{dwp}\) is the estimated
fraction of carcasses lying in the searched area.

Note that
\(\hat{M} = X/(\hat{g}\cdot\widehat{dwp}) = \frac{X/\hat{g}}{\widehat{dwp}} = \hat{m}_\textrm{in}/\widehat{dwp}\),
which incorporates both the uncertainty in estimating the number of
carcasses within the searched area (\(m_\textrm{in}\)) and the
uncertainty in the fraction of carcasses lying in the searched area
(\(\widehat{dwp}\)). Thus, \(\hat{M}\) accounts for the two major
sources of uncertainty discussed previously, and in theory it should
give accurate confidence intervals, so that a 90\% confidence interval
should have a 90\% chance of including \(M\).

\begin{figure}
\centering
\includegraphics{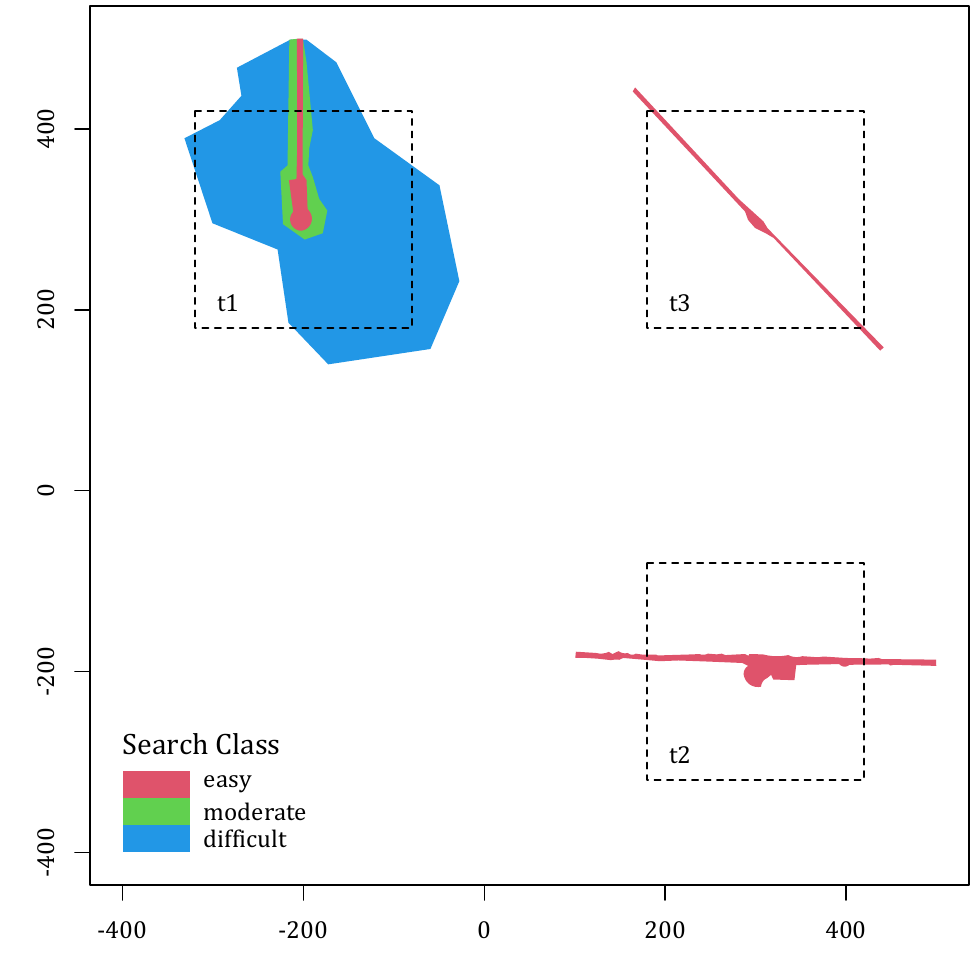}
\caption{\label{fig:layout}Maps of search areas for three turbines.
Dashed lines indicate search boundaries.}
\end{figure}

To verify the accuracy, we ran a simulation of the processes of carcass
deposition and discovery, field trials for estimating \(g\), and
distance modeling to estimate \(dwp\). Then, we used \texttt{GenEst} to
estimate mortality for each turbine and the total for the site. We
tallied the number of 90\% confidence intervals that included \(M\),
calculated the \emph{coverage} as the fraction of the CIs that included
\(M\), and compared the coverages with the target coverage of 90\%. The
parameters for the simulation scenarios:\\
Simulation:

\begin{enumerate}
  \item generate carcass distances for $M = 300$ or 1000 carcasses as 
    $~\textrm{gamma}(\alpha = 1.774, \beta = 28.17)$
  \item arrival at site with three turbines, with layouts as in fig. \ref{fig:layout}
  \item searches in within squares with radius 120 meters, days = 0, 5, ..., 150
  \item Weibull persistence with parameters: 
    \begin{itemize}
      \item Easy visibility: $\alpha = 0.64$, $\beta = 1.705$
      \item Moderate visibility: $\alpha = 0.64$, $\beta = 7.37$
      \item Difficult visibility: $\alpha = 0.64$, $\beta = 9.47$
    \end{itemize}
  \item Searcher efficency with parameters:
    \begin{itemize}
      \item Easy visibility: $p = 0.8$, $k = 0.75$
      \item Moderate visibility: $p = 0.3$, $k = 0.75$
      \item Difficult visibility: $p = 0.15$, $k = 0.75$
    \end{itemize}
  \item summary parameters
    \begin{itemize}
      \item $\psi$ = 0.898, 0.254, 0.0967 for turbines t1, t2, and t3, resp.
      \item $g$ = 0.283, 0.285, 0.179 for search classes E, M, and D, resp.
    \end{itemize}
\end{enumerate}

\begin{figure}
\centering
\includegraphics{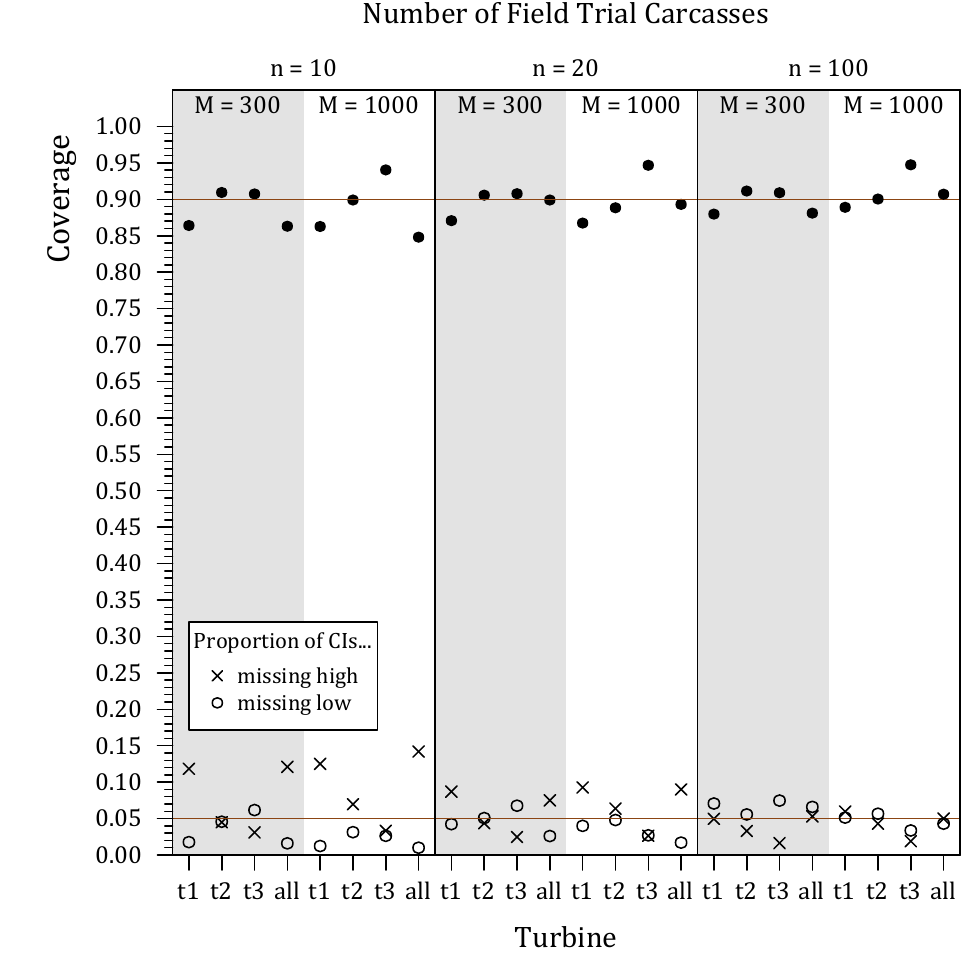}
\caption{\label{fig:MCI}Coverages for nominal 90\% confidence intervals
for the three turbines at a site and for the total (solid circles) and
the proportion of confidence intervals that fail to cover M, missing
high (\(\hat{M}_{lwr} > M\), indicated by \(\times\)) or missing low
(\(\hat{M}_{upr} < M\), indicated by \(\bigcirc\)).}
\end{figure}

The estimated coverage probabilities range from 85\% and 95\% for all
combinations of mortality (\(M = 300,\, 1000\)) and field trial
carcasses (\(n = 10, \, 20,\, 100\)) at all turbines and the site as a
whole (fig.~\ref{fig:MCI}), with a single exception. Coverage for total
mortality for the site as a whole when \(n = 10\) and \(M = 1000\) was
84.8\%, with 14.8\% of the confidence intervals missing high (that is,
\(\hat{M}_{lwr} > M\)) and 1.8\% missing low. This indicates a tendency
to slightly overestimate the total under these conditions. With discrete
random variables (\(M\) and \(n\)), coverage probabilities for 90\% CIs
are guaranteed to differ from the nominal 90\% except for in rare
circumstances, so the small deviations from the target coverage is
neither a surprise nor cause for alarm, especially when \(n = 10\)
(Madsen et al.~2019). When \(M = 1000\), the CIs appear to be slightly
too wide at turbine 3, as coverages approach 95\%, with approximately
equal probability of the CI missing high or low (fig \ref{fig:MCI}).

\hypertarget{acknowledgements}{%
\section*{Acknowledgements}\label{acknowledgements}}
\addcontentsline{toc}{section}{Acknowledgements}

We thank Bat Conservation International, and Avangrid Renewables for
providing the carcass location data used in our bat examples and
Norwegian Institute of Nature Research for carcass location data used in
our eagle examples. We appreciate the statistical advice provided by L.
Madsen, although if there are any errors, they are entirely our own.
Funding for this research was provided by the U. S. Geological Survey's
Ecosystems Mission Area Wildlife Program, the U. S. Department of
Energy, Bat Conservation International and the American Wind Wildlife
Institute. Any use of trade, firm, or product names is for descriptive
purposes only and does not imply endorsement by the U.S. Government.

\hypertarget{references}{%
\section*{References}\label{references}}
\addcontentsline{toc}{section}{References}

Akaike, H. 1974. A new look at the statistical model identification.
IEEE Transactions on Automatic Control, 19 (6): 716--723,
\url{doi:10.1109/TAC.1974.1100705}.

Allen, DM 1974. The Relationship between Variable Selection and Data
Augmentation and a Method for Prediction. Technometrics. 16 (1):
125--127.

Andersen, E. 1970. Sufficiency and Exponential Families for Discrete
Sample Spaces. Journal of the American Statistical Association. 65
(331): 1248--1255.

Arnett, E. B., M. Schirmacher, M. M. P. Huso, and J. P. Hayes. 2009a.
Effectiveness of changing wind turbine cut-in speed to reduce bat
fatalities at wind facilities. An annual report submitted to the Bats
and Wind Energy Cooperative. Bat Conservation International. Austin,
Texas, USA.

Arnett, EB, MR Schirmacher, MMP Huso, and JP Hayes. 2009b. Patterns of
bat fatality at the Casselman Wind Project in south-central
Pennsylvania. An annual report submitted to the Bats and Wind Energy
Cooperative and the Pennsylvania Game Commission. Bat Conservation
International. Austin, Texas, USA.

Bañuelos-Ruedas, Francisco, César Ángeles Camacho, and Sebastián
Rios-Marcuello. 2011. Methodologies Used in the Extrapolation of Wind
Speed Data at Different Heights and Its Impact in the Wind Energy
Resource Assessment in a Region, in Wind Farm - Technical Regulations,
Potential Estimation and Siting Assessment, ed.~by Gaston Orlando
Suvire. InTech, Rijeka, Croatia.

Berger, J.O., Bernardo, J.M., and Sun, D. 2012, Objective priors for
discrete parameter spaces: Journal of the American Statistical
Association: v. 107, no. 498, p.~636--648,
\url{DOI:10.1080/01621459.2012.682538}

Burnham, K.P. and Anderson, D.R. 2002. Model Selection and Multimodel
Inference: A Practical Information-Theoretic Approach. Springer-Verlag,
New York. 488 pp.

Canty, A. and B. Ripley. 2021. boot: Bootstrap R (S-Plus) Functions. R
package version 1.3-27.

Cavanaugh, J. E. 1997. Unifying the derivations of the Akaike and
corrected Akaike information criteria. Statistics \& Probability
Letters, 31 (2): 201--208,
\url{https://doi.org/10.1016/S0167-7152(96)00128-9}.

Counihan, J. 1975. Adiabatic atmospheric boundary layers: A review and
analysis of data from the period 1880--1972. Atmospheric Environment
9(10): 871-905.

Dalthorp, D., Huso, M., and Dail, D. 2017. Evidence of absence (v2.0)
software and user guide: US Geological Survey Data Series 1055,
\url{https://doi.org/10.3133/ds1055}

Dalthorp D., Simonis, J, Madsen, L, Huso, H, Rabie, P, Mintz, J,
Wolpert, R, Studyvin, J, and Korner-Nievergelt, F. 2018a. GenEst:
Generalized Mortality Estimator. R package version 1.4.4.
\url{https://CRAN.R-project.org/package=GenEst}

Dalthorp, D., Madsen, L., Huso, M., Rabie, P., Wolpert, R., Studyvin,
J., Simonis, J., and Mintz, J., 2018b. GenEst statistical models---A
generalized estimator of mortality: U.S. Geological Survey Techniques
and Methods, book 7, chap.~A2, 13 p.,
\url{https://doi.org/10.3133/tm7A2}.

Davison, A.C. and Hinkley, D.V. 1997. Bootstrap Methods and Their
Applications. Cambridge University Press, Cambridge. 582 pp.

Efroymson, M.A.~1960. Multiple regression analysis. In Ralston, A. and
Wilf, H.S., eds., Mathematical Methods for Digital Computers. Wiley, New
York.

Grodsky, SM, Behr, MJ, Gendler, A, Drake, D, Dieterle, BD, Rudd, RJ, and
Walrath, NL. 2011. Investigating the causes of death for wind
turbine-associated bat fatalities. Journal of Mammalogy 92(5): 917-925.

He, Y, Monahan, AH, McFarlane, NA. 2013. Diurnal variations of land
surface wind speed probability distributions under clear-sky and
low-cloud conditions. Geophysical Research Letters 40:1-7.
\url{doi:10.1002/grl.50575}

Huso, MMP, Som, N, and Ladd, L. 2012. Fatality estimator software and
user's guide. US Geological Survey Data Series 729.
\url{https://doi.org/10.3133/ds729}

Huso, MMP and Dalthorp, D. 2014. Accounting for unsearched areas in
estimating wind turbine-caused fatality. Journal of Wildlife Management
78: 347-358. DOI: 10.1002/jwmg.663

Huso, MMP, Wilson, Z, and Arnett, EB. 2011. Density-weighted Area
Estimation. Paper presented at \emph{Conference on Wind Energy and
Wildlife Impacts}, Trondheim, Norway.

Kaltschmitt, M. and Wiese, A. 2007. Wind Energy, \emph{in} Renewable
energy: technology, economics, and environment, ed.~by Kaltschmitt, M.,
Streicher, W. and Wiese, A. Springer, Berlin, pp.~49-65.

Madsen, L, Dalthorp, D, Huso, MMH, and Aderman, A. 2019. Estimating
population size with imperfect detection using a parametric bootstrap.
Environmetrics 31:e2603. \url{https://doi.org/10.1002/env.2603}

Maurer, J., Huso, H., Dalthorp, D., Madsen, L., and Fuentes, C. 2020.
Comparing methods to estimate the proportion of turbine-induced bird and
bat mortality that occurred within the searched area under a road and
pad search protocol. Environmental and Ecological Statistics 27:
769-801.

McCullagh, P and Nelder, JA. 1983. Generalized Linear Models, 2nd
Edition. Chapman \& Hall CRC. Boca Raton, Florida.

Nelder, JA and Wedderburn, RWM. 1972. Generalized linear models. Journal
of the Royal Statistaical Society, Series A, 135(3): 370-384.

Prakash, S and Markfort, CD. 2020. Experimental investigation of
aerodynamic characteristics of bat carcasses after collision with a wind
turbine. Wind Energy Science 5: 745-758.

Prakash, S and Markfort, CD. 2021. Development and testing of a
three-dimensional ballistics model for bat strikes on wind turbines.
Wind Energy. 01 April 2021. \url{https://doi.org/10.1002/we.2638}

Richards SA. 2005. Testing ecological theory using the
informationtheoretic approach: examples and cautionary results. Ecology
86:2805--2814.

Studyvin, J, and Rabie, P. 2019. windAC: Area Correction Methods. R
package version 1.0.0. \url{https://CRAN.R-project.org/package=windAC}

Vuong, Q.H. 1989. Likelihood Ratio Tests for Model Selection and
Non-Nested Hypotheses Econometrica 57(2): 307-333

Whitney, WO and Mehlhaff, CJ. 1987. High-rise syndrome in cats. J Am Vet
Med Assoc 191(11): 1399-1403.

\end{document}